# Testing Explanations of Short Baseline Neutrino Anomalies

A DISSERTATION PRESENTED
BY
NICOLÒ FOPPIANI
TO
THE DEPARTMENT OF PHYSICS

IN PARTIAL FULFILLMENT OF THE REQUIREMENTS
FOR THE DEGREE OF
DOCTOR OF PHILOSOPHY
IN THE SUBJECT OF
PHYSICS

HARVARD UNIVERSITY
CAMBRIDGE, MASSACHUSETTS
AUGUST 2022

Dissertation Advisors: Professor Roxanne Guenette
and Professor Carlos Argüelles-Delgado

Nicolò Foppiani

# Testing Explanations of Short Baseline Neutrino Anomalies

## Abstract

The experimental observation of neutrino oscillations profoundly impacted the physics of neutrinos, from being well understood theoretically to requiring new physics beyond the standard model of particle physics. Indeed, the mystery of neutrino masses implies the presence of new particles never observed before, often called sterile neutrinos, as they would not undergo standard weak interactions. And while neutrino oscillation measurements entered the precision era, reaching percent-level precision, many experimental results show significant discrepancies with the standard model, at baselines much shorter than typical oscillation baselines. Such experimental measurements include LSND, MiniBooNE, gallium experiments, and reactor antineutrino measurements. These *short baseline anomalies* seem to be explainable by the addition of a light sterile neutrino, with mass in the $1 - 10\,\mathrm{eV}$ range. However, this hypothesis is in strong tension with many *null* experimental observations. Other explanations that rely on models containing sterile states with masses in the $1 - 500\,\mathrm{MeV}$ could resolve the tension. In this thesis, we test both classes of models. On the one hand, we look for datasets collected at a short baseline which can constrain heavy sterile neutrino models. We find that the minimal model is fully constrained, but several extensions of this model could weaken the current constraint and be tested with current and future datasets. On the other hand, we test the presence of neutrino oscillations at short baselines, induced by a light sterile state, with the data collected by the MicroBooNE experiment, a liquid argon time projection chamber specifically designed to resolve the details of each neutrino interaction. We report null results from both analyses, further constraining the space of possible explanations for the short baseline anomalies. If new physics lies behind the short baseline anomaly puzzle, it is definitely not described by a simple model.

# Contents

*To my family,*
*who shared with me*
*every emotion*
*of this journey.*

# Acknowledgments

This Ph.D. was a turbulent journey, which could not be possible without the support of the many friends, colleagues, and mentors I found along the way. Ph.D. advisors are fundamental pillars during graduate studies.

I would like to thank Roxanne Guenette for welcoming me into her group during my first year and always supporting my exploratory mindset and ideas, often off the beaten path. She made me realize that scientific endeavors can be very complex not simply because of the science but because of the many human interactions behind them. She created a humane environment where students are respected as individuals, acknowledging our needs, spare time, and lives outside of physics.

I would like to acknowledge Carlos Argüelles-Delgado for being first a friend, then a respected colleague and collaborator, and eventually a thesis advisor. He always encouraged me to think broadly, outside the box, and constantly innovate in every aspect of my scientific research. I will genuinely miss our physics chats, always full of enthusiasm, passion, and room for unconventional ideas. At the same time, I also appreciated all our deep conversations about the other aspects of life, always putting our well-being first, recognizing that, without it doing good science is challenging.

I also would like to acknowledge Melissa Franklin and Matt Reece for serving on my committee. Melissa convinced me to join Harvard in the first place, and she has always been a reference point for me. She often challenged my ideas, pushing me to think broadly about options and opportunities, both science- and career-wise. I always appreciated Matt's humility and thoughtfulness in our physics discussions. Since my first year, when I was just a QFT student, he never dismissed my questions, always finding some exciting and nontrivial aspects.

I thank all my collaborators in Guenette's group, especially Wouter, Marco, Roberto, Justo, and Lars. We shared many aspects of the MicroBooNE experience and many trips to Fermilab. Wouter has also been a special friend, sharing many outdoor adventures and trips during these past years. I thank the MicroBooNE collaboration for allowing me to work on the flagship analysis, and the Pandora eLEE team for sharing the endeavor of the electron neutrino searches. I especially thank David and Giuseppe for collaborating, following me closely, and mentoring me during this time. I thank the *Neutrino Penguins'* group (Argüelles' group) for being a new home after the pandemic. It also broadened my horizons to the edge of the universe, where astrophysical neutrinos originate. Within this group, I would like to especially thank Matheus for the fun projects we developed in the last two years. Although mainly working remotely, our meetings always brought joy, thanks to the friendly and enjoyable atmosphere we created.

If I am here writing this thesis, it is because my journey started some time ago when Gigi Rolandi caught me with his enthusiasm and passion during the first particle physics class at Scuola Normale in Pisa. He supervised me at CERN for both my Bachelor's and Master's theses, where the collaborative environment of the CMS experiment made my interest in particle physics grow further. It is thanks to his mentorship and all the time he dedicated to me that I decided to embark on the journey of a physics Ph.D.

Even if science was the main ingredient of this journey, it was definitely not the only one. I am grateful that I could be part of many *families* during these few years. The Harvard physics department has been a homy environment, where I found friends among the other students, especially within my cohort. I especially would like to thank Sepehr, Qianshu, Cari, Brendon, Nick, Zoe, Grace, Ruihua, Tim, Paloma, and Rashmish for the many fun and celebratory moments we shared in these few years. I would also like to acknowledge Lisa, Jacob, and Carol for always supporting and helping me navigate the troubles of the Ph.D.

Harvard also gave me opportunities to challenge myself in different subjects. The class in Human-Computer-Interaction was an exciting experience that allowed me to apply my knowledge of statistics in a different field. The mentorship of Professor Elena Glassman and the great teamwork with Sana, Sophie, and Ziv culminated in a publication that I feel very proud of because it is outside of my primary field. How to Make Almost Anything was an extraordinary class. For a semester, the fabrication lab and the MIT Media Lab have been my second home. I had a chance to dive into digital fabrication and meet fantastic classmates and mentors. I would like to acknowledge Nathan for running the lab and ensuring everyone could complete their final projects, and Ibrahim, Gabby, Rosalie, Alfonso, Ian, and Camron for sharing many nights fighting with CAD and Arduino. The challenges I encountered in my scientific collaborations made me appreciate the importance of human relationships and led me to the Negotiation Workshop at Harvard Law School. Since the class is about human interactions, I could grow significantly thanks to the people I negotiated, debriefed, and argued with. I would like to thank the fantastic "group blue": especially Vi, for helping us stu-dents be always ready for the following case, and Professor Michael Chaffers for leading us through all the challenges posed during the class. After all our conversations reconciling my analytical skills with his negotiation experience, I am very grateful that I can consider him a mentor.

Challenges do not appear only in academia: the Harvard Mountaineering Club was a place for embarking on new challenges every week. I learned new mountaineering skills, improved my rock climbing, and even started ice climbing, thanks to the cold New England winters. Still, most importantly, I learned how to tackle mental challenges when finding myself in difficult and uncomfortable situations. Among my fantastic climbing partners and friends, I would like to acknowledge Enrico, Larissa, Dom, and Walter for the many adventures we shared across multiple continents. I especially would like to thank Carlo for being not just a climbing partner or a friend but also a mentor: our conversations when returning from a long climbing trip always gave me perspective and clarity on my future steps during my Ph.D.

I am genuinely grateful to have had the privilege to join the Harvard Horizons program. The fantastic team at the Bok Center, Pamela, Erika, Marlon, Jordan, and Casey, helped me develop better outreach skills that, I am sure, improved both this manuscript and my thesis defense. My fellow scholars, Vanessa, Harry, Hannah, Karina, Chika, and Juliana, have been an essential element of this journey. All the time we spent together, from the very first meeting to the final day of the Symposium in Sanders Theater, made me not just better at communicating my research but made me also understand it more deeply.

I am also thankful to the Harvard Italian Student Society for reminding me of home whenever I felt nostalgic of Italy, between joyful glasses of Aperol Spritz and pasta competitions. I would like to acknowledge the many friends I met within this family: Alberto, Davide, Greta, Giuseppe, Gabriele, Benedetta, Elena, Eleonora, Gaia, and the many others I cannot name. Our shared experience in Boston is a bond we will keep in the future.

Meeting new friends from entirely different backgrounds is always a great opportunity to learn, and my friends from the Master in Public Administration and International Development (MPA/ID) at the Harvard Kennedy School were definitely not an exception. I found myself being caught in conversations about economics, thinking about introspective leadership, and reflecting on global poverty. This group definitely made me a little more humble. I especially acknowledge Alex, Jossie, Bia, Christian, Beto, Mechi, Nicole, and Nico, and all the other people I met during many trips, events, and parties.

Within this family, I am immensely grateful to Ruth for being my partner and supporting me throughout the last two years. We did not simply share the many pleasant moments of our jour-neys at Harvard. She truly supported me over all the difficult patches of my Ph.D., always showing empathy, compassion, and care.

You cannot say you moved to a new place until you find a new home, which I found in a cozy and quiet house in Somerville, at 19A Harvard Street. What made it home were my roommates, especially Sepehr, with whom we shared the house from the very beginning, and Sean, Mark, Diego, Ali, Brendon, Barton, and all the other roommates who have spent part of the journey at Harvard St. I will conserve all the lovely memories from the parties and meals we had during these few years. All the challenges we faced living together, from finding compromises between different cultures and habits to sharing the space during the pandemic, forged our friendships and made us grow as people.

Not all the people who supported me were around Cambridge: many of them were actually across the ocean, back in Europe. I am grateful to have been part of Officine Italia, a group of friends that gave me hope and energy at the beginning of the pandemic when we felt there was no way for us to help the world: I am looking forward to many more projects to come.

It always feels special to walk through the streets of Pisa, stop by the University, and meet my former classmates Marco, Federica, Olmo, Tommaso, Francesca, Marianna, and Agnese, either in person during our reunions or for a phone call now that we live far away. I feel thankful that I can always rely on this special group of friends with whom I shared one of the most meaningful experiences of my life.

And, going back to Chiavari, my hometown, always gives me pleasant feelings, evoking many lovely memories from the past. I am always happy to share new moments with my more historical friends, Giacomo, Sara, Luca, Erik, Riccardo, Emanuele, Serena, Gaia, and Francesco. I am thankful Chiavari will always be the place to come back to our roots and share our stories and experiences.

And eventually, my family is what I will always call home, always waiting for me throughout my adventures: my parents, Adriana and Luciano, and my grandmother, Nonna Lidia. Despite not fully grasping what a Ph.D. is about and what living in the US looks like, they lived this adventure with me, sharing all the joyful moments and all the difficult patches. Although this time has been challenging for them, I am grateful that they have and always will support me.

# 0

## Prologue

It is astonishing that all the complex structures we observe in nature, from stars and galaxies to living organisms and molecules, emerge from the same underlying elementary pieces: the quarks up and down, and the electron. It is through their fundamental interactions that these three particles can combine together, giving rise to an exponentially large amount of combinations and possibilities. This is what particle physics is about: understanding nature at its fundamental level. Particle physics is often referred to as high energy physics, because large energies correspond to small scales, which, in turn, require experiments at high energy to be investigated. Among the elementary particles there is one more type: the elusive neutrino. Since neutrinos only undergo the weak interaction, it does not bind with the other particles, and its probability of interacting with matter is very small for the typical energies of neutrinos around the Earth, which made it difficult to discover, and makes it both hard to study and unknown to most people.

However, neutrinos have brought many surprises over the past few decades. For example, we know they come in three distinct types, or *flavors*, and that their interaction violates symmetries under parity and charge conjugations, which are instead respected by other forces. More importantly, neutrinos were first theorized as massless particles because of the lack of experimental evidence. We now know neutrinos are massive because the three flavors mix among each other. This discovery represents a crack into the standard model of particle physics, which requires the existence of new physics and additional particles to be explained. Moreover, in a moment in which most measurements performed in the lab agree with the theoretical predictions, several neutrino experiments report disagreements. These discrepant experimental results are often referred to as *short baseline anomalies*, and form a truly unclear puzzle, which might hide new physics discoveries.

Such motivations are what pushed me to pursue research in neutrino physics as a Ph.D. student. This thesis represents the summary and the completion of the work I performed between 2017 and 2022. All the original work in this thesis has been either published in peer-reviewed journals or is currently under review. The work is not organized in chronological order but rather in a logical order, hoping that the reader would agree with the rationale behind this scheme.

Part I sets the background for the thesis work, introducing the current understanding of neutrinos and the theoretical and experimental motivations that justify the search for new particles. While it does not contain any original work, I explained the physics in my preferred way and included all the insights I find effective in understanding neutrinos. Although I started working on short base-line anomalies five years ago, I described the theoretical and experimental status at the time of writing the thesis. I want to acknowledge, however, that several things have changed, and new insights were brought in by both the experimental and theoretical communities.

Part II contains some more *phenomenological* work I performed, considering possible solutions to the short baseline anomalies in terms of heavy sterile neutrino models and looking for datasets and experiments that could test them. Chapter 4 is adapted from [1], which was published in Physical Review D. Chapter 5 is adapted from [2], which is currently under review in Physical Review X. Both works have been performed together with Matheus Hostert and Carlos Argüelles-Delgado. Together with testing physical models, we also developed a new statistical approach to test models in an ample parameter space, which explains why, although started a year before the first paper, the second was made public only recently.

Part III summarizes most of my contribution in the MicroBooNE collaboration. I joined MicroBooNE in 2018 and have been an active member until the electron neutrino search was finalized and published. Chapter 6 introduces the MicroBooNE experiment, describing the most important features relevant to this thesis. Chapter 7 describes how particles are measured and identified in MicroBooNE. A large chunk of the chapter is adapted from [3], published in the Journal of High Energy Physics, which describes a new method to improve the identification of track-like particles in MicroBooNE substantially. This new methodology was not only essential for the electron neutrino searches I was directly involved in, but it is also the basis of many detailed cross-section measurements published by MicroBooNE. The rest of the chapter complements the paper and describes other unpublished work. Chapter 8 and chapter 9 are adapted from and expand the electron neutrino search published in [4,5], in Physical Review Letters and Physical Review D, respectively.

# Part I

# Active and Sterile Neutrinos

*I admit that my remedy may seem almost improbable because one probably would have seen those neutrons, if they exist, for a long time. But nothing ventured, nothing gained, and the seriousness of the situation, due to the continuous structure of the beta spectrum, is illuminated by a remark of my honored predecessor, Mr Debye, who told me recently in Bruxelles: "Oh, It's better not to think about this at all, like new taxes." Therefore one should seriously discuss every way of rescue. Thus, dear radioactive people, scrutinize and judge.*

Wolfgang Pauli, Dear Radioactive Ladies and Gentlemen[6]

# 1

# Neutrinos within the Standard Model

Neutrino, literally the "little neutral one," is the Italian diminutive of neutral. This name, born during conversations at the Institute of Physics of Via Panisperna in Rome, was introduced in the scientific language by Enrico Fermi. He used it for the first time during conferences in 1932 and 1933. It indicated, and still does, an almost massless and almost invisible elementary particle, which took twenty-five years to be discovered from its first theoretical proposal in

1930.

Neutrino physics is now a well-established field of particle physics, with tens of experiments running worldwide and thousands of physicists working on the theory, the experiments, and the interpretation of the data. We now review some of the most salient aspects of the history of neutrinos[*], the current understanding of the physics of neutrinos, and its current limitations.

## 1.1  A brief history of the physics of neutrinos

Back in 1930, before the neutron was even discovered, protons and electrons were the only established particles. At that point, Wolfgang Pauli proposed[6] the existence of a new, basically invisible particle to explain the apparent non-conservation of energy and angular momentum of nuclear beta decays[†]. Indeed, suppose a beta decay was simply a two-body decay. In that case, the electron energy should always be at the same well-defined value, while the total angular momentum would be missing half a unit, impossible to compensate with orbital angular momentum, which is also how we experimentally know that neutrinos are spin-1/2 fermions. Fermi incorporated the neutrino in its theory of beta decay[8], modeling weak forces through a point-like interaction between four fermions: proton, neutron, electron, and neutrino. It become clear that the interaction constant $G_F = 1.166 \times 10^{-5}\,\mathrm{GeV}^{-2}$ was too small to hope to ever measure neutrinos, justifying the word *weak* for these interactions. About twenty-five years later, in 1956, Reines and Cowan finally measured anti-neutrinos produced by the reactor at Savannah River[9], observing inverse beta decay for the first time. This technology will be the protagonist of several experiments in neutrino physics. In the same year, the most fundamental aspects of the weak interactions started to be unveiled. With one of the experiments that would later become a cornerstone of particle physics, Chien-Shiung Wu demonstrated that beta decay violates parity as a symmetry of nature [10], established instead in electromagnetic and strong interactions. The final decay products, namely the electron and the unobserved neutrino, would be emitted primarily in the preferential direction of the spin of the parent nucleus, thus breaking the reflection symmetry. Just a year after, through a careful experiment involving several decays of isotopes, Maurice Goldhaber determined that beta decay also violates charge conjugation [11]. His experiment measured the helicity of neutrinos by measuring the other spins in the decay and assuming angular momentum. He observed about five-hundred decays containing left-handed neutrinos and not a single right-handed neutrino, concluding that all neutrinos are left-handed.[‡] These fundamental ingredients eventually led to the first formulation of the Standard Model (SM) in the seminal paper by Weinberg in 1967 [12]. Weak interactions are mediated by the heavy $W$ and $Z$ bosons, and only states with left-chirality can interact with the $W$ boson, explaining the previous observations of Wu and Goldhaber. The fact that the four-fermion theory was only a low-energy approximation was already evident. While the calculations were accurate for small momentum transferred, amplitudes diverge at energies $\sim 100$ GeV. The new theory cured the divergent calculations through the on-shell production of the $W$ boson. And indeed, this made it clear why weak interactions are weak: the Fermi constant

$$G_F = \frac{\sqrt{2}}{8} \frac{g^2}{M_W^2} \qquad\qquad (1.1)$$

is small not because the coupling constant $g$ is small, its actual value is $\sim 0.65$, but rather because $M_W$ is large. This reasoning also explains why the passage of neutrinos through matter differs significantly from that of charge leptons. While electromagnetic interactions at low momentum trans-ferred are favorable because the cross section diverges, the converse is true for neutrinos, making them impossible to slow down in a controlled way.

---

[*]This historical *excursus* is far from being complete, and highlights some of the aspects that the author finds more exciting and undervalued.

[†]Pauli named this particle *neutron*, which was supposed to explain these phenomena and be the actual neutron. When Chadwick discovered the actual neutron[7], it was clear that its mass was too large to be also the actual neutrino, but the hypothesis was not discarded, instead picked up by Fermi.

---

[‡] I always wonder what the same paper, published these days instead, would look like. It would probably contain exclusion regions in the parameter space of different models predicting the production of right-handed neutrinos...

In the following decades scientists learned that neutrinos are not all equal, rather they come in three different *flavors*, which are associated with the three leptons: electron, muon, and tau. While the muon neutrino was discovered in 1962 at the AGS experiment at Brookhaven[13], it was only in the new millennium that the tau neutrino was finally observed by the DONUT collaboration[14]. However, by the early 90s it was already well understood that three different flavors of neutrinos existed. Precision measurements of the $Z$ resonance carried out at the Large Electron Positron (LEP) collider showed that its invisible width is compatible only with three active light§ neutrinos[15].

While the SM looks beautifully elegant, as we will discuss in section 1.2, this was not the end of the story. In those years, Ray Davis, who previously attempted measuring neutrinos, performed the first measurements of the solar neutrino flux[16] in what would be called the "Homestake experiment," set in the dismissed Homestake goldmine, finding a discrepancy with the theoretical prediction[17] of a factor larger than two. It took almost twenty years to solve the solar neutrino problem, proving that neutrinos change flavor as they travel and establishing neutrino oscillations. In the 90s, the Super-Kamioka Neutrino Detection Experiment (Super-Kamiokande) observed a zenith-dependent deficit of atmospheric muon neutrinos[18]: the rate of neutrinos coming from below, which traveled through the entire Earth, was smaller than the rate of neutrinos coming from right above the experiment, which only traveled through the atmosphere. This result brought the first clear evidence for neutrino oscillations: neutrinos which traveled a longer distance, must have changed flavor along the way, explaining the observed discrepancy. While dedicated solar neutrino experiments confirmed Davis' measurements[19,20], it was not until Sudbury Neutrino Observatory measured the flavor-agnostic neutral current rate[21] that the battle was settled. The neutral current rate agreed with the theoretical calculation of the solar neutrino flux, while the rate of charged current electron neutrino interactions agreed with Ray Davis' measurement. It turned out that the solar neutrino flux predictions and Ray Davis' measurements were both correct. What was missing was that neutrinos can change flavour while traveling from the core of the Sun to the detector on the Earth. The solar neutrino puzzle brought to the discovery of neutrino oscillations, which, in turn, require massive neutrinos. In fact, massless neutrinos could not mix with each other, implying the conservation of the neutrino flavor, which is explicitly violated by neutrino oscillations. However, oscillations can only reveal the differences between the squares of the values of the masses, because this combination appears in the oscillation frequency. The last twenty years of history have been devoted to understanding the nature of this mass, which was not included in the first formulation of the SM and does not have a unique and straightforward interpretation. Experiments have measured the mass mixing parameters, set limits on the absolute scale of the neutrino mass, tried to determine the ordering of the neutrino mass states, and looked for evidence for the nature of the neutrino mass (Dirac or Majorana). Recent neutrino physics was devoted to search for more compelling and direct evidence of BSM physics through measurements of deviations from the SM and direct searches of additional neutrino states.

---

§Active because they need to undergo weak interaction, otherwise the $Z$ boson could not decay into the, and lighter than half of the $Z$ mass, otherwise the decay would be kinematically forbidden.

## 1.2    The theory of neutrinos

Neutrinos are an essential ingredient of the SM, the theory that describes our current understanding of elementary particles and fundamental forces. Although they were first introduced as massless states with left chirality only, explaining neutrino masses requires some additional ingredients, thus pointing to physics Beyond the Standard Model (BSM). However, because neutrino masses are small and their effect is tiny, all current aspects of neutrino physics can be embedded in the SM through an effective field theory (EFT) approach. What is still unclear is the completion of this EFT and what lies beyond some of the still unclear experimental results in neutrino physics, like short baseline anomalies, as discussed in chapter 2.

The rest of the chapter is not an overview of neutrino physics and phenomenology: the careful reader is welcome to refer to [22],[23].

## Zooming in the Standard Model

The SM of particle physics is a Quantum Field Theory (QFT) that describes nature at its most fundamental level. It incorporates electromagnetism, the weak and strong nuclear forces. Like classical electromagnetism, the SM is a gauge theory, meaning that the Lagrangian is invariant under local transformations. These transformations are generated by the group $SU(3)_C \times SU(2)_L \times U(1)_Y$, where $C$ stands for color, the charge of the strong force, $L$ stands for *left*, because the weak force involves only fields with left chirality, and $Y$ refers to the hypercharge. The global versions of these transformations are fundamental symmetries resulting in conserved currents to which the gauge bosons couple. The local transformations express the redundancy of the description adopted in QFT: in order to use a manifestly Lorentz covariant description, meaning that the laws of physics have the same form in every reference system, we need to use four-vectors and four-tensors, with more degrees of freedom than the physical ones. For example, although the electromagnetic field has two degrees of freedom (polarizations), we describe it with a four-vector, which has four components. Gauge *redundancy* expresses the equivalence between different gauges or descriptions. In this theory, eight gluons mediate the strong force, three weak bosons responsible for the weak force, and one boson for the hypercharge. The other particles in the SM are spin-1/2 fermions. The charged leptons and the neutrinos, called leptons, interact only through the electroweak force, while the various quarks, called hadrons, also experience the strong force. The left-chiral parts of the charged leptons ($\ell$) and the neutrino ($\nu$) are arranged into $L_L$, a doublet of $SU(2)_L$ that transforms in the fundamental representation, while the charged lepton part with right-chirality $\ell_R$ is a singlet of $SU(2)_L$. Similarly, the left-chiral parts of the quarks ($u$ and $d$) are arranged into the doublet $Q_L$, while the right-chiral parts, $u_R$ and $d_R$, are singlets.

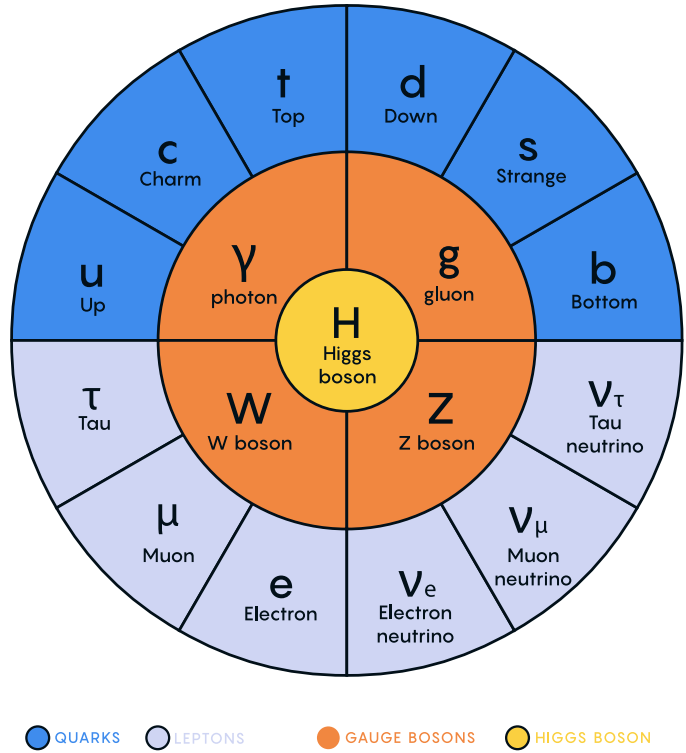

**Figure 1.1:** The fundamental particles in the Standard Model are illustrated in this artistic cartoon.

All quarks transform in the fundamental representation of $SU(3)_C$, while leptons are singlets. To complete the charge assignment, as shown in table 1.1, hypercharge values are obtained by requiring anomalies to cancel to keep the theory gauge invariant at the quantum level. All fermions are repeated in three copies (like the electron, muon, and tau), called generations, with identical properties but different masses, making only the first and lightest generation stable. This theory does not include any right-chiral neutrino because of the lack of experimental evidence of such fields, for example, as seen in the Goldhaber experiment. Even if such fields existed, they would not be detectable

| | $Q_L = (u_L, d_L)$ | $u_R$ | $d_R$ | $L_L = (\nu_L, \ell_L)$ | $\ell_R$ | $G$ | $W$ | $B$ | $H$ |
|---|---|---|---|---|---|---|---|---|---|
| $SU(3)_C$ | 3 | 3 | 3 | 1 | 1 | 8 | 1 | 1 | 1 |
| $SU(2)_L$ | 2 | 1 | 1 | 2 | 1 | 1 | 3 | 1 | 2 |
| $I_3$ | (1/2, -1/2) | 0 | 0 | (1/2, -1/2) | 0 | 0 | (1, 0, -1) | 0 | (1/2, -1/2) |
| $U(1)_Y$ | 1/3 | 4/3 | -2/3 | -1 | -2 | 0 | 0 | 0 | 1 |
| $U(1)_{em}$ | (2/3, -1/3) | 2/3 | -1/3 | (0, -1) | -1 | 0 | (1, 0, -1) | 0 | (1, 0) |

**Table 1.1:** Charge assignment of all the fields in the Standard Model. For $SU(3)_C$ and $SU(2)_L$, we indicate the dimension of the representation under which the fields transform, while for $U(1)_Y$, we show the charge. The electromagnetic charge, that generates transformations under $U(1)_{EM}$, is obtained through the Gell-Mann–Nishijima formula as $Q_{EM} = I_3 + Y/2$, where $I_3$ is "the third component" of the weak isospin, *i.e.* the analogous of $m$ if these were normal spins. For example, $I_3 = 1$ for $\nu_L$, resulting in $Q_{EM}(\nu_L) = 0$, and $I_3 = -1$ for $\ell_L$, resulting in $Q_{EM}(\ell_L) = -1$.

since neutrinos only interact through $SU(2)_L$.

This theory so far would not be able to explain the massive gauge bosons and the fermion masses, as mass terms would not be gauge invariant. The Higgs mechanism explains masses through electroweak spontaneous symmetry breaking (SSB). The Higgs field $H$, a complex scalar that transforms as a doublet of $SU(2)_L$, is included with a Mexican-hat-like potential. $SU(2)_L \times U(1)_Y$ is spontaneously broken when the Higgs field acquires a vacuum expectation value (vev) $v \sim 247$ GeV. The remaining symmetry is the electromagnetic $U(1)_{EM}$, and a combination of the weak charge and the hypercharge results in the conserved electromagnetic charge. Three out of the four degrees of freedom of $H$ are absorbed by the broken generators, becoming their longitudinal parts and turning the massless weak and hypercharge bosons into the massive $Z$ and $W$ bosons. Mass terms for the fermions arise thanks to the Yukawa interactions between the Higgs field and the fermions. However, because no right-chiral neutrino is present, this theory does not include neutrino masses, leaving this question for the next section.

The SM also includes some accidental global symmetries, meaning they emerge without requiring the theory to respect them. The baryon number $B$, which is the total number of quarks minus anti-quarks, and the lepton number $L$, which is the total number of leptons minus anti-leptons, are conserved quantities at tree level. While $B$ and $L$ are only conserved at the tree level, and broken at the quantum level, interestingly, the combination $B - L$ is anomaly-free; thus, many scientists speculated it might be an additional fundamental symmetry of nature. This statement is supported by quantum gravity conjectures that forbid global symmetries, suggesting that it is either gauged, meaning there is a gauge boson associated with it, or broken by some new physics at a larger scale. Lepton flavor is another accidental symmetry of the SM: meaning that the lepton number is conserved separately for each flavor. It turns out that neutrino masses explicitly break this conservation law through phenomena like neutrino oscillations and neutrino decay.

### The mystery of neutrino masses

We described the SM without including mass terms for the neutrinos. Indeed, until the "recent" discovery of neutrino oscillations, there was no evidence of neutrino masses. We now discuss the general ideas to include massive neutrinos in the SM and the implications for future experiments trying to test them.

We can introduce neutrino masses $M_{\alpha\beta}$ in an EFT fashion by considering the so-called Weinberg operator [24,25]:

$$\mathcal{O}^{Weinberg} = \frac{y_{\alpha\beta}}{\Lambda}(\overline{L_\alpha^c}\tilde{H}^*)(\tilde{H}^\dagger L_\beta) \rightarrow M_{\alpha\beta}\overline{\nu^c_{L,\alpha}}\nu_{L,\beta} \quad \text{with} \quad M_{\alpha\beta} = \frac{y_{\alpha\beta}v^2}{\sqrt{2}\Lambda}, \tag{1.2}$$

where $y_{\alpha\beta}$ is an arbitrary matrix of coefficients, $\alpha$, and $\beta$ are flavor indices, and $\Lambda$ is the scale that suppresses this dim-5 operator, where we expect this description to fail and new physics to appear. $L^c = C\overline{L}^T$ is the charge conjugated spinor, where $C = i\gamma^2\gamma^0$ in the Dirac and Weyl representations. Notably, the Weinberg operator is the only gauge-invariant dim-5 operator allowed in the SM, corroborating the evidence that neutrino masses point towards BSM physics. This mass term has a Majorana nature: contrary to the other SM fermions, the mass term mixes the neutrino with its charged conjugate spinor. The consequence is that neutrinos and neutrinos are simply related by flipping the helicity: therefore lepton number cannot be conserved. Indeed, the Weinberg operator breaks $L$ and $B - L$ explicitly. While these are accidental symmetries of the SM, one might argue that it is *technically natural* that the coefficients $y_{\alpha\beta}$ should be small, as they restore these symmetries in the limit in which they vanish. In fact, if the coefficients $y_{\alpha\beta}$ are $\mathcal{O}(1)$, and $m_\nu \sim 0.1\,\text{eV}$, we need $\Lambda \sim 10^{14}\,\text{GeV}$, making yet another hopeless prediction to test. The dim-5 Weinberg operator can be completed in many different ways by introducing additional particles. The tree-level completions are called seesaw mechanisms, and we will discuss them in more detail in section 3.1. Radiative neutrino masses, which generate neutrino masses at loop level, have also been considered in the literature [26,27].

However, there are other solutions: neutrinos masses might be of Dirac nature. If right-handed neutrinos exist, their masses could be generated in the same way as the other fermions in the SM:

$$\mathscr{L}_{\nu-\text{mass}} \supset \overline{\nu_R^\alpha} i\slashed{\partial}\nu_R^\alpha - y_{\alpha\beta}^\nu (\overline{L^\alpha}\tilde{H})\nu_R^\beta - (y_{\alpha\beta}^\nu)^* \overline{\nu_R^\beta}(\tilde{H}^T L^\alpha) \tag{1.3}$$

$$\Longrightarrow \overline{\nu_R^\alpha} i\slashed{\partial}\nu_R^\alpha - y_{\alpha\beta}^\nu \frac{v}{\sqrt{2}} \overline{\nu_L^\alpha}\nu_R^\beta - (y_{\alpha\beta}^\nu)^* \frac{v}{\sqrt{2}} \overline{\nu_R^\beta}\nu_L^\alpha, \tag{1.4}$$

where $\nu_R$ are the right-handed fields, which are completely uncharged under the SM, and would therefore be impossible to detect. However, they also admit a Majorana mass term

$$\mathscr{L}_{\nu-\text{mass}} \supset -M_{\alpha\beta}\overline{\nu_R^\alpha}\nu_R^\beta, \tag{1.5}$$

where the scale $M_{\alpha\beta}$ is completely unrelated to the EW scale and can assume any value as far as we know. There are now several scenarios. If $\nu_R$ is assigned a lepton number, similarly to the Majorana case, we can require $M_{\alpha\beta}$ to be small to restore the conservation of $L$ and $B - L$. We can then explain neutrino masses with a basic Higgs mechanism, requiring Yukawa terms of the order of $10^{-12}$, which would be theoretically very small and significantly smaller than the other Yukawa coefficients.

A second option is not to assign a lepton number to $\nu_R$. In this case, the Yukawa interaction term explicitly breaks $L$ and $B - L$. $M_{\alpha\beta}$ can be arbitrarily large, and the difference in scales between $M_{\alpha\beta}$ and $v$ could explain the smallness of the neutrino masses. This mechanism is called Type-I seesaw[28] and turns out to be the most straightforward tree-level completion of the Weinberg operator. Another option is that $B - L$ is gauged and spontaneously broken at a scale $v_{B-L}$[29]

$$\mathscr{L}^{B-L} \sim y_{B-L}\phi_{B-L}\overline{\nu_R}\nu_R \implies y_{B-L}v_{B-L}\overline{\nu_R}\nu_R, \tag{1.6}$$

where $\phi_{B-L}$ is the gauge boson of the $B - L$ force, $y_{B-L}$ is a coupling constant, and $v_{B-L}$ is the vev acquired by field $\phi_{B-L}$, which would generate $M \sim y_{B-L}v_{B-L}$.

These are only some of the options to generate neutrino masses. Most of them, like the Dirac mass we discussed earlier, requires the presence of new fields, typically called sterile neutrinos, as they would not interact through the weak force.

Experimental tests of the nature of neutrino masses are complex. Any experiment trying to test the Majorana nature of neutrino mass requires observing lepton number violations, which is thus suppressed by $(m_\nu/E)^2$ for ultrarelativistic neutrinos. Given that all neutrinos are produced ultrarelativistic, and we cannot slow them down in a controlled way, any experiment that tries to be sensitive to neutrino masses needs to confront this small factor. The current most promising road is the search for neutrinoless double beta decay in nuclei that undergo double beta decay[30]. If neutrinos are Majorana in nature, the two neutrinos produced in double-$\beta$ decay could annihilate and never leave the nucleus. Another idea would appear in the so-called neutrino-antineutrino oscillations. For Majorana neutrinos, the mass term could flip the helicity of the neutrino, turning it into an antineutrino. Therefore, an antineutrino produced with a lepton could turn to a neutrino while propagating and produce another lepton in the final state when re-interacting through the weak

interaction. The process

$$W^- \to \bar{\nu}\ell^- \to \nu\ell^- \to W^+\ell^-\ell^- \tag{1.7}$$

violates lepton number by two units, and would still be suppressed by

$$(m_\nu/E)^2 \sim (0.1\,\text{eV}/500\,\text{KeV})^2 \sim 2 \times 10^{-19}. \tag{1.8}$$

On the other hand, a Dirac mass term would flip a neutrino to its right-handed partner, which cannot interact through weak interaction and thus would not be detectable.

Lastly, independently of the mechanism behind neutrino masses, there is no reason why we should expect $M_{\alpha\beta}$ to be diagonal, and this is what is in fact happening. Therefore, neutrino masses explicitly break lepton flavor conservation and allow the conversion between different generations. This mechanism is the basis of neutrino mixing, which find its largest experimental evidence in neutrino oscillations.

## Neutrino mixing

The three neutrino states can be described in two different interesting bases: the flavor and mass bases. The flavor basis that we have used so far, is the most natural one, because, in case of massless neutrinos, it would be the only relevant one. In the flavor basis the weak interaction term is diagonal:

$$\mathscr{L}_{weak\ interaction} \supset \sum_{\alpha=e,\mu,\tau} \overline{\ell_\alpha} W \frac{1-\gamma_5}{2} \nu_\alpha + h.c., \tag{1.9}$$

where $\frac{1-\gamma_5}{2}$ is the projector on the left-chiral state, and $\nu_e$, $\nu_\mu$, and $\nu_\tau$ are the states produced or detected through the weak interaction. However, any mass term would result in a mass matrix $M_{\alpha\beta}$, like in eq. (1.2), which is generally nondiagonal in flavor basis. It can be diagonalized with a rota-

tion¶ defined by the unitary matrix $U_{PMNS}$, the Pontecorvo-Maki-Nakagawa-Sakata (PMNS) matrix[31,32], also called lepton-mixing matrix, giving origin to the mass basis $\nu_{i=1,2,3}$:

$$
\begin{pmatrix} \nu_e \\ \nu_\mu \\ \nu_\tau \end{pmatrix} = U_{PMNS} \times \begin{pmatrix} \nu_1 \\ \nu_2 \\ \nu_3 \end{pmatrix} .
\tag{1.10}
$$

Not all the nine complex entries of this matrix are independent: analogously to the quark mixing case, we can describe the matrix with three angles and one complex phase. In the Majorana case, there are fewer fermions we can re-phase, leaving two additional phases as free parameters. However, in contrast with the hadronic case, there is no hierarchy in the magnitude of the mixing matrix elements. For this reason, the typical parametrization is:

$$
U_{PMNS} = \begin{pmatrix} 1 & 0 & 0 \\ 0 & c_{23} & s_{23} \\ 0 & -s_{23} & c_{23} \end{pmatrix} \times \begin{pmatrix} c_{13} & 0 & s_{13}e^{-i\delta} \\ 0 & 1 & 0 \\ -s_{13}e^{i\delta} & 0 & c_{13} \end{pmatrix} \times \begin{pmatrix} c_{12} & s_{12} & 0 \\ -s_{12} & c_{12} & 0 \\ 0 & 0 & 1 \end{pmatrix} \times \begin{pmatrix} 1 & 0 & 0 \\ 0 & e^{i\alpha_2/2} & 0 \\ 0 & 0 & e^{i\alpha_3/2} \end{pmatrix} ,
\tag{1.11}
$$

where $c_{ij} = \cos\theta_{ij}$ and $s_{ij} = \sin\theta_{ij}$, $\delta$ is the phase, and $\alpha_2$ and $\alpha_3$ can be different from zero only for Majorana neutrinos. Neutrino oscillations are the most powerful tool to measure these mixing parameters, as we will discuss in detail in the next paragraph.

---

¶This procedure, which is analogous to the Cabibbo-Kobayashi-Maskawa (CKM) matrix, is a little more complex than discussed here. It differs slightly between the Majorana case, where $M$ is a complex symmetric matrix, and the Dirac case, where $M$ is diagonalized by its singular value decomposition. In this latter case, $\nu_L$ and $\nu_R$ can be rotated independently.

Despite their rare and weak interactions, several physical phenomena are currently studied in physics laboratories. Many experiments are measuring neutrino oscillations, the footprints of neutrino mixing, in many different contexts: we will discuss this phenomenon in detail as it is relevant for this thesis, especially as a tool to possibly discover additional types of neutrinos. On the other hand, the quest for observing neutrinoless double beta decay ($0\nu\beta\beta$), the holy grail for discovering the Majorana nature of neutrinos, has been ongoing for tens of years, setting more and more stringent limits. At the same time, careful measurements of beta decay try to see the small effect of neutrino masses in a deformation of the beta decay spectrum, slowly becoming sensitive to the sub-eV mass region. This question is closely related to the mass ordering question, often called "normal hierarchy or inverted hierarchy?". Indeed, the measurements of the mass splittings leave doubts on which neutrinos are the heaviest and lightest. Moreover, the nucleus-neutrino interactions are essential to analyze oscillation experiments' data properly: we will provide an overview in section 8.1.

## Neutrino oscillations

There are many similarities between the mixing among generations in the quark and the lepton sector. However, there is one crucial difference: the quarks we talk about are mass eigenstates, while we always refer to neutrinos in the flavor basis. The reason is that we can only produce and detect neutrinos through the weak interaction. While the case of neutral current (NC) interactions, mediated by the $Z$ boson, is basis agnostic as it is proportional to the identity matrix, charged current (CC) interactions, mediated by the $W$ boson, differ by flavor. A typical neutrino experiment will look like the cartoon in fig. 1.2. A neutrino is produced through a CC interaction, propagates along some distance, and is detected through another CC interaction. The flavor of the neutrino at production ($\alpha$) and detection ($\beta$) is determined by the lepton's flavor in the interaction. We can summarize it

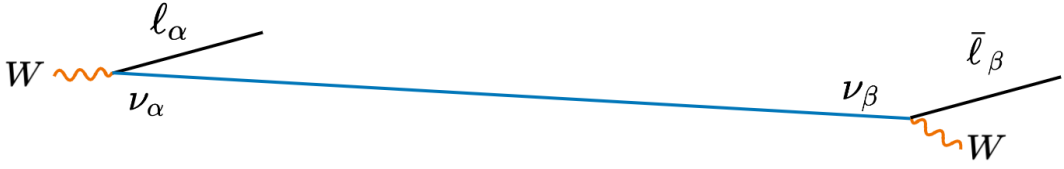

**Figure 1.2:** Neutrino oscillations happen when a neutrino, produced through weak interaction with flavor $\alpha$ determined by the associated lepton, propagates for a certain distance and is detected, resulting in a lepton of different flavor $\beta$.

with the process

$$W^+ \to l_\alpha^+ \nu_\alpha \to \text{propagation} \to l_\alpha^+ \nu_\beta \to l_\alpha^+ l_\beta^- W^+, \tag{1.12}$$

with a final-state pair $l_\alpha^+ l^- \beta$. If the PMNS matrix were equal to the identity, then $\alpha = \beta$ always. However, because of the mixing, the initial state $|\psi(t=0)\rangle = |\nu_\alpha\rangle$ undergoes a non trivial temporal evolution. The different mass states present in the flavor state $|\psi(0)\rangle = \sum_{i=1}^{3} U_{\alpha i} |\nu_i\rangle$ gain different phases while propagating $|\psi(t)\rangle = \sum_{i=1}^{3} e^{i\phi_i(t)} U_{\alpha i} |\nu_i\rangle$, resulting in a different state at detection $|\psi(t)\rangle \neq |\psi(t=0)\rangle$, which has overlap with the other flavor states $\langle \nu_\beta | \psi(t) \rangle \neq 0$. The probability of transition between states $\alpha$ and $\beta$, after propagating for a length $L$, in vacuum, at energy $E$ is given by:

$$P(\nu_\alpha \to \nu_\beta) = |\langle \nu_\beta | \psi(t) \rangle|^2 \tag{1.13}$$

$$= \sum_{i,j} U_{\alpha i}^* U_{\alpha j} U_{\beta j} U_{\beta i}^* \exp\left(i \frac{\Delta m_{ij}^2 L}{2E}\right) = \tag{1.14}$$

$$= \delta_{\alpha\beta} - 4 \sum_{i>j} \text{Re}\left(U_{\alpha i}^* U_{\alpha j} U_{\beta i} U_{\beta j}^*\right) \sin^2\left(\frac{\Delta m_{ij}^2 L}{4E}\right) \tag{1.15}$$

$$- 2 \sum_{i>j} \text{Im}\left(U_{\alpha i}^* U_{\alpha j} U_{\beta i} U_{\beta j}^*\right) \sin\left(\frac{\Delta m_{ij}^2 L}{2E}\right), \tag{1.16}$$

where $U$ is the PMNS matrix and $\Delta m_{ij}^2 = m_i^2 - m_j^2$ is the difference between the square of the mass of $\nu_i$ and $\nu_j$. If $\Delta m_{ij}^2 L / E \ll 1$ for a given $i, j$, then the contribution from $\Delta m_{ij}^2$ to oscillation is negligible. In the opposite limit, $\Delta m_{ij}^2 L / E \gg 1$, oscillations are still allowed, but they average to $1/2$ because of energy and distance resolution.

A handy formula for the oscillation phase can be obtained by turning it from natural units to typical experimental units:

$$\phi_{osc} = 1.27 \, \frac{\Delta m^2}{\text{eV}^2} \, \frac{L}{\text{km}} \, \frac{\text{GeV}}{E}, \tag{1.17}$$

where 1.27 is unitless.

Often, the oscillation problem can be approximated as a two-states case, for example, when the experiment is sensitive to only one value of $\Delta m^2$. The typical situation is when $\Delta m_{ij}^2 L / E \sim 1$ for one specific combination of $i, j$, while $\Delta m_{ij}^2 L / E \ll$ or $\gg 1$ for all the other values. In this case, the formula reduces to

$$P(\nu_\alpha \rightarrow \nu_\beta) = \left| \delta_{\alpha\beta} - \sin^2 2\theta_{\alpha\beta} \sin^2 \left( \frac{\Delta m_{ij}^2 L}{4E} \right) \right|, \tag{1.18}$$

where $\theta_{\alpha\beta}$ is an effective mixing angle for the specific values of $\alpha, \beta, i, j$. This formula is helpful in a variety of contexts. For example, it is appropriate for atmospheric oscillations between $\nu_\mu$ and $\nu_\tau$, and for solar oscillations between $\nu_e$ and a superposition of $\nu_\mu$ and $\nu_\tau$. These approximations are justified by the fact that $\sin^2 \theta_{13}$ is much smaller with respect to the other mixing angles and because two of the three mass states are almost degenerate because of the hierarchy between mass splittings. A third case where this formula is helpful is central for this thesis: oscillations at shorter baseline induced by a fourth neutrino with $\Delta m_{41}^2 \gg \Delta m_{12,23,13}^2$. Oscillations are induced between the light states and the sterile state at a single frequency dictated by $\Delta m_{41}^2$. We discuss this case in greater detail in section 3.3.

The oscillation formula in eq. (1.13) is often obtained with an approximate derivation based on plane waves. While leading to the correct answer, this derivation is intrinsically problematic because plane-wave states have infinite uncertainty on the position, implying that the probability of being detected in the experiment would be zero. A proper derivation, based on the internal wave packet approach[33] or more sophisticated treatments in Quantum Field Theory called the external wave packet approach[34,35,36,37] relies on a finite-size wave packet. The finite size of the wave packet allows oscillations only if the mass states are in a coherent superposition at detection, meaning that they still overlap after propagating for a finite size. This observation has significant consequences: for example, cosmological neutrinos produced in the early Universe are most likely decohered, and they would be detected as mass eigenstates. The case of a fourth neutrino state in the MeV mass range is more relevant to this thesis: it would mix with the standard neutrinos in a typical neutrino experiment. However, it would decohere immediately from the beam, resulting in a separate component of the beam, as discussed in section 3.2. Moreover, all derivations assume the ultrarelativistic approximation, which is valid for basically any neutrino experiment.[‖] In fact, in the case of non-relativistic neutrinos, oscillations would not be possible because the different mass states would decohere way before oscillating.

If neutrinos travel long distances through dense matter, the oscillation probabilities must be corrected for the coherent forward scattering with matter through CC and NC interactions[38,39]. Similar to photons, they effectively acquire a refractive index. These *matter effects* can be observed only if the interaction is not degenerate among the three flavors, which is the case because of CC interaction between $\nu_e$ and atomic $e^-$. We do not discuss these effects further because they are not relevant to this thesis. They often need to be considered in oscillation experiments at long baselines, for solar neutrinos traveling the dense core of the Sun, astrophysical neutrinos going through supernovae, atmospheric neutrinos traveling through the Earth, and for neutrinos propagating in the dense plasma of the early Universe.

Let us go back to the parallel in the hadronic case: why are there neutrino oscillations and not quark oscillations? If we could design an experiment for which a quark would be produced through

---

[‖]The only example of non-relativistic neutrinos come from cosmological neutrinos, which the expansion of the Universe has slowed down.

weak interaction and detected later through weak interaction, provided the mass states do not decohere in between, we could observe quark oscillations. However, nature does not allow this phenomenon: the strong force is confined, impeding quarks from freely traveling distances $\gtrsim 1/\Lambda_{QCD} \sim 1\,\text{fm}$. Therefore, quark mass mixing results in different branching ratios of the $W$ boson into different pairs of quarks rather than quark oscillations. However, oscillations can happen between states that can freely propagate, like neutral mesons. For example, the propagation eigenstates of the $K_0 - \overline{K_0}$ system differ from the detection eigenstates. Lastly, why are there neutrino oscillations and no charged lepton oscillations? Charged leptons are mass eigenstates: it is enough to rotate one of the two fermions involved in the interaction to diagonalize the interaction term. However, suppose we were to produce a coherent superposition of ultrarelativistic charged leptons through a scattering of a perfect neutrino mass eigenstate. Furthermore, suppose we could detect it downstream by measuring another neutrino mass eigenstate. In that case, we could observe oscillations between the different leptons. However, such an experiment is barely possible to imagine and would not be particularly helpful, making its way only into the imagination of a Ph.D. student.

## Measurement of neutrino oscillations

Neutrino oscillations have been measured in many experiments over the last twenty years. Figure 1.3 shows the experimental precision on the six parameters relevant for oscillation physics from 1998 till now: the three mixing angles, the two $\Delta m^2$, and the complex phase $\delta$. The graph also emphasizes the experiments that contributed the most to the sharp reduction of the uncertainties in each parameter. Remarkably, the precision on most quantities is small, at the percent level, which is why people often say that "neutrino physics entered the precision era." There is one exception, which is the phase $\delta$, that is hard to determine experimentally. It enters in CP violation measurements, like

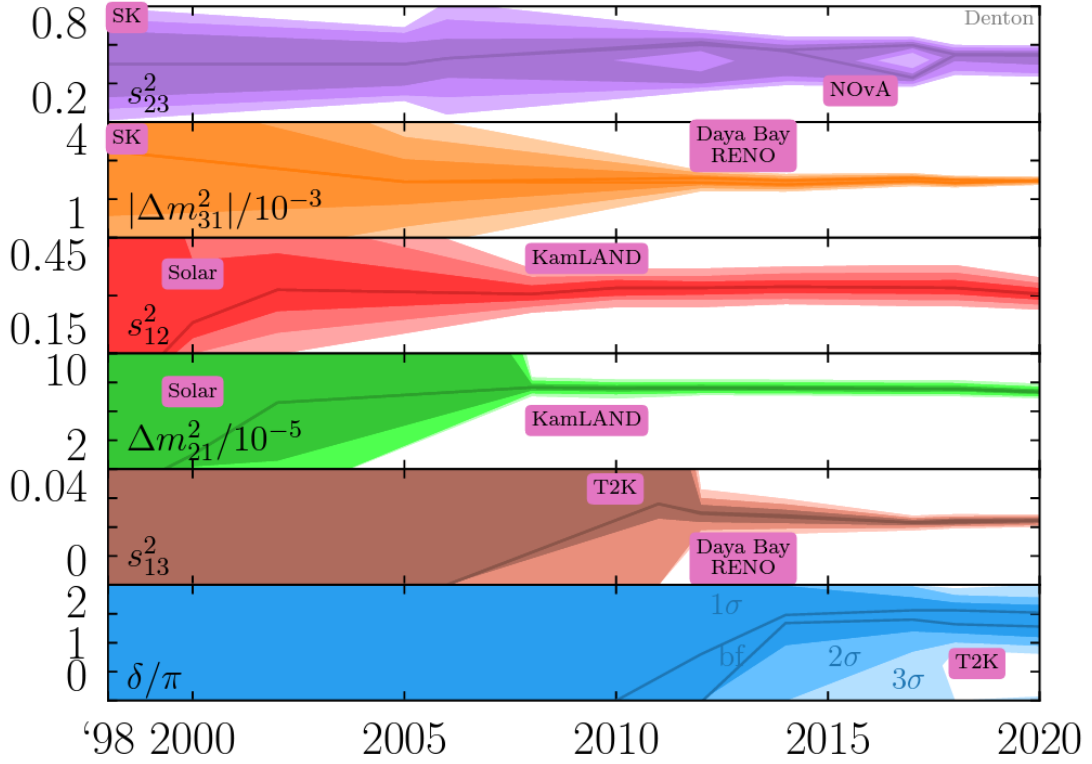

**Figure 1.3:** The experimental uncertainties on the six parameters determining neutrino oscillation physics have decreased significantly in the last 20 years. Super-Kamiokande, the solar neutrino experiments, KamLand, T2K, Daya Bay, Reno, and Nova, are the experiments that contributed the most to the reduction of the uncertainties. The only parameter for which the precision is not yet at the percent level is the $\delta$ phase. Figure taken from [40].

the difference in the rate of $\nu_\mu$ and $\overline{\nu_\mu}$ disappearance, through the Jarlskog invariant [41], defined as:

$$J_{CP} \sum_{\gamma,l} \varepsilon_{\alpha\beta\gamma} \varepsilon_{jkl} = \text{Im}\left( U_{\alpha j}^* U_{\alpha k} U_{\beta k} U_{\beta j}^* \right),$$ (1.19)

where $\varepsilon$ is the Levi-Civita symbol in three dimensions. It can also be written in terms of the mixing

matrix parameters

$$J_{CP} = s_{12} c_{12} s_{13} c_{13}^2 s_{23} c_{23} \sin \delta.$$ (1.20)

Other processes, like $\nu_\mu$ disappearance itself, are sensitive to $\cos\delta$. Notice that there is no complex phase in the two-neutrino case, so there is no CP violation. Therefore experiments for which the two-neutrino approximation is valid are not sensitive to $\delta$ and CP violation. Proving that $\delta \neq 0$ would imply $J \neq 0$, fueling explanations of baryogenesis through leptogenesis. However, this last process depends on more ingredients than just $\delta$, making it $\delta \neq 0$ only a necessary condition rather than sufficient.

These measurements can be re-interpreted in terms of the values of the PMNS matrix. The precision on the entries of the matrix reached the percent level, as illustrated by the most recent values from global fit[42], shown in fig. 1.4, making neutrino physics fully inside the era of precision measurements. However, while most of the experimental observations fit very well in this framework,

$$\boxed{\text{NuFIT 5.1 (2021)}}$$

$$|U|_{3\sigma}^{\text{w/o SK-atm}} = \begin{pmatrix} 0.801 \to 0.845 & 0.513 \to 0.579 & 0.143 \to 0.156 \\ 0.232 \to 0.507 & 0.459 \to 0.694 & 0.629 \to 0.779 \\ 0.260 \to 0.526 & 0.470 \to 0.702 & 0.609 \to 0.763 \end{pmatrix}$$

$$|U|_{3\sigma}^{\text{with SK-atm}} = \begin{pmatrix} 0.801 \to 0.845 & 0.513 \to 0.579 & 0.144 \to 0.156 \\ 0.244 \to 0.499 & 0.505 \to 0.693 & 0.631 \to 0.768 \\ 0.272 \to 0.518 & 0.471 \to 0.669 & 0.623 \to 0.761 \end{pmatrix}$$

**Figure 1.4:** These confidence intervals for the entries of the PMNS matrix were obtained by propagating the uncertainties from the mixing parameters. The first row shows the result without the atmospheric data from the Super-Kamiokande experiment, while the second one includes it. While this dataset tightens the confidence intervals, the lack of proper systematic uncertainties released by the collaboration makes the combination less trustworthy. These results are published in the latest analysis from the NuFit collaboration[42].

several interesting ones do not, providing yet another experimental indication that this is not the end of the story.

# 2

## Short baseline anomalies

In contrast with the consistent picture of neutrino oscillations that we discussed in section 1.3, several experimental measurements do not fit consistently in the SM with massive neutrinos, which are commonly known as short-baseline (SBL) anomalies[*]. What is common to all these anomalous measurements is that the experiments detect neutrinos at SBL, much shorter than well-measured neutrino oscillations:

$$\left( \frac{L/\mathrm{km}}{E/\mathrm{GeV}} \right)^{SBL} \ll \frac{1}{1.27 \Delta m_{atm}^2 / \mathrm{eV}^2} \sim 300, \qquad (2.1)$$

where $L$ is the distance between production and detection, $E$ is the neutrino energy, and $\Delta m_{atm}^2 = 2.51 \times 10^{-3} \, \mathrm{eV}^2$ is the atmospheric squared-mass splitting, which induces faster oscillations than the solar mass splitting $\Delta m_{solar}^2 = 7.42 \times 10^{-5} \, \mathrm{eV}^2$ [42].

---

[*]Here anomaly means that the measurement differs from the SM predictions, not to be confused with Quantum Field Theory anomalies.

This puzzle, with experimental results dating back to more than twenty years ago, sets the context of this thesis. All the original work described in part II and part III aims at testing possible explanations of these anomalies with more recent experiments and datasets. It is arguably one of the most interesting open problems in neutrino physics. First, it consists of many different experimental results observed in different channels, with different neutrinos at different energies, with significances varying from roughly $2\,\sigma$ to almost $5\,\sigma$. Moreover, new experiments meant to test previous anomalies with slightly different conditions often produce inconsistent results. While there are probably many systematic effects involved, it is hard to believe that all these discrepancies are related to systematic errors and underestimated uncertainties. Second, there is currently no theory that explains all the anomalies consistently. The most straightforward one is the light-sterile neutrino hypothesis, which would explain the anomalies as oscillations at a frequency faster than SM oscillations induced by an additional neutrino with mass in the range $m_4 \sim 1-10$ eV. This hypothesis is fascinating because new states, like sterile neutrinos, are needed to explain neutrino masses, and SBL oscillations could be a powerful way to discover them. Moreover, since so many anomalies point towards this idea, it is hard to disentangle SBL anomalies from light sterile neutrinos. However, the picture is certainly more complex than that: not all anomalous results fit well with the light sterile neutrino interpretation, even in richer modifications of this theory, and many more experiments disfavoring this theory.

This chapter provides an overview of the most striking anomalies and their possible interpretations. For a complete overview of the experimental panorama, we refer the interested reader to Ref.[43]. Chapter 3 contains an overview of the sterile neutrino models invoked to explain some of the SBL anomalies tested in this thesis work.

## 2.1 LSND and MiniBooNE

LSND and MiniBooNE are the anomalies that captured the most attention during these years because their significance is relatively high $\sim 4-5\sigma$, and the straightforward interpretation of light sterile neutrino oscillations fits them reasonably well.

## LSND

The Liquid Scintillator Neutrino Detector (LSND), operated at Los Alamos between 1993 and 1998, was looking for neutrino oscillations and was designed before the $\Delta m^2$ values were well-established. An 800 MeV proton beam shot against a target produced pions, later stopped in a beam dump. While $\pi^-$ bind with nuclei and are absorbed without neutrino emission on timescales much smaller than the charged $\pi$ lifetime, $\pi^+$ decay at rest into $\nu_\mu\mu^+$. Muons also stop in the absorber, decaying to $e^+\nu_e\bar{\nu}_\mu$. The beam contains $\nu_e, \nu_\mu, \bar{\nu}_\mu$, with minor contamination of $\bar{\nu}_e$, at the $10^{-3}$ level with respect to the other flux.

The experiment measured the $\bar{\nu}_e$ component of the beam by measuring the inverse $\beta$ decay (IDB) process $\bar{\nu}_e p \rightarrow e^+ n$. This powerful signature, schematically illustrated in the left panel of fig. 2.1, consists of the $e^+$ Cherenkov light, followed by the two photons from the $e^+$ annihilation, and about 200 μs later, the light emitted the captured neutron. The result[46,47,48,49,45], right plot of fig. 2.1, shows a significant excess 3.8 $\sigma$ of the data (black dots) with respect to the expectation (red and green). It can be explained by neutrinos oscillation $\bar{\nu}_\mu \to \bar{\nu}_e$ with $\Delta m^2 \gtrsim 1\,\mathrm{eV}^2$ (blue). This anomaly, which started the long journey of searches for light sterile neutrinos oscillations, is highly robust. Because of the double signature, interpreting these events differently than $\bar{\nu}_e$ is difficult, which is why many theories addressing SBL anomalies fail to explain the LSND anomaly.

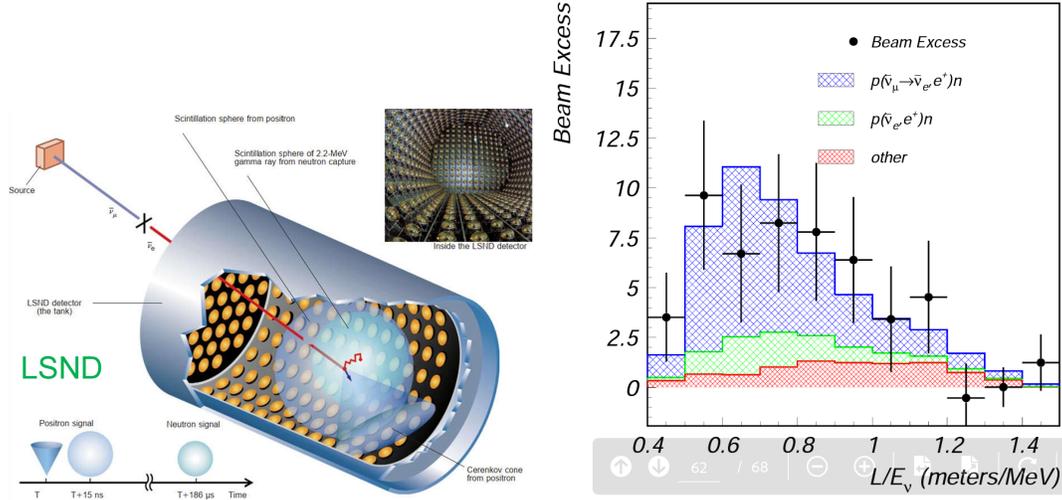

**Figure 2.1:** Left: The LSND experiment, consisting of a cylindrical liquid scintillator tiled with over a thousand photo-tubes, can measure $\bar{\nu}_e$ thanks to the double signature produced by the $e^+$ first and by the neutron afterward. Figure taken from [44]. Right: the reconstructed $L/E$ spectrum after subtracting all the backgrounds not traceable to $\bar{\nu}_e$. The red and green histograms show the beam's expected contamination of $\bar{\nu}_e$. At the same time, the blue represents the distribution under the light sterile neutrino hypothesis for the best-fit value to the data (black dots). Figure taken from [45].

## MiniBooNE

Built to test the oscillatory interpretation of the LSND result, the Mini Booster Neutrino Experiment (MiniBooNE) used a different $L$ and a different $E$ to look for oscillations at the same $L/E$. MiniBooNE operated from 2002 to 2019 in the Fermilab Booster Neutrino Beam (BNB), consisting of neutrinos with an average energy of 800 MeV produced from mesons' decay in flight, originating from the 8 GeV proton beam against a Beryllium target. Figure 6.1 shows the location of MiniBooNE in the context of the SBL experiments at Fermilab. MiniBooNE was a Cherenkov detector shaped as a sphere of roughly 5 m of radius, filled with about 800 ton of mineral oil. Surrounded by about 1000 photomultipliers, MiniBooNE could detect different final states through different types of Cherenkov "rings", as illustrate in fig. 2.2. A clear and full ring is the signature of a muon, while electrons produce fuzzy rings. Individual photons result in electromagnetic showers indistinguishable from electrons, while di-photon events, which originated from $\pi^0$ decays, produce a broader signature that can be fit by two rings. MiniBooNE was not only insensitive to the differ-ence between electrons and photons but was also insensitive to protons and neutrons. Therefore, to reconstruct the incoming neutrino energy, neutrino interactions are assumed to be quasi-elastic, meaning elastic on a single, unbounded nucleon, which is mostly the case at the MiniBooNE energy range (more details in section 8.1).

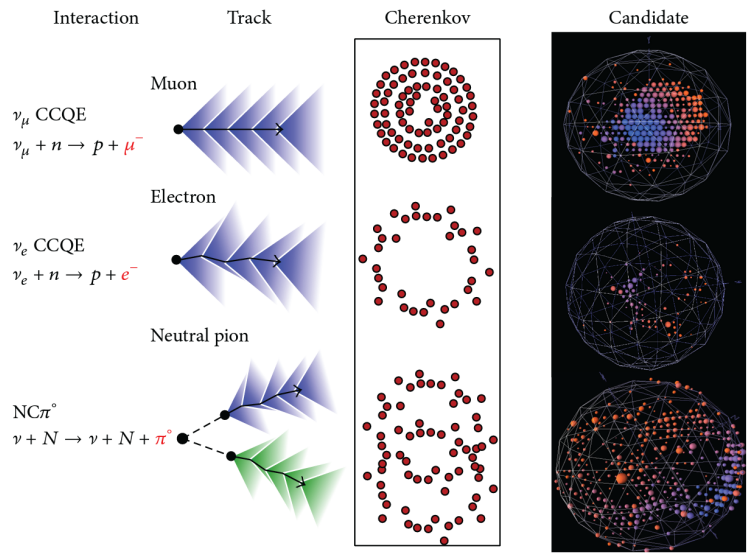

**Figure 2.2:** The spherical MiniBooNE liquid Cherenkov detector can measure three signatures, traceable to the footprints of muons, electrons, and $\pi^0$, produced through specific neutrino interactions. Figure taken from [50].

The plot on the left of fig. 2.3 shows the distribution of the reconstructed neutrino energy in the search for electron neutrinos - the fuzzy rings - on the latest iteration of the analysis [51]. This graph shows the so-called *low energy excess* (LEE): the excess of electromagnetic events in the 200 − 600

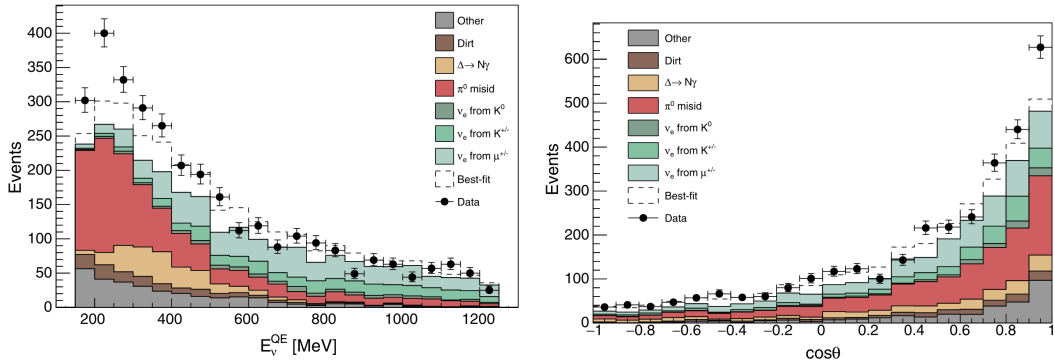

**Figure 2.3:** The reconstructed neutrino energy spectrum (left) and polar angle (right) in the search for single electromagnetic shower events show an excess concentrated at low energy. This plot shows the data collected in neutrino mode, but the excess is also visible in anti-neutrino mode. The most considerable background (red) comes from misidentified $\pi^0$ events, which has been constrained from a dedicated *in-situ* measurement of the $\pi^0$ background. Single photons produced in radiative decays of the $\Delta$ resonance (yellow) also contribute to the low energy region. The rest of the background (green) is induced by genuine electrons, produced by interactions of the minor contamination (0.5%) of $\nu_e$ in the beam. The other backgrounds (gray and brown) are also constrained with specific sidebands. While many interpretations can reproduce the energy spectrum, only a few are also able to simultaneously fit the wide angular distribution of the excess. Figures taken from[51].

MeV energy range. When looking at the angular distribution (right plot), the excess appears to be spread over the entire angular range, although most events are forward-going. The excess has been observed in both neutrino and anti-neutrino mode, and each analysis iteration[52,53,54,55,51,56] only strengthened the significance, now about $5\sigma$.

It is also clear that the excess is related to the beam and is not simply a background effect. The plot on the left of fig. 2.4 shows the radial coordinate of the interaction vertex. The excess is mostly in the inner part of the detector, similar to standard neutrino interactions, excluding its possible relationship with backgrounds close to the boundary of the detector. Similarly, the plot on the right shows the distribution of the timing of the event with respect to the beginning of the bunch. The excess perfectly aligns with the beam structure, below 8 ns, while the data agrees perfectly with the simulation in the region containing non-beam related background. Moreover, no evidence for an excess was found in the beam-dump run[57], where mesons were not focused along the beamline, and only possible heavier BSM particles were expected to reach the detector.

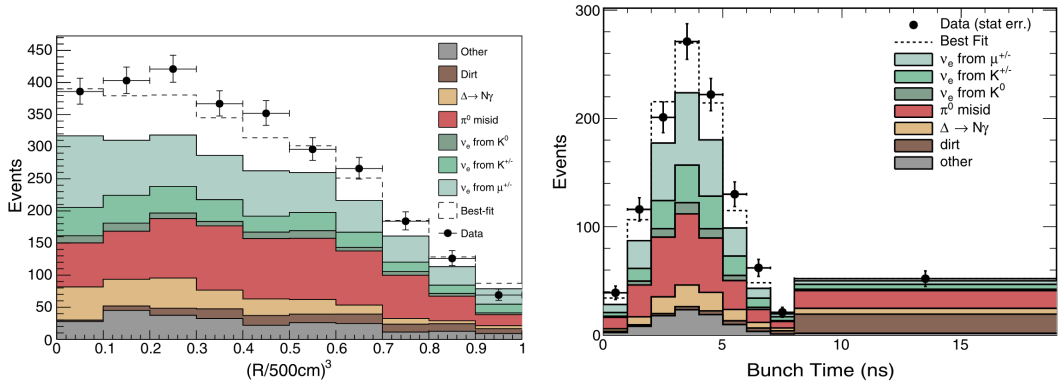

**Figure 2.4:** The same data shown in fig. 2.3 is now plotted in the radial coordinate of the neutrino interaction vertex (left) and in the time of the interaction with respect to the beginning of the bunch. The excess closely follows the distribution of the neutrino-induced backgrounds, supporting the interpretation of the excess as part of the neutrino beam. Figures taken from [51].

There are three interpretations for the MiniBooNE excess, as shown in fig. 2.5: electrons from electron neutrino interactions, single photons from neutrino interactions, and $e^+e^-$ collimated pairs that could not be distinguished from single electromagnetic showers. While all these explanations could be generated by new physics BSM, the first and the second explanations might also be related to mismodeling of the neutrino flux or nuclear cross sections. The $e^+e^-$ hypothesis, however, requires physics BSM, such as models discussed in chapter 4.

While it seems that the light-sterile neutrino hypothesis fits the LSND energy spectrum quite well, this is not precisely true for MiniBooNE. The best-fit point underestimates the data, clearly visible in figure fig. 2.6, where the number of excess events is plotted against the reconstructed energy, together with the best-fit lines. Even large modifications of the best-fit parameters within uncertainties do not significantly improve the agreement in the lower energy bins.

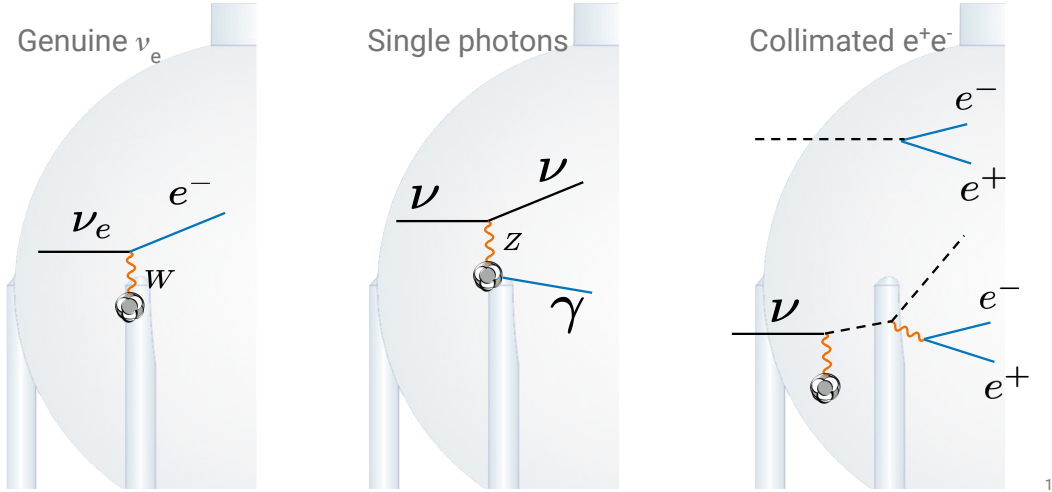

**Figure 2.5:** The MiniBooNE low energy excess of electromagnetic events can be interpreted in three ways: pure electrons, induced by CC neutrino interactions, pure photons, produced through NC neutrino interactions, and $e^+e^-$ collimated pair. These last ones require the presence of BSM particles, either part of the beam and decaying in flight in the detector or produced directly in the detector through neutrino scattering. Figures taken from [51].

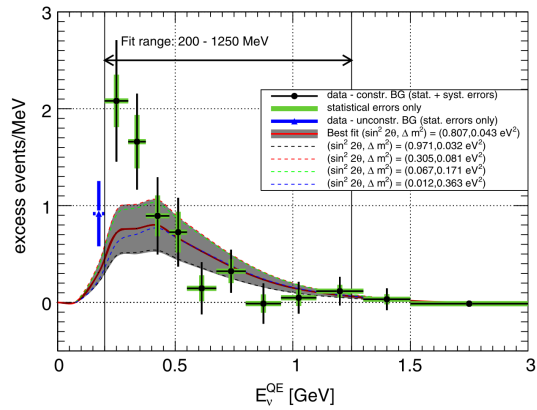

**Figure 2.6:** The best-fit value to the light sterile neutrino model (solid red line) does not explain the excess completely, especially in the lower energy bins. Even varying the model parameters does not improve the agreement in the lower energy bins. Figure taken from [51].

The SAGE[58,59] and GALLEX (later called GNO)[60,61] experiments were designed to confirm neutrino oscillations by measuring solar neutrinos through the reaction:

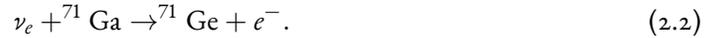

$$\nu_e + {}^{71}\mathrm{Ga} \rightarrow {}^{71}\mathrm{Ge} + e^-. \tag{2.2}$$

The germanium is extracted and counted through its decay back to gallium, with a typical timescale of $\sim 11$ days. To double check the observed deficit of solar neutrinos, which was indeed induced by oscillations, they design intense sources of ${}^{51}\mathrm{Cr}$ and ${}^{37}\mathrm{Ar}$, which emit $\nu_e$ from electron capture[62]. Both experiments observed a deficit with respect to the expectation with both isotopes[63,64,65,66]. The value of the ratio between observed and expected, averaged among both experiments, was $R = 0.86 \pm 0.05$[67].

The BEST experiment was later built to follow up on these results, with a larger detector with increased sensitivity. A schematic representation is shown in the left plot of fig. 2.7. They found[68,69] a deficit $R_{in} = 0.791 \pm 0.05$ and $R_{out} = 0.766 \pm 0.05$, for the inner and outer part of the detector, respectively, consistent with SAGE and GALLEX. These results have sensitivities between 4 and $5\sigma$, which grow well above $5\sigma$ when combined with SAGE and GALLEX. Given the short baseline $L/E \sim 1\,\mathrm{m/MeV}$, this result can be interpreted as electron neutrino disappearance from oscillations at $\Delta m^2 \gtrsim 1\,\mathrm{eV}^2$. However, these experiments are not sensitive to oscillations aside from the inner-vs-outer part of the detector. No dependence on oscillation length was seen, favoring large oscillation frequencies, but leaving room for other interpretations, like systematic errors in the cross-section calculation and efficiency determination.

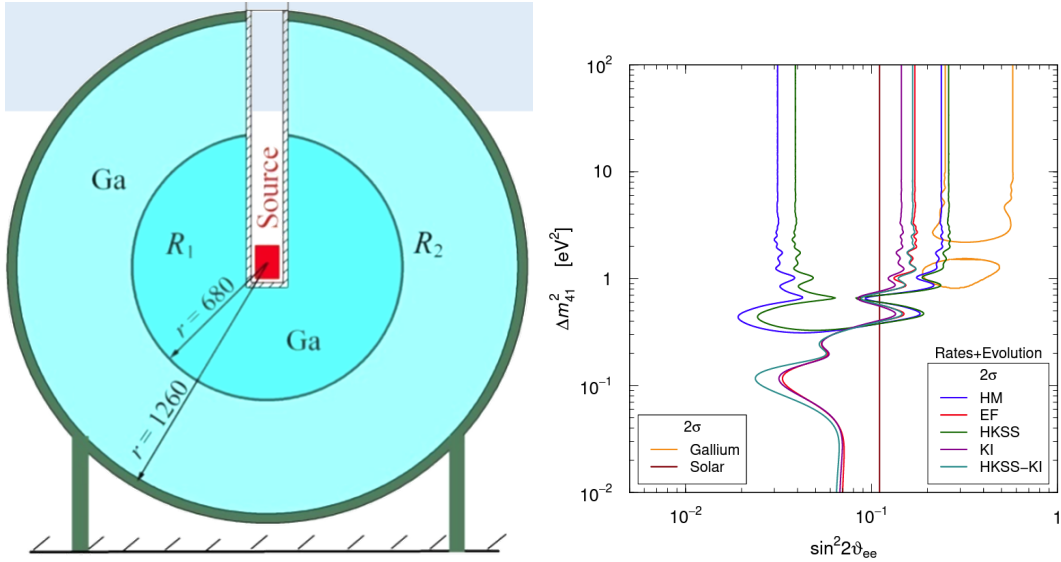

**Figure 2.7:** Left: The BEST experiment consists of a sphere full of Gallium, divided into two sections ($r < 68$ cm and $68$ cm $< r < 126$ cm). The source is inserted in the center through a small vertical hole. Right: Parameter space of the light sterile neutrino model for what concerns the radiochemical and the reactor antineutrino anomalies, showing the oscillation frequency $\Delta m^2$ on the vertical axis and the effective mixing angle $\sin^2 2\theta_{ee}$ on the horizontal one. The graph demonstrates how older flux predictions (HM and HKSS) produce a closed best-fit region in parameter space, in contrast with the latest predictions (EF, KI, HKSS+KI), which result in exclusion regions. Moreover, this graph emphasizes the tension between Solar neutrinos, which exclude everything above $\sin^2 2\theta_{ee} \gtrsim 0.1$, and the Gallium measurements, which prefer larger values. The figure is taken from[70].

## 2.3 REACTOR ANTINEUTRINO ANOMALIES

For about ten years, people believed in the reactor antineutrino anomalies (RAA), and this story is finally reaching an end. Reactor antineutrinos are produced in the fission of $^{235}$U, with energy $\sim 3$ MeV. They are detected through the IDB process, and their energy is inferred from the electron energy in the final state. Reactor neutrino experiments, like Daya Bay, RENO, and Double Chooz, were initially designed to measure $\sin^2\theta_{13}$ in conventional neutrino oscillations and extended their analysis to look for faster oscillations. These very well-understood detectors measure antineutrinos at $L \sim \mathcal{O}(1\,\text{km})$, with very large samples ($\gtrsim 10^6$), very low background from the reactor, providing sensitivity up to $\Delta m^2 \lesssim 0.1\,\text{eV}^2$. Because of the large distance, the resolution on $L$ smears out the oscillation wiggles to the average value of the mixing angle involved in the disappearance (see eq. (1.18)). The RAA started in 2011[71]: after a careful reevaluation of the reactor neutrino flux[72,73] (called HM model), most reactor experiments found discrepancies in the ratio of the expected to the predicted IDB yield. Such discrepancy is quantified with the average ratio across experiments: with the updated HM model, this was found to be $R = 0.925^{+0.025}_{-0.023}$, with a significance of $2.9\sigma$.

However, this discrepancy is now dismantled for three reasons. First, newer calculations relying on more recent measurements of the decays of $^{235}$U emerged, obtaining a better agreement with the experimental data. The most up-to-date calculation with the KI model results in $R = 0.975^{+0.022}_{-0.021}$[70,74,75], basically compatible with unity within the uncertainty, as shown in fig. 2.8. Moreover, they produce open regions in the light sterile neutrino parameter space, in contrast with closed regions produced by older flux models, as seen in the right plot of fig. 2.7. Second, it has

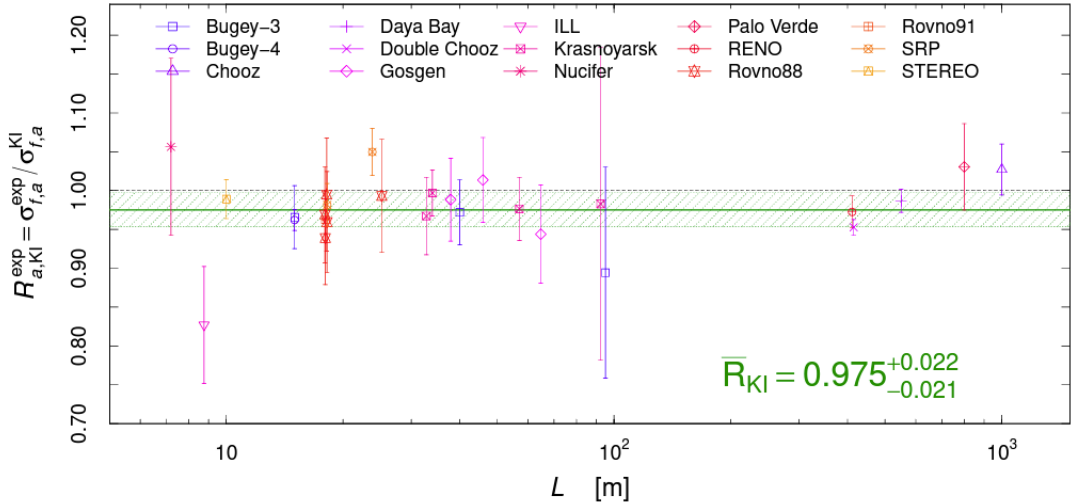

(c) KI model [24]: no RAA (1.1 $\sigma$).

**Figure 2.8:** The reactor antineutrino anomaly, quantified by the average ratio of the observed to the predicted inverse beta decay yields across experiments, is cured by the most up-to-date flux calculations, providing a value for the ratio compatible with one within uncertainties. Notably, this effect is independent of the distance between the source and the detector in the range $10\,\mathrm{m} < L < 1\,\mathrm{km}$ because oscillations average to their mean value. The figure is taken from[70].

been noticed that using a proper statistical treatment, using Feldman-Cousin confidence interval rather than relying on Wilks' theorem, reduces the significance of most anomalies, especially when interpreted as light sterile neutrinos[76]. Third, the newer generation of experiments at very-short-baseline $L \sim \mathcal{O}(10\,\text{m})$ attempted measurements of the IDB yield at different distances through segmented or movable detectors. By performing relative comparisons, these experiments are sensitive to baseline-dependent spectral distortions induced by oscillations without relying on neutrino flux predictions. DANSS[77], NEOS[78], PROSPECT[79], and STEREO[80] all reported null results, disfavoring and weakening the RAA further. Among this class of experiments, only Neutrino-4 reported evidence for oscillations at the $2.9\sigma$ level[81]. However, the community claimed this result was controversial because of the analysis technique employed[82]. The best-fit region is also incompatible with the results from PROSPECT[79] and STEREO[80].

## 2.4 Null results

What makes the saga of SBL anomalies so complex and never-ending is the wide variety of null results. The most straightforward interpretation of the anomalies in terms of light sterile neutrinos predicts oscillations in a wide variety of channels and experiments. However, many of these do not report any disagreement with the SM, interpreted as a null result for the light sterile neutrino model. If, on the one hand, these results made it clear that the vanilla oscillation hypothesis is not sufficient, on the other hand, they pushed the community to come up with many new theories to explain the SBL anomalies.

### Accelerator neutrinos

The KARMEN (KArlsruhe Rutherford Medium Energy Neutrino) experiment searched for transitions $\bar{\nu}_\mu \to \bar{\nu}_e$, from pion decays at rest, analogously to LSND. It exploited the time structure of the beam to isolate $\bar{\nu}_\mu$ flux and a segmented liquid scintillator calorimeter to measure the resulting $e^+$, coated with gadolinium to enlarge the sensitivity to neutrons. It did not observe any disagreement from the expectation[83], excluding a large portion of the LSND best-fit region, especially at large $\Delta m^2$.

While the search for electromagnetic events led to observing the low energy excess, MiniBooNE's measurements of $\nu_\mu$ and $\bar{\nu}_\mu$[84] reported results in agreement with the SM, both with Booster Neutrino Beam and with the NuMI beam. The lack of muon-neutrino disappearance disfavors the oscillation hypothesis, with the caveat that, without a near detector, a high-frequency oscillation could average out to a too small value to be distinguished from systematic uncertainties. For this reason, MiniBooNE performed additional searches combined with SciBooNE[85], still leading to null results. These results have been further corroborated by many additional experiments, most remarkably by the MINOS/MINOS+[†] analyses[86]. MINOS was located along the NuMI beamline, which now serves the NoVA experiment, to measure neutrino oscillations, using a near detector located 1 km downstream and a far detector 735 km away. The most recent analysis combines data from both detectors and rules out the parameter space that could explain LSND and MiniBooNE in the light sterile neutrino model.

### Atmospheric neutrinos

Together with MINOS, the IceCube measurement of atmospheric neutrinos is the most stringent constraint of the light sterile neutrino explanation of the SBL anomalies. IceCube, a cubic kilometer of the Antarctica glacier instrumented with photomultipliers, can test oscillations through the matter effects. For $\Delta m^2 \sim 0.1 - 10 \, \mathrm{eV}^2$, a resonant matter effect leads to a dramatic enhancement of $\nu_\mu$ disappearance at energies $\sim 1 \, \mathrm{TeV}$, allowing very strong constraints from atmospheric neutrinos, as shown in fig. 2.9. The latest analysis using eight years of data[87,88] sets some of the strongest limits on $\nu_\mu$ disappearance induced by light sterile neutrinos, and it is in agreement with the SM with a p-value of 8%[‡]. This analysis, already powerful because testing SBL anomalies through a different mechanism, is complemented by another analysis at smaller energies, where sterile neutrinos would leave an imprint on standard atmospheric oscillations[89].

---

[†]MINOS+ is the name given to the second phase of the experiment, benefitting from the upgraded neutrino beam.

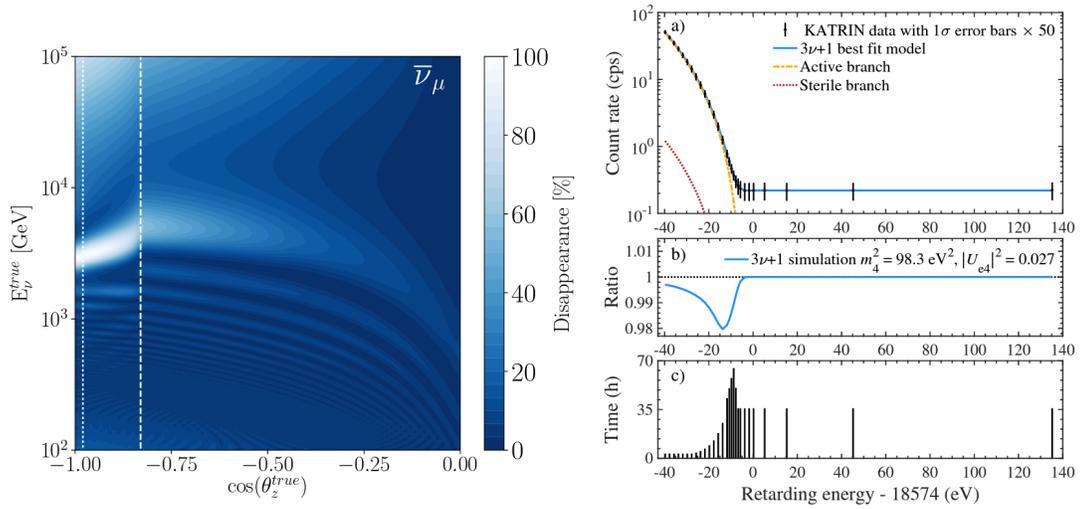

**Figure 2.9:** Left: this graph is known in IceCube as the *oscillogram*. It shows the $\overline{\nu}_\mu$ disappearance probability as a function of the neutrino energy and the zenith angle with respect to IceCube, which impacts the distance traveled through the Earth. Around $\sim 1$ TeV and for small angles, matter effects result in a resonant depletion of the $\overline{\nu}_\mu$, which are fully transitioned into the sterile state (called MSW effect). The vertical white lines indicate the transition between the inner and outer core of the Earth and between the core and the mantle. The sterile neutrino parameters are set to the global best-fit value from[90]. Figure taken from[87]. Right: KATRIN simulation of the measured tritium beta decay spectrum in the presence of sterile neutrinos with arbitrary parameter values. The first plot shows the contributions of the individual contributions of the active and the sterile neutrinos. The second plot shows the ratio of the spectra with and without sterile neutrinos, while the third shows the integrated time distribution. Figure taken from[91].

---

[‡]Some physicists interpreted the closed contour at 90% as mild evidence for sterile neutrinos. However, technically speaking, the result agrees with the SM within less than $2\sigma$.

## Solar neutrinos

While they have been measured by the same class of experiments, radiochemical anomalies are in strong tension with solar neutrino measurements. Indeed, the radiochemical anomalies require strong mixing angles to explain the observed deficit, which should have been observed on top of neutrino oscillations in solar neutrinos, which have similar energies. A combined analysis of solar neutrino measurements firmly excludes the SBL probability of electron neutrino disappearance $> 0.0168$, independently of the oscillation frequency[92].

## Neutrino masses

The Karlsruhe Tritium Neutrino experiment (KATRIN) is designed to improve the sensitivity to the effective neutrino masses by measuring the tritium beta decay spectrum with high precision. Given the sensitivity reached with their latest result[93], claiming $m_\nu \lesssim 0.8\,\mathrm{eV}$, KATRIN can effectively constraint light sterile neutrinos at masses $\sim 1\,\mathrm{eV}$. An additional state would modify the endpoint of the beta decay spectrum with an extra kink produced by the sterile state, as shown in the right plot of fig. 2.9. The results[94,91] exclude interesting regions of the parameter space that could explain the gallium anomalies, especially in the large $\Delta m^2$ region.

## Coherent neutrino-nucleus scattering

Prospects for robust tests of the oscillatory interpretations of MiniBooNE and LSND come from coherent neutrino-nucleus scattering measurements. Coherent neutrino-nucleus scattering (CE$\nu$NS[§]) is a neutral current interaction with the entire nucleus, which produces a nuclear recoil as the only signature. It is the dominant process at small neutrino energy, with a considerable enhancement of the cross section proportional to the square of the number of neutrons, thanks to the coherent interaction. Even if the CE$\nu$NS process was observed only recently [95,96,97], a dedicated search for neutrino disappearance could already prove very powerful in the next few years. Measuring the flux of neutrinos from pion decay at rest, with fixed and well-known energy, at different distances from the source, varying from a few to a few tens of meters, could resolve oscillation wiggles if light sterile neutrinos exist. An upgrade to the COHERENT experiment, using a 10-kg CsI detector at cryogenic temperature could test the best-fit regions of MiniBooNE and LSND and 90% confidence level. The Coherent CAPTAIN-Mills experiment [98] will also test the anomalies with similar techniques.

---

[§]While I would tend to read this acronym like "ce-nuns," people in the field read it as "se-vens" - yes, like the number 7 at plural.

## Cosmology

Any sterile neutrino would be produced, at least through mixing, in the early universe. If sterile neutrinos are in thermal equilibrium when the temperature is at the MeV scale, they would affect the history of the universe and are thus strongly constrained by cosmological observations. At temperature around the MeV, about one second after the big bang, neutrinos decouple from the thermal bath, and their population freezes out. This additional extra radiation contributes to the rate at which the universe expands and effectively modifies the Hubble parameter [99]. This effect is quantified through the measurement of the effective extra relativistic degrees of freedom species, which was measured to be $N_{eff}^{exp} = 2.99 \pm 0.17$ [100], in agreement with the Standard Model prediction $N_{eff}^{SM} = 3.046$ for the three active neutrinos. This argument does not directly extend to heavier sterile neutrinos. We will see that MeV sterile neutrinos affect the rate at which deuterium is formed during Big Bang Nucleosynthesis in section 4.1, while KeV sterile neutrinos would typically require very small mixings that they would not be in thermal equilibrium with the bath when neutrinos decouple from the bath.

# 3

# Neutrinos beyond the Standard Model

Neutrinos are a rather special ingredient of the Standard Model (SM). The absence of electric charge and their extremely small but non-vanishing mass implies that, contrary to all other fermions, neutrinos do not have their properties uniquely determined by the SM gauge group, $G = SU(3) \times SU(2) \times U(1)$. Indeed, in order to uniquely determine the origin of neutrino masses, we need additional ingredients, like new symmetries, such as $U(1)_{B-L}$, or new scales, such as the Majorana mass of their right-handed partners. Indeed, this second route is the central topic of investigation in this thesis. However, any new model should also provide explanations for the anomalous observations at short baselines. All the models discussed here are tailored to explain the MiniBooNE excess and, with some additional tweaks, can explain other anomalies too. This chapter provides a phenomenological discussion of models containing sterile neutrinos, distinguishing the case of heavy sterile neutrinos, which do not produce oscillations in most accelerator-based experiments, from light sterile neutrinos, which induce oscillations at a much shorter baseline than standard neutrinos. The first case is the topic of study of the two works discussed in part II, where we show how a phenomenologically minimal model is excluded within a certain, interesting mass range and how some variations of this model can be constrained by present and future experiments. The second model is investigated with the MicroBooNE experiment: the full experimental analysis and interpretation is the central topic of part III.

## 3.1 Mixing between active and sterile neutrinos

We introduce additional states together with all the allowed mass terms. We call these states sterile, as they are typically uncharged* under $SU(2)_L$, thus not undergoing the typical weak interaction with the $Z$ and $W$. Among the other states, there are some right-handed ones that allow the construction of mass terms for neutrinos in the Lagrangian. In the simplest model, these right-handed partners are uncharged under any group and therefore allow a new Majorana mass term, which could be very large, making neutrinos masses effectively small. This mechanism is called Seesaw, as it produces neutrino masses as a ratio between the electroweak and the new Majorana mass scale. Moreover, as a complete singlet under $G$, right-handed neutrinos could also provide unique insight into the

---

*In type-II [101] and type-III [102] seesaw among these states, there are triplets of $SU(2)_L$. However, this thesis only focuses on the classical type-I seesaw mechanism, which involves additional neutrinos uncharged under $SU(2)_L$.

possible existence of other hypothetical particles, such as dark matter, the dark photon, or additional Higgs bosons.

After diagonalization, the three lightest neutrinos are identified as the active neutrinos, which make up almost entirely the three left-handed flavor states that interact through the weak force. The flavor states $\nu_e$, $\nu_\mu$, and $\nu_\tau$ would contain a small fraction of the heavier mass states, $\lesssim 3\%$ as constrained by the unitarity of the PMNS matrix. In the most *minimal* model, these mixing terms are the only tools to possibly discover sterile neutrinos.

## Seesaw and more

In the canonical Type-I Seesaw mechanism [103,28,104,105,106,107,108,109,102] right-handed neutrinos $\nu_R$ are introduced. Generally, we expect these neutrinos to be more than one, as one is insufficient to fix at least two non-zero neutrino masses. $\nu_R$ transform as complete singlets in the SM, allowing a Majorana mass term and a Yukawa coupling with the Higgs:

$$\mathscr{L}_{\text{masses}} \supset M_M \overline{\nu_R} \nu_R + y_\nu (\overline{L} \tilde{H}) \nu_R + h.c., \tag{3.1}$$

where $M_M$ is the Majorana mass matrix, $y_\nu$ is a general Yukawa coupling matrix, $L$ is the lepton doublet, and $H$ is the Higgs doublet. After electroweak symmetry breaking, the Higgs gets a vev $v$, leading to:

$$\mathscr{L}_{\text{masses}} \supset M_M \overline{\nu_R} \nu_R + M_D \overline{\nu_L} \nu_R + h.c., \tag{3.2}$$

where $M_D = y_\nu v/\sqrt{2}$ is the resulting Dirac mass matrix. After diagonalizing these mass terms, we found that there are two different sets of mass eigenstates with masses:

$$m_{1,2} = \frac{1}{2}(M_M \pm \sqrt{M_M^2 - 4M_D^2}), \tag{3.3}$$

which, in the Seesaw limit ($M_D \ll M_M$), leads to

$$m_1 \simeq \frac{M_D^2}{M_M} = \frac{(y_\nu v)^2}{2M_M}, \quad m_2 \simeq M_M, \tag{3.4}$$

where the first formula reads as $m_1 = M_D^T M_M^{-1} M_D$ in case of a nontrivial structure of these mass matrices. Although the SM forbids any renormalizable Majorana mass term for the left-handed neutrinos, by integrating out $\nu_R$, we can generate an effective dim-5 operator, the so-called Weinberg operator [24,25], as shown in fig. 3.1:

$$\mathscr{L}_{d=5} = \frac{y_\nu^2}{M_M}(\overline{L^c}\tilde{H}^*)(\tilde{H}^\dagger L) \rightarrow \frac{(y_\nu v)^2}{M_M}\overline{\nu_L^c}\nu_L, \tag{3.5}$$

which, after electroweak symmetry breaking, explicitly violates the lepton number through an effective Majorana mass for the light neutrinos. However, any effect involving this operator will be proportional to $(m_\nu/E)^2$, where $E$ is the energy scale of the experiment under consideration. Since neutrino masses have not been measured yet, it is clear that experiments are not yet sensitive to this operator.

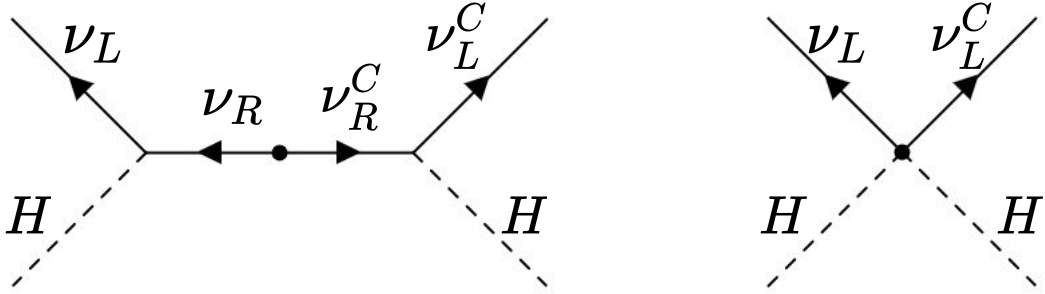

**Figure 3.1:** The Type-I seesaw mechanism requires a Yukawa interaction between the Higgs, the active left-handed, and the right-handed neutrinos, and a Majorana mass term for the right-handed neutrinos (left). If the sterile neutrino is much more massive than the electroweak scale, it can be integrated out, producing an effective Weinberg operator of dimension five (right). When the Higgs takes a vev, the Weinberg operator results in a Majorana mass term for the active neutrinos.

We can now generate small neutrino masses by a combination of a large Majorana mass scale $M_M$, which is theoretically completely disconnected from the electroweak scale, and by small Yukawa couplings $y_\nu$. If we assume couplings of order one, to get neutrino masses of around 0.1 eV, we need $M_M \sim 10^{15}$ GeV, conventionally known as high-scale Seesaw, with heavy neutrinos partners impossible to be produced in the laboratory. However, this is not the case in low-scale variations of the model [110,111,112,113,114,115,116], which are both ubiquitous and well-motivated theoretically, even if less predictive. Among the most interesting cases is the inverse Seesaw, where approximate conservation of lepton number guarantees the smallness of neutrino masses in a technically natural way [117,118]. This variation requires the introduction of additional fermions in the spectrum, with a lepton number opposite to the heavy neutrinos. As a result, the additional state combines into a pseudo-Dirac pair when diagonalizing the mass matrix. In the limit in which the parameters that violate the lepton number, *i.e.* the Majorana mass terms for the heavy neutrinos, are small, the lepton number is conserved, and the pseudo-Dirac pair becomes exactly Dirac. Neutrino masses are now proportional to the lepton number violating terms, which are now allowed to be small, below the electroweak scale, as they restore the lepton number in the limit in which they vanish. This framework is behind the models for heavy sterile neutrinos considered in this thesis.

However, there are other ways to realize a low-scale Seesaw mechanism: Linear Seesaw [112,113,114] and Extended Seesaw [119,115,116]. Lastly, here we discussed only the Type-I Seesaw mechanism, where the Weinberg operator is UV completed by a new fermion which transforms as a singlet under $SU(2)_L$. Two other types of Seesaw mechanisms exist, exhausting all possibilities in which the Weinberg operator could be UV completed. Type-II Seesaw introduces a scalar that transforms as a triplet under $SU(2)_L$ [120], while Type-III Seesaw requires a fermion transforming as a triplet of $SU(2)_L$ [102].

## Heavy or light?

From a phenomenological or even experimental point of view, an interesting distinction is between heavy and light sterile neutrinos. Let's consider a typical accelerator neutrino experiment, where neutrinos are produced in decays of pions and kaons, and sterile neutrinos with a mass smaller than the kaon mass $m_K \simeq 494$ MeV so that they can be produced in conventional neutrino beams. While pions and kaons decay into pure flavor states, these flavor states contain a sterile neutrino component. Depending on the mass of the sterile neutrino with respect to the light neutrinos, two different regimes exist, as illustrated in fig. 3.2. Suppose the sterile neutrino is light enough so that the difference in velocity between the sterile neutrino and the active one is small enough. In that case, the wave packets will significantly overlap at the detection moment, producing a coherent oscillation pattern between at least four flavors. However, if the sterile neutrino is heavy, it will immediately lose coherence from the rest of the wave packet, turning into an additional component to the beam. Roughly speaking, we will talk about light sterile neutrinos when

$$\frac{pL}{c}\left(\frac{1}{m_{\text{active}}} - \frac{1}{m_{\text{sterile}}}\right) << \sigma_z, \tag{3.6}$$

where $p$ is the momentum of the neutrino, $L$ is the distance traveled between production and detection, $c$ is the speed of light, and $\sigma_z$ is the wave packet size along the axis of propagation. This last quantity is difficult to compute as it requires taking into account the *quantum* uncertainty in the location of the production and detection of the neutrino [37,121,122].

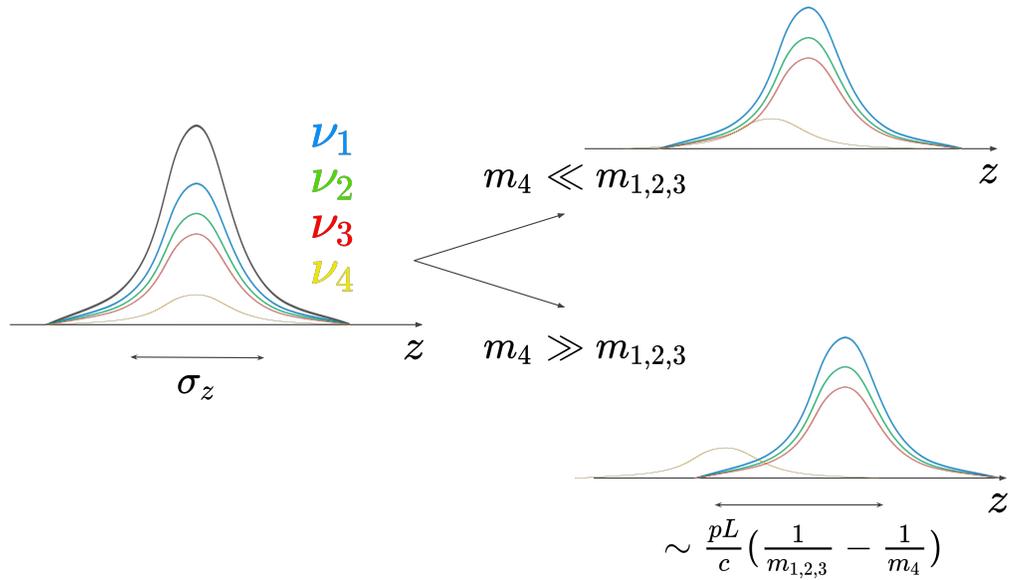

**Figure 3.2:** Sterile neutrinos can be classified into heavy and light in a phenomenological way. Light sterile neutrinos stay coherent with the rest of the wave packet on length scales much longer than the typical accelerator-based neutrino experiments. Heavy sterile neutrinos decohere from the other mass states as soon as they are produced, effectively resulting in an additional component to the neutrino beam.

A sterile neutrino is often indicated as $\nu_4$ when is light, producing a new oscillation frequency $\Delta m_{41}^2$, and as $N$ when is heavy. Across this thesis the notation will be clarified case by case, and will be determined by the context.

The typical mass range of interest for a heavy sterile neutrino case is between $\sim 1\,\text{MeV}$ so that it can decay at least into an $e^+e^-$ pair and $\sim 500\,\text{MeV}$ so that it can be produced in conventional neutrino beams through kaon decays, while for the light case is around $\sim 1 - 10\,\text{MeV}$, which results in oscillations at short baseline of the order of the baselines of the experiments which observed anomalous results. In this thesis we do not discuss the intermediate regime of sterile neutrinos with KeV masses, because it does not provide an interesting solution to the short baseline anomaly puzzle. KeV sterile neutrinos are interesting as they could be a dark matter candidates. However, the minimal version of this model is mostly excluded by x-ray searches looking for their decay into photons.

Heavy sterile neutrinos in the MeV-GeV range, often called *Heavy neutral leptons* (HNL) or, simply, heavy neutrinos, are an interesting solution to some of the short baseline anomalies, which is also typically very predictive and thus testable with different experiments. This section discusses only the so-called minimal scenario, the ones in which there is only one (or one set of) heavy sterile neutrino(s). As discussed before, a more realistic model is likely to contain additional non-minimal ingredients. Different ingredients change the phenomenology in different ways, and this will be discussed in more detail in chapter 4 and chapter 5.

### Weaker-than-weak

Heavy neutrinos are produced and decay at the tree level through mixing with the active flavors. Figure 3.3 shows the Feynman diagrams for the heavy neutrinos' weaker-than-weak interactions: they undergo weak interactions as active neutrinos, further suppressed by small mixing values in the amplitude.

### Production and decay through mixing

Typical production of heavy neutrinos happens through pions and kaons decays, while decay can happen into multiple channels, through neutral current or charged current, depending on the heavy neutrino mass and flavor structure *i.e.* combination of mixing values with the active neutrinos. Figure 3.4 illustrates a mode of production through charged current and decay through neutral current for a heavy neutrino that mixes with the muon flavor. Heavy neutrinos mixing with muon flavor are extremely interesting as their production from meson decays is enhanced by the large branching ratios into muon neutrinos (for helicity suppression). Heavy neutrinos that mix primarily with the tau flavor are less constrained because they can only be produced in the decay of heavier mesons, like $D$ or $B$ mesons.

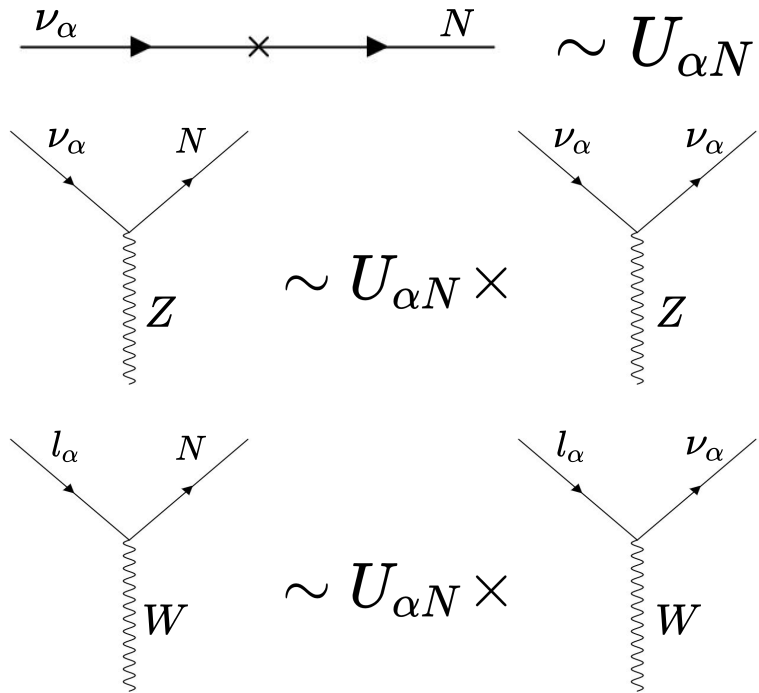

**Figure 3.3:** The tree level amplitudes for a sterile neutrino $N$ undergoing neutral or charged current interaction can be obtained by multiplying the corresponding amplitude with active neutrino $\alpha$ by the mixing factor $U_{\alpha N}$. These amplitudes can be at most $\sim 10^{-2}$, so these interactions are often called weaker-than-weak.

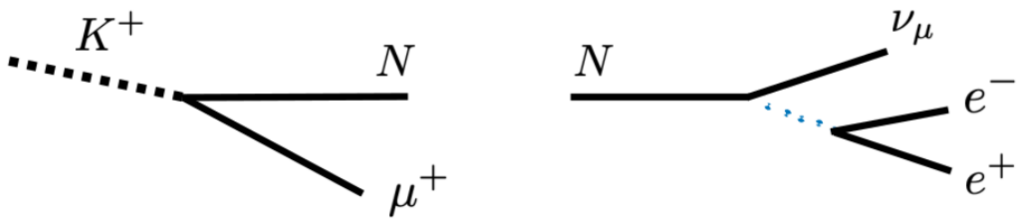

**Figure 3.4:** Heavy sterile neutrinos in the range $1 - 500\,\mathrm{MeV}$ are produced through the decay of charged mesons, like kaons, through charged current processes. They can decay into multiple final states, among which the $\nu e^+ e^-$ has a mass threshold of $m_N \gtrsim 1\,\mathrm{MeV}$, through neutral current, and through charged current, if $U_{\alpha N} \neq 0$.

Decay modes depend on the heavy neutrino mass and the mixing with the active flavor. Decay modes and branching ratios have been computed in the literature for arbitrary mixing values[123]. The total decay width increases with the lifetime, and, mode by mode, it is proportional to the value of the mixing square. The top left plot in fig. 3.5 shows the lifetime multiplied by the mixing square as a function of the mass for heavy neutrinos that mix primarily with the electron, muon, and tau flavor, respectively, *i.e.* assuming only one mixing value is different from zero. The lifetime curve is similar for the three different cases because most decay modes are in common: however, the case of $|U_{eN}| \neq 0$ shows a faster decrease because of the decay into $e\pi$, which is accessible before the decay into $\mu\pi$ or before the decay into $\tau\pi$ which would be out of the range of this plot. The significant kink between 0.1 and 0.2 GeV is related to the opening of the decay mode into $\nu\pi$. The lifetime curve spans values between $c\tau^0 \sim 10^9$ cm and $c\tau^0 \sim 10$ cm for mixing values of 1. If we consider more realistic values of the order of $|U|^2 \sim 10^{-8}$, we see that lifetimes are extremely long, making them survive for long distances if produced in a neutrino experiment. The other three plots show the branching ratios into the different channels as a function of the heavy neutrino mass for the three different cases of mixing with electron, muon, or tau only. Among the general feature of the branching ratio curves, the invisible decay mode into three neutrinos is always allowed. At larger masses, other decay channels become available, like $\nu e^+ e^-$, $e\pi$, $e\mu$, $\mu\pi$, etc. Combining searches into multiple decay modes is powerful enough to set strong constraints in the whole mass range. However, even a single decay mode can set powerful constraints, as shown in part II.

While the case of a single mixing larger than zero might seem artificial, constraining one mixing at a time is a powerful way to constrain this model, as specific experiments are much more sensitive to one mixing value than to the others. And more generally, unless it is protected by a specific symmetry, we do not expect any of the mixing values to be exactly zero. Thus, ruling out the parameter

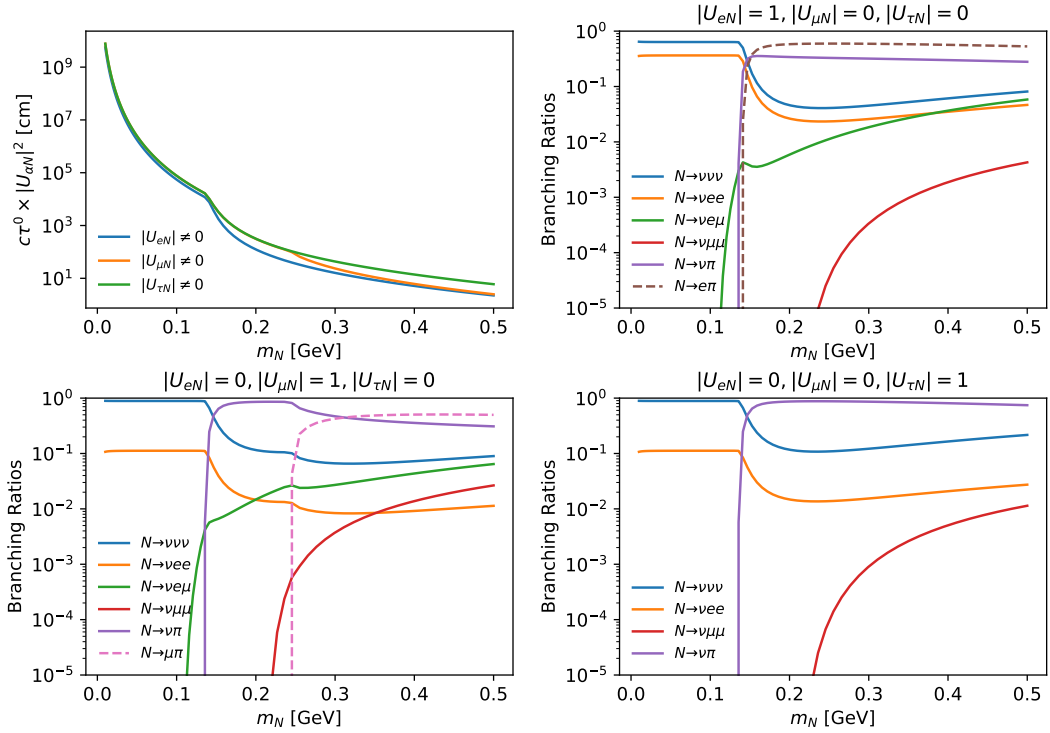

**Figure 3.5:** Top left: Lifetimes times mixing square as a function of the heavy neutrino mass, for heavy neutrinos mixing exclusively with the electron, muon, and tau flavor. This product is independent of the mixing value because the lifetime is proportional to one over the mixing square. Cases of interest require much smaller mixing values, making the lifetime in a realistic model much longer than what is displayed on the vertical axis. Top right and bottom row: Branching ratios for a heavy sterile neutrino that mixes exclusively with the electron, muon, and tau flavor as a function of its mass. The curves are quite similar and are modified by the channels with electrons or muons in the final state, which can proceed through neutral or charged current depending on the flavor structure.

space for one specific mixing constrains the model significantly.

## 3.3 Phenomenology of Light Sterile Neutrinos

Light sterile neutrinos are sterile neutrinos that induce oscillations in experiments with baselines of the order of $L/E \sim 1\,\mathrm{km/GeV}$. Despite their decay channel into three neutrinos, their lifetime is too long to be of any interest, and the only way to test this model is through oscillations.

Although this model is interesting for explaining short baseline anomalies, it is severely con-strained by cosmology [100], particularly by the measurement of $N_{eff} \sim 3$. A light sterile neutrino would be produced in the early universe, have decoupled from the plasma, and become a light relic, which would have affected the expansion rate of the universe. Variations of this model that avoid this problem are possible, although not very favorable.

Lastly, a seesaw mechanism that generates neutrino masses with eV sterile neutrinos is hard to realize. It is possible to find a fit to the standard seesaw, resulting in Yukawa couplings of the order of $10^{-12}$ [124]. Alternatively, it can be realized as an accident of a full seesaw mechanism, in which the structure of the mass matrices results in cancellations between different mass terms [125].

## Fast oscillations

Essentially, light sterile neutrinos induce neutrino oscillations at a much faster frequency than standard neutrino oscillations. Assuming small values of the mixings between active and sterile, as constrained by the unitarity of the PMNS matrix, at first order, the observable effect is neutrino disappearance, while, at second order, we have the appearance of active neutrinos. The survival and transition probabilities at short baseline (SBL) can be written as

$$P_{\nu_\alpha \to \nu_\alpha}^{SBL} \simeq 1 - \sin^2 2\theta_{\alpha\alpha} \sin^2 \Delta_{41}, \qquad P_{\nu_\alpha \to \nu_\beta}^{SBL} \simeq \sin^2 2\theta_{\alpha\beta} \sin^2 \Delta_{41} \quad (\alpha \neq \beta), \qquad (3.7)$$

where

$$\sin^2 2\theta_{\alpha\alpha} = 4|U_{\alpha 4}|^2(1 - |U_{\alpha 4}|^2), \quad \sin^2 2\theta_{\alpha\beta} = 4|U_{\alpha 4}|^2|U_{\beta 4}|^2 \ (\alpha \neq \beta), \quad \Delta_{41} = \frac{(m_4^2 - m_1^2)L}{4E},$$
$$(3.8)$$

where $\alpha$ and $\beta$ are active flavors, $U$ is the PMNS matrix, $L$ is the distance the neutrino traveled, $E$ is the neutrino energy, and $\Delta_{41}$ contains the difference in the squares of the masses between the sterile and the active ones. While we indicate this last quantity with $\Delta_{41}$, it does not matter which of the active mass states we consider, because $m_4 \gg m_{1,2,3}$ and, therefore, $(m_4^2 - m_1^2) \simeq m_4^2$. In principle, neutrinos could be more massive than their mass differences and almost degenerate, generating standard oscillations through this special combination of mass values. This pretty fine-tuned possibility is discussed, together with other options, in this complete review of the topic[126]. Notice that these amplitudes do not distinguish neutrinos and antineutrinos, as, in this approximation, this is a two-neutrino oscillation problem. To be compatible with standard neutrino oscillations measurements, the amplitude of these oscillations must be smaller than a few percent, as constrained by measurements of the neutral current rate at long baselines. In the long baseline regime, the oscillations induced by light sterile neutrinos would average out to the constant value $\sin^2 2\theta_{\alpha\alpha}$.

Light sterile neutrinos would induce electron neutrino disappearance, which could be responsible for radio-chemical anomalies, like the ones observed by GALLEX, SAGE, and BEST, while anti-electron neutrino disappearance might be related to some of the reactor antineutrino anomalies, although this option has been disfavored recently. Light sterile neutrinos would induce electron neutrino appearance, which might explain the LSND and MiniBooNE observations, as observed in the most recent MiniBooNE analysis and illustrated in fig. 3.6.

However, when analyzed through a global fit, this model is in strong tension with the *null results* discussed in section 2.4. Figure 3.7 shows the best-fit regions to the light sterile neutrino model for appearance and disappearance separately. The left plot shows the best-fit regions by several experiments searching for $\nu_\mu \to \nu_e$ appearance, at 99% CL. The best-fit regions overlaps and are therefore combinable in the overall best-fit region (red shape). However, this same region does not overlap at all with the best-fit region from experiments searching for $\nu_\mu$ disappearance (blue line in the right plot) at 99.73% CL, creating strong tension between different experimental results when interpreted under the light sterile neutrino model. All the details and perils of this analysis are discuss in comprehensive reviews of the topic[43,127,90].

This strong tension fueled the study of extensions of the *vanilla* light sterile neutrino scenario.

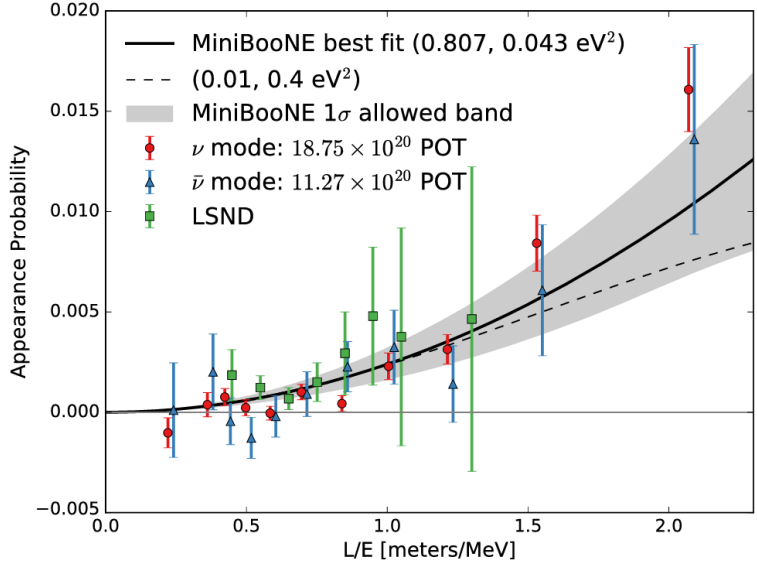

**Figure 3.6:** The left plot shows the appearance probability of $\nu_e$, in a beam primarily made of $\nu_\mu$, such as the ones employed by the LSND and MiniBooNE experiments, as a function of $L/E$ [MeV/m]. Each datapoint represents a bin from the LSND and MiniBooNE analyses, where $L$ is estimated as the average distance neutrinos travel along the beamline and $E$ from the reconstructed energy. The appearance probability is estimated by looking at the excess of events in a given energy bin compared to the $\nu_\mu$ beam content in the same energy bin. The datapoints show a clear increase with $L/E$, although they do not necessarily lie perfectly on the best fit line. The uncertainty of the fit is significant, as it can accommodate values of $\sin^2 2\vartheta_{e\mu}$, $\Delta m^2_{41}$ that differ even by a factor of 10 from the best fit values.

These extensions include an arbitrary number of sterile states[128,129], non-standard interactions which induce additional matter effects[130,131,132,133] or additional[†] decay modes that makes the sterile neutrino decay over distances typical of SBL experiments, to either visible[90,134,135] or invisible[136,137,138,139] final states. Moreover, it has been found that considering the finite wave-packet size might relax the significance of the light sterile neutrino explanation of some of the anomalies[140]. However, none of these alternatives seems to fully satisfy the experimental puzzle. Moreover, some of these alternatives, especially the ones including additional decay modes, require proper models that generate which are not always easy to build.

---

[†]Sterile neutrinos would decay into active neutrinos, but the rate would be too small to affect SBL experiments.

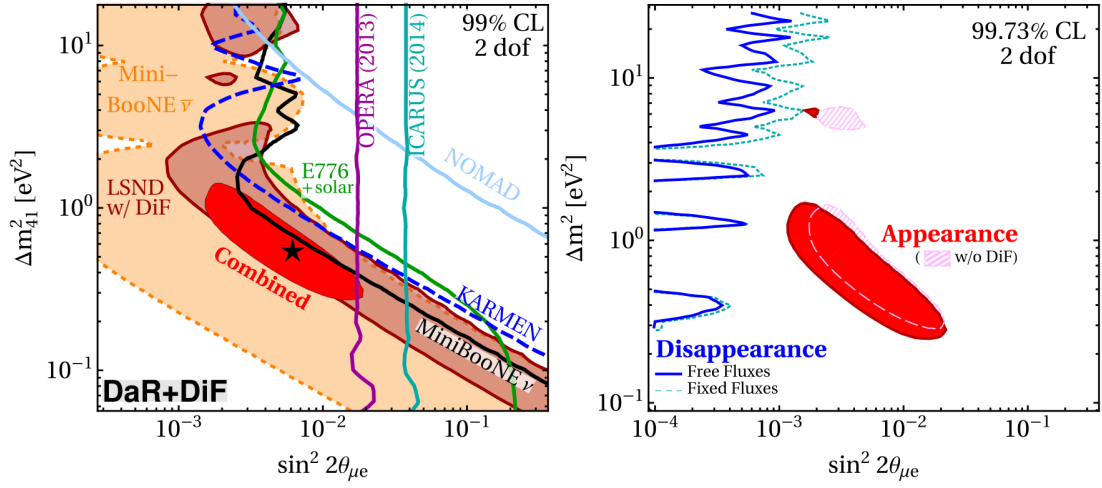

**Figure 3.7:** Best-fit regions on the light sterile neutrino parameter space defined by $\Delta m^2_{41}$ (y-axis) and $\sin^2 2\theta_{\mu e}$ (x-axis). All the experiments looking for $\nu_\mu \rightarrow \nu_e$ appearance seems to have overlapping best-fit regions, and can therefore be combined to obtain the global best-fit region (red region, 99% CL) (left). This region is, however, incompatible with the global best-fit region from experiments looking for disappearance (blue), showing no overlap between the two at 99.73% CL. Both figures are taken from [127].

In summary, while the light sterile neutrino model seems hard to fit in the full panorama, it is the most natural interpretation of the SBL anomalies because it would induce SBL oscillations. For this reason, part III is dedicated to the investigation of a proxy of this model using MicroBooNE data, especially targeting the regime that would best explain the MiniBooNE anomaly.

If anything, this preamble teaches us that the puzzles of neutrino masses and short-baseline anomalies are complex, and, very likely, the physics that explains them is complex too. However, to learn about the right complex solution, it is a good idea to start testing simpler models combining different experimental datasets and theoretical insights.

# Part II

# Heavy Sterile Neutrino Explanations

# 4

# Minimal and non-minimal models

The small nonzero neutrino masses challenge the conservation laws and particle content of the Standard Model (SM). As introduced in chapter 3, the existence of right-handed neutrinos $N_R$, singlets under the SM gauge symmetries, is an appealing solution to this puzzle. After diagonalization of the mass matrix, a heavy sterile neutrino $N$ would be present in the spectrum. It would interact via the weak force suppressed by a small mixing element with SM neu-trinos, the so-called *weaker-than-weak* interactions discussed in section 3.2. In this work, we build up on previous research [141,142,143,144,145], showing that this mixing is strongly constrained in the region between 10 MeV and $m_K \simeq 494$ MeV, thanks to laboratory-based searches, which provide upper bounds, and cosmological limits, which constrain the lifetime of $N$ to be $\tau_N < \mathcal{O}(0.1)$ sec, and therefore give a lower bound. However, the physics of heavy sterile neutrinos might be more complex, with additional interactions contributing to their production or decays. Additional forces can significantly modify their decay widths even for couplings that would be otherwise very difficult to probe experimentally. Of particular interest are scenarios wherein $N$ is shorter-lived than $\tau_N \lesssim 0.1$ s, to escape cosmological limits but still sufficiently long-lived that it could survive $c\tau_N \gtrsim 100$ m, the typical distance from production to detection at beam-dump experiments, also called accelerator-neutrino experiments. These scenarios are most effectively constrained with these experiments, where $N$ could be copiously produced in meson decays and observed through its decay products inside large-volume detectors typically used for neutrino detection.

In this work, we consider decay-in-flight (DIF) searches at hodoscopic neutrino detectors for $N \rightarrow \nu e^+ e^-$ and derive new bounds on the mixing between $N$ and muon-neutrinos. Hodoscopic — from the Greek *hodos* meaning path and *scopos*, observer — describes detectors that precisely reconstruct, track, and identify charged particles. This capability is essential for low-background searches for heavy neutrinos and other long-lived particles. We consider three detectors: the T2K near detector ND280 [146], MicroBooNE [147], and PS191 [148,149]. We revisit constraints from PS191, thought to be the strongest, showing that they have been significantly overestimated. We then extend the DIF search at T2K to heavy neutrinos lighter than the pion, showing that T2K data provides the leading lab-based constraints in that region of the minimal model. We show that these limits are enough to rule out all parameter space in this mass region for heavy neutrinos that mix predominantly with the muon flavor. These limits are then re-interpreted under three new scenarios with additional interactions between $N$ and the SM: a transition magnetic moment (TMM), a four-fermion leptonic interaction, and a leptophilic axion-like-particle ($\ell$ALP) portal. While the minimal model is too strongly constrained, these non-minimal models are exciting as they provide interpretations of the excess of electron-like events in the MiniBooNE experiment[55,51]. By re-interpreting the previous analyses under these model variations, we obtain some of the strongest limits, ruling out significant portions of the interesting parameter space to explain the MiniBooNE excess. The code to obtain our limits and simulate heavy neutrino decays is open source, and can be found on GitHub.[*]

## 4.1 From minimal to non-minimal

In this first section, we discuss the models considered in this analysis. The description of the minimal model builds on the basic concepts of heavy sterile neutrinos illustrated in chapter 3. The additional models we consider are built to provide extensions of the theoretically well-motivated models in an effective field theory approach or by considering axion-like particles. The interesting parameter space for these models is defined by where they can explain the MiniBooNE anomaly.

### Minimal model

The minimal model with a single heavy neutrino is defined by the low-energy Lagrangian describing its *weaker-than-weak* interactions

$$-\mathscr{L}_{\text{int}} \supset \frac{g}{2c_W} U_{\alpha N}^* \overline{\nu_\alpha} \not{Z} P_L N + \frac{g}{\sqrt{2}} U_{\alpha N}^* \overline{\ell}_\alpha \not{W} P_L N + \text{ h.c.}, \tag{4.1}$$

where $N$ is the heavy mass eigenstate, which may be a Majorana or (pseudo-)Dirac particle, while $\alpha$ denotes any of the three SM flavors. Although mixing with all three SM flavors is expected, we focus on dominant mixing with the muon-neutrinos, $|U_{eN}|, |U_{\tau N}| \ll |U_{\mu N}|$. Our conclusions will be analogous in the case of dominant $|U_{eN}|$. The case of dominant $|U_{\tau N}|$ is constrained by DIF searches at high-energy experiments, such as CHARM[150] and NOMAD[151], see also[152]. This work focuses on the decay $N \rightarrow \nu_\mu e^+ e^-$ that proceeds via the neutral-current (NC). We assume $N$ as a Dirac particle, as it provides a more interesting theoretical model in the low-scale inverse seesaw scenario.

### Heavy neutrinos and Big Bang Nucleosynthesis

$N$ lifetimes in the minimal model are very long, as illustrated in fig. 3.5, and, for the channel under consideration $N \rightarrow \nu e^+ e^-$ can be expressed as $\tau^0 \sim 1 \sec \times (10^{-6}/|U_{\mu N}|^2)(100 \text{ MeV}/m_N)^5$. This observation has significant consequences for cosmology, which thus result in strong upper bounds. In the early Universe, $N$ will be thermally produced and, if it survives to the onset of Big Bang Nucleosynthesis (BBN), will impact the abundance of light elements[153,154,155,156]. This happens in two ways: $N$ and its decay products upset the neutron-to-proton ratio, especially if $N$ can decay hadronically, in which case the known Helium abundance requires $\tau^0 < 0.023 \sec$[157]. Moreover, its electromagnetic decay products heat the plasma, changing the baryon-to-photon ratio and impacting the deuterium abundance. Throughout this work, we use the detailed limits found in[158], neglecting effects from modified branching ratios.

### Non-minimal models

Additional contributions to the heavy neutrino decay rate could make it decay before BBN, avoiding cosmological constraints and opening up some unconstrained parameter space. We consider enhancements to the dilepton channel $N \rightarrow \nu e^+ e^-$ from low-energy operators at dimensions five and six, as well as from a low-energy extension with a light axion-like particle.

## Transition magnetic moment

We start with the dimension-5, TMM operator

$$-\mathcal{L}_{\text{int}} \supset \frac{\mu_{\text{tr}}}{2} \overline{\nu_\alpha} \sigma^{\mu\nu} N F_{\mu\nu} + \text{h.c.}. \tag{4.2}$$

We set again the flavor index $\alpha = \mu$ motivated by phenomenological applications [159,160,161,162,163,164,135]. If $|\mu_{\text{tr}}| \gg (G_F m_N / 2\sqrt{3}\pi)$, $N$ predominantly decays electromagnetically. The left plot of fig. 4.1 shows the diagram for the process that enhances the width of $N \to \nu e^+ e^-$, while the plot on the right shows the branching ratios as a function of $m_N$, for a choice of parameters at the edge of the constraints from supernovae. The primary decay channel is into a real photon $N \to \nu_\mu \gamma$, which can be observed in high-density detectors, where the photon converts into an observable $e^+ e^-$ pair.

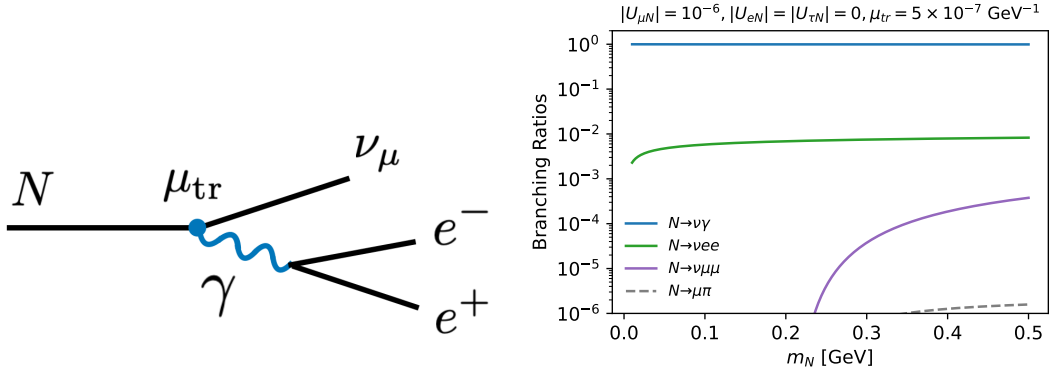

**Figure 4.1:** If the heavy neutrino possesses a transition magnetic moment, heavy neutrinos can decay electromagnetically into photons $N \to \nu\gamma$. While this is the dominant decay channel, the additional interaction would enhance the decay rate $N \to \nu e^+ e^-$ through the diagram containing a virtual photon (left). For a choice of parameter space at the edge of the bounds from supernovae, the primary decay rates predicted by this model are electromagnetic, into a single photon (blue), $e^+ e^-$ (green), and $\mu^+ \mu^-$ (purple) (right plot).

However, low-density detectors can measure the smaller rate to virtual photon, $\mathcal{B}(N \to \nu\gamma^* \to \nu e^+ e^-) \sim 0.7\%$ at $m_N \sim 250$ MeV, benefiting from small neutrino-interaction backgrounds.

The operator in eq. (4.2) is not invariant under $SU(2)_L$, and we discuss its possible UV completion. Depending on the underlying model, this may bear consequences for the masses and the mixing of neutral leptons. If the completion of eq. (4.2) contains charged particles that couple only to heavy neutrinos, then $\mu_{\mathrm{tr}}$ can only be generated via mixing. In particular, a pseudo-Dirac pair of fermions $N_{L,R}$ with a large magnetic moment $(\mu_N/2)\overline{N_L}\sigma^{\mu\nu}N_R F_{\mu\nu}$ will generate a light-heavy transition moment in the mass basis, $\mu_{\mathrm{tr}} \sim U^*_{\alpha N}\mu_N$. Heavy neutrino decay rates are suppressed by mixing in this case. On the other hand, if the new charged particle content couples to light neutrinos, it may directly generate $\mu_{\mathrm{tr}}$. To avoid a relation between eq. (4.2) and the Dirac mass term $m_D\overline{\nu_L}N_R$ (and therefore to $U_{\alpha N}$), one may borrow several results from the literature on light Dirac neutrinos with large magnetic moments [165,166,167,168,169]. Following Voloshin's mechanism [165], for instance, if the completion of eq. (4.2) respects some approximate SU(2) symmetry under which $(\nu_\mu, N^c_R)^T$ transforms as a doublet, then the dim-5 operator could be significant, and the mixing would be protected by the symmetry. We assume this case, rendering the lifetime of $N$ from eq. (4.2) independent of $|U_{\mu N}|^2$.

The TMM operator also generates a corresponding magnetic moment for $\nu_\mu$ due to the mixing $U_{\mu N}$. The $\nu_\mu - \nu_\mu$ magnetic moment is of the order of

$$|\mu_\nu| = |\mu_{tr} U_{\mu N}| \sim 3 \times 10^{-11}\mu_B \times \left(\frac{|U_{\mu N}|}{10^{-2}}\right)\left(\frac{|\mu_{\mathrm{tr}}|}{1\,\mathrm{PeV}^{-1}}\right) \tag{4.3}$$

where $\mu_B$ is the Bohr magneton. This value is within the range of the XENONnT results [170], and therefore provides an upper limit on the mixing parameter in our plots.

Four-fermion interaction  At dimension six, we consider a vectorial four-lepton interaction

$$-\mathscr{L}_{\mathrm{int}} \supset \frac{G_X}{\sqrt{2}}\left(\overline{N}\gamma^\mu N\right)\left(\overline{\ell_\beta}\gamma_\mu \ell_\beta\right) + \mathrm{h.c.}, \tag{4.4}$$

where $\beta \in \{e, \mu, \tau\}$. For $G_F/G_X \ll 1$, heavy neutrinos decay primarily via $N \to \nu\ell^+_\beta\ell^-_\beta$ through mixing with light neutrinos. This decay mode is analogous to the weak-decay case: the Feynman diagram is displayed on the left plot of fig. 4.2. The plot on the right shows the branching ratios as a function of $m_N$ for an interesting choice of parameters. The decay modes into electrons and muons (above the threshold $m_N \gtrsim 2m_\mu$) completely dominate the rate.

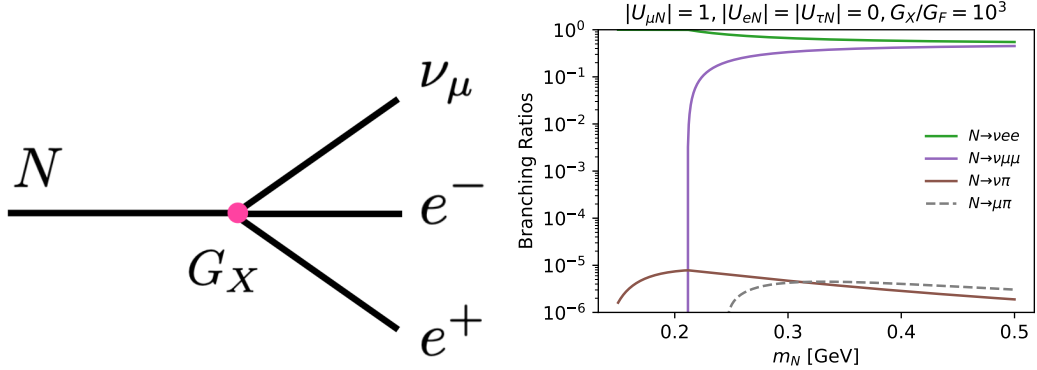

**Figure 4.2:** An effective four-fermion interaction enhances the decay rate into $e^+e^-$ (left plot) and $\mu^+\mu^-$, overcoming all other decay modes completely (right plot). Contrary to the TMM model, it requires mixing a heavy neutrino into an active one, making all amplitudes proportional to the mixing.

For large $G_X/G_F$, the effective operator in eq. (4.4) is only valid up to scales around the mediator mass. Therefore, constraints at high energies, for instance, from the branching ratio of $Z \rightarrow NN(e^+e^-)$ [171], do not apply. Ultraviolet completions of eq. (4.4) include gauge extensions of the Standard Model, such as $U(1)_{B-L}$ and dark $U(1)_X$ gauge symmetries [172,173]. In the latter, the dark photon $A'$ couples to dark leptons, $g_X \overline{N_D} \slashed{A}' N_D$, and to charged SM particles via kinetic mixing with the photon, $(\varepsilon/2)F_{\mu\nu}X^{\mu\nu}$. Thus eq. (4.4) is independent of flavor, resulting in

$$\frac{G_X}{\sqrt{2}} \sim \frac{g_X e \varepsilon}{m_{A'}^2}, \tag{4.5}$$

where $g_X$ is the gauge coupling. As a result, the amplitude for $N \rightarrow \nu e^+ e^-$ is proportional to $G_X U_{\mu N}$, shortening the heavy neutrino lifetime by a factor $\kappa \sim (G_F/G_X)^2$ with respect to the minimal model. Model-independent limits on kinetic mixing constrain $\kappa \gtrsim 10^{-4}$ and require the mediator to be relatively light, around the GeV scale. Above the dimuon threshold, $N \rightarrow \nu \mu^+ \mu^-$ is allowed and would further strengthen our constraints. In addition, for theoretical consistency, such hidden sectors typically contain several heavy neutrinos, which may also be produced in the decay of their heavier partners. Even faster decays like $N \rightarrow N'\ell^+\ell^-$ could dominate since the rate would be independent of $|U_{\mu N}|^2$. We do not comment further on this possibility and assume that decay rates mediated by eq. (4.4) are always proportional to $|U_{\mu N}|^2$.

Finally, light neutrinos also interact via the dim-6 operator above due to mixing. There are two types of processes to consider: i) $\nu_\mu e \rightarrow Ne$, and ii) $\nu_\mu e \rightarrow \nu_\mu e$. Process i) is kinematically forbidden for neutrino energies below $E_\nu^{\text{th}} = m_N + m_N^2/2m_e$, which at the smallest masses we consider, $m_N = 20$ MeV, takes values of $E \sim 420$ MeV. This process would create a single electron shower inside experiments like MINER$\nu$A [174,175] and CHARM-II [176], which have measured the SM rate for neutrino-electron scattering at a precision of $\mathcal{O}(10\%)$ and $\mathcal{O}(3\%)$, respectively. Since signal i) comes from an inelastic scattering, the electron would be less forward, and the signal efficiency due to stringent experimental cuts on $E_e \theta_e^2$ would be reduced. Nevertheless, requiring that the rate for process i) be less than 10% of the weak rate and neglecting the effects of $m_N$, we find $|U_{\mu N}|G_X/G_F \lesssim 0.1$, which for our benchmark of $G_X/G_F = 10^3$ gives $|U_{\mu N}| < 10^{-4}$. For process ii), the scattering on electrons is elastic, and therefore there is no threshold. Given that the new operator interferes with the SM amplitude, a naive scaling provides a limit of $|U_{\mu N}|^2 < 10^{-4}$ for our benchmark, again requiring the interference term to be below 10% of the SM cross section.

LIGHT MEDIATORS    We now consider a low-energy extension of the SM where heavy neutrinos can decay to a light mediator, which in turn decays to $e^+e^-$. A straightforward example was already alluded to in the discussion above, where a dark photon mediator $X$ was proposed as a completion of the four-fermion interaction. While we focused on the case of $m_X > m_N$, justifying the effective operator approach, it may very well be possible that $X$ is lighter than $N$ so that it can be produced in two-body decays of the heavy neutrinos, $N \rightarrow \nu X$, to subsequently decay promptly into $e^+e^-$ via $X \rightarrow e^+e^-$. While a dark photon is attractive from a model-building perspective, it is certainly not the only one.

In Ref. [177], the authors proposed an extension of the SM with a leptophilic axion-like particle ($\ell$ALP). The simplified Lagrangian used was given by,

$$-\mathcal{L} \supset \frac{\partial_\mu a}{2f_a} \left( c_N \overline{N} \gamma^\mu \gamma^5 N + c_e \overline{e} \gamma^\mu \gamma^5 e \right) \tag{4.6}$$

where $f_a$ is the axion decay constant. The mixing of $N$ with active-neutrinos is then responsible for the decay of $N \to \nu a$, which overwhelms the branching ratios of $N$ for the parameter space of interest. The ALP decays promptly into $e^+ e^-$ via the leptonic coupling, as shown in the plot on the left of fig. 4.3. The plot on the right shows the branching ratios as a function of $m_N$, emphasizing how the $e^+ e^-$ final state dominates all the decay modes. There are also loop-induced decays with $\mathcal{B}(a \to \gamma\gamma) \lesssim 10\%$. We account for these loops without considering them as part of our signal definition, making our result a conservative estimate. The interactions in Eq. (4.6) are again not gauge-invariant and require a UV completion. As discussed in more detail in Ref. [177], the ALP $a$ can be identified with the pseudo-Goldstone boson of a global axial symmetry $U(1)_A$, under which $N$ is charged. It is also assumed that $a$ couples most strongly with electrons. While this assumption is not strictly necessary for our current study, it can be satisfied if the $U(1)_A$ symmetry has some non-trivial flavor structure, for example, in Froggatt-Nielsen models [178].

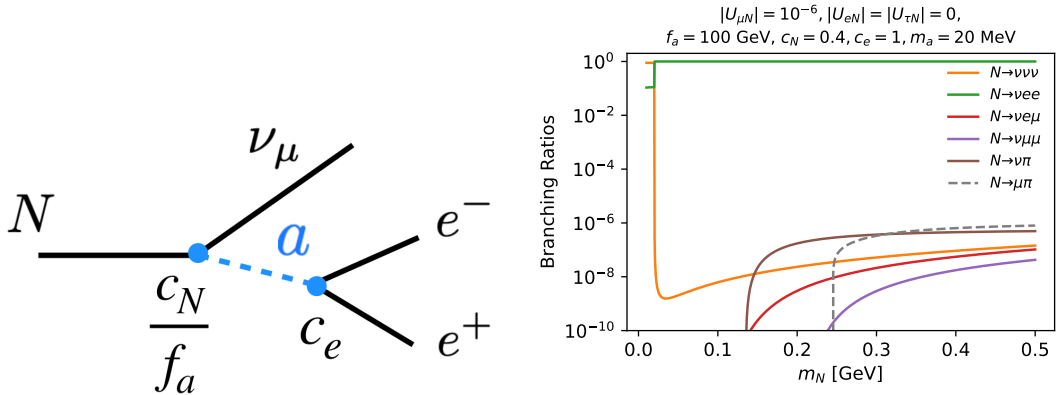

**Figure 4.3:** In the extension of the heavy neutrino model with an ALP-like particle, the decay rate into the $e^+ e^-$ final state is enhanced by the diagram mediated by the ALP-like particle (left plot). As soon as the ALP-like particle is produced on-shell, the branching ratio into $e^+ e^-$ dominates all the other decay channels (right plot).

Finally, as before, the new interactions also mediate neutrino-electron scattering. For $E_\nu \gg m_a, m_N$, the inelastic scattering cross section decreases with energy,

$$\sigma_{\nu_\mu e \to Ne}(E_\nu \gg E_\nu^{\text{th}}) \sim \frac{|U_{\mu N}|^2 |c_e c_N|^2 m_N^2 m_e^2}{256 \pi f_a^4} \frac{1}{2 E_\nu m_e} \tag{4.7}$$
$$\sim 8 \times 10^{-49} \, \text{cm}^2 \left(\frac{1 \, \text{GeV}}{E_\nu}\right) \left(\frac{|U_{\mu N}|^2}{10^{-2}}\right) \left(\frac{100 \, \text{GeV}}{f_a}\right)^4,$$

for our benchmark of $m_N = 20 \, \text{MeV}$, $c_N = 0.4$ and $c_e = 1$, so safely below weak-interaction cross sections even for such large values of mixing. For $m_N$ values above $\mathcal{O}(100) \, \text{MeV}$, the threshold becomes too large for $N$ to be produced in accelerator neutrino experiments. Elastic cross sections vanish within the limit of massless neutrinos.

### Comparing the different models

Although all models illustrated previously predict heavy neutrinos that decay into $e^+ e^-$, the experimental signatures may differ because of different differential decay rates. However, what matters the most in our analysis is that the selection efficiency does not differ significantly between different models. While computing efficiencies as a function of experimentally relevant variables is not easy, we will focus on showing that experimentally relevant variables that impact the efficiency curves are similar among the models under consideration.

For the minimal model, while we discussed lifetimes, decay channels, and branching ratios in chapter 3, here we show the differential rate to explicitly clarify the differences between Dirac and Majorana and decays occurring purely via CC, NC, or both. These points are particularly relevant because previous limits, like the ones from PS191, are easily misinterpreted because of their assump-tions about the nature of the heavy neutrino and the decay process. Neglecting the final lepton masses, we find

$$\frac{\mathrm{d}\Gamma^{\mathrm{Dir}}}{\mathrm{d}E_+\mathrm{d}E_-} = |U_{\alpha N}|^2 \frac{G_F^2 m_N}{2\pi^3} \left[ g_R^2 E_-(m_N - 2E_-) + (1 - g_L)^2 E_+(m_N - 2E_+) \right], \quad (4.8)$$

$$\frac{\mathrm{d}\Gamma^{\mathrm{Maj}}}{\mathrm{d}E_+\mathrm{d}E_-} = |U_{\alpha N}|^2 \frac{G_F^2 M_N}{2\pi^3} \left[ (1 + g_L)^2 + g_R^2 \right] \left[ m_N(E_+ + E_-) - 2(E_+^2 + E_-^2) \right], \quad (4.9)$$

where $g_L = \sin^2(\theta_W) - 1/2$, $g_R = \sin^2 \theta_W$, and $E_+$ and $E_-$ are the positive and negative charged lepton energies in the center-of-mass frame, respectively. The expression holds for both left-handed and right-handed polarized heavy neutrinos. Note that the above rates in a purely CC decay mode can be recovered with $g_R, g_L \to 0$, and in a purely NC decay mode with $(1 + g_L) \to g_L$. The expression used by the PS191 collaboration in [148] agrees with our calculation in the CC-only limit. For a full discussion regarding the difference between Dirac and Majorana heavy neutrinos, see [179]; in the following, we will discuss and assume Dirac heavy neutrinos. In the Dirac case, the shape of the differential rate is sensitive to the interference between NC and CC. However, this effect is primarily irrelevant for variables of experimental interest. Figure 4.4 illustrates the distribution of a few kinematical variables for heavy neutrinos decaying inside of the T2K near detector, as an example of the experiments under consideration, comparing the cases of NC only, CC only, and the complete and correct amplitude. These experimentally relevant variables, namely the lowest energy of the $e^+e^-$ pair, its opening angle, and the distance between the trajectories bent by the magnetic field after 20 cm are insensitive to the diagram responsible for the decay.

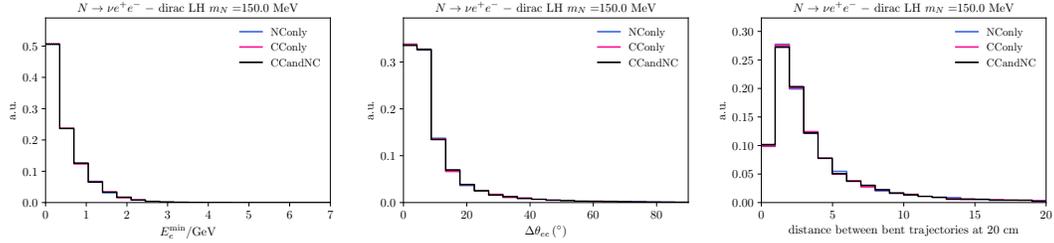

**Figure 4.4:** Kinematic distributions for different assumptions on the diagram leading to the decay: NC only (blue), CC only (pink), NC and CC with interference (black). The first is the decay of the mixing with the muon or tau flavor only, while the full amplitude applies to the case with the mixing with the electron flavor. Left: the energy of the lowest energy particle of the electron-positron pair. Center: the opening angle between the $e^+e^-$ pair. Right: the distance between $e^+e^-$ pair at $20$ cm from the decay point along the direction of the total $e^+e^-$ momentum. We consider the bent trajectories of pairs assuming a constant $B = 0.2\,\mathrm{T}$ magnetic field. Monte-Carlo errors are too small to be seen. The kinematics is mostly independent of the nature of the decay.

We performed a similar study comparing the three new physics models considered, in several experimental variables, as shown in fig. 4.5. The general comparison shows some mild differences that squeeze the distributions of the experimentally relevant variables towards smaller values. However, the bulk of the distribution lies above values where we expect efficiencies to plateau.

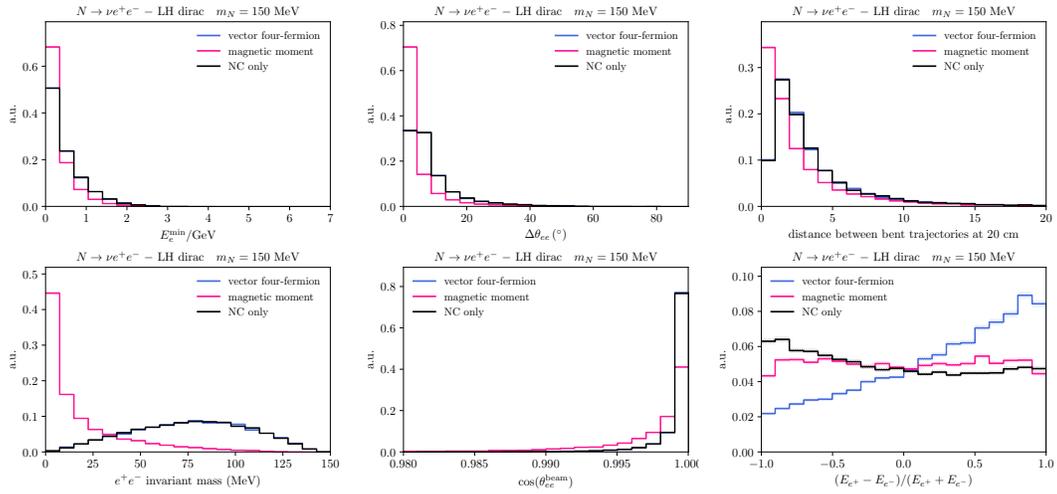

**Figure 4.5:** Decay kinematics in different models. In addition to the main variables also shown in fig. 4.4, we show the $e^+e^-$ invariant mass, the angle of the sum of the $e^+e^-$ momenta with respect to the neutrino beam, and the energy asymmetry of the pair.

We now provide a more detailed study of the differential decay rates for the TMM and four-fermion vector interactions.

In the TMM model, the main decay channel is $N \to \nu_\mu \gamma$, for which the decay rate of Dirac heavy neutrinos is

$$\Gamma_{N \to \nu \gamma}^{\text{Dir}} = \frac{|\mu_{\text{tr}}|^2 m_N^3}{16\pi}, \tag{4.10}$$

recalling that $\mu_{\text{tr}} = 2d$, where $d$ is the dipole parameter often used in part of the literature[162]. However, the decay rate into virtual photons into $e^+e^-$ is significant for our analysis. Due to the small electron mass, the branching ratio is enhanced by large logs and is larger than a naive factor of $\alpha/4\pi \times \mathcal{B}(N \to \nu\gamma)$. In fact, the differential rate will peak at the smallest lepton energies, as follows

$$\frac{\mathrm{d}\Gamma_{N\to\nu\gamma^*\to\nu\ell^+\ell^-}^{\text{Dir}}}{\mathrm{d}E_+\,\mathrm{d}E_-} = \frac{\alpha|\mu_{\text{tr}}|^2}{8\pi^2} \frac{m_N(E_+ + E_-) - 4E_+E_-}{(E_+ + E_-) - m_N/2}, \tag{4.11}$$

where we have taken the massless limit for the amplitude. In the massive case, the rate is regulated by the lepton mass to give

$$\Gamma_{N\to\nu\gamma^*\to\nu\ell^+\ell^-}^{\text{Dir}} = \frac{\alpha|\mu_{\text{tr}}|^2}{48\pi^2} m_N^3 L\left(\frac{m_\ell}{m_N}\right), \tag{4.12}$$

with

$$\begin{aligned}
L(r) &= \left(2 - \frac{r^6}{8}\right)\operatorname{sech}^{-1}(r) - \frac{24 - 10r^2 + r^4}{8}\sqrt{1 - 4r^2} \\
&\simeq 2\log\left(\frac{1}{r}\right) - 3 + \mathcal{O}(r^2),
\end{aligned} \tag{4.13}$$

where we show the leading-log approximation for small $r$ in the second line. As an example, for a heavy neutrino with $m_N = 100$ MeV and negligible mixing with active neutrinos, we get $\mathcal{B}(N \to \nu e^+e^-) = 0.68\%$. Above the dimuon threshold, we find $\mathcal{B}(N \to \nu e^+e^-) = 0.75\%$ and $\mathcal{B}(N \to \nu\mu^+\mu^-) = 3.8 \times 10^{-3}$ for $m_N = 300$ MeV. At large masses, vector meson dominance dictates that $N$ will decay to final $\rho$, $\omega$, and $\phi$ mesons, which produce primarily pions and kaons accompanied by a final state neutrino.

The rate peaks at the lowest dilepton invariant masses, implying that the signal will be very similar to real photons that convert inside the detector material. The dependence of the rate on an artificial cut on $m_{\ell\ell}^2 = (p_+ + p_-)^2$ can be understood analytically. At small values of $r_{ee} = m_{ee}^{\min}/m_N$, the rate will decrease as $\log(1/r_{ee})$, while for large $r_{ee}$, we expand in $a = 1 - r_{ee}^2$ to find a rate as in eq. (4.12) with the replacement

$$L(r) \to a^3 \sqrt{1 - 4r^2} \left( \frac{1}{2} + r^2 \right).$$ (4.14)

In the case of the four-fermion vector interactions, the decay rate into leptons is given by

$$\Gamma_{N \to \nu(Z')^* \to \nu e^+ e^-}^{\mathrm{Dir}} = \frac{G_X^2 M_N^5}{192\pi^3}.$$ (4.15)

The differential rate, in this case, differs from the weak-decays only due to the lack of axial-vector couplings,

$$\frac{\mathrm{d}\Gamma_{N \to \nu \ell^+ \ell^-}^{\mathrm{Dir}}}{\mathrm{d}E_+ \mathrm{d}E_-} = \frac{G_X^2 m_N}{32\pi^3} \left( m_N(E_+ + E_-) - 2(E_+^2 + E_-^2) \right).$$ (4.16)

It is easy to see that this is the limit of eq. (4.8) where CC is absent and $g_L = g_R$.

## 4.2 Laboratory-based searches for decays in flight

In this section, we discuss the methodology employed for setting limits on the models described in section 4.1 using accelerator-based experiments. We first discuss how to estimate the heavy neutrino flux and the limitations of this method. We then describe the experiments and datasets used in our analysis before moving to the new bounds.

### General methodology

Accelerator neutrino beams are obtained from the DIF of magnetically-focused mesons. If heavy neutrinos exist, they are part of the neutrino beam produced through mixing. The flux of heavy neutrinos in a given experiment can be estimated from the known neutrino flux per parent meson by rescaling it by [180,181]

$$\frac{\Gamma_{P \to N\ell}}{\Gamma_{P \to \nu\ell}} = \rho(r_\ell, r_N) = \frac{|U_{\ell 4}|^2(r_\ell + r_N - (r_\ell - r_N)^2)\sqrt{\lambda(1, r_\ell, r_N)}}{r_\ell(1 - r_\ell)^2}, \qquad (4.17)$$

where $P$ is the associated parent meson, $r_\ell = (m_\ell/m_M)^2$, $r_N = (m_N/m_P)^2$, and $\lambda$ is the Källén or triangle function:

$$\lambda(x, y, z) = x^2 + y^2 + z^2 - 2xy - 2yz - 2xz. \qquad (4.18)$$

The left plot of fig. 4.6 shows $\rho$ as a function of $r_N$, for pions and kaons, decaying to electron and muons. When the mass of the heavy neutrino is close to the parent mass, the rescaling can reach values of $10^4$. In this work, we include only the contribution from kaon decays, neglecting the additional production from pions and muons. Estimating contributions from pions and muons would require significant effort and a detailed experimental simulation, as the T2K analysis estimated efficiencies and limits using the flux from kaons only. This procedure yields conservative results. However, they are sufficient in the low heavy neutrino mass region where $m_N \lesssim (m_M - m_\ell)/2$. This procedure also automatically considers the effect of the magnetic focusing at the production point and other geometrical effects. For larger $m_N$ values, our approach underestimates the heavy neutrino flux and yields conservative results. Heavier heavy neutrinos are produced with lower transverse momentum with respect to parent particles than light neutrinos. Therefore, they are more collimated with the beam direction, increasing the angular acceptance of on-axis detectors to heavy neutrinos, particularly at low energies where light neutrinos would otherwise have a larger angular spread. This behaviour can be understood in the right plot of fig. 4.6, showing the neutrino flux from kaons at the T2K near detector (black), the official heavy neutrinos flux (dashed) for masses of 150 MeV (blue) and 250 MeV (red), compared to our rescaling (solid). Our rescaling works well at large energies, but undershoots the more accurate fluxes obtained by the T2K collaboration in the low energy region. The flux is more underestimated at larger masses than at lower masses. Sec-tion 4.4 discusses how underestimating the heavy neutrino flux reflects on the limits in the heavy neutrino parameter space. When considering new forces that shorten the lifetime while keeping $N$ long-lived enough to reach the detector, the probability for $N$ to decay to some final state $X$ inside the detector is independent of the total lifetime $(1/\Gamma)$, and proportional only to the partial decay width in the signal channel $(\Gamma_{N\to X})$

$$P_{N\to X} = e^{-L\Gamma/\beta\gamma}(1 - e^{-\ell_{\mathrm{det}}\Gamma/\beta\gamma}) \, \mathcal{B}(N\to X) \qquad (4.19)$$
$$\simeq \frac{\ell_{\mathrm{det}}}{\gamma\beta} \, \Gamma_{N\to X},$$

where $\ell_d$ is the detector's length, $L$ the distance between the production and decay points, $\gamma\beta = p_N/m_N$, and $\mathcal{B}$ denotes the branching ratio. Ultimately, in the long-lifetime and single-flavor-dominance limits, the new upper bound is given by $|U_{\alpha N}^{\mathrm{new}}|^2 = |U_{\alpha N}|^4 \times \widehat{\Gamma}_{N\to X}/\Gamma_{N\to X}^{\mathrm{new}}$, where $\widehat{\Gamma} = \Gamma/|U_{\alpha N}|^2$. The argument is analogous when relaxing the assumption of single-flavor domi-nance. If $\Gamma^{\text{new}}_{N \to X} \propto |U^{\text{new}}_{\alpha N}|^2$, the new upper-bound is proportional to the square root of the ratio of signal decay rates, while the new lower bound if it exists, will be linearly dependent on the ratio of the total rates. This statement is also approximately true for the lower bounds posed by cosmology.

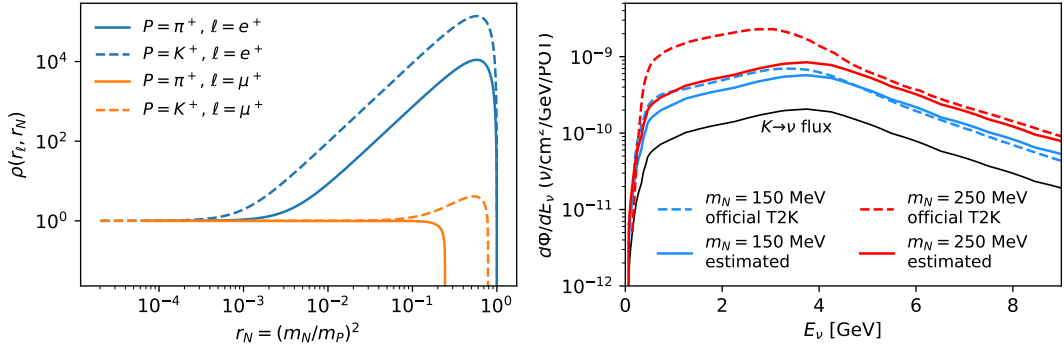

**Figure 4.6:** Left: the function $\rho$ is the multiplicative factor to approximately convert the neutrino flux into heavy neutrino flux. For small values of $r_N = (m_N/m_P)^2$, the kinematics of the decay into heavy neutrinos is equivalent to the one into active neutrinos, resulting in $\rho \simeq 1$. At larger masses, the heavy neutrino flux can be enhanced significantly with respect to the standard neutrino as it is not limited by helicity suppression. Right: Comparison between the official heavy neutrino flux employed by T2K (dashed) and our estimate (solid) for two different values of the heavy neutrino mass ($150\,\mathrm{MeV}$ in blue and $250\,\mathrm{MeV}$ in red), as a function of the beam energy, for particles originated from kaon decays. Our estimate is a simple rescaling of the standard neutrino flux from kaons (black), and it is a better approximation at lower values of the mass and higher energy. Our estimate always under-estimates the actual flux, resulting in more conservative bounds than achievable with a complete analysis.

## Angular distributions

In order to further understand the heavy neutrino flux at ND280, we performed some studies regarding the relationship between energy, angle, and heavy neutrino mass. While a full simulation of the heavy neutrino flux at ND280 is possible [182], it is pretty resource expensive and susceptible to mismodeling of the geometry of the experiment. The approach followed in this work by rescaling the neutrino flux by the function in eq. (4.17) turns out to be conservative and sufficient to support strong bounds on the heavy neutrino parameter space. These studies, however, emphasize how the flux is generated and some of the limitations of the previous method, which works exactly only in the case of an on-axis experiment, with very forwardly collimated beam. The kinematics of heavy neutrinos in the lab frame depends only on the decay angle in the center of mass because of the isotropic two-body decay of kaons and pions, which are scalar particles. Figure 4.7 shows the relationship between the angle in the lab frame and the angle in the center of mass frame, and between the energy in the lab frame and the angle in the center of mass frame, for different heavy neutrino masses and different kaon momentum. The ND280 off-axis detector selects a flux around $\theta_N^{lab} \sim 2°$, which implies a well-defined angle in the center of mass and thus well-defined energy in the lab frame for a given mass and kaon momentum. A conclusion is that, when fixing the parent momentum, light neutrinos and heavy neutrinos contribute to substantially different regions of the energy spectrum. This relationship is emphasized further in fig. 4.8, where we neglect the angle in the lab frame in order to show the relationship between angle and energy in the lab frame, losing information about the probability density function along the line, which depends on the derivatives of the functions shown in fig. 4.7. Figure 4.9 and fig. 4.10 illustrate the effect of the heavy neutrino mass more strikingly, for kaon and pion decays, respectively. They illustrate the same relationships as in fig. 4.9, but comparing different masses for the same parent momentum. Remarkably, when slicing the plot at a given angle, for example 2° for ND280, the same parents can contribute significantly to the flux at a certain mass, and negligibly at a different one. For pions, because the available energy in the center of mass is more limited, the difference between the lightest and heaviest possible heavy neutrinos is not as striking as in the kaon case.

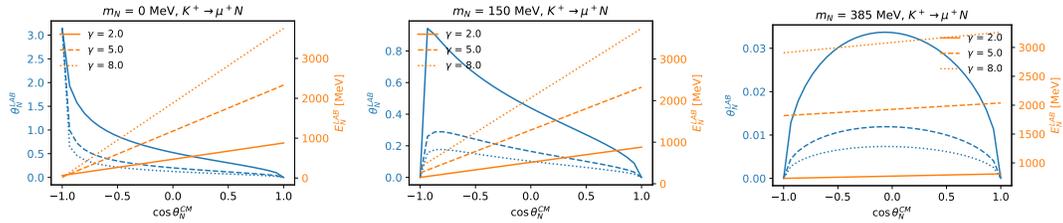

**Figure 4.7:** The kinematics of heavy neutrinos from kaon decays $K \to \mu N$ depends on its mass (three plots) and on the kaon momentum (different line styles for three $\gamma$ factors). The angle (blue) and energy (orange) in the lab frame are functions of the angle in the center of mass (x-axis), which is uniformly distributed because kaons are scalars. These plots emphasize how heavy neutrinos at different masses cover very different regions in energy and angle, making the experiment, ND280, which is off-axis, sensitive to a specific part of the phase space.

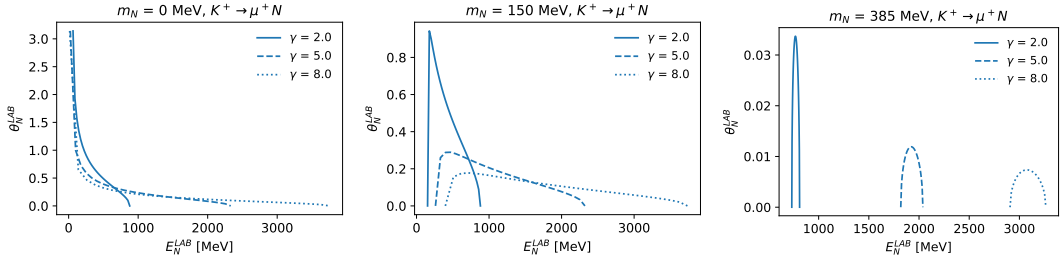

**Figure 4.8:** Analogous series of plots to fig. 4.7, this time showing directly the relationship between the angle in the lab-frame (y-axis) and the energy in the lab-frame (x-axis). These plots allow a direct understanding of what part of the energy spectrum is produced at what angles, but do not make explicit the distribution across the line, which is not uniform.

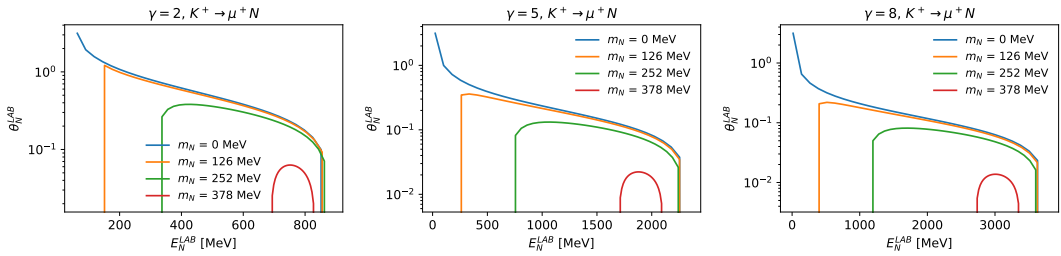

**Figure 4.9:** Similarly to fig. 4.8, these plots shows the relationship between the angle in the lab-frame (y-axis) and the energy in the lab-frame (x-axis) for different masses (different colors) and different kaon momenta (different plots).

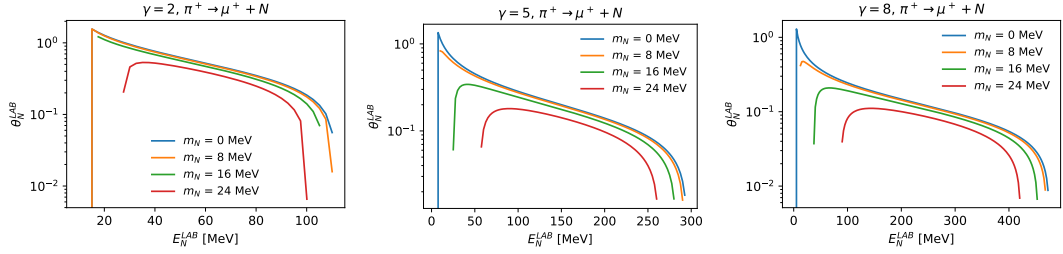

**Figure 4.10:** Analogous series of plots to fig. 4.9, but for heavy neutrinos resulting from pion decays $\pi \rightarrow \mu N$. In this case, the maximum heavy neutrino mass is about 29 MeV, and the difference between different masses is less pronounced than in the kaon case.

### T2K ND280

The T2K collaboration searched for the DIF of heavy neutrinos in the three Gaseous Argon Time Projection Chambers (GArTPC) of the off-axis near detector ND280 [183], as illustrated in the left plot of fig. 4.11. Because of the low density of the argon gas, this search has a minimal background from neutrino interactions, and the gas allows excellent tracking and identification of the $e^+e^-$ final state. The analysis observes no event in all channels and provides some of the strongest limits in the mass region 140 MeV $\leq m_N \leq$ 493 MeV. We use their null results and extrapolate the experimental efficiencies, shown on the right plot of fig. 4.11, to estimate the constraint on light heavy neutrinos with 20 MeV $\leq m_N \leq$ 140 MeV. We neglect systematic uncertainties and backgrounds, as they provide negligible contributions to the limits. We reasonably accurately reproduce the official T2K result above the pion mass. ND280 is currently being upgraded to a new configuration [184], with the replacement of the $\pi^0$ detector, currently made of lead and scintillator, with two new GArTPCs. DIF searches will benefit from the larger GAr volume and the reduced number of backgrounds from coherent neutrino interactions upstream of the TPCs. We estimate the sensitivity of a future search with this upgrade by considering the increased volume, and a total of $2 \times 10^{22}$ POT [185], $4 \times 10^{21}$ POT before (already collected), and $16 \times 10^{21}$ POT after the upgrade. This conservative estimate neglects improvements to reconstruction and background rejection.

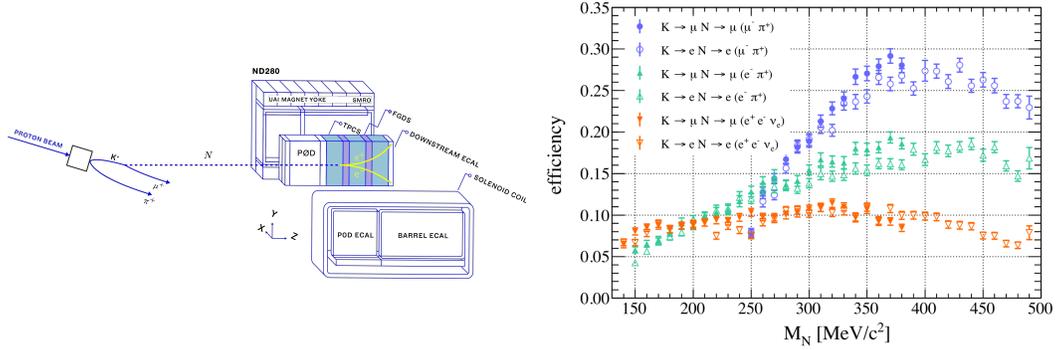

**Figure 4.11:** Left: The T2K near detector, ND280, and the heavy neutrino decay-in-flight signature. The detector is located at an angle of $2.042°$ with respect to the proton beam and a distance of $284.9$ m from the center of the production target. Thanks to the artist and graphic designer Jackapan Pairin for the drawing. Right: Reconstruction and selection efficiencies for different heavy neutrino production and decay modes, as a function of the heavy neutrino mass, as employed in the official T2K analysis [183]. The fact that the orange lines do not extend below the pion mass, despite being kinematically allowed, sparked this work, particularly the extension of this limit to smaller masses in the minimal model.

## PS191

PS191 was a low-density detector located at CERN and designed to search for the decays of long-lived particles. The 19.2 GeV proton beam from the Proton Synchrotron (PS) provided a neutrino beam for the on-axis BEBC experiment and could also be used by PS191 at 40 mrad off-axis location, where $\langle E_\nu \rangle \sim 1$ GeV. The sketch on the left panel of fig. 4.12 shows the schematics of PS191 and the other experiments related to the PS beam dump. The detector is comprised of helium bags separated by scintillator planes followed by a dense electromagnetic calorimeter downstream, as shown in the schematic in the right panel of fig. 4.12. The calorimeter was used in a search for

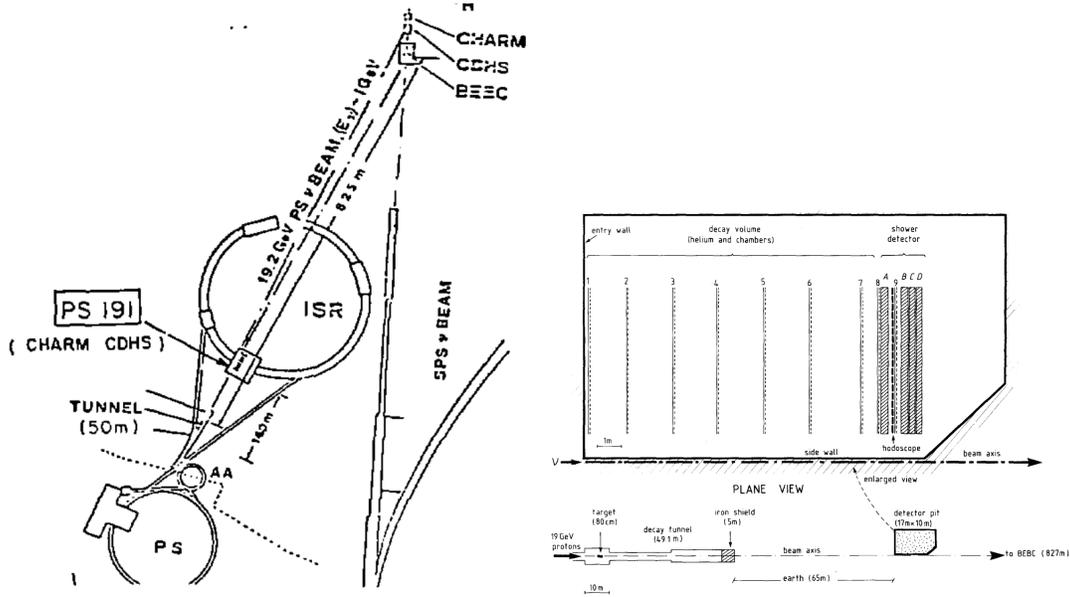

**Figure 4.12:** Left: PS191 was a neutrino detector located at CERN in the neutrino beam generated by protons from the PS storage ring. PS191 was a short baseline detector, located only 140 m from the production point, while the same beam served CHARM, CCHS, and BEEC about 825 m away. These last experiments were also sensitive to neutrinos produced by the SPS beam. Figure taken from [186]. Right: a sketch of the PS191 detector, which consisted of a large decay region made of six helium bags separated by scintillator planes and followed by an electromagnetic calorimeter. Figure taken from [148].

$\nu_\mu \to \nu_e$ oscillations, where an excess was observed [187], but not for the searches for heavy neutrinos discussed here. Their final results considered only the CC decays of Dirac heavy neutrinos [188,149]. [†]. As a consequence, the search for $K^+ \to \mu^+ (N \to \pi^+ \mu^-)$ and $K^+ \to \mu^+ (N \to \nu_e \mu^- e^+)$ was used to constrain only the $|U_{\mu N}|^2$-dominant case [‡] and the search $K^+ \to \mu^+ (N \to e^+ e^-)$ was used to constrain $|U_{\mu N} U_{eN}|$. We are concerned with the latter case, as even in a $|U_{\mu N}|^2$-dominant case, heavy neutrinos still decay to $\nu e^+ e^-$ via NC, and a constraint can be derived. This point was first appreciated in [189] and later discussed in [190,143]. These constraints were thought to be the strongest lab-based in this mass region for the $|U_{\mu N}|^2$-dominant case. With our simulation, we show that this is not the case. The bound on $|U_{\mu N}|^2$ is a factor of 6 weaker than the published ones, corresponding to an event rate 36 times smaller [§], which is corroborated by the fact that T2K and PS191 have very similar total exposures and neutrino fluxes, noting that PS191 ran for only a month.

---

[†]In the first publication [148] it was incorrectly stated that the limits are independent of the Dirac or Majorana nature of the heavy neutrinos, which was later corrected in [188,149].

[‡]We note that the channel $K^+ \to \mu^+ (N \to \nu_\mu \mu^+ \mu^-)$ was not considered even though it also proceeds via CC diagrams and could also constrain $|U_{\mu N}|^2$.

### Heavy neutrino production from pion and muon decays

As previously stated, one of the limitations of this analysis is the lack of heavy neutrino flux from pion and muon decay. We aim to give insights into this flux at ND280 and PS191 in this subsection.

The left panel of Fig. 4.13 shows the neutrino fluxes used in our primary analysis. The neutrino fluxes separated by parent particle are obtained from [191] for T2K and [192] for PS191, and they agree reasonably well with those provided in Refs. [186] and [187].

T2K did not include the target's heavy neutrino production from pion and muon decays. For this reason, we leave these decay channels out of our analysis since the final signal efficiencies can vary significantly between these channels due to the energy distribution and acceptance. Nevertheless, we perform a naive comparison between the event rate in these channels and the one from kaon production before efficiencies. Our results are shown on the right panel of fig. 4.13. To compute the heavy neutrino flux from muon decays, we take the approximation that the heavy neutrino flux is given by $\Phi_{\mu \to \overline{N} e \nu_e} \sim \Phi_{\mu \to \overline{\nu_\mu} e \nu_e} \times |U_{\mu 4}|^2 \times (\Gamma_{\mu \to \overline{N} e \nu_e} / \Gamma_{\mu \to \overline{\nu_\mu} e \nu_e})$, which is expected to be less accurate than in the case of the two-body meson decays used in the main text, and discussed in eq. (4.17).

### MicroBooNE

MicroBooNE is the first ton-scale liquid argon (LAr) TPC operated in a neutrino beam. It can perform searches for the DIF of new particles [193,194,195]. However, since LAr is a high-density material, providing both the target and the detector material, one needs to rely on extra schemes to reject neutrino-induced background. The delayed arrival of heavy neutrinos with respect to neutrinos was explored in the search for heavy neutrinos in the $\mu\pi$ decay channel [196]. Another attractive option has been employed in the search for light scalars [197] This search successfully rejects background utilizing the directionality and time of arrival of new particles, as produced in decays-at-rest of kaons at the beam dump of the NuMI beam. The left plot of fig. 4.14 shows the schematic of the detector location and orientation relative to the NuMI beam and the location of the absorber. The analysis did not observe any excess with respect to the background, providing firm limits in the scalar parameter space of mass versus coupling, as shown in the right plot of fig. 4.14.

---

[§]We have found a similar factor in the channels $N \to \mu\pi$ and $N \to \nu e\mu$. For the latter, we find a discrepancy by a factor of 6 at the largest heavy neutrino masses, which decreases for lower heavy neutrino masses. We attribute this effect to us overestimating the signal efficiency. For $\mu\pi$, the discrepancy is a constant factor of 7.5.

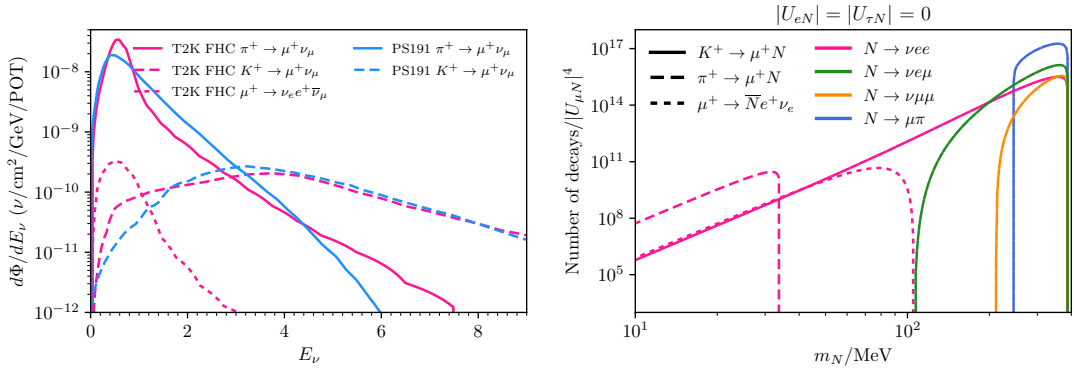

**Figure 4.13:** On the left panel, we show the $\nu_\mu$ and $\bar{\nu}_\mu$ flux separated by parent particle at T2K in neutrino mode and PS-191 as a function of the neutrino energy. On the right, we show the number of decays in flight in the T2K ND280 detector for $18.63 \times 10^{20}$ POT as a function of the heavy neutrino mass. The number of decays shown assumes the long-lifetime limit and is normalized by $|U_{\mu 4}|^4$. We include an estimate of the number of decays from heavy neutrinos produced in pion and muon decays for comparison.

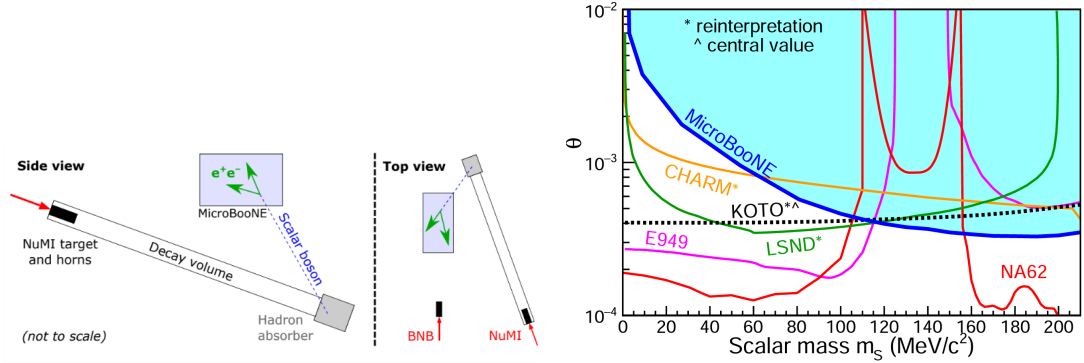

**Figure 4.14:** Left: a sketch of the location and orientation of MicroBooNE with respect to the main axis of the NuMI beamline, its decay volume, and absorber. In the search for light scalar bosons [197], the new particles are produced in the absorber from the kaons decaying at rest, fly upwards and backward, and decay into $e^+e^-$ inside the detector. Right: MicroBooNE limits on the light scalar boson parameter space exclude all the interesting regions to explain the KOTO anomaly. We recast this result to models of heavy neutrinos decaying into $e^+e^-$, which differs from the signature MicroBooNE searched for only for the kinematic distribution of the $e^+e^-$ pair. Both figures are taken from [197].

The authors of [198] recast this null result into limits on the heavy neutrino mixing for the minimal model. We further re-interpret this constraint as bounds on the non-minimal models considered. These results use the reconstruction described in [199] and could be improved in MicroBooNE and future LArTPC by using new and innovative reconstruction techniques [200,201,3,202]. Additionally, MicroBooNE performed searches for signatures that could explain the MiniBooNE excess in terms of electron [203,204,205,206] and single photon [207] production in the detector. Although these results do not report excesses with respect to the standard model expectation, they do not entirely exclude that the MiniBooNE anomaly is related to new particles in the beam, as discussed in [208]. Moreover, these analyses do not significantly constrain the models we are considering in this paper because they do not include a tailored search for $e^+e^-$ from decay in flight. In contrast, the search for single photons with no protons in the final state is not sensitive to event rates compatible with the MiniBooNE excess.

At the time of writing this thesis, about a year after the publication of this work, the same strategy employed in the search for $e^+e^-$ from the NuMI absorber [197] has been employed to search for heavy neutrino and heavy scalar decaying into $\mu^+\mu^-$ and $\mu\pi$ pairs [196,209], which are more easily reconstructed and measured in MicroBooNE than $e^+e^-$ pairs. Although less powerful than T2K ones, this analysis provides interesting constraints above the $\mu\pi$ threshold $m_N \gtrsim 250$ MeV.

## Comparison between the different experiments

Table 4.1 provides a summary of the comparison between the three different experiments considered in our analysis.

| | T2K | PS-191 | MicroBooNE (NuMI KDAR) |
|---|---|---|---|
| $\langle E_{\nu^{\pi-\text{decay}}} \rangle$ | 0.9 GeV | 1 GeV | — |
| $\langle E_{\nu^{K-\text{decay}}} \rangle$ | 4 GeV | 4 GeV | 234 MeV |
| $\nu^{\pi-\text{decay}}/\text{cm}^2/\text{POT}$ | $1.8 \times 10^{-8}$ | $1.7 \times 10^{-8}$ | — |
| $\nu^{K-\text{decay}}/\text{cm}^2/\text{POT}$ | $9.1 \times 10^{-10}$ | $1.0 \times 10^{-9}$ | $6.6 \times 10^{-11}$ |
| POT | $(12.34 + 6.29) \times 10^{20}$ | $0.89 \times 10^{19}$ | $1.93 \times 10^{20}$ |
| area | $1.7\,\text{m} \times 1.96\,\text{m}$ | $3\,\text{m} \times 6\,\text{m}$ | $10.36 m \times 2.56\,\text{m}$ |
| length | 1.68 m | 12 m | 3 m |
| baseline | 280 m | 128 m | 102 m |
| signal efficiency | $6\% - 12\%$ | $\lesssim 30\%$ | $\sim 14\%$ |
| target | beryllium | carbon | — |
| target length | 80 cm | 91.4 cm | — |
| baseline | 280 m | 128 m | — |
| decay tunnel | 96 m | 49.1 m | — |
| proton energy | 30 GeV | 19.2 GeV | — |
| off-axis angle | $2.042°$ | $2.29°$ | — |

**Table 4.1:** Comparison between T2K and PS191 experimental design. The top rows show numbers that enter directly in the overall normalization of the event rate. The quantity $\langle E_{K \to \nu} \rangle$ is defined as the average energy of neutrinos produced in kaon decays and $\nu/\text{cm}^2/\text{POT}$ is defined as the total neutrino flux integrated over all energies.

The naive estimates of the flux-averaged heavy neutrino decay rates at PS191 and T2K are of the same order, showing that we should expect limits of the same order. The total number of heavy neutrinos crossing the detector in the long-lifetime and ultra-relativistic limit, up to experiment-independent factors, is

$$\text{norm} \simeq \frac{n_{\text{POT}} \times \Phi_N \times \text{Area} \times \text{Length} \times \langle \varepsilon_{\text{sig}} \rangle}{\langle E_N \rangle}, \qquad (4.20)$$

where $\langle \ldots \rangle$ denotes an average over the heavy neutrino spectrum, and $\varepsilon$ the signal efficiency, and $\Phi_N$ the total flux of heavy neutrinos. A simple ratio between the two experiments using the information in table 4.1 is

$$\frac{(\text{norm})_{\text{PS-191}}}{(\text{norm})_{\text{T2K}}} \simeq 0.5. \tag{4.21}$$

The numbers used for PS191 are obtained from Refs. [186,210,148,149]. We also note that T2K explores several analysis channels, including the efficiency to select $e^+e^-$ final states in other selection channels such as $\mu\pi$ and $e\pi$. This combination provides additional statistical power to the T2K analysis.

The most uncertain ingredient in our calculation of the normalization is the PS191 efficiency. From [186], we know that the detection efficiency is at most 70%. In addition, we are given the geometrical acceptance for $\pi$ final states, $\sim 40\%$, which in the low-density detector cannot be much different from the efficiency to detect electron final states. Therefore, 30% is the largest efficiency we allow for PS-191 to have, assuming it remains constant for all heavy neutrino masses. The actual efficiency is likely to be smaller, especially at low heavy neutrino masses where the $e^+e^-$ final state is more collimated. In any case, even for 100% efficiencies, we do not find reasonable agreement with the published bounds.

We have performed this check for the $\mu\pi$ and $\nu e\mu$ channels, finding similar discrepancies. For low heavy neutrino masses, we observe that the ratio between our bounds and the published PS191 ones goes from $\sim 6$ to lower values, which is most likely due to our assumption that the efficiencies remain constant.

We now compare our limits using the T2K dataset and those set by the authors of [198] using the MicroBooNE analysis in [197]. In the long-lived heavy neutrino limit, the ratio between the naive normalization factors in T2K and MicroBooNE is approximately

$$\frac{(\text{norm})_{\text{MicroBooNE}}}{(\text{norm})_{\text{T2K}}} \simeq 1.8 \times \frac{\varepsilon_{\text{MicroBooNE}}}{\varepsilon_{\text{T2K}}}, \tag{4.22}$$

where the larger neutrino flux compensates for the difference in detector size at T2K. As seen in

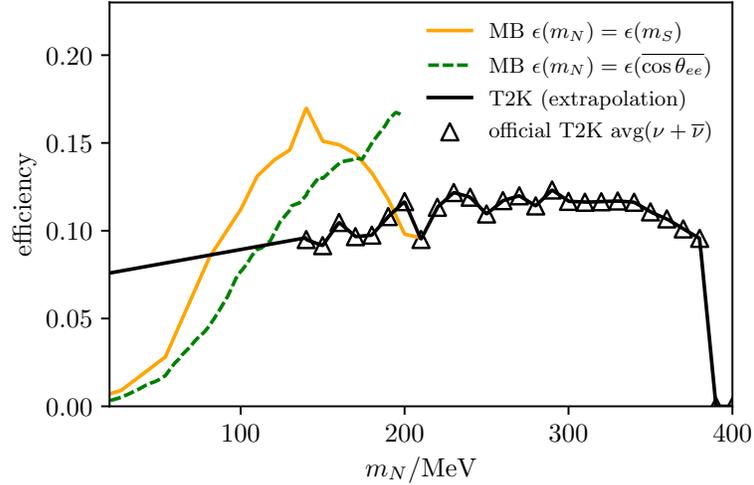

**Figure 4.15:** Comparison between the MicroBooNE and T2K efficiencies used in this work. Black triangles indicate the official efficiencies quoted in [146], averaged according to the POT in neutrino and antineutrino mode. The black line shows our extrapolation of the efficiency. The green curve has been obtained from the analysis of Ref. [198].

fig. 4.15, the MicroBooNE efficiencies fall rapidly at low heavy neutrino masses. While the T2K efficiencies we use at low masses are obtained from an extrapolation, it is clear that their dependence on $m_N$ is much milder, as is also the case in other reconstruction channels, such as in $\mu^+ \mu^-$ or $\mu^\pm e^\mp$. This is due to the low-density material, which prevents showering of the electrons, and the magnetic field, which splits the $e^+ e^-$ pairs with measurable angles. A great example of the capability of ND280 to reconstruct the highly-collinear dilepton pairs is the single photon selection in the $\nu_e$CC measurements [211,212] and the single photon search [213]. In the former, the selection efficiency for photons converted in the Fine-grained detector was 12%. While it might appear odd that the signal efficiency does not vanish at $m_N \sim 0$, this is because of the magnetic field inside the ND280 detector. In MicroBooNE, on the other hand, low-mass heavy neutrinos that decay into overlapping $e^+ e^-$ pairs are much less likely to be reconstructed as two separate objects. More detailed analyses by the collaborations can refine our estimates for the efficiencies at T2K and improve on the numbers shown in table 4.1.

This section shows the bound in parameter space for the minimal heavy neutrino model and its non-minimal extensions, as described in section 4.1, obtained using the methodology and the experimental datasets discussed in section 4.2. We conclude the chapter with some considerations for future searches and some additional cross-checks.

### Exploring the parameter space of the minimal model

Our results for the minimal model are shown in fig. 4.16. Our bounds and sensitivity estimates rule out heavy neutrinos below the kaon mass with dominant muon mixing in the minimal model. This result also puts more strain on models that could also account for the baryon asymmetry of the Universe[214]. As discussed previously, official limits from PS191 have been overestimated by one order of magnitude, and T2K provides the leading constraints at the lowest masses. Future data can improve the leading constraints below 200 MeV, where the kaon peak searches become insensitive. We note that our results are based on extrapolated efficiencies and conservative flux simulation and that a complete simulation within the collaboration will provide improved results.

### Constraining the parameter space of the TMM and four-fermion interaction models

The new constraints with decays via the dimension five, eq. (4.2), are shown in fig. 4.17. while for dimension six, eq. (4.4), in fig. 4.18. In these scenarios, the combination of lab-based and cosmological constraints does not exclude heavy neutrinos below the kaon mass. Our work complements searches for neutrino upscattering at CHARM-II, which provides stronger constraints for new physics scales below 1 PeV[216,162], and constraints from supernovae, which dominate above $\sim 1$ EeV. For $G_X/G_F = 10^3$, BBN constraints still exclude the smallest mixings, but our lab-based results pro-vide the best upper limits in the newly allowed parameter space. For this choice of parameters, one expects the existence of a new vector mediator with a mass of a few GeV, which can be searched for in collider experiments[217].

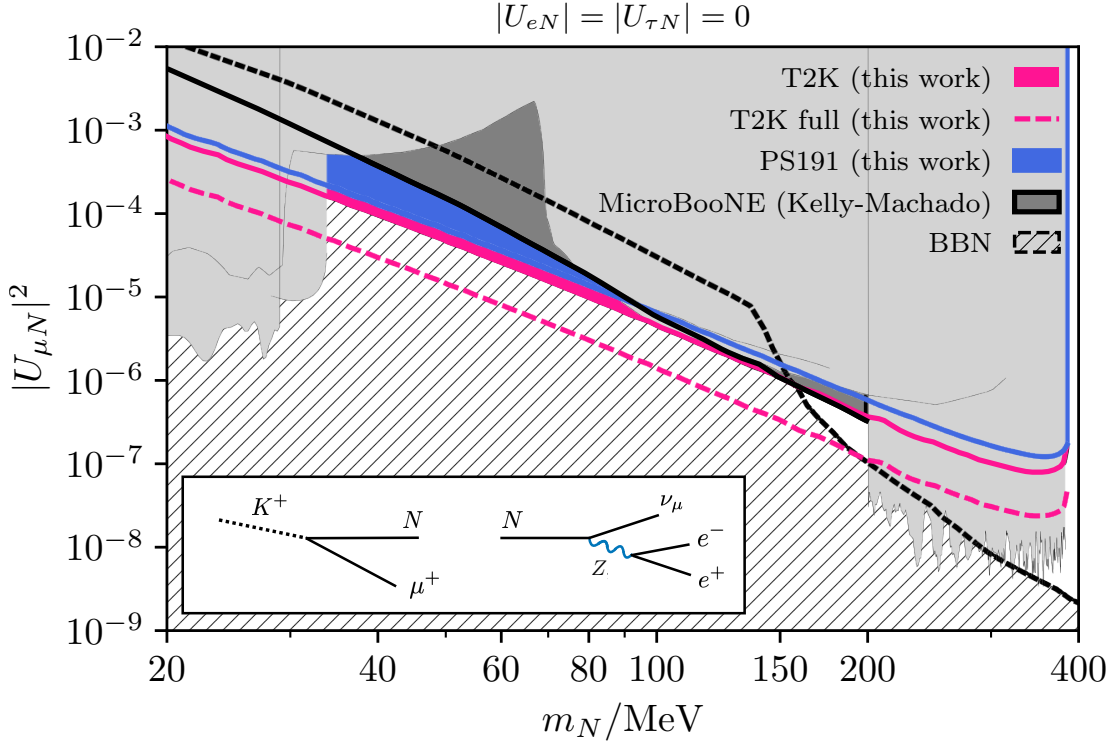

**Figure 4.16:** Constraints on the mixing of heavy neutrinos with the muon flavor as a function of its mass for a minimal heavy neutrino model at 90% C.L., considering only the production and decay mode: $K \rightarrow \nu_\mu N \rightarrow \nu_\mu(e^+e^-\bar{\nu}_\mu)$. For MicroBooNE, T2K, and PS191, the regions above the lines are excluded, while BBN excludes the region below the line. In gray, we show other model-independent constraints. T2K full refers to the projected sensitivity of T2K with the final dataset, which will be collected by the end of the experiment.

Our constraints are relevant to new physics explanations of the MiniBooNE excess of electron-like events[55,51]. The authors of[215] proposed that the excess can be explained by the DIF of heavy neutrinos with a TMM. This observation can be generalized to any model with enhanced $N \rightarrow \nu e^+e^-$ or $\nu\gamma$ rates. Our limits are derived using off-shell photon decays, therefore suppressing the $e^+e^-$ signal rate at T2K, PS191, and MicroBooNE with respect to the $\gamma$ signal rate at MiniBooNE by approximately $\alpha/4/\pi \times \log\left(m_N^2/m_e^2\right) \approx 6 \times 10^{-3}$ in the parameter space of interest. However, our study shows that MicroBooNE and T2K already constrain interesting parameter space for the TMM model.

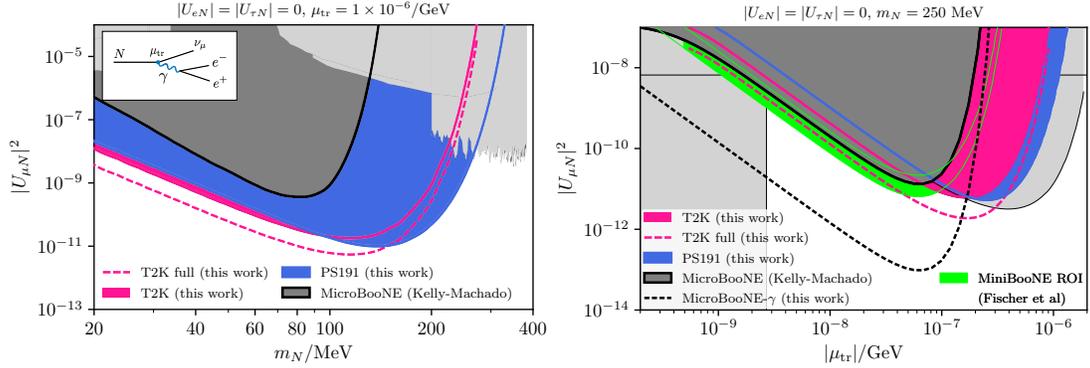

**Figure 4.17:** Left: Same as fig. 4.16 but for heavy neutrinos with a TMM $\mu_{tr} = 1\,\mathrm{PeV}^{-1}$. Right: The same constraints as above but shown as a function of $\mu_{tr}$ for a fixed value of the heavy neutrino mass. The region of interest to explain the MiniBooNE excess is shown in green [215]. In fine dashed black, we show an optimistic estimate of a NuMI-neutrino single-photon search at MicroBooNE. The shape of the constraints is dictated by the combination of the rate and lifetime. A very small $\mu_{tr}$ results in a low rate and a long lifetime, while a large $\mu_{tr}$ increases the rate and shortens the lifetime. There is an optimal point where the strongest constraint in mixing can be placed. Heavy neutrinos become very short-lived for larger $\mu_{tr}$ values, thus decaying before reaching the detector.

We acknowledge that the distribution of the $e^+e^-$ invariant mass in the TMM model is significantly different than in the other models, as it peaks at masses close to zero because highly off-shell photons are disfavored. We believe that applying the same efficiency to this model as to the other two models might lead to overestimated limits; therefore, only detailed analyses from the experimental collaborations can provide the most accurate bounds. However, MicroBooNE could exploit a complementary strategy to test this model. Because of the high density, photons convert in the detector. Thus a search for the single-photon decay channel rather than the $e^+e^-$ branching ratio could be extremely powerful, as seen by the additional curve (black dashed) drawn on fig. 4.17. Here we assume the same efficiencies, backgrounds, and exposure as the $e^+e^-$ search at MicroBooNE. While this projection relies on optimistic assumptions on the reconstruction and selection effi-ciency for a photon converting in the detector, which is likely to be lower than for a genuine $e^+e^-$ pair and to have higher backgrounds, even if the rate estimation would be higher by a factor of ten, lowermost of the MiniBooNE preferred region will be tested. Moreover, with a volume of 0.66 and 4.5 times the MicroBooNE volume, respectively, and a similar beam exposure, SBND[218] and ICARUS[219] could complement the MicroBooNE constraints, thanks to the different distances from the beam target and absorber.

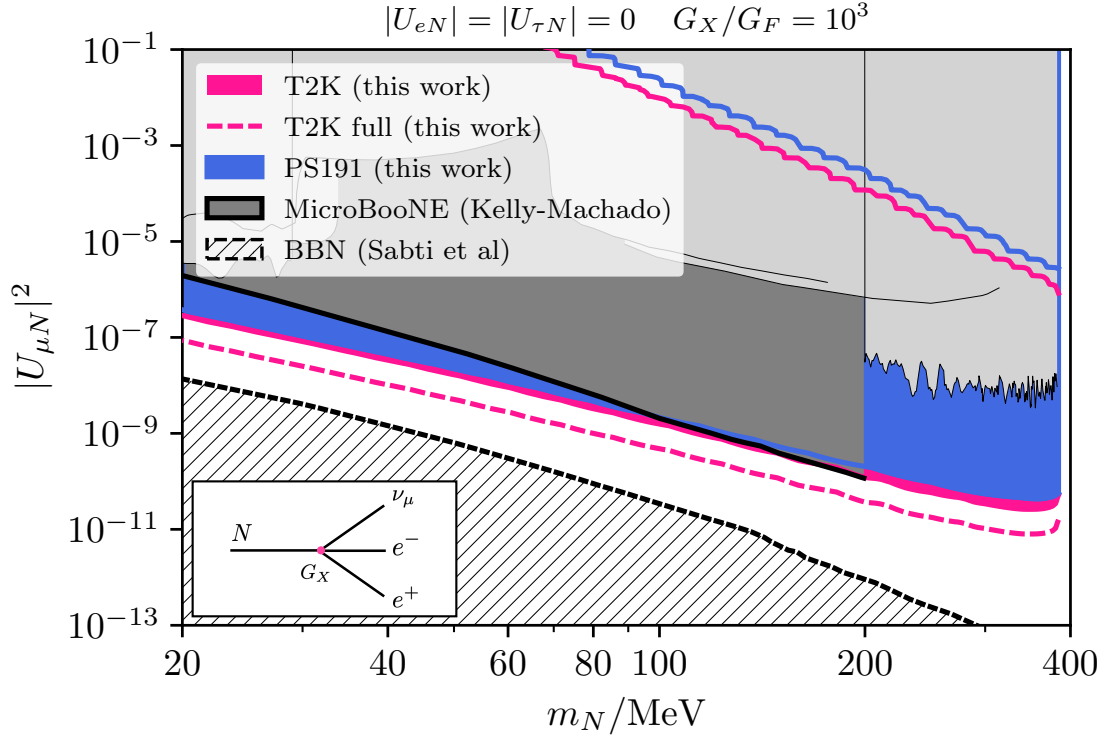

**Figure 4.18:** Same as fig. 4.16 but for heavy neutrinos with a four-fermion leptonic interaction with $G_X = 10^3 G_F$. The inset diagram shows the dominant decay process in the model.

At the time of writing this thesis, about a year after the publication of this work, new constraints on this model from MINERvA have been derived, probing most parts of the region of interest for MiniBooNE[220]. A dedicated analysis from the MINERvA collaboration will likely be sensitive to the entire best-fit region to the MiniBooNE excess.

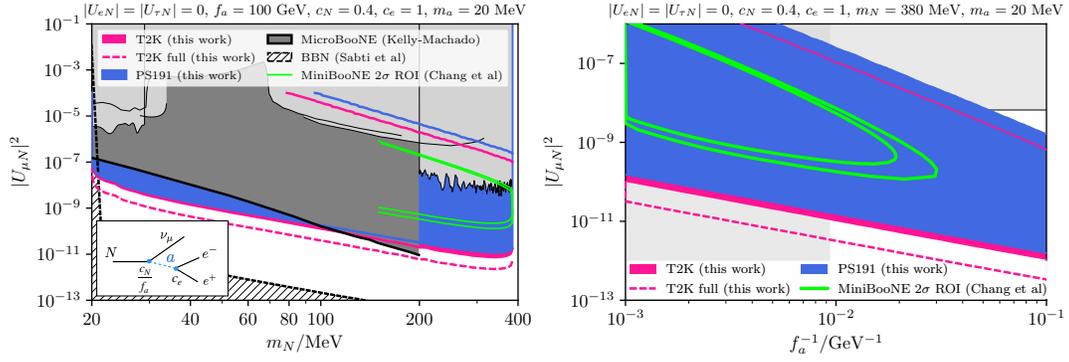

**Figure 4.19:** Same as fig. 4.16 but for heavy neutrinos that decay to a light leptophilic ALP of $m_a = 20$ MeV. The two solid green lines define the MiniBooNE region of preference at $2\sigma$.

Constraints on the heavy neutrino decay to $\ell$ALP are shown in fig. 4.19. On the left panel, we show the limits in mass and mixing for fixed values of the heavy neutrino-$\ell$ALP couplings and $\ell$ALP mass. On the right, we fix the heavy neutrino mass to be 380 MeV and vary the heavy neutrino mixing and the $\ell$ALP decay constant, $f_a$. The limits from T2K and PS191 fully cover the region of preference to explain the MiniBooNE excess in both panels, excluding that the excess is due to decays in flight to a great significance. On the right panel, we do not show the MicroBooNE limits as the heavy neutrino mass is beyond the range considered by the experiment. However, inspecting the left panel, one can deduce that MicroBooNE would also strongly constrain the large mass region of the explanation. Similar to the TMM model, the invariant mass of the $e^+e^-$ produced in heavy neutrino decays is tiny; in fact, $m_{e^+e^-} = m_a$. For MicroBooNE, separating the lepton pair would be more challenging due to the absence of magnetic fields. When setting our limits in this model, we assign the final signal efficiency to be that of a standard heavy neutrino model with $m_N = m_a$, that is, we set $\varepsilon_{\ell\text{ALP}} = \varepsilon(m_N = m_a)$.

## Outlook

Lastly, when heavy neutrinos are too short-lived to be probed in DIF searches, they may be produced by coherent neutrino-nucleus upscattering in the detector itself[160,161,216,135,221,222,223], before traveling a distance of the order of the detector size and decay within the detector. The dark neutrino model is an example and constitutes a possible UV completion of the four-fermion interaction model. The next chapter of this thesis thoroughly explores this model and the constraints derived with ND280 by exploiting the upscattering in the dense layer of lead of the calorimeter preceding the GArTPCs.

Prompt heavy neutrino decays in pion and kaon factories, such as PIENU[224,225] and NA62[226,227], should also be searched for. The channel $K^+ \to \ell^+(N \to \nu e^+ e^-)$, proposed in[223], would be sensitive to light dark sector models and, to a lesser extent, to TMM.

Low- and high-density hodoscopic detectors like ND280 and MicroBooNE can play a central role in the search for long-lived particles. Due to its hybrid design, ND280 can place firm limits on the upscattering production of light particles, like dark neutrinos[221,222,223,2] and co-annihilating dark matter[228,229]. Future detectors, such as the planned DUNE near detector, could also benefit from a hybrid detector design since they would be sensitive to both charged-track and single photon final states while having a region of low neutrino-induced backgrounds.

## 4.4 Cross-checks

Here we discussed some simple cross-checks that validate our analysis and bounds.

### Reproducing the T2K official limits

We now discuss the consistency between our limits in fig. 4.16 and the official T2K result[183]. A direct comparison between our limit and the figures in[183] would not be fair because of the differ-ent physical and methodological assumptions. Figure 4.20 illustrates the individual effect of each different ingredient and to what extent we reproduce the T2K official result correctly. First, T2K performs a Bayesian analysis for Majorana neutrinos, while we are showing a frequentist analysis for Dirac neutrinos. Moreover, T2K shows both marginalized limits, where the posterior is integrated over the parameters not shown in the current figure ($U_{eN}$ and $U_{\tau N}$ for our case), and profiled limits where the posterior density is shown after conditioning on the values of the parameters not shown equal to zero ($U_{eN} = U_{\tau N} = 0$). The second case is appropriate for our comparison, and we were able to plot it from the data release (black solid line). From the data release, we also obtained the flux and the efficiency in the mass range considered for the analysis, from which we extracted a frequen-tist limit without considering systematics and backgrounds (orange dashed line), which play a minor role in this analysis. The difference between a Dirac and a Majorana (green dashed curve) analysis is a factor of two in the heavy neutrino rate — aside from a small difference in efficiency between charge conjugate channels for $e\mu$, $e\pi$, and $\pi\mu$ channels — and becomes exactly a factor of two for the $e^+e^-$ channel. We reproduce the T2K heavy neutrino flux using eq. (4.17), which largely underestimates the flux at large masses, but becomes accurate within 10% at $m_N = 150$ MeV (blue dashed line curve). Eventually, considering only the $e^+e^-$ decay mode gives the most conservative limit as we are not adding contributions from $e\mu$, $e\pi$, $\pi\mu$, and $\mu\mu$. However, below the muon mass, since $e^+e^-$ is the only kinematically allowed decay mode, the limit matches the case where all decay modes are considered, as shown by comparing the solid pink line with the blue dashed line.

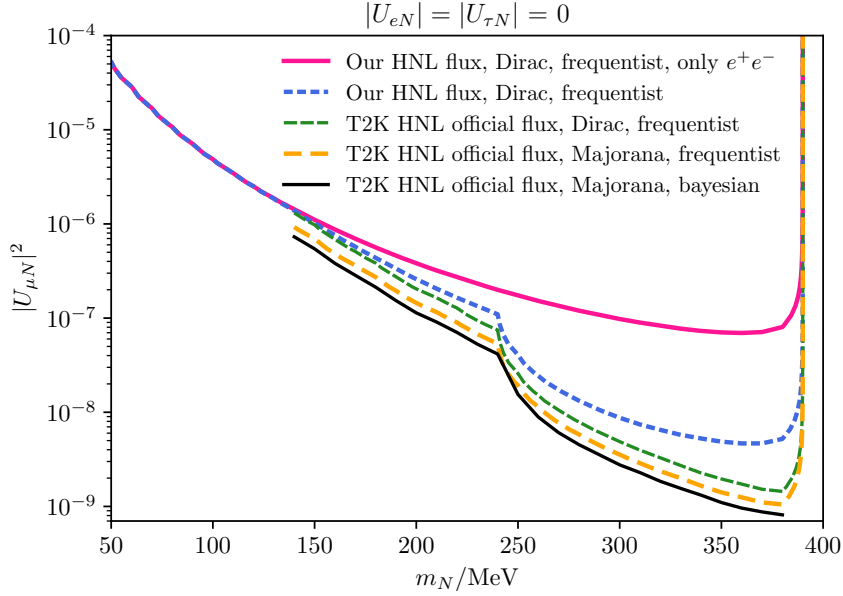

**Figure 4.20:** Differences between our analysis and the T2K official result lead to a conservative limit. The T2K official result sets a limit for Majorana neutrinos in a Bayesian fashion (black solid), which is stronger than using a frequentist limit (orange dashed), and is about a factor of $\sqrt{2}$ stronger than the limit for Dirac neutrinos (green dashed). Our flux estimation is conservative at large heavy neutrino mass and matches the full simulation at lower masses (blue dashed). Excluding all decay modes aside for $e^+e^-$ (pink line) leads to a conservative limit above the muon mass and matches well with the equivalent T2K constraints below the pion mass.

### Comparison in the light scalar case

If constraints on light scalars can be translated to constraints on heavy neutrinos, as illustrated in [198], it is also possible to translate between bounds on the heavy neutrino mixing to bounds on the mixing of light scalars with the SM Higgs. In particular, due to the proximity between the muon and pion masses, the bounds on the scalar mixing can be approximately related to those on the heavy neutrino mixing under a muon-dominance assumption as follows:

$$\theta_{\text{bound}}^4 = |U_{\mu N}|_{\text{bound}}^4 \times \frac{\mathcal{B}(K^+ \to \mu^+ N)}{\mathcal{B}(K^+ \to \pi^+ \phi)} \frac{\hat{\Gamma}_{N \to \nu e^+ e^-}}{\hat{\Gamma}_{\phi \to e^+ e^-}}, \tag{4.23}$$

where $\hat{\Gamma}$ is the decay rate normalized by the relevant mixing angle. This estimate neglects contributions from $K_L$ and $K^-$ decays and hadron regeneration, providing a conservative estimate for the bound on $\theta$. These effects can contribute as much as a factor of 2 to the total rate since 0.065 $K^+$, 0.032 $K^-$, and 0.044 $K_L$ are produced per proton on target according to [230]. For our PS-191 constraint, Eq. (4.23) gives $\theta^2 < 2.8 \times 10^{-7}$ for $m_\phi = 150$ MeV. Including a naive factor of 2 in the rate for the other kaon sources translates to $\theta^2 < 2 \times 10^{-7}$. These values are not far from the constraints found in [230]. Using the naively rescaled PS191 constraints on heavy neutrinos, we would have found $\theta < 4 \times 10^{-8}$, which is much stronger than the quoted value in [230].

Recasting our results below for current (future) T2K data, we find a constraint on the scalar mixing of $\theta < 2.3\,(1.5) \times 10^{-4}$ for $m_\phi = 150\,\text{MeV}$, which is the leading constraint in the "pion gap" [231].

### Existing limits from PS191

We revisit the constraints set by the PS191 experiment on the in-flight decays of heavy neutral leptons (heavy neutrino). We are primarily interested in the search for $K \to \mu(N \to \nu e^+ e^-)$. As discussed previously, PS191 omitted neutral-current (NC) decays of heavy neutrinos, so this channel was used to constrain only the product $|U_{eN}U_{\mu N}|$. Including neutral currents, however, this limit can be naively translated to a limit on $|U_{\mu N}|^2$ as follows

$$|U_{\mu N}|^2_{\text{new-limit}} = \left( \frac{1 - 4s_{\text{W}}^2 + 8s_{\text{W}}^4}{4} \right)^{1/2} \times |U_{eN}U_{\mu N}|_{\text{PS191}}, \qquad (4.24)$$

where $s_{\text{W}} = \sin\theta_{\text{W}}$ is the sine of the weak mixing angle. We expect a similar selection efficiency between charged-current (CC) and NC decays. The bound obtained with this rescaling procedure differs from ours by a factor of $\approx 6$, corresponding to a discrepancy of 36 in the event rate. This estimate agrees with what was found in [230] for the scalar case.

# 5

# Dark Neutrinos

While the canonical Type-I Seesaw mechanism is an exciting solution to the puzzle of neutrino masses, if heavy neutrinos are in the mass range for explaining short-baseline anomalies ($1 - 500$ MeV), a minimal model is not enough. We explored several effective operators that can shorten the lifetime of heavy neutrinos to avoid cosmological constraints from BBN. However, as a complete singlet under $G$, right-handed neutrinos could also provide unique insight into the possible existence of other hypothetical particles, such as dark matter, the dark photon, or additional Higgs bosons.

In fact, theoretical activity in low-scale dark sectors has intensified and, unsurprisingly, new dark-sector solutions to the short-baseline puzzle have been brought to light [160,161,232,233,234,162,135,235,133] and [221,172,222,236,223,237,173,238,239,240,241,242,243,215,177,244].

Among them is the possibility that light dark particles are produced in the scattering of neutrinos with matter, and misidentified as electron-neutrinos due to their electromagnetic decays. These models have been popularized by their connection to neutrino masses in low-scale seesaw models, the possibility to explain other low-energy anomalies, and, chiefly, due to their falsifiability.

Among these, dark sectors containing a dark photon are interesting because mixing with the electromagnetic photon provides a portal to the dark sector. However, these dark sector solutions often involve several new particles and, therefore, several independent parameters. While the experimental signatures are straightforward to identify, the coverage of the model parameter space that is needed can quickly become unmanageable. This curse of dimensionality is especially burdensome due to the need to repeat sophisticated simulations of the model predictions in an experimental setting.

In this work, we solve this problem using a re-weighing method. We do so in the context of a model of a dark neutrino sector, one of the working new physics explanations of the Mini-BooNE excess based on heavy neutrinos coupled to a dark photon. We apply our technique to the background-free search for $e^+e^-$ in the multi-component near detector of T2K, ND280 [183], that we also used in chapter 4. The signatures arise from the upscattering of neutrinos inside the high-density region of the detector, followed by the decay of the heavy neutrino into $e^+e^-$ pairs inside the gaseous Argon (GAr) Time Projection Chambers (TPC) of ND280. By leveraging the power of our re-weighing technique, we can take advantage of our detailed detector simulation throughout a much broader parameter space.

The novelty of our technique lies in the usage of a single Monte Carlo simulation that simulta-neously samples physical quantities, like phase space variables, and model parameters, such as the masses of heavy neutrinos and the dark photon. With sufficient statistics, these samples can be used to construct a Kernel Density Estimator (KDE), which computes the model prediction and corresponding likelihood for any choice of model parameters within the boundaries of the simulation. The result is a fast interpolation of the posterior probability of the model and greater flexibility in determining confidence intervals in various slices of parameter space.

The code to obtain our limits and simulate dark neutrinos in the ND280 detector is open source, and can be found on GitHub.[*]

## 5.1   When heavy sterile neutrinos meet a dark sector

The idea that low-scale seesaw extensions of the SM can co-exist with new gauge symmetries, most famously with $B - L$ [245,246,247,248,249,250,251,252,253,254], has been discussed throughout the literature also in the context of baryonic [255], leptonic [256,257], or completely hidden gauge symmetries [258,259,260,261,262,263,264]. These models present a complicated mass spectrum and a self-interacting dark sector that can be challenging to identify experimentally. Nevertheless, they remain viable and testable examples of low-scale neutrino mass mechanisms and deserve experimental scrutiny.

This work focuses on a low-scale dark sector containing heavy neutrino states and a broken $U(1)_D$ gauge symmetry [221,172,222,236,223,237,173]. The heavy neutrinos interact with the mediator of the dark gauge group, the dark photon, and SM neutrinos via mixing. Through a combination of neutrino and kinetic mixing between the SM photon and the dark photon, this dark sector leads to new interactions of neutrinos with charged particles and contains heavy neutrino states that decay primarily through the new force. This model is exciting in the context of short-baseline anomalies as it predicts the production of heavy neutrinos inside detectors, which subsequently decay to electro-magnetic final states, mimicking $\nu_\mu \to \nu_e$ appearance signatures.

---

[*] github.com/mhostert/dark-neutrinos-at-T2K.

We start with the definition of a simplified, low-energy Lagrangian that will be used throughout this work. The minimal particle content we consider contains a single mediator, corresponding to a kinetically-mixed dark photon $Z'$, and heavy neutrino states $\nu_{i \geq 4}$. We provide further details on the UV completions of the model at the end of this section. In terms of the physical fields, our Lagrangian reads

$$\mathcal{L} \supset \mathcal{L}_{\nu\text{-mass}} + \frac{m_{Z'}^2}{2} Z'^\mu Z'_\mu + Z'_\mu \left( e\varepsilon j^\mu_{\text{EM}} + g_D j^\mu_D \right) , \tag{5.1}$$

where $\mathcal{L}_{\nu\text{-mass}}$ contains all the mass terms for the neutrino fields after proper diagonalization. The dark photon interacts with the electromagnetic current of the SM, $j^\mu_{\text{EM}}$, proportionally to the electric charge $e$ and kinetic mixing $\varepsilon$, as well as with the neutral leptons in the dark current $j^\mu_D$ to the gauge coupling $g_D$. The above Lagrangian includes all interactions of interest in the limit of small $\varepsilon$ and $(m_{Z'}/m_Z)^2$.

In terms of an interaction matrix $V$, the dark current in the mass basis is given by

$$j^\mu_D = \sum_{i,j}^{n+3} V_{ij} \overline{\nu_i} \gamma^\mu \nu_j, \tag{5.2}$$

where $n$ is the number of heavy neutrino states. Here, $\nu_{1,2,3}$ are the mostly-SM-flavor light neutrinos, and $\nu_{i \geq 4}$ are the heavy neutrinos that contain small admixtures of SM flavors. We can express $V_{ij}$ in terms of the mixing matrix $U_{di}$ between the mass eigenstates $i$ and dark flavor states $d$ as $V_{ij} = \sum_d Q_d U^*_{di} U_{dj}$, where the index $d$ runs over all dark-neutrino flavors $\nu_d$ of $U(1)_D$ charge $Q_d$. Assuming $Q_d = 1$ for all flavors, $|V_{ij}| \leq 1$ for all $i$ and $j$ due to the unitarity of the full neutrino mixing matrix. Since we are not interested in the specifics of the flavor structure of the full model, we stick with the generic notation of Equation (5.2), noting that experimental constraints require $V_{ij} \ll 1$ if either $i$ or $j$ are in $\{1, 2, 3\}$. We now consider the upscattering of light neutrinos into one of the $n$ heavy neutrinos states $\nu_N$, hereafter referred to as $N$ for brevity, and its subsequent decay into lighter neutrinos and an $e^+ e^-$ pair. The Feynman diagrams responsible for the processes are shown in fig. 5.1. Specifically,

$$\widehat{\nu}_\mu + A \rightarrow (N \rightarrow \nu e^+ e^-) + A, \tag{5.3}$$

where $\widehat{\nu}_\mu$ stands for the coherent superposition of $\nu_{1,2,3}$ produced in the neutrino beam, and $\nu$ for all possible daughter neutrinos. This work considers only coherent scattering on the nucleus $A$.

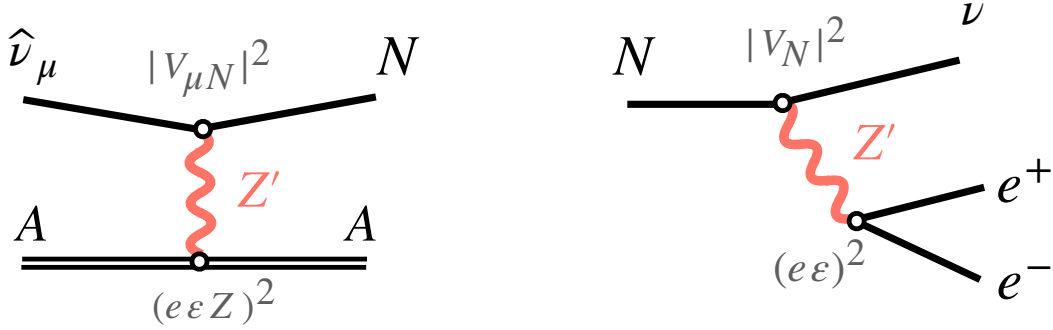

**Figure 5.1:** The diagrams for coherent neutrino-nucleus upscattering ($\widehat{\nu}_\mu A \rightarrow NA$) and heavy neutrino decays ($N \rightarrow \nu e^+ e^-$) considered in this work. We indicate the relevant parametrization of each interaction vertex.

UPSCATTERING AND DECAY

The coherent neutrino-nucleus scattering is mediated by the dark photon with amplitude

$$\mathcal{M}_{\mathrm{ups}}^{Z'} = \frac{e\varepsilon g_D}{q^2 - m_{Z'}^2} \, \ell_\mu b^\mu, \tag{5.4}$$

where $q^2$ is the momentum exchange with the nucleus, $h^\mu = \langle A | \hat{J}^\mu_{\rm EM} | A \rangle$ is the elastic electromagnetic transition amplitude for the nuclear ground state of $A$, and $\ell^\mu$ is the leptonic current

$$
\ell^\mu = \langle N | \hat{J}^\mu_D | \hat{\nu}_\mu \rangle = \frac{\sum_{i \leq 3} U^*_{\mu i} V_{iN} \langle N | \overline{N} \gamma^\mu \nu_i | \nu_i \rangle}{\left( \sum_{k \leq 3} |U_{ki}|^2 \right)^{1/2}},
$$

$$
= V_{\mu N} \langle N | \overline{N} \gamma^\mu \nu_i | \nu_i \rangle, \tag{5.5}
$$

where we defined the vertex factor

$$
V_{\alpha N} \equiv \frac{\sum_{i \leq 3} U^*_{\alpha i} V_{iN}}{\left( \sum_{k \leq 3} |U_{ki}|^2 \right)^{1/2}}. \tag{5.6}
$$

In a model with a single dark flavor $d = D$ and one heavy neutrino $N = \nu_4$, it is possible to show that $V_{\alpha N} = U_{\alpha 4} |U_{D4}|^2 \simeq U_{\alpha 4}$, which is small and directly constrained by laboratory experiments. The complete cross section is then computed in the usual fashion. We have implemented a data-driven Fourier-Bessel parametrization for the nuclear form factors[265].

The decay process can be computed similarly, now summing over the daughter neutrinos incoherently,

$$
|\mathcal{M}^{Z'}_{\rm dec}|^2 \equiv \sum_{i < N} |V_{Ni} \mathcal{M}(m_i)|^2 \simeq |\mathcal{M}(0)|^2 \sum_{i < N} |V_{Ni}|^2, \tag{5.7}
$$

where we factorize the matrix elements assuming that all daughter neutrinos have a negligible mass with respect to $m_N$. We define the remaining vertex factor as

$$
|V_N|^2 = \sum_{i < N} |V_{iN}|^2. \tag{5.8}
$$

As before, for a model with a single dark flavor and one heavy neutrino, $|V_N|^2 = |U_{D4}|^2 (1 -$

$|U_{D4}|^2) \simeq |U_{e4}|^2 + |U_{\mu 4}|^2 + |U_{\tau 4}|^2$. $|V_N|^2$ may be similar in size or much larger than $|V_{\mu N}|^2$, depending on the flavor structure of the model.[†] In this way, the production cross section is effectively decoupled from the lifetime of $N$. This effect is irrelevant for light dark photons ($m_{Z'} < m_N$) because the decay is always prompt, rendering most signatures independent of $|V_N|^2$. However, in MiniBooNE explanations where $N$ decays via an off-shell dark photon, the requirement $|V_N| > |V_{\mu N}|$ helps ensure that the production of $N$, as well as its decays, happen inside the detector.

We also keep the number of daughter neutrinos unspecified to effectively cover models where $N$ does not decay only into $\nu_{1,2,3}$, but also into other heavy neutrinos $\nu_j$ with $3 < j < N$. In this case, $V_{jN}$ is mainly insensitive to the direct limits on the mixing of active and heavy neutrinos, $|U_{\alpha 4}|^2$, and can be of order one. The properties of these new states are rather model-dependent, so we conservatively consider them invisible and not observable. For simplicity, we require that $\nu_j$ be light enough such that $(m_N - m_j)/m_N \ll 1$. Therefore, we do not consider scenarios with small mass splittings between the upscattered and the daughter neutrinos.

The relevant decay rate for a Dirac $N$ with off-shell $Z'$ is

$$\Gamma_{N \to \nu e^+ e^-} = \frac{\alpha \alpha_D \varepsilon^2 |V_N|^2}{48\pi} \frac{m_N^5}{m_{Z'}^4} L(m_N^2/m_{Z'}^2),$$

(5.9)

where $L(x) = \frac{12}{x^4} \left( x - \frac{x^2}{2} - \frac{x^3}{6} - (1-x) \log \frac{1}{1-x} \right)$, with $L(0) = 1$. For a light, on-shell $Z'$ we need only compute $N \to \nu Z'$ since $Z' \to e^+ e^-$ is always prompt,

$$\Gamma_{N \to \nu Z'} = \frac{\alpha_D |V_N|^2}{4} \frac{m_N^3}{m_{Z'}^2} \left( 1 - \frac{m_{Z'}^2}{m_N^2} \right)^2 \left( \frac{1}{2} + \frac{m_{Z'}^2}{m_N^2} \right).$$

(5.10)

---

[†]This was the idea proposed in [222], where by virtue of $|U_{\tau 4}|^2 \gg |U_{\mu 4}|^2$, the daughter neutrino produced had a significant admixture of the tau flavor.

Note that the decay rate is bounded from above and below by

$$\Gamma_N(|V_N| = |V_{\mu N}|) < \Gamma_N < \Gamma_N(|V_N| = 1), \tag{5.11}$$

where MiniBooNE explanations prefer to saturate the right-most inequality whenever $m_N < m_{Z'}$. The lifetimes of $N$ and $Z'$ are always prompt in the light $Z'$ case. For the heavy case, we assume $Z'$ to be prompt, while $N$ can be prompt or longer lived. For instance, taking $|V_N| = 1$ and $m_{Z'} = 1.25$ GeV, we find

$$c\tau_{\min}^0 \simeq 1 \, \text{cm} \times \left(\frac{10^{-2}}{\varepsilon}\right)^2 \left(\frac{100 \, \text{MeV}}{m_N}\right)^5 \left(\frac{m_{Z'}}{1.25 \, \text{GeV}}\right)^4. \tag{5.12}$$

ANALYTICAL APPROXIMATION FOR UPSCATTERING CROSS SECTION

For convenience, a crude approximation for the upscattering cross sections above $E_\nu > 1$ GeV is given below. These have been obtained assuming a box function for the coherent and dipole form factors with cut-offs around the QCD scale and vector mass, respectively. For upscattering on nuclei

$$\sigma_{\text{coh}}^{\nu_\alpha \to N} \simeq \frac{|V_{\alpha b}|^2 (Z e \, \varepsilon)^2 \alpha_D}{4 E_\nu^2 m_{Z'}^4} \left[2(M^4 + s^2) - sM^2(x_A^2 + 4)\right], \tag{5.13}$$

where $Z$ is the atomic number of the nucleus with mass $M$, $x_A = 2\Lambda_{\text{QCD}}/A^{1/3}$ with $\Lambda_{\text{QCD}}$ from 100 to 200 MeV, and $A$ the atomic mass number. The dependence of the total cross section on $\Lambda_{\text{QCD}}$ is stronger for lower energies.

EXPLAINING THE MINIBOONE ANOMALY

We now comment on the broader context of the MiniBooNE anomaly and present two choices of model parameters that will help us benchmark the MiniBooNE explanation in the context of dark photon models. Our model interprets the MiniBooNE anomaly as a sequence of upscatter-ing and decay in the active volume of the detector, as shown in the plot on the left of fig. 5.2. The collimated $e^+e^-$ pair is then misidentified as a single shower, producing the excess of events at low energy. While several models exploring this mechanism have been studied, only a small subset of those can also explain LSND[244] due to the much harder-to-fake inverse-beta-decay signature. In that case, a neutral mediator in the neutrino-nucleus scattering process should kick out a neutron from inside the Carbon nucleus. Not only is this a negligible effect for a dark photon mediator, but it also requires larger neutrino energies to produce the heavy neutrinos and remain above the $E_e > 20$ MeV analysis threshold. Because of this, we proceed to present benchmark points that are compatible with the MiniBooNE observation only.

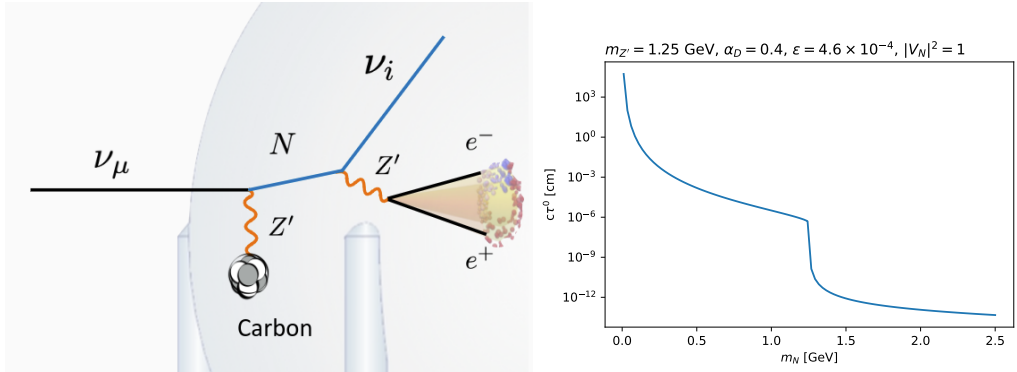

**Figure 5.2:** Left: Dark Neutrinos provide a new interpretation of the MiniBooNE anomaly. Because their lifetime can be tiny, they would be produced directly inside the detector by upscattering an active neutrino into a heavy neutrino. It would travel a distance smaller than the detector size and decay into an $e^+e^-$ pair, collimated enough not to be resolved and thus identified as a single shower by MiniBooNE. Right: The heavy neutrino lifetime spans several orders of magnitude when varying the heavy neutrino mass $m_N$. For very small values of $m_N$, the decay is a three-body decay analogous to the muon decay, as the mediator is much more massive than the heavy neutrino and goes like $c\tau^0 \sim m_{Z'}^4/m_N^5$. A significant jump is visible around $m_N \sim m_{Z'}$: the decay is now a two-body decay with an on-shell $Z'$, which proceeds much faster. In the heavy $Z'$ case, $N$ can be either long- or short-lived, while in the light $Z'$ case, $N$ is always assumed to be prompt.

BENCHMARK A, LIGHT $Z'$.— Benchmark points based on a best-fit to MiniBooNE data have been derived in the literature in the past[221], and have been constrained by the neutrino-electron elastic scattering measurement performed using MINERvA and CHARM data[236]. This work targets the low $N$ mass region of the parameter space, which is not constrained very effectively by MINER$\nu$A due to systematic uncertainties in the background.

BENCHMARK B, HEAVY $Z'$.— This benchmark is inspired by the benchmarks provided in Refs.[223] and[173]. It illustrates the case of a heavy dark photon, where the coherent upscattering contribution is not dominant but still significant.

| Benchmark | (A) Light $Z'$ | (B) Heavy $Z'$ |
|---|---|---|
| $m_N$ (MeV) | 100 | 100 |
| $m_{Z'}$ (GeV) | 0.03 | 1.25 |
| $|V_{\mu N}|^2$ | $8 \times 10^{-9}$ | $2.2 \times 10^{-7}$ |
| $\alpha_D$ | 1/4 | 0.4 |
| $\varepsilon$ | $1.7 \times 10^{-4}$ | $2 \times 10^{-2}$ |
| $|V_N|^2$ | $|V_{\mu N}|^2$ | 1 |
| $c\tau_N^0$ (cm) | $7 \times 10^{-5}$ | 0.54 |
| $N_{\text{upscattering}}^{\text{T2K}}$ | 15.5 | 560 |
| $N_{\text{decay}}^{\text{T2K}}$ | 15.5 | 4.6 |

**Table 5.1:** Parameters of our two benchmark points. These choices are compatible with the excess of events observed at MiniBooNE.

Both choices of parameters above, outlined in table 5.1, can explain the MiniBooNE energy spectrum but not the angular spectrum. A fascinating difference between the two models, light and heavy $Z'$, comes from the resulting lifetime of the heavy neutrino $N$. For the heavy mediator case, the decay $N \to \nu e^+ e^-$ happens through an off-shell $Z'$. Depending on the choice of parameters, the decay rate can be fast or slow, with lifetime changing by several orders of magnitude. However, for the light mediator case, the decay is a sequence of two-body decays $N \to \nu Z' \to e^+ e^-$, which proceed much faster than the heavy case resulting in microscopic lifetimes and prompt decays. The right plot of fig. 5.2 illustrates this effect: the kink at $m_N \sim m_{Z'}$ represents the transition between the off-shell and on-shell decays, with a significant decrease in the lifetime of the model.

The exchange of a dark photon with the nucleus gives rise to very low-$Q^2$ processes and, therefore, very forward $e^+ e^-$ final states. This is in apparent contradiction with the MiniBooNE observation. Quantifying this tension, however, is not currently possible due to the lack of public information on the background systematics in $\cos\theta$. Since systematic uncertainties in the background prediction dominate the significance of the MiniBooNE excess, a proper fit should include the systematic uncertainties in the angular bins and, most importantly, their correlations.

Models with helicity-flipping interactions and heavy mediators, such as scalar mediator models [244], have a better chance of describing the angular spectrum. These models also have the advantage of having larger cross sections with neutrons and being attractive in the context of LSND. Due to their broader angular spectrum, we expect these models to have smaller selection efficiencies at T2K than the ones we find for the dark photon model. We leave the exploration of these models for future work after proper fits to the MiniBooNE excess have been performed.

Many constraints have been posed on the model above, from accelerator neutrino experiments to kaon decays. We note the study of Ref. [266], where the authors point out a large set of experimental observables that can be used to constrain MiniBooNE explanations, among which the ND280 data that we make use of. In our analysis, we properly take into account detector effects and systematics, carefully describing the detector's geometry, which is an essential ingredient to correctly determine the bounds on the heavy $Z'$ case, which are extremely sensitive to the lifetime of the heavy neutrino.

## UV completions

Possible UV completions of Eq. (5.1) have been discussed in Refs. [172,223,173]. The general idea is to consider new fermions $\nu_D$ charged under the new gauge symmetry, which mix with SM neutrinos upon symmetry breaking. Two main categories can be identified, depending on the pattern of the

$U(1)_D$ breaking. Schematically, they make use of the following operators,

$$(\text{I}): (\overline{L}\tilde{H}_D)\nu_D, \tag{5.14}$$

$$(\text{II}): (\overline{L}\tilde{H})(\Phi\nu_D). \tag{5.15}$$

The first route requires new $SU(2)_L$ scalar doublets, $H_D$, also charged under the dark symmetry. The mixing between SM neutrinos and the dark leptons $\nu_D$ is then generated by the expectation value of $H_D$, which breaks the $U(1)_D$, together with the SM Higgs, also the electroweak symmetry. The second method considers instead an SM-singlet dark scalar $\Phi$, whose expectation value breaks only the dark symmetry. The main idea is illustrated by the dimension-five operator in Equation (5.15), which induces a mixing term between $\nu_D$ and the SM neutrinos after symmetry breaking. This effective operator can be easily generated by the exchange of a singlet (sterile) neutrino $\nu_s$, which serves as a bridge between the SM and the dark sector via $(\overline{L}\tilde{H})\nu_s$ and $\overline{\nu}_s(\nu_D\Phi)$.

Note that in both cases, the dark photon gets a mass from breaking the $U(1)_D$, while the masses of the neutral leptons and the additional scalar degrees of freedom will depend on the specifics of the model. The interplay between the expectation values of the new scalars, Yukawa couplings, and new arbitrary Majorana mass terms will determine the coupling vertices $V_{ij}$ in Equation (5.2) and should also generate the correct value for the light neutrino masses. In particular, both model types are flexible enough for $N$ to be a (pseudo-)Dirac or Majorana particle. However, to generate small neutrino masses due to approximate conservation of lepton number, pseudo-Dirac states are preferred. Any additional fermion in the model can be heavier than a few GeV, where their interactions with the SM would be poorly constrained at the values of neutrino mixing we consider.

ND280 is the off-axis near detector of T2K located at 280 m from the target at an angle of ∼ 2.042° with respect to the beam[267]. The mean neutrino energy at this location is very similar to that of the Booster Neutrino Beam, where MiniBooNE is located. The comparison of the two fluxes in Figure 5.3 clearly shows that the flux seen by ND280 is significantly larger than that seen by MiniBooNE for the same exposure. The active mass, however, is much smaller – MiniBooNE contains a total of 818 t of mineral liquid scintillator ($CH_2$) compared with 18 t of total active mass. Considering these two elements, we expect a similar number of upscattering events to happen in the two detectors, enabling T2K to test the dark neutrino interpretation of the MiniBooNE excess directly. Our analysis reinterprets two public results by T2K: the search for the in-flight decays of long-lived heavy neutrinos[183] and the $\nu_e$CCQE cross-section measurement[212]. The former directly searched for appearing $e^+e^-$ pairs inside the low-density region of the detector. At the same time, the latter measured the rate of single photons that convert into $e^+e^-$ pairs inside one of the tracking components of the detector. We discuss the detector components below to then discuss the two analyses.

Our analysis is based on a simplified detector simulation: we implemented the analysis selection criteria and detector geometry on upscattering events generated by our own modified version of the DarkNews generator[268]. More details are given in section 5.3.

### The ND280 detector

ND280 is a highly segmented and magnetized detector. The detector modules that will be used in our work are shown in Figure 5.4, and constitute most of the detector's active volume. The first three modules constitute the PØD detector, a layered arrangement of high-Z material such as lead and brass, intertwined with plastic scintillators and water bags. The latter ones serve as an elec-

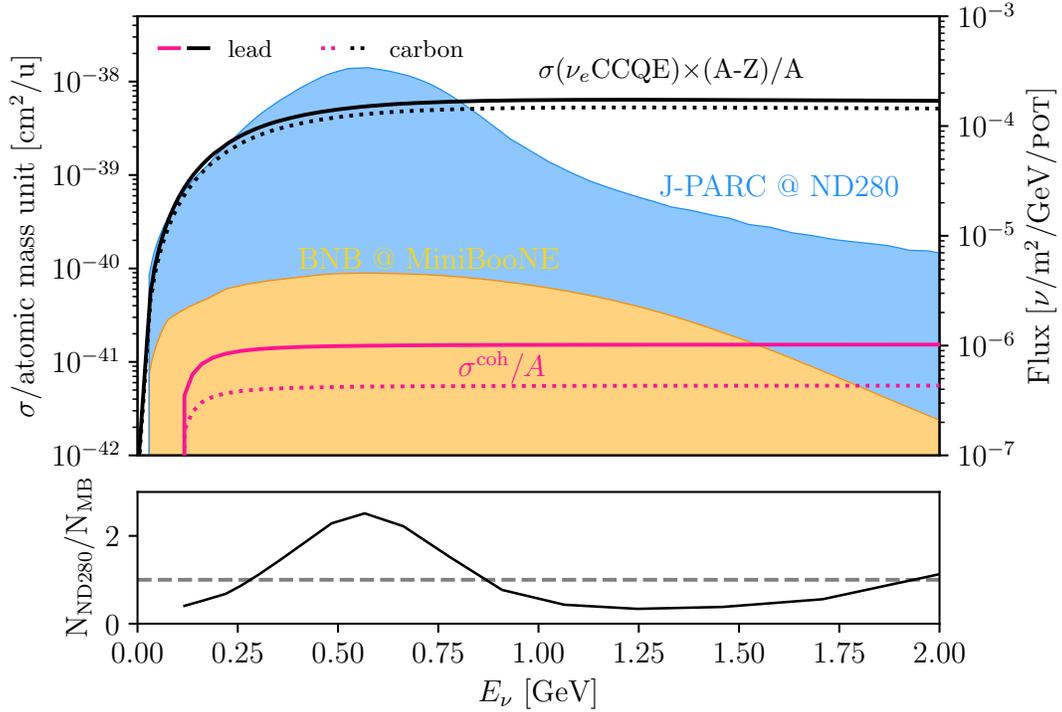

**Figure 5.3:** Comparison between relevant neutrino fluxes and cross sections. Although MiniBooNE has a much larger mass, ND280 benefits from a significantly larger neutrino flux and higher Z materials. The solid and dotted lines show the cross section per atomic mass unit for lead and carbon, respectively. The $\nu_e$CCQE cross sections are shown as black lines, while the coherent upscattering cross sections for our heavy dark photon benchmark (B) as pink lines. Considering all the active material in ND280 and MiniBooNE, we found that the ratio of upscattering between the two experiments is $\mathcal{O}(1)$ across the energy spectrum.

tromagnetic calorimeter (ECAL) and an active water target. This active water target is specially designed to study $\pi^0$ production and neutrino cross sections in water, both critical inputs for the oscillation analyses using the Super-Kamiokande far detector. The first and third ECALs contain only layers of lead and scintillator plates. The module in between contains the water bags and layers of brass and scintillator plates. Downstream we have the tracking modules composed of three gaseous Argon time projection chambers (GArTPC), separated by fine-grained scintillator detectors (FGD). These components have fewer neutrino interactions and provide a better environment for particle identification. The whole detector is in a 0.2 T magnetic field, surrounded by other side and back ECAL and side muon detectors. The dimensions and composition of each region are summarized in Table 5.2 (see Refs. [269,270,271] for more information on ND280). Our simulation assumes that the lead, brass, water, and scintillator layers are distributed uniformly inside each module. Each

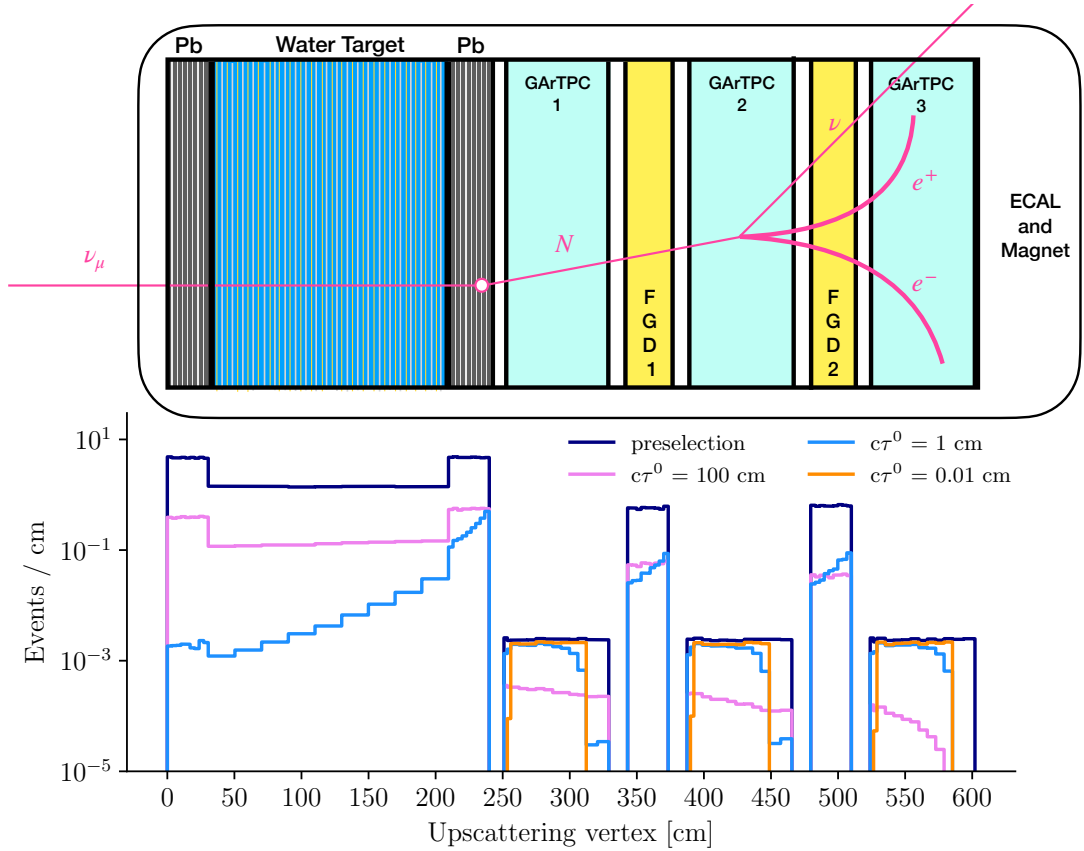

**Figure 5.4:** Diagram of the T2K near detector, ND280, showing all the active components of the detector and the new-physics signature we are interested in. Below, we show the event rate distribution as a function of the upscattering position $z$ before and after geometrical and analysis selection. For long lifetimes ($c\tau^0 \gtrsim 1$ cm), the event rate is dominated by upscattering on lead, while for much smaller values, only upscattering on the gaseous argon modules contributes.

GArTPC is enclosed in a plastic and aluminum cage. The cage, in turn, is composed of an external wall (10.1 kg), an internal volume of $CO_2$ gas, and an internal wall (6.9 kg). Inside the TPC, on the Y-Z plane, one can find the cathode (6 kg, made of 34% C, 25% O, 17% Cu, 17% Si). Even though these materials are the closest to the active volume of the TPC, we neglect them as the total number of neutrino interactions recorded in them of order thousand times smaller than in the other targets.

| ND280 module | Active volume $X \times Y \times Z$ [cm] | $z_{\text{begin}}^{\text{active}}$ | $M_{\text{tot}}^{\text{active}}$ | Fiducial volume $X \times Y \times Z$ [cm] | $z_{\text{begin}}^{\text{fiducial}}$ | $M_{\text{tot}}^{\text{fiducial}}$ | composition (in mass) |
|---|---|---|---|---|---|---|---|
| PoD-ECAL1 | $210 \times 224 \times 30.5$ | 0 | 2.9 | | | | 6.5% H, 40% C, 53.5% Pb |
| PoD-water | $210 \times 224 \times 179$ | 30.5 | 10 | | | | 10% H, 43% C, 22% O, 16% Cu, 9% Zn |
| PoD-ECAL2 | $210 \times 224 \times 30.4$ | 209.6 | 2.9 | | | | 6.5% H, 40% C, 53.5% Pb |
| GArTPC1 | $186 \times 206 \times 78$ | 251 | 0.016 | $170 \times 196 \times 56$ | 256 | 0.010 | 100% Ar |
| FGD1 | $186 \times 186 \times 30$ | 343 | 1.1 | $175 \times 175 \times 29$ | 344 | 0.92 | 8% H, 88% C, 4% O |
| GArTPC2 | $186 \times 206 \times 78$ | 387 | 0.016 | $170 \times 196 \times 56$ | 256 | 0.010 | 100% Ar |
| FGD2 | $186 \times 186 \times 30$ | 480 | 1.1 | $175 \times 175 \times 29$ | 481 | 0.92 | 9% H, 50% C, 41% O |
| GArTPC3 | $186 \times 206 \times 78$ | 524 | 0.016 | $170 \times 196 \times 56$ | 256 | 0.010 | 100% Ar |

**Table 5.2:** The active and fiducial volume dimensions, mass, and composition of each ND280 module simulated in our analysis. No fiducial volume is shown for the P∅D because we do not employ this detector for measuring heavy neutrino decays but only for its production. For the GAr modules, we show each TPC's fiducial volume and mass, taking $\rho_{\text{GAr}} = 1.78 \text{ g/cm}^3$.

## Detector description

We simulate the three subdetectors of ND280: the P∅D, the two FGDs, and the three GAr TPCs. In Table 5.2, we report the sizes of the active volume, where upscaterring occurs, and the fiducial volume, where $e^+e^-$ pairs are detected. We also report total active and fiducial mass, as well as the material composition in mass. We account for gaps between the detector volumes and report the $z$ coordinate along the beam axis, where the active or fiducial volume begins.

Figure 5.5 shows the 2d distribution density of upscattering vertices along the z and x axes for the heavy $Z'$ case, using our benchmark point in parameter space. The three different sections of the P∅D, the three different TPCs, and the two FGDs are distinguishable, with the gaps between volumes.

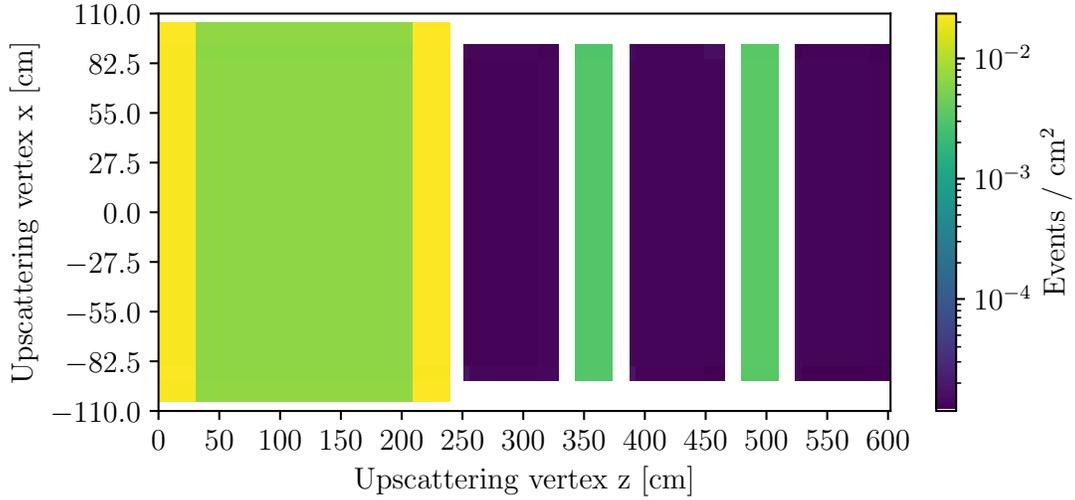

**Figure 5.5:** Distribution of the upscattering vertex across the z-x plane, for the heavy $Z'$ case, using our benchmark point. The color bar shows the number of upscattering events per bin

### Analysis I – heavy neutrino searches in the GAr TPC

The analysis in [183] looked for the decay in flight (DIF) of heavy neutrinos inside the three GAr TPCs. Heavy neutrinos are produced in the target through mixing between active and sterile neutrinos, and, after propagating from the target to the detector, they decay in the detector TPCs. They look for multiple final states, although the relevant one for our analysis is $N \to \nu e^+ e^-$, which also extends to the lowest masses, with a threshold of $m_N \sim 1\,\mathrm{MeV}$. This analysis benefits from a clear signature with zero background, as the total Argon mass is so tiny that neutrino interactions inside the TPC do not produce a relevant background. In order to achieve the zero background, the original selection imposes a tight fiducial volume cut in the TPC, with a requirement of no additional visible energy deposition in the detector in addition to the charged particles produced in the TPC. Our model can be tested with this analysis because it predicts a sizeable coherent cross section, resulting in a very low-energy nuclear recoil that is invisible in these detectors.

This analysis is a counting experiment performed over $12.34 \times 10^{20}$ proton on target (POT)

in neutrino mode and $6.29 \times 10^{20}$ POT in anti-neutrino mode. Observing zero $e^+e^-$ events over the neutrino background expectation of 0.563 in neutrino mode and 0.015 in anti-neutrino mode sets firm limits on new physics. A histogram-like representation of the data for the data collected in neutrino mode is shown in the plot on the left of fig. 5.8. The expected signal is shown for a larger value of $\varepsilon$ with respect to our best-fit point to make the signal more clearly visible. Limits on long-lived heavy neutrinos produced in kaon decays at the target have been discussed in Ref. [183,1].

For dark neutrino models, those can be recast, considering the production of $N$ via upscattering inside the detector. In particular, we will focus on parameters such that the lifetimes of $N$ are not much larger than $\mathcal{O}(10)$ m, as otherwise, these particles would not provide an excellent fit to MiniBooNE as well. If $N$ propagates more than $\mathcal{O}(15)$ cm, it can be produced via upscattering in the dense material of the PØD, where the cross section is significantly enhanced due to the coherent scaling with proton number, $Z^2$. It would then decay into a visible $e^+e^-$ pair inside one of the TPCs. This particular signature is present in the most interesting parameter space for the heavy mediator case. When the lab-frame lifetime becomes shorter than $\mathcal{O}(15)$ cm, the heavy neutrinos produced in the PØD decay before entering the TPCs, and therefore the $e^+e^-$ are rejected by the selection to avoid large neutrino-induced backgrounds. Nevertheless, the upscattering can happen inside the TPCs, where they would be visible. Despite the relatively small number of targets in the TPC fiducial volume, a handful of events is enough to constrain the model due to the absence of backgrounds. Such fast decays always happen in the light dark photon case ($m_{Z'} < m_N$), but it can also happen in regions of large $|V_N|^2$ values of the heavy dark photon parameter space. The left plot of fig. 5.6 shows the fraction of heavy neutrino decaying in one of the three TPCs as a function of the proper lifetime, for the case the upscattering happens in the lead or the argon.

We expect differences in the reconstruction and selection efficiencies for upscattering with respect to the decay-in-flight signatures considered in Refs. [183,1]. Figure 5.7 shows the comparison between standard heavy neutrino signatures and our scattering-induced signatures for both the heavy and the light mediator case. The reconstruction efficiency depends on the kinematics of the heavy neutrino decay products, including the detector capability to separate the $e^+e^-$ tracks as a function of their opening angle and their distance of closest proximity. The selection efficiency relies on the cuts described in eq. (5.16), which we implemented in our analysis. No smearing of the kinematics of the electron and positron has been applied at this stage; however, the resolution effect in the GArTPC measurement would likely have a small impact on a zero-background search.

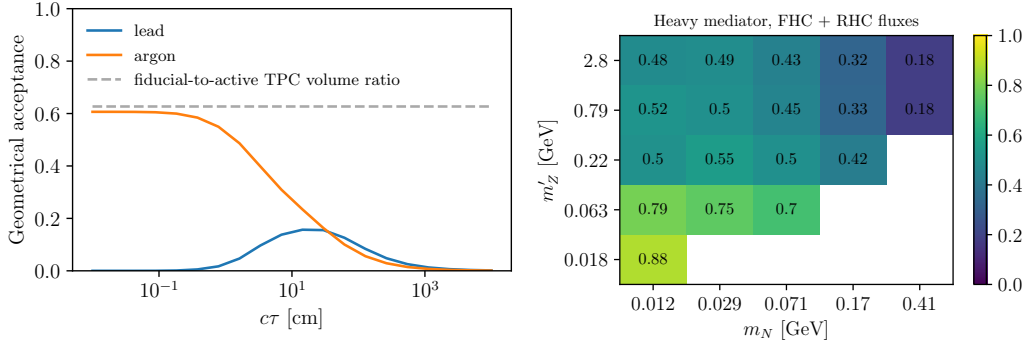

**Figure 5.6:** Left: the geometrical acceptance of ND280 as a function of the $N$ proper lifetime. For the smallest lifetimes, only the GArTPCs can pick up the decays. Right: Selection efficiency resulting from the kinematical cuts in eq. (5.16), as a function of the parameter space $m_{Z'}$ - $m_N$. The efficiency varies between $0.88$ and $0.18$ in the parameter space under consideration. This variation does not have a significant impact on the final result.

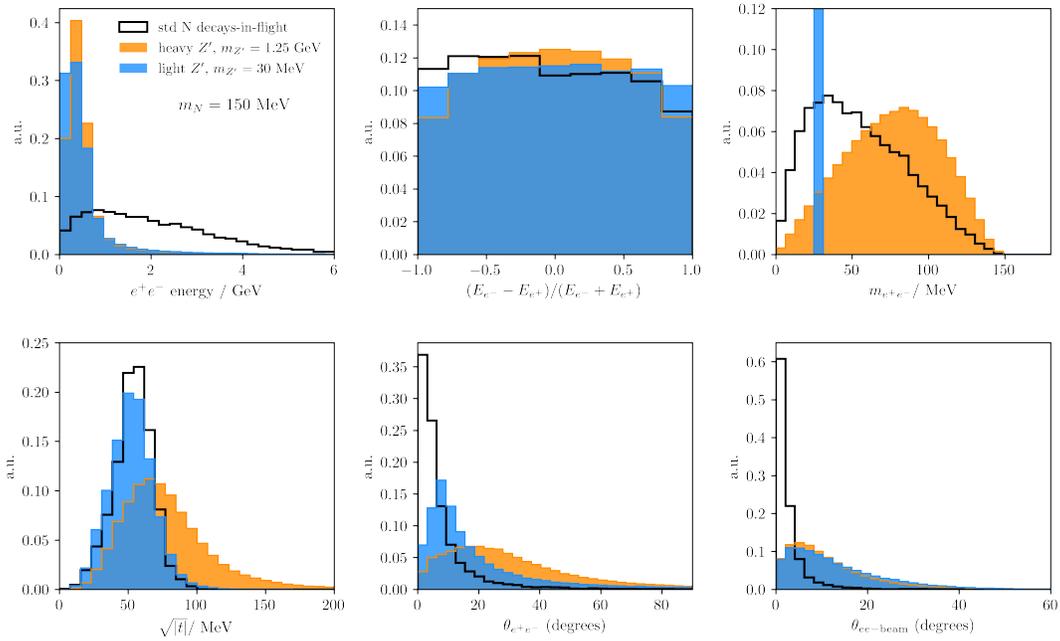

**Figure 5.7:** Comparison between standard decay-in-flight $N \rightarrow \nu e^+ e^-$ signatures of heavy neutral leptons produced at the target (solid black) and that of heavy neutrino decays initiated by coherent scattering in a light dark photon (filled blue) and heavy dark photon (filled orange) model. Histograms are area-normalized. The variables are defined in eq. (5.16).

$$E_{e^+e^-} \equiv E_{e^+} + E_{e^-} > 0.150 \text{ GeV}, \tag{5.16a}$$

$$m_{e^+e^-} \equiv \sqrt{(p_{e^+} + p_{e^-})^2} < 0.7 \text{ GeV}, \tag{5.16b}$$

$$|t| \equiv (E_{e^+e^-} - p_{e^+e^-}^z)^2 - |\vec{p}_{e^+e^-}^T|^2 < 0.03 \text{ GeV}^2, \tag{5.16c}$$

$$\cos\theta_{e^+e^-} \equiv \frac{\vec{p}_{e^+} \cdot \vec{p}_{e^-}}{|\vec{p}_{e^+}||\vec{p}_{e^-}|} > 0, \tag{5.16d}$$

$$\cos\theta_{ee-\text{beam}} \equiv p_{e^+e^-}^z / p_{e^+e^-} > 0.99. \tag{5.16e}$$

The plot on the right of fig. 5.6 shows a map of the efficiency for these cuts by varying $m_N$ and $m_{Z'}$ across parameter space, which are the only two parameters that impact the kinematics and not simply the total rate. The efficiency is about 50% for our benchmark point, and it is independent of $m_{Z'}$ for large values of $m_{Z'}$, where $m_{Z'}$ only determines the total rate, similar to the $W$ mass in the muon decay. However, it grows to almost 90% at lower values of $m_{Z'}$, while it decreases to $\sim 20\%$ at larger $m_N$. All these effects are taken into account when scanning the parameter space. Moreover, given that the efficiency in the original heavy neutrino analysis is of the order $10 - 15\%$, we applied an additional 10% efficiency factor to consider reconstruction effects conservatively.

We also perform a sensitivity study, projecting this analysis's status by the end of the T2K data. The current analysis could be extended to about $4 \times 10^{21}$ POT which the experiment has already collected. Moreover, ND280 is currently being upgraded to a new configuration [184]: the PØD is being replaced by two new GArTPCs and a Super FGD module. A future search post-upgrade, looking only at upscattering inside the Argon, could be performed on the 3 TPCs, plus the two new TPCs, on a forecast of $16 \times 10^{21}$ POT[185]. This conservative estimate neglects improvements to reconstruction and background rejection and a benefit of a tailored analysis for this model.

### Analysis II – photons in the FGD

The second analysis uses the photon-like control sample of the $\nu_e$CCQE cross-section measurement in the first FGD. Even though it focuses on a very different measurement, it can provide a helpful constraint for our model. The largest background for this analysis comes from photons that convert inside of the FGD and for which one of the two particles has not been reconstructed. In order to better measure this background, they look at a specific sideband, selecting $e^+e^-$ in the FGD in the same way they select single electrons or positrons for the primary measurement. The $e^+e^-$ invariant mass is a helpful quantity for them to select real photons. It can be used to constrain the dark neutrino signal in the case of a light mediator. The $Z'$ is produced on-shell and decays promptly to an $e^+e^-$ pair, which, if reconstructed correctly, shows a peak in the invariant mass spectrum at $m_{ee} = m_{Z'}$. The right plot of fig. 5.8 shows an example of the measured $m_{ee}$ spectrum. We implement the smearing of momenta and zenith angles based on the 2D histograms provided in Ref.[272]. Figure 5.9 shows the resulting smearing matrices for electron/positron momentum, angle, and the invariant mass of the pair. In this case, the resolution is significantly larger than for the GArTPCs. However, it is essential for a correct sensitivity estimate because of the extensive background from single-photon conversions that spreads on the entire spectrum. We consider a flat 10% reconstruction and selection efficiency, which considers the 30% efficiency for the $\nu_e$CCQE analysis, squared to account for the two leptons. We also estimate the sensitivity of a projection of this analysis by expanding to two FGDs, with a larger dataset and including the SuperFGD after the upgrade.

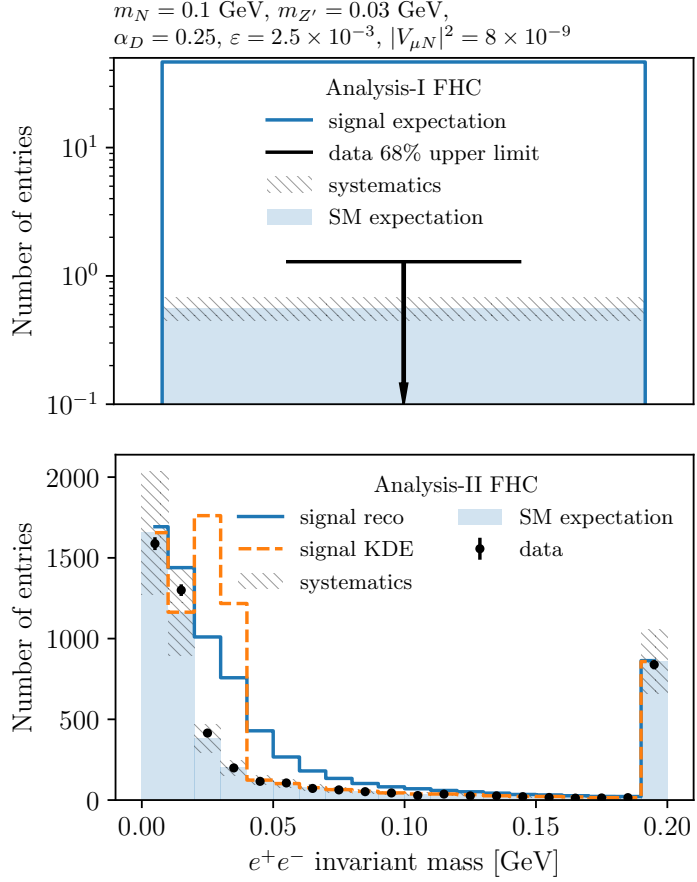

**Figure 5.8:** The two plots show the data used in Analysis-I (first panel) and Analysis-II (second panel) in neutrino mode (the plots for the anti-neutrino mode are analogous). The expected signal is shown for a larger value of $\varepsilon$ with respect to our best-fit point to make it more clearly visible. Analysis-I is a one-bin experiment, while Analysis-II is a search for a resonance on the $e^+e^-$ invariant mass spectrum. The shaded blue region represents the expected SM background, while the black points show the observed data. No event was observed in Analysis-I; therefore, we display the upper limit at 68% CL. The blue lines show the signal we expect to observe for the light $Z'$ benchmark point, modified with a larger $\varepsilon$. The orange line illustrates the signal at the true level, smeared only by our KDE interpolation. It cannot be compared with the data but gives a sense of the effect of the experimental resolution in measuring $e^+e^-$ pairs in the FGD.

## 5.3 Exploring the parameter space

Extended dark sectors like the ones considered in this work typically involve a large number of independent parameters. This observation poses a challenge from the phenomenology point of view.

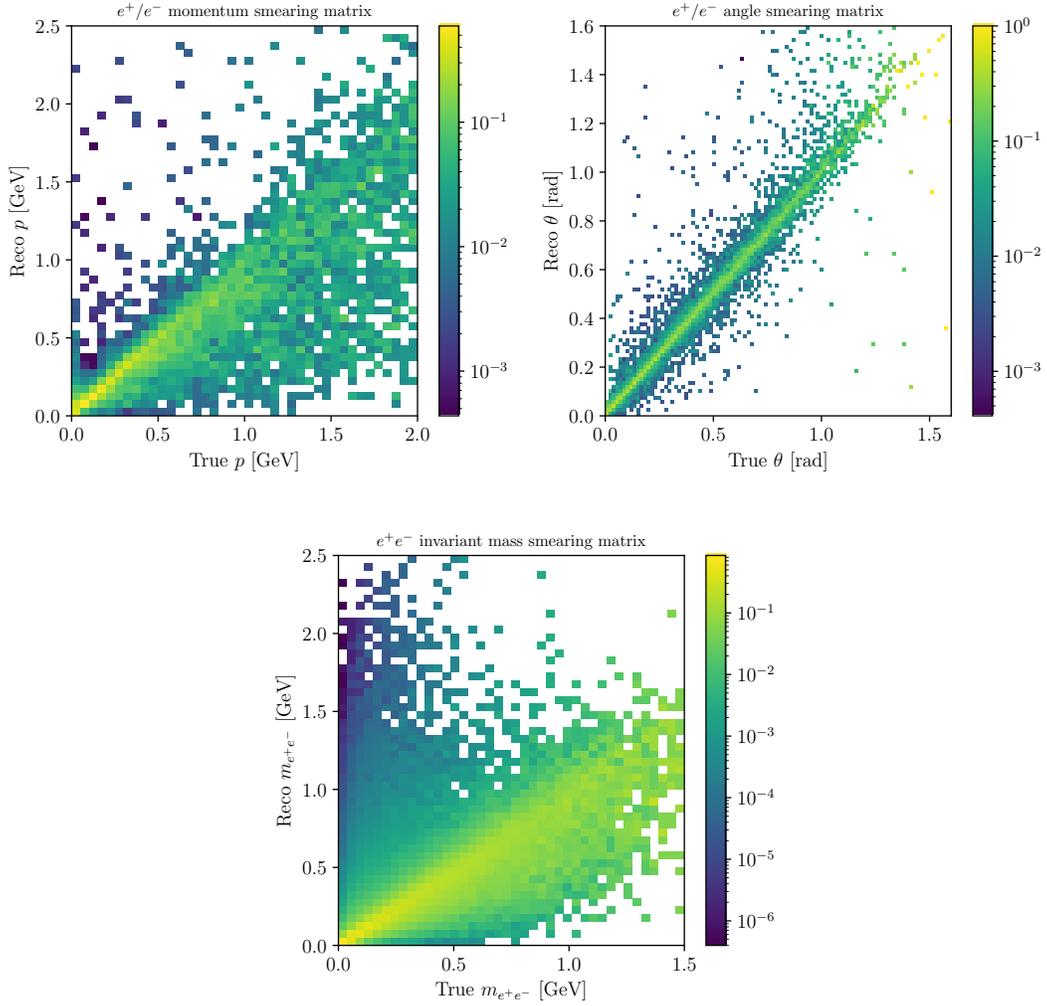

**Figure 5.9:** These smearing matrices show the distributions of a reconstructed quantity as a function of its actual value for the electron (or positron) momentum, the polar angle with respect to the beam axis, and the $e^+e^-$ invariant mass which incorporates both smearing effects. These smearing matrices are normalized such that the content of each vertical column sums to one. All color scales are logarithmic, emphasizing the tails of the distributions; however, the resolution is excellent, and the bulk of these distributions lie along the diagonal lines. 2D distributions of reconstructed and true electron momentum $p$ (left) and polar angle $\hat{\vartheta}$ (right), as employed by T2K in the sideband of analysis-II[212]. These smearing matrices are obtained from the distribution shown in [212] and provided as open data in [272], and employed in our recast of analysis-II. No smearing on the azimuthal angle $\phi$ and no correlations between the momentum and the angular resolutions are considered; however, we expect these contributions to be small.

Performing inference in this ample parameter space requires predicting distributions of observables for all possible choices of parameters. It can be costly and quickly become unfeasible for more detailed simulations. For this reason, promptly predicting our signal is crucial to improving our coverage of the model parameter space. We apply existing statistical methods to interpolate model predictions across the parameter space, allowing the computation of the model prediction from a single simulated sample. Fast interpolation of physical predictions across the parameter space has been discussed in the context of event generators for colliders[273,274] as well as in fast generators of dark matter direct detection signatures[275]. However, these methods rely on the prediction's parametrization in terms of analytic functions. Our technique complements these methods by deriving a non-parametric estimate of the observables. A similar approach to the one discussed here has been proposed for treating nuisance parameters and systematic uncertainties in IceCube[276]. The IceCube scheme overcomes the curse of dimensionality of the production of many distinct Monte Carlo samples, sometimes described as the "multiple Universes" approach. While the IceCube method derives a non-parametric estimate of the observables as a function of the nuisance parameters in the neighborhood of the central value, our method applies to the full parameter space.

## General idea

We can think of a model as a family of probability density functions (PDF) $p(x|\theta)$, where $\theta$ are the physical parameters of the theory over which we want to perform inference, like masses and couplings, whereas $x$ are observables, like particle momenta. The model also predicts a normalization factor $\mathcal{N}(\theta)$: not just the observable distribution depends on the parameter, but also the total rate. Both $\theta$ and $x$ are multi-dimensional, varying from several to $\mathcal{O}(10)$ dimensions.

Inference is performed by computing the expectation value $E_\theta[T(x)]$ of a test statistic $T$ for each value of $\theta$. The typical approach proceeds as follows: *i)* start from an initial definition of a multi-dimensional grid of a total of $m$ points in the parameter space $\theta_{j=1,\ldots,m}$, *ii)* run a simulation for each $\theta_j$, *i.e.*, draw $n_j$ samples $x_{i=1,...,n^j} \sim p(x|\theta_j)$, *iii)* compute the expectation value for each $\theta_j$ as:

$$\mathrm{E}_{\theta_j}[T(x)] = \sum_{i=1}^{n^j} w_i^j T(x_i^j),  \tag{5.17}$$

where $w_i^j$ are weights associated with the sampling, such as importance-sampling weights. Finally, *iv)* one eventually interpolate these values across the parameter space, in order to predict $\mathrm{E}_{\bar{\theta}}[T(x)]$ for a $\bar{\theta}$ which has not been simulated.

While the method above works, our procedure provides a more efficient way to interpolate the expectation. It allows us to rapidly compute multiple and more complex test statistics, like histograms, using a single set of samples, *i.e.*, running only one simulation. We promote $\theta$ to a random variable by considering $p(x, \theta) = p(x|\theta)\mathcal{N}(\theta)q(\theta)$ where $q(\theta)$ is a prior over $\theta$, and we sample $x_i, \theta_i \sim p(x, \theta)$ with weights $w_i$, for $i = 1, ..., n$. Using these samples, we obtain $\mathrm{E}_{\bar{\theta}}[T(x)]$ by interpolating across the parameter space using Kernel Density Estimation:

$$\begin{aligned}
\mathrm{E}_{\bar{\theta}}[T(x)] &= \sum_i^n w_i T(x_i) \frac{w(\bar{\theta}, \theta_i)}{q(\theta_i)} \\
&= \sum_i w_i T(x_i) \frac{K(d(\bar{\theta}, \theta_i), \hat{\partial})}{q(\theta_i)},
\end{aligned} \tag{5.18}$$

where $K(d, \hat{\partial})$ is a Kernel function, $d(\bar{\theta}, \theta_i)$ is a distance in parameter space, and $\hat{\partial}$ is the bandwidth or smoothing parameter. By sampling over parameter space, we exploit the fact that neighbor parameters will produce similar observable distributions. Using importance adaptive sampling or Markov Chain Monte Carlo, this method will guarantee to sample observables with significant contribution to any test statistic, *i.e.* large weights. However, the adaptation over the parameter space will also make to sample parameters where they result in larger weights. The function $q(\theta)$ allows us to control this effect and skew the distribution of samples towards our preferences. For exam-ple, if performing inference by conditioning on the posterior, *e.g.* when setting limits on a slice of the parameter space, fixing a subset of the total parameters, and varying the other ones, we might be in a region where there are no samples, as that slides contains a small probability with respect to the total model. A sketch of this method is given on the left of fig. 5.10, where we summarized the many parameters $\theta$ in a single axis and the many experimental variables in the single axis $x$ in order to illustrate the method in a with a three-dimensional graph.

Finally, despite the power of this interpolation, it introduces some statistical uncertainty related to the finite sample size. Whenever a close formula for $w(\bar{\theta}, \theta_i)$ is present, it is more effective to use that. For this reason, we split the parameter space into $\theta^\alpha$, for which a KDE weight is computed, and $\theta^\beta$, for which the weight is computed through an analytical formula.

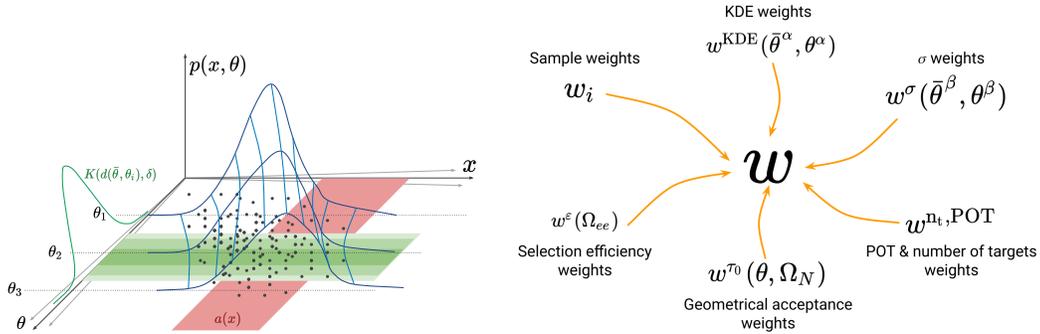

**Figure 5.10:** Left: The method employed in this work requires extending the probability density function $p(x|\theta)$ (dark blue, for different values of $\theta$) to a 2D distribution $p(x, \theta)$ (light blue). Using this sample, we can now sample events from this 2D distribution (black dots) and compute expectations at different values of $\theta$. We interpolate across parameter space using a kernel (green) that weighs events depending on their proximity to the target value. At the same time, we apply an acceptance or selection function to the observables (red). Right: To quickly compute the model prediction across the parameter space, we implement a set of weights that accounts for the simulation of the cross section, the kinematics, and detector effects. By taking the product of all these weights, we obtain a single weight that can be used to compute the final model prediction for any observable.

We use this framework in the context of the dark neutrino model discussed in the previous sections. In this model, $\theta$ is a 5-dimensional parameter space for the light mediator case and 6-dimensional for the heavy case since $|V_N|$ is only a relevant variable for the latter. More precisely, $\theta^\alpha = \{m_N, m_{Z'}\}$, while $\theta^{\beta,\text{light}} = \{|V_{\mu N}|^2, \alpha_D, \varepsilon\}$, while $\theta^{\beta,\text{heavy}} = \theta^{\beta,\text{light}} \cup \{|V_N|^2\}$. The product of the differential cross section in the observable space $x = \{\Omega_N, \Omega_{ee}\}$ times the neutrino flux $\phi(E_\nu)$ is our $p(x|\theta)$, as

$$p(x|\theta) = \frac{d\sigma}{d\Omega}\phi(E_\nu) = \phi(E_\nu)\frac{d\sigma}{d\Omega_N}(\Omega_N|\theta, E_\nu)\frac{1}{\Gamma(\theta)}\frac{d\Gamma}{d\Omega_{ee}}(\Omega_{ee}|\Omega_N, \theta), \quad (5.19)$$

where $\Omega_N$ describes the kinematics of the heavy neutrino and the nuclear recoil, while $\Omega_{ee}$ describes the kinematics of the $e^+e^-$ pair and the final state neutrino. However, we are not interested in the degrees of freedom of the recoil and the final state neutrino, so we implicitly integrate over those variables. Here, $\Gamma$ is the total decay width of the heavy neutrino and is given by

$$\Gamma(\theta) = \int d\Omega_{ee}\frac{d\Gamma}{d\Omega_{ee}}(\Omega_{ee}|\Omega_N, \theta), \quad (5.20)$$

and can be computed analytically using Equation (5.9) and Equation (5.10) for the heavy and light scenarios, respectively. Finally, an essential parameter for the simulation of the events in the detector is the heavy neutrino lifetime in its rest frame $\tau^0(\theta) = \hbar/\Gamma(\theta)$.

## Monte Carlo Event Generator

We implemented the physics matrix elements in a Monte Carlo event generator, and we sample events using the Vegas Monte Carlo algorithm [277,278] with its Python implementation [279].

In a typical simulation, we would use Vegas to sample the following integral

$$\int dE\phi(E_\nu)\sigma(\theta, E_\nu) = \int E_\nu\phi(E_\nu) \int d\Omega_N \frac{d\sigma}{d\Omega_N}(\Omega_N|\theta, E_\nu) \int d\Omega_{ee} \frac{1}{\Gamma(\theta)} \frac{d\Gamma}{d\Omega_{ee}}(\Omega_{ee}|\Omega_N), \quad (5.21)$$

that we now extend to

$$\int d\theta^\alpha q(\theta^\alpha) \int E_\nu\phi(E_\nu) \int d\Omega_N \frac{d\sigma}{d\Omega_N}(\Omega_N|\theta, E_\nu) \int d\Omega_{ee} \frac{1}{\Gamma(\theta)} \frac{d\Gamma}{d\Omega_{ee}}(\Omega_{ee}|\Omega_N), \quad (5.22)$$

where $\theta^\alpha = \{m_N, m_{Z'}\}$. We use $q(\theta^\alpha)^{\text{light}} = m_{Z'}^2/m_N^{3.5}$ and $q(\theta^\alpha)^{\text{heavy}} = m_{Z'}^8/m_N^5$, designed to provide samples distributed as uniformly as possible in the $\{m_N, m_{Z'}\}$ plane.

We employ the total number of selected events in a single-bin analysis as test statistics,

$$\mu(\theta) = \sum_k n_t^k \times \text{POT} \times$$
$$\times \phi(E_\nu) d\Omega_N \frac{d\sigma}{d\Omega_N}(\Omega_N|\theta, E_\nu) \times$$
$$\times \frac{1}{\Gamma(\theta)} d\Omega_{ee}(\Omega_{ee}|\Omega_N)\varepsilon(\Omega_{ee})a(\Omega_N, \Gamma(\theta)), \quad (5.23)$$

where the cross section has been multiplied by $n_t^k$, the number of targets for each material, indexed by $k$, and by the collected beam exposure in terms of protons on target (POT). We also folded in the selection efficiency $\varepsilon(\Omega_{ee})$ and detector acceptance $a(\Omega_N, \Gamma(\theta))$, which depends on the kinematics as well as on the lifetime of the heavy neutrino. Both functions can be computed as multidimensional cuts on the observables.

By introducing weights for parameters $\theta^\alpha$ and $\theta^\beta$ such that:

$$E_{\bar\theta}[T] = \int dE_\nu d\Omega_N d\Omega_{ee} T(x) \phi(E_\nu) \frac{d\sigma}{d\Omega_N}(\Omega_N | \bar\theta, E_\nu) \frac{1}{\Gamma(\bar\theta)}(\Omega_{ee}|\Omega_N)$$

$$= \sum_i^n w_i T(x_i) w_i^{\mathrm{KDE}}(\bar\theta^\alpha, \theta_i^\alpha) w_i^\sigma(\bar\theta^\beta, \theta_i^\beta), \tag{5.24}$$

and by defining $\varepsilon(\Omega_{ee,i}) = w_i^\varepsilon(\Omega_{ee,i})$, $a(\Omega_{N,i}, \Gamma(\theta_i)) = w_i^{\tau_0}(\theta_i, \Omega_{N,i})$, and $n_t^{k_i} \times \mathrm{POT} = w_i^{n_t, \mathrm{POT}}$ if event $i$ is generated on material $k_i$, we can rewrite Equation (5.23) as a product of weights:

$$\mu(\theta) \simeq \sum_i^n w_i w_i^{\mathrm{KDE}}(\bar\theta^\alpha, \theta_i^\alpha) w_i^\sigma(\bar\theta^\beta, \theta_i^\beta)$$

$$w_i^\varepsilon(\Omega_{ee,i}) w_i^{\tau_0}(\theta_i, \Omega_{N,i}) w_i^{n_t, \mathrm{POT}}. \tag{5.25}$$

For simplicity, we will discuss the method by writing the expectation for a single-bin analysis, which is the case for Analysis-I. Binned analyses, like Analysis-II, represent a trivial extension.

### Multidimensional re-weighting scheme

We now discuss the weights appearing in Equation (5.25). The plot on the right of fig. 5.10 summarizes the different aspects of the re-weighting scheme.

Cross-section KDE weights.— The cross section has a non-trivial dependence on $\theta^\alpha = \{m_N, m_{Z'}\}$. We define the KDE weight as

$$w_i^{\mathrm{KDE}}(\bar\theta^\alpha, \theta_i^\alpha) = K(d(\bar\theta^\alpha, \theta_i^\alpha), \delta)/q(\theta_i^\alpha). \tag{5.26}$$

We studied the accuracy of different kernels, distance functions, and smoothing parameters by comparing the interpolation on a benchmark grid with a dedicated, high-statistics sample for different values of $\theta$. In the rest of the work, we used the Epanechnikov kernel, a logarithmic distance, and $\hat{\delta} = 0.005$ along the direction of both parameters, in an uncorrelated way.

CROSS-SECTION ANALYTICAL WEIGHTS.— The up-scattering cross section is proportional to $|V_{\mu N}|^2 \alpha_D (e\varepsilon)^2$, as seen by squaring the amplitude in Equation (5.4). We implement a trivial scaling along with the parameters $\theta^{\cancel{\theta}}$ allowing a quick re-weight of the events in this parameter space. In this case, we define:

$$w_i^\sigma(\overline{\theta^{\cancel{\theta}}}, \theta_i^{\cancel{\theta}}) = \frac{|V_{\mu N}|^2 \alpha_D (e\varepsilon)^2}{(|V_{\mu N}|^2 \alpha_D (e\varepsilon)^2)_i}, \tag{5.27}$$

where the parameters $\overline{\theta^{\cancel{\theta}}}$ are fixed in the entire simulation and so independent of $i$, but could, in principle, be varied as well.

RECONSTRUCTION AND SELECTION EFFICIENCY WEIGHTS.— We implemented $\varepsilon(\Omega_{ee})$ as a function that is 0 for the events which are not selected and $\overline{\varepsilon}$ for the events that are selected. For Analysis-I, this weight is $\overline{\varepsilon} = 10\%$ and the selection follows Equation (5.16). For Analysis-II, this weight is a flat $\overline{\varepsilon} = 10\%$ for every event.

LIFETIME RE-WEIGHTING.— This weight applies only to the heavy case, where lifetimes span multiple orders of magnitude, while the light case always leads to a prompt decay ($c\tau^0 \leq 0.1$ cm). In this latter case, we simulate interactions directly in the fiducial volume. The easiest way to compute the acceptance for different lifetimes is to sample a number from the exponential distribution with scale parameter equal to the $N$ lab-frame lifetime, propagate $N$ to the detector, and accept or reject the event if the decay point happens within the TPC fiducial volume. However, this method has an important drawback as it produce small effective sample sizes, especially at short lifetimes, where most interactions from the P∅D will not make it to the detector. To avoid this issue, we instead account for the geometrical acceptance by multiplying by a lifetime-weight, which is equal to the integral of the trajectory within the TPC weighted by the exponential distribution. The trajectory of the heavy neutrino in the lab frame in the event $i$ enters and exits each of the three different TPC at points $(a_i^j, b_i^j)$, where $j = 0, 1, 2$ is the TPC index. If the heavy neutrino never enters a given TPC, we can take both numbers as infinity. For each event, we can compute $(\beta\gamma)_i = p_i/m_N$, and given a value of the lifetime in the proper frame $c\tau_0$, the lifetime weight is computed as:

$$w_i^{\tau_0}(\theta, \Omega_N) = \sum_j \int_{a_i^j}^{b_i^j} \frac{ds}{(\beta\gamma)_i c\tau_0} e^{-s/(\beta\gamma)_i c\tau_0}$$
$$= \sum_j (e^{-a_i^j/(\beta\gamma)_i c\tau_0} - e^{-b_i^j/(\beta\gamma)_i c\tau_0}). \tag{5.28}$$

POT and number of targets weights.— This re-weighting is the most trivial , as it depends only on these multiplicative factors

$$w_i^{n_t, POT} = n_t \times POT, \tag{5.29}$$

and can be computed on the fly, in order to easily change beam exposure and target material and mass.

## Likelihood evaluation

We compute a Poisson likelihood of the observed data ($N_{obs}$) given the expectation, summing up the expected background $b$ and the signal $\mu(\theta)$ across the parameter space. We account for systematic uncertainties by using the *effective likelihood* framework [280], which provides an analytic formula to marginalize over systematic uncertainties:

$$\mathcal{L}(\theta) = \mathcal{L}^{eff}(\theta | N_{obs}, b + \mu(\theta), \sigma^2(b, \mu(\theta))), \tag{5.30}$$

where

$$\sigma^2(b, \mu(\theta)) = \sum_i w_i^2 + (b + \mu(\theta))^2 * \eta^2, \qquad (5.31)$$

accounts for systematic uncertainties. The first addend accounts for the finite sample size. In contrast, the second includes the analysis systematics (*e.g.*, flux, and cross section), using the fractional systematic uncertainties published with the analysis, which are typically close to a flat 20%. This formula can be easily extended to a multi-bin analysis by taking the product of the likelihood for each bin. We sum the likelihood together when combining different analyses, like the TPC search and the FGD sideband. When computing projections, we scale the signal and the background proportional to the number of targets and the POT, and we assume $N_{obs} = int(b + \mu(\theta))$, where $int()$ is just approximating to an integer number.

## 5.4 DARK-NEW CONSTRAINTS

Given that neither analysis observed any excess of events with respect to the background prediction, they can be used to constrain the parameter space of the dark neutrino model. Analysis-I allows to set constraints on the long-lived heavy case by looking at upscattering in the PØD and decay in the TPCs, and on the short-lived case (both heavy and light) when the upscattering and decay happen directly in the TPCs. On the contrary, Analysis-II constrains only the light case by looking for peaks in the $e^+e^-$ invariant mass spectrum. We show limits on interesting slices of the parameter space, discuss the contribution of each analysis, and conclude with prospects for the future, both of this model and the methodology.

## LIMITS IN THE RELEVANT PARAMETER SPACE

Our novel method allows us to interpolate the prediction of physical observables across parameter space using a single batch of simulated events and set limits in arbitrary slices of the parameter space. While our methodology allows setting constraints on arbitrary slices of the parameter space, we show our limits for two particular projections: the $m_N - |U_{\mu N}|^2$ plane, describing the heavy neutrino properties, and the $m_{Z'} - \varepsilon$ plane, describing the dark photon properties. We compute the likelihood on the plane by summing up the negative log-likelihoods of the relevant analyses, subtracting the minimum, and tracing the contour at constant likelihood $= 2.3$, which produces regions of exclusion at 90% C.L.

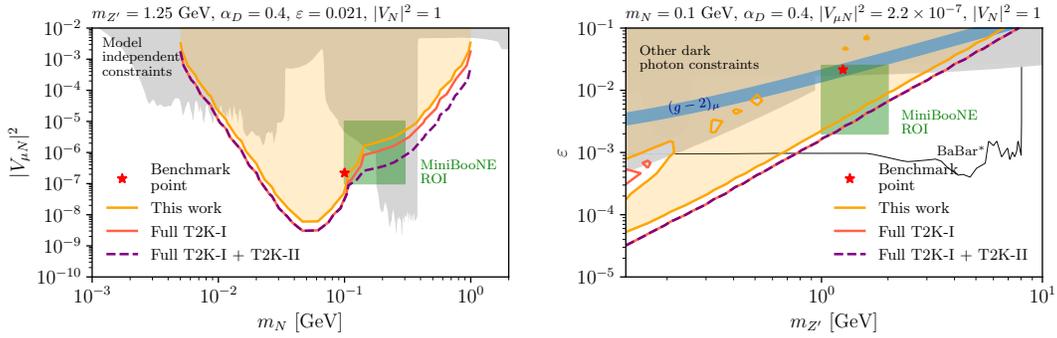

**Figure 5.11:** Limits on the dark neutrino model for a scenario with a heavy dark photon, where $c\tau_N^0$ is typically greater than centimeters. On the left, we show the limits on $|V_{\mu N}|^2$ as a function of $m_N$ and on the right on $\varepsilon$ as a function of $m_{Z'}$, choosing the remaining parameters according to benchmark (B). The MiniBooNE region of interest (ROI) is shown as a large green area surrounding the benchmark point in [223].

Figure 5.11 shows the limits for the heavy case. When the two parameters on the plot are varied, all the others are kept fixed at their benchmark values. In the plot on the left, vertical lines have a constant lifetime, while horizontal lines have a constant upscattering rate. By folding in the geometrical acceptance seen in fig. 5.6, we can explain the shape of the curves seen in the plot. Given that no fit of this model to the MiniBooNE data has been performed for the heavy case, we consider a region of interest around the benchmark point, while for the light case, we consider the best-fit region from [221]. T2K data strongly constrains these particles as explanations of the MiniBooNE excess but does not entirely rule them out for sufficiently short lifetimes. Similar conclusions can be drawn from the plot on the right: diagonal lines parallel to the edge of the excluded region have a constant lifetime, while lines more or less perpendicular have a constant upscattering rate (which scales like $\sim \varepsilon^2/m_{Z'}^4$). A complete discussion on the implications of models predicting this type of signature as a function of the production rate and the lifetime is extrapolated in section 5.4.

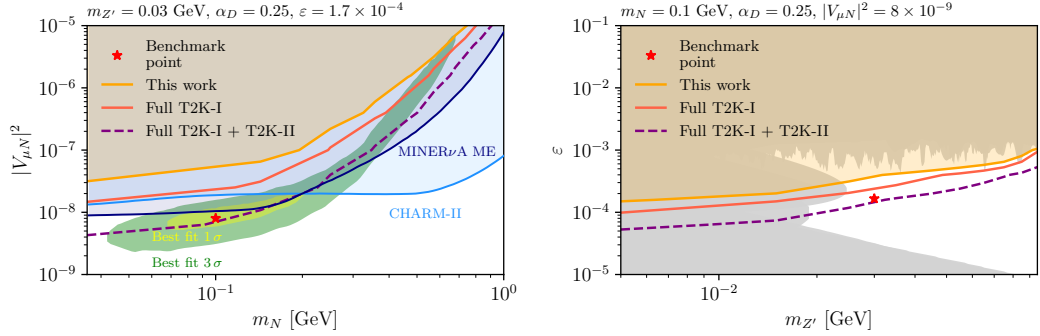

**Figure 5.12:** Limits on the dark neutrino model for a scenario with a light dark photon, where $N$ decays are prompt. On the left we show the limits on $|V_{\mu N}|^2$ as a function of $m_N$ for $m_{Z'} = 30$ MeV and on the right on $\varepsilon$ as a function of $m_{Z'}$ for $m_N = 100$ MeV. We show the allowed region from Ref.[221] on the left. Limits on the light and visible dark photon have been obtained from[281]. Sentence about the little islands

Constraints on the light case are much less powerful with this analysis because Analysis-II is background-dominated and Analysis-I at short lifetimes requires scattering in the GArTPCs, which have a tiny active mass. Nevertheless, these constraints are interesting and show how even a tiny mass of argon, about 17 kg, can provide limits with a zero-background analysis. Future iterations of these searches from the T2K collaboration will be able to set much more powerful constraints, benefitting from a larger dataset, an upgraded detector, and dedicated analyses.

### Complementarity of the ND280 sub-detectors

In the heavy mediator case, both the scattering in the PØD and in the argon contribute significantly to the constraints, but in different regions of the parameter space. Scattering in the gaseous argon is rare because of the low density. However, it is the most powerful component in constraining the shortest lifetimes since it is where the fiducial volume of the analysis is contained. Figure 5.13 shows our constraints, as in Figure 5.11, splitting the limits into the contribution from the GArTPCs and from the PØD. Between $m_N = 0.1\,\mathrm{GeV}$ and $m_N = 0.2\,\mathrm{GeV}$, the model is very short lived, and all heavy neutrinos produced in the PØD decay before reaching the TPCs. This region is constrained only by prompt decays of heavy neutrinos produced inside the argon and is, therefore, less constrained. In the right plot, we show the dark photon parameter space, where, despite the larger upscattering rate at smaller $Z'$ masses, the model cannot be constrained by the PØD events due to the short lifetimes. However, in several regions of parameter space, the model predicts a significant number of events in the argon, which allows for a robust exclusion of the largest values of $\varepsilon$.

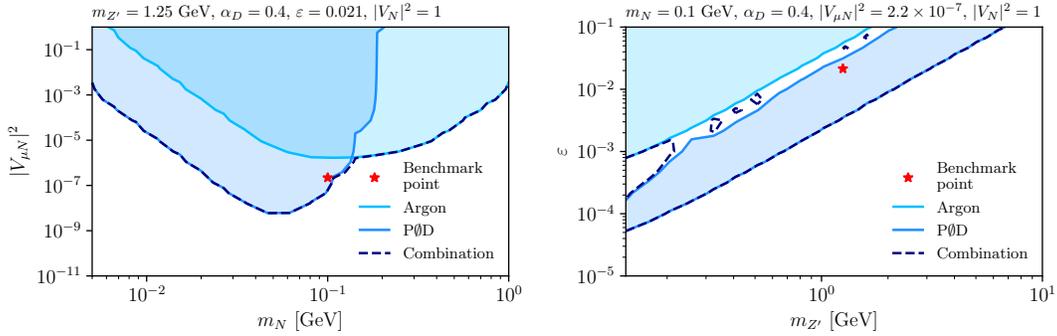

**Figure 5.13:** Limits for the heavy $Z'$ case, as shown in Figure 5.11, splitting into the two regions where upscattering happens in the GArTPCs or the PØD. On the left, the argon provides wider limits because it is sensitive to shorter lifetimes. These limits, however, do not go as deep into mixings as the PØD because of a smaller active mass. On the right, the PØD excludes a well-defined range of lifetimes, complementary to the top-left corner excluded by the argon, at a shorter lifetime and larger rate.

In the light mediator case, we combine Analysis-I and Analysis-II, considering upscattering happening in the GArTPCs, for the first case, and in the FGDs, in the second case. The two analyses contribute similarly to the limit, as shown in Figure 5.14.

### Model-independent constraints

The reasoning sparked by studying fig. 5.11 motivates a more model-independent study by probing constraints from Analysis-I on the heavy $Z'$ case as a function of the two central ingredients: the upscattering rate and the proper lifetime. First, we look at the exclusion as a function of the proper lifetime $c\tau^0$, shown in the top-left plot of fig. 5.15. In this study, all parameters are fixed to the heavy benchmark point, while we vary $|V_N|^2$, or $c\tau^0$ equivalently. We exclude the lifetimes $c\tau_N^0 \lesssim 3$ cm and $c\tau_N^0 \gtrsim 4 \times 10^3$ cm, although the latter limit is not interesting as it would be too long-lived to explain MiniBooNE. The first threshold of $c\tau_N^0 \lesssim 3$ cm would require $|V_N|^2 > 1$, which is unphysical. However, a different model predicting the same final state with the same number of events and this lifetime would be excluded.

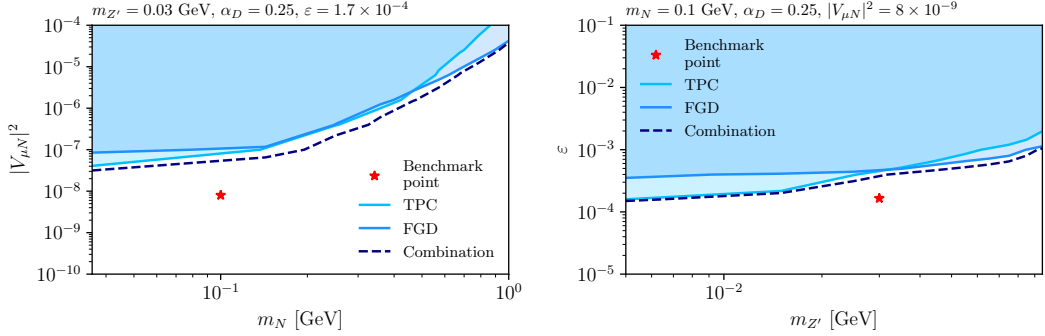

**Figure 5.14:** Limits for the heavy $Z'$ case, as shown in Figure 5.12, splitting into the two regions where upscattering happens in the FGD or the GArTPCs.

We extend this statement to constraints on the plane upscattering rate (x-axis) and proper lifetime (y-axis), shown in the top-right plot of fig. 5.15. This upscattering rate contains all the efficiencies except for the geometrical effect, meaning all events that, if decayed into one of the TPCs, would be measured in Analysis-I. The region on the right side of the curves is excluded. The shape and location of the curves depend exclusively on $m_N$, as heavier masses result in a lower rate because of the higher threshold and a smaller boost, probing only longer lifetimes. The bottom plot shows a zoom of the lower-left portion of the exclusions curves. The dots here corresponds to the lines of the previous figure, while the dashed lines show best-fit curves using a well-motivated functional form:

$$c\tau_{min}^0 = \frac{\alpha[\text{cm}]}{\log_{10}(\text{event rate}) - \beta}, \tag{5.32}$$

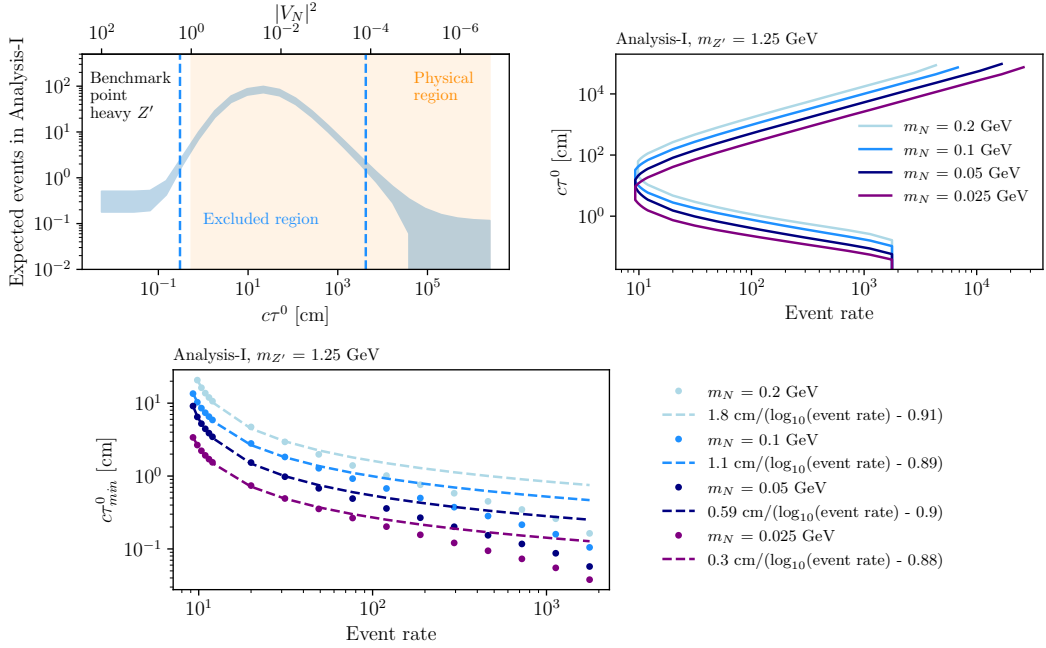

**Figure 5.15:** Top left: The expected number of events in Analysis-I after the complete selection as a function of the proper decay length of $N$, $c\tau^0$, for the heavy $Z'$ benchmark point. The blue band represents the uncertainty as obtained using Equation (5.31). Once we fix all the other parameters in the model, there is a bijection between $c\tau^0$ and $|V_N|^2$, as shown on the top x-axis. However, $|V_N|^2$ is physical only in the region allowed by Equation (5.11), shaded in orange. Besides very long-lived heavy neutrinos, we exclude most of the physical region. The minimum lifetime we exclude is 0.3 cm, although our model cannot generate such short lifetimes for this combination of parameters. Top right: Model-independent constraints in the 2D plane of proper lifetime $c\tau^0$ and upscattering rate after all selections but the geometrical one, for different values of the heavy neutrino mass and fixed value of $m_{Z'} = 1.25$ GeV. The excluded region is on the right of the curves. At first order approximation, different $m_{Z'}$ values result in different event rates without affecting the kinematics of the decay. However, different values of $m_N$ result in different production thresholds and lab-frame lifetimes, explaining the difference between the different curves. Bottom: Zoom of the previous plot in the lower-left part of the graph. The dots represent the curves obtained with the analysis, while the dashed lines are the best-fit values obtained with a simplified model. Although the functional form cannot perfectly fit all the points, it describes the curves reasonably accurately, providing model-independent constraints which can be applied to models predicting a similar phenomenology.

where $\alpha$ and $\beta$ are free parameters. It can be derived by assuming a point-like source of heavy neutrinos decaying in a detector of size $d$ at a distance $z$ and assuming that the process happens in one dimension only. The number of decays in the detector is given by event rate $\times$ ($e^{-z/\beta\gamma c\tau^0} - e^{-(z+d)/\beta\gamma c\tau^0}$). We are considering the region of the smallest lifetimes we exclude, so $c\tau^0 \ll z$, and the second term of the exponential is negligible. Finally, the limit is set when the event rate is equal to a constant value $C$, so the equation defining the limit can be re-written as $C = e^{-z/\beta\gamma c\tau^0}$, with $\alpha = z/\beta\gamma$ and $\beta = f(C)$ as free parameters, with $f$ a certain functional form. It can now be inverted to obtain eq. (5.32). The best fit to the different curves accurately describes the points at small event rates, underestimating the limit at smaller lifetimes. As expected, the values of $\beta$ are roughly constant, while $\alpha$ varies with $m_N$, roughly in a proportional way. The reason is that the momentum available in the heavy neutrino production is constant, so $m_N\beta\gamma$ is roughly constant. While this functional form is limited and cannot replace the complete study, it provides a model-independent benchmark, which can easily show the size of the constraints resulting from this analysis for an arbitrary model without needing a complete analysis.

## Outlook

As previously discussed in [182,183,1], the gaseous Argon (GAr) Time-Projection-Chambers (TPCs) of the T2K near detector, ND280, provide a powerful probe of long-lived particles. Visible decays inside the low-density volume of the TPCs, where neutrino-induced backgrounds are negligible, are identified. In this work, we showed that combining the high-density material in the P∅D detector with the low-density TPCs downstream is even more powerful. The former enhances the production of new particles in neutrino-nucleus scattering due to the large mass of lead. At the same time, the latter provides a desirable volume to search for charged final states. In addition, the magnetic field allows for improved identification of $\ell^+\ell^-$ pairs even at the smallest opening angles and energies.

We note that we have not exhausted the list of models, having not covered cases with small mass splittings between $N$ and the daughter neutrinos, scalar mediator models, and other $2 \to 3$ scattering signatures involving the emission of on-shell dark bosons [282,238,239,240,241,242,243]. We expect different signal selection efficiencies for these models, especially those that better fit the MiniBooNE angular spectrum. We encourage the T2K collaboration to pursue a dedicated search for all such upscattering signatures, including the one discussed in this paper, leveraging the full power of their detector simulation. In particular, we expect a full reconstruction simulation by the collaboration to overcome the simplifying assumption in this work of energy-independent signal efficiencies. In addition, further public data on the reconstruction efficiencies as a function of physical observables, like energies and angles, rather than model parameters, would be incredibly beneficial to the phenomenology community.

In the context of modern accelerator experiments, this method will constitute a valuable tool for phenomenologists to explore rich dark sectors. It will also be synergetic with the latest progress in building neutrino-nucleus Monte Carlo generators capable of simulating new physics processes [283,284,285]. With these latest tools, the user can proceed to build their kernel density estimator and perform confidence interval studies across a much broader region of parameter space with relative ease.

Interesting future directions include applying our methodology to searches for new physics outside the context of short-baseline anomalies. Attractive models include decay-in-flight signatures of multiple light dark particles. These are predicted in models such as higgsed low-scale $U(1)$ symmetries, co-annihilating dark matter models, and other heavy neutrino sectors with new interactions. We hope our method will speed up the exploration of the large-dimensionality of these models when searching for their experimental signatures at experiments like the Short-Baseline Neutrino program at Fermilab [286,287], atmospheric neutrino experiments like IceCube and KM3NET [288], as well as future high-intensity long-baseline experiments like DUNE [289] and Hyper-Kamiokande [290].

# Part III

# Light Sterile Neutrino Explanations

# 6

# The Micro Booster Neutrino Beam Experiment

THE EXISTENCE OF A LIGHT STERILE NEUTRINO with a mass of $\sim 1\,\mathrm{eV}$ has been proposed as the most straightforward interpretation of $\nu_e$ appearance and disappearance at baselines much shorter than usual neutrino oscillations. However, there are strong tensions between the interpretations of different experimental results under this hypothesis. Even for the case of MiniBooNE, the best fit under the light sterile neutrino model does not explain the data sufficiently well. For this reason, the community considers the light sterile neutrino model an essential benchmark model. However, if realized in nature, it must contain some additional ingredients.

I spent most of my Ph.D. as part of the MicroBooNE (Micro Booster Neutrino Experiment) collaboration. MicroBooNE is a neutrino experiment located at Fermilab. Its main physics goal is the investigation of the excess of events above the standard model expectation, as observed by Mini-BooNE, referred to as Low Energy Excess (LEE). In this chapter, I will introduce the MicroBooNE experiment, its technology, and setting up the stage for the LEE searches.

## 6.1    From Mini to Micro

MiniBooNE and MicroBooNE differ by more than just a name, but maybe less than three orders of magnitude. They are placed along the same beamline, the Booster Neutrino Beam [291,292] (BNB), and at about the same distance from the target, where neutrinos are produced. Both experiments also benefit from the off-axis beam from the Neutrino from the Main Injector (NuMI) beamline. Figure 6.1 shows a picture of the neutrino facility, with superimposed lines illustrating the proton storage rings and the neutrino beamlines, as well as the location of the experiments and the office building. While MiniBooNE set out to reproduce LSND excess with the same baseline and detector technology while varying the distance and the energy, MicroBooNE sets out to investigate Mini-BooNE excess with the same beam and distance while using a different technology. MicroBooNE is a liquid argon time projection chamber (LArTPC), which makes use of argon at cryogenic temperature as the target and detector material. Thanks to its mm-level resolution, MicroBooNE should be able to gain more insights into the MiniBooNE excess by reconstructing and distinguishing individual final state particles. However, to compromise on the cost of this new technology, the design empowers a smaller detector than MiniBooNE, hence the name *Micro*. While MiniBooNE is a spherical Cherenkov detector consisting of 818 tons of active material (mineral oil, mostly $CH_2$), the MicroBooNE detector is an 85-ton active mass, measuring 2.56 m, 2.33 m, and 10.36 m, along the drift direction ($x$), the vertical direction ($y$), and the beam direction ($z$), respectively.

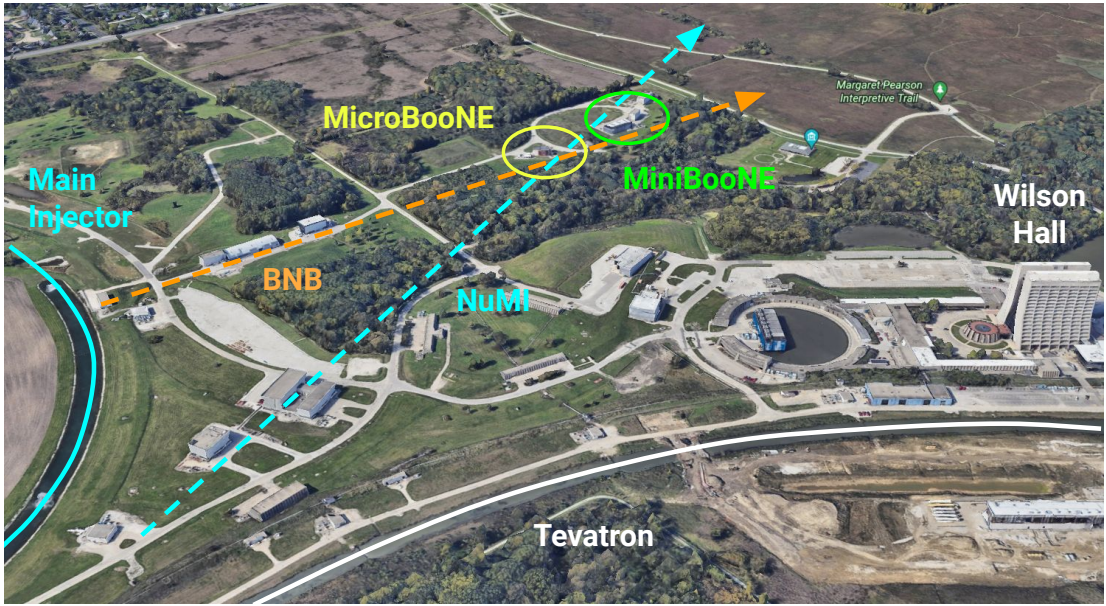

**Figure 6.1:** An aerial view of the Fermi national laboratory, its accelerator complex, and the short baseline experiments MiniBooNE, circled in green, and MicroBooNE, circled in yellow. The Booster Neutrino Beam (orange) originates from the Booster storage ring and serves MiniBooNE and MicroBooNE on-axis. While serving primarily long-baseline neutrino experiments, the Neutrino from the Main Injector beamline (light blue) provides some off-axis flux to MiniBooNE and MicroBooNE. Most neutrino scientists are lodged in the offices in Wilson Hall, a peculiar building that stands out against the other buildings in the lab. Picture taken using Google Earth.

The two pictures in fig. 6.2, obtained during the assembly in 2014, show the cylindrical cryostat, with the photomultiplier (PMT) system on the side (left picture), and the moment when the TPC was inserted in the cryostat, before sealing and filling it with argon. The following sections describe the process from the production of neutrinos to the data used for the analysis.

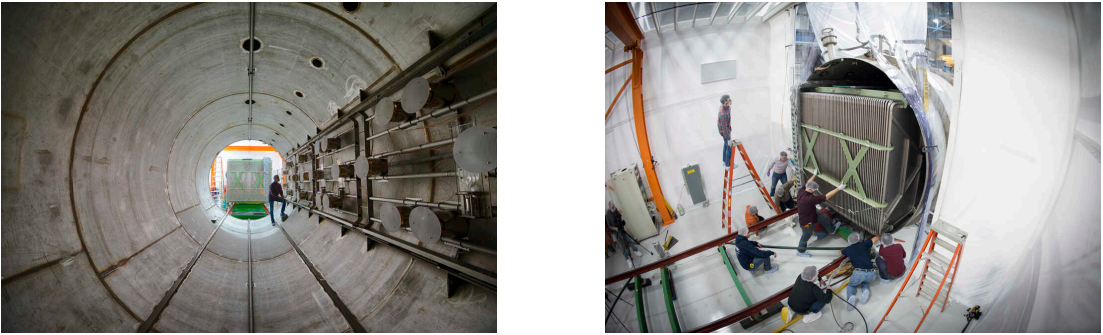

**Figure 6.2:** After being sealed in 2015, the cryostat has never been opened again. The picture on the left shows the inner part of the cylindrical vessel, with the PMT system sitting on the right side. In contrast, the picture on the right crystallizes the delicate moment of the insertion of the TPC and the field cage into the cryostat.

## 6.2 Turning protons into neutrinos

MicroBooNE is located along the BNB, while it can also benefit from the off-axis NuMI, which primarily targets MINOS, MINERvA, and NOνA. The cartoon in fig. 6.3 describes the process of turning protons into neutrinos. The Booster is a 450 m-circumference storage ring, accelerating protons to an energy of 8 GeV. Protons are then extracted and made to collide with a beryllium target of 71.2 cm, about 1.75 interaction lengths. As a result of this collision, secondary mesons, mostly pions and kaons, are produced. A focusing horn produces a magnetic field that focuses charged particles of one sign while deflecting out the opposite charge. Focused mesons are allowed to decay in an empty region of about 100 m, producing neutrinos. Typically, charged pions are the most abundant mesons, and because of helicity suppression, $\nu_\mu$ are the most abundant neutrinos. The sign of the mesons' charge determines if the beam is mostly made of neutrinos or antineutrinos. Typically, the running mode selecting mostly neutrinos is called $\nu$−mode or forward-horn-current (FHC), while the opposite is called $\bar{\nu}$−mode or reverse-horn-current (RHC). The flux is given in units of neutrinos per surface area, energy bin, and protons of target (POT), which is a measurement of the integrated exposure collected to the neutrino beam collected by a given experiment, analogous to the concept of luminosity for colliders. The resulting flux is shown in the left plot of fig. 6.4 for the case of FHC, which is what MicroBooNE employed for the data taking. The flux is dominated by $\nu_\mu$ and $\bar{\nu}_\mu$, with contamination of $\nu_e$ at the 0.5% level. Despite being small, the $\nu_e$ flux constitutes a major background for any search for $\nu_e$ above the standard model expectation. The second important point is that the flux peaks at energies of the order of 800 MeV. As discussed in more detail later, at this energy, the most probable charged-current interaction process is quasi-elastic, scattering against a single nucleon inside a nucleus. The right plot shows the off-axis flux from the NuMI beam in FHC mode. Interestingly, the $\nu_e$ contamination is significantly larger. This is because most $\nu_e$ are produced by decays of charged kaons $K^\pm \rightarrow \bar{\nu}_e e^\pm$, which is a larger Q value than the pion decay. This Q value can be used as transverse momentum with respect to the meson direction, resulting in a larger angular spread than for $\nu_\mu$ for charged pion decay. Therefore, selecting the off-axis flux enhances the relative contribution of $\nu_e$ with respect to $\nu_\mu$. Moreover, the typical neutrino energy of the off-axis flux from the NuMI beam is similar or smaller to the BNB case, which seems counter-intuitive because the NuMI beam relies on more energetic protons (20 GeV), which, in turn, should result in more energetic mesons and neutrinos. While this statement is valid on-axis, the fact that MicroBooNE is significantly off-axis significantly reduces the observed energy spectrum, making it comparable to the BNB and valuable for additional sideband studies.

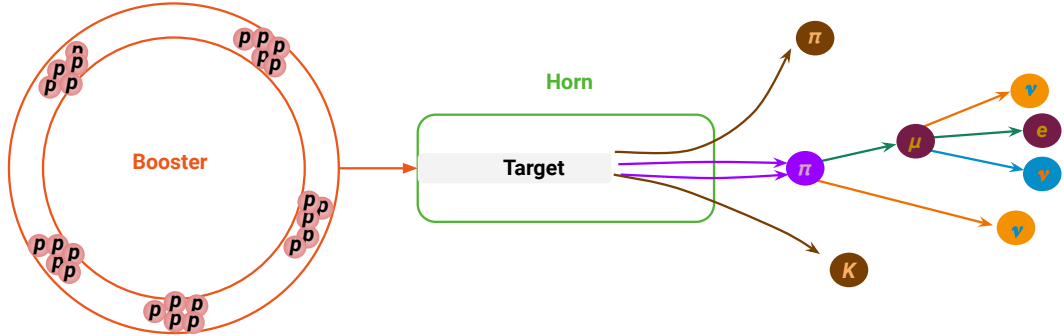

**Figure 6.3:** The Booster accelerates protons up to 8 GeV and stores them until they hit a beryllium target, where mesons, mainly pions, and kaons, are produced. The magnetic horns focus positively charged particles into the decay pipe while deflecting away mesons with opposite charges. Meson decays result in neutrinos which travel through the dirt undisturbed, while an absorber stops mesons at the end of the decay pipe.

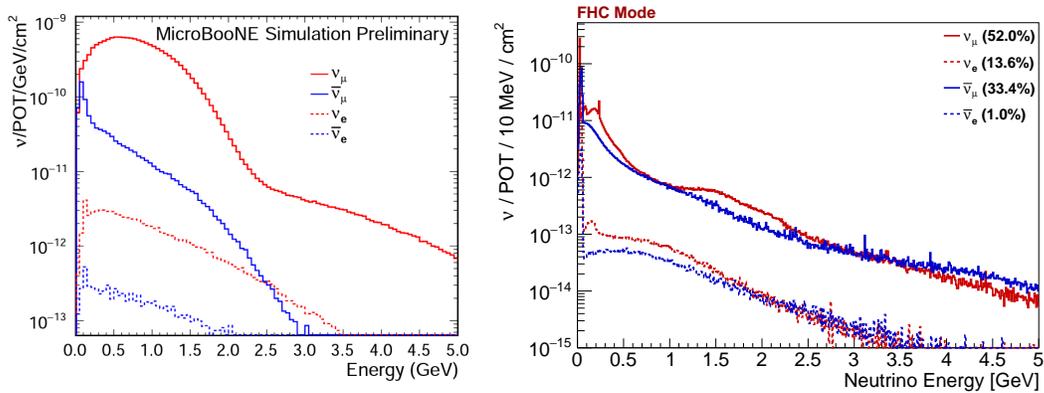

**Figure 6.4:** Fluxes of neutrinos per POT, energy, and surface area, as seen by MicroBooNE from the BNB (left) and NuMI (right) beamlines, in neutrino mode. Different lines refer to different neutrino components. Despite being in neutrino mode, the muon antineutrino component is more predominant than the electron neutrino component.

## 6.3 From invisible to visible

Neutrinos are invisible until they undergo a hard scattering interaction, whose resulting particles can be detected and measured.

Let's start with the description of all the processes happening in MicroBooNE, from the fastest to the slowest, following the cartoon in fig. 6.5. First, the hard scattering interaction takes place in a super-short timescale of a zs = $10^{-21}$ s, roughly the timescale associated with a transferred momentum of 1 MeV. Any highly unstable particle decays on the spot, like $\pi^0$ or $K^7$, while metastable particles, like muons, charged pions, and $K^l$, can propagate through the detector together with stable particles. After exiting the nuclear medium, they travel at speeds of the order of the speed of light, covering distances of $\mathcal{O}(10\,\text{cm})$ in timescales of $\mathcal{O}(1\,\text{ns})$. While traveling through the detector, charged particles release energy to argon atoms, exciting atomic levels or ionizing them. In $\mathcal{O}(1\,\text{ns})$, argon atoms create metastable states called dimers. Dimers are also produced by the recombination of free electrons and ions. When decaying, dimers produce the first detectable signal of the neutrino interaction: scintillation light. Depending on the spin configuration, the dimer decays in $\mathcal{O}(1\,\text{ns})$ (fast light) or $\mathcal{O}(1\,\mu\text{s})$ (slow light). This light has a wavelength of 100 nm and is shifted to visible light by a coating applied to the PMTs. While the slow light is emitted, the other neutrinos, pro-duced in the same bunch, cross the detector. The typical spread of a bunch is at the scale of 1.6 μs. However, because only one of every 600 bunch crossings results in a neutrino interaction in Mi- croBooNE, it is improbable that two neutrinos from the same bunch interact. The superposition of multiple neutrino interactions is much more likely to happen in the future experiment SBND, which is about four times closer to the beam production. Finally, the remaining electrons that did not recombine with nearby ions drift towards the anode through an electric field of 273.9 V/cm. The drift velocity is of the order 0.11 cm/s, which is why this process happens on a timescale of sev- eral ms. From the moment of a trigger, data is recorded for 3.2 ms after and an additional 1.6 ms before to fully measure cosmic rays that intercept the detector. While drifting, the electrons induce signals on the wires of the two induction planes and are collected by the collection plane. The beam has an average repetition rate of about 5 Hz: eventually, about 200 ms later, another bunch will cross the detector, and this entire process will start again if another neutrino happens to interact with the argon.

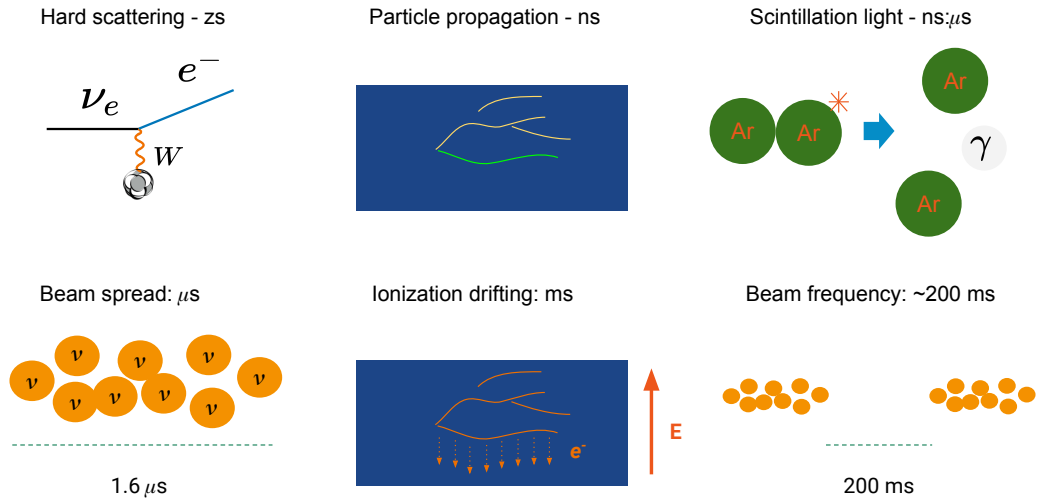

**Figure 6.5:** Between a neutrino interaction and the next time a beam spill crosses the detector, many different processes happen in the detector over different timescales. While scattering happens over a zs, particles take some ns to propagate through the detector and ionize argon atoms. Scintillation light has two components: a fast one, of the order of a ns, and the slow one, of the order of a $\mu$s. A $\mu$s is also the distance in time between the first and the last neutrinos crossing the detector within the same bunch, also known as beam spread. The electrons produced through ionization require a timescale of ms to drift to the wire planes. 200 ms afterward, another bunch of neutrinos will cross the detector again.

## 6.4   From 3D to 2D, and from 2D to 3D

### Projecting neutrino interactions in 2D

Figure 6.6 illustrates how the TPC measurement projects the deposited charge onto three two- dimensional pictures (only two projections are shown for simplicity). The drifted ionization charge from particle interactions is read out by three wire planes spaced 0.3 cm apart, with a 0.3 cm wire spacing, oriented vertically for the collection plane (Y plane), at 60° for the first induction plane (U plane), and −60° for the second induction plane (V plane).

The projection happens because each wire integrates all the charge deposited in the direction par- allel to the wire, and samples in the perpendicular direction, the drift direction, at different times. Let's consider the case of the collection plane, with wire oriented along the vertical direction $\hat{y}$. Any charge deposited simultaneously, on the same wire, but at different $y$ will be integrated and cannot be distinguished. Thus, if we look at the 2D plot of the deposited charge as a function of wire number (horizontal axis) and time (vertical axis), we can see a 2D projection of the 3D charge onto the $\hat{z} - \hat{x}$ plane. It is the case of the third image of the first row of fig. 6.7. Blue means no charge, while a redder color corresponds to a larger charge. Let's consider the corresponding diagram in the second row. We see the schematic representation of the two particles (blue and orange) and their measurement in the $\hat{z} - \hat{x}$ plane. The two induction planes, oriented at $\pm 60°$ with respect to the collection plane provide projections over the $1/2\hat{y} \pm \sqrt{3}/2\hat{z} - \hat{x}$ planes. Notice that the three blue and orange traces share the same vertical coordinates, from start to end, because the drift axis $\hat{x}$ is shared among the three planes. This simple observation is the seed for the 3D reconstruction.

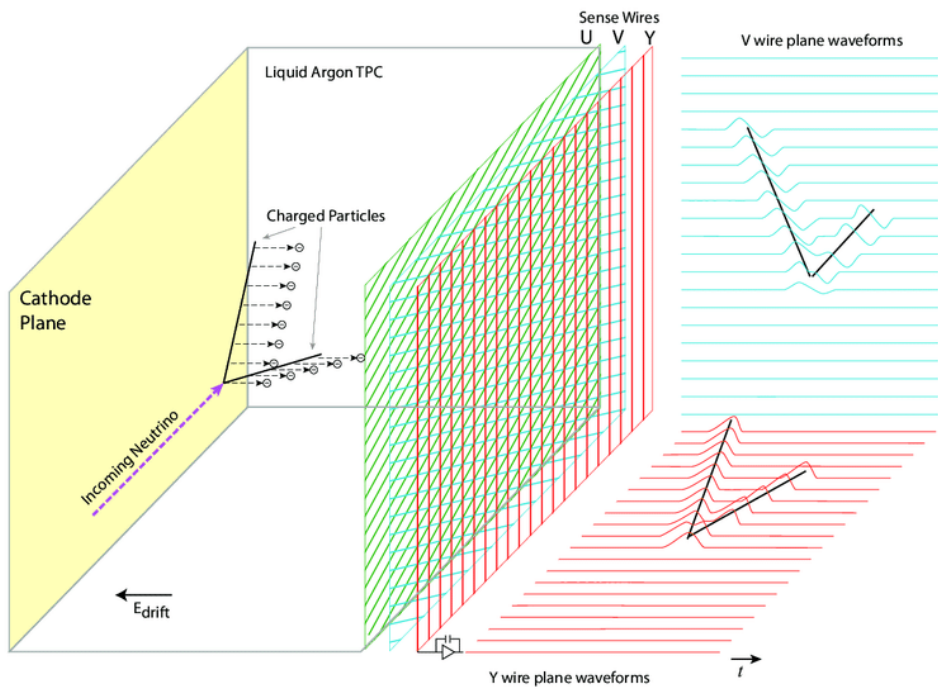

**Figure 6.6:** In this schematic representation of the MicroBooNE LArTPC, a neutrino interacts in the detector, producing two charged particles in the final state. While drifting towards the anode, the ionized electrons induce waveforms on the wire planes, displayed for the V and Y planes. Figure taken from [147].

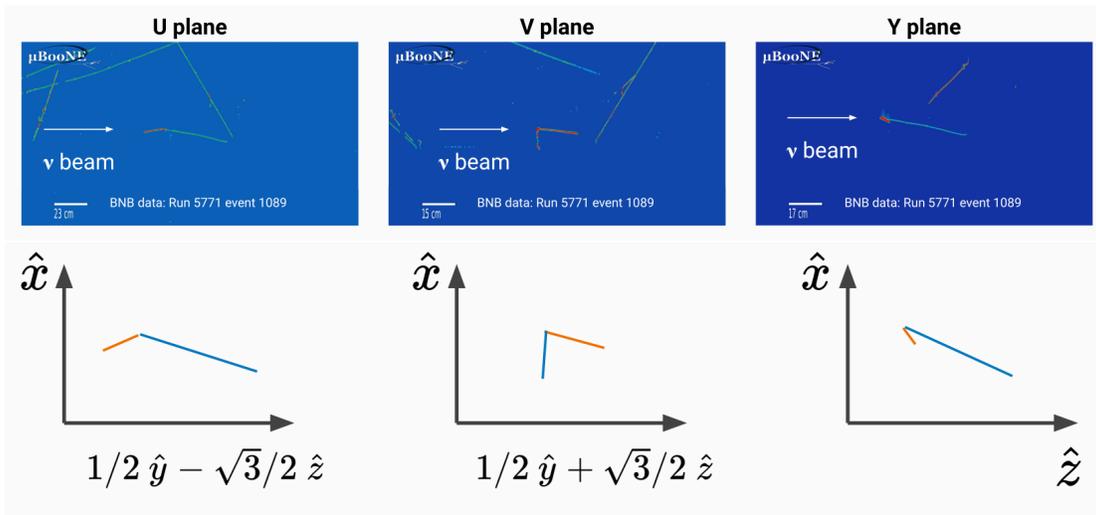

**Figure 6.7:** The three wire planes record the deposited charge in the form of images, projecting the 3D space onto three different planes. The three planes share the vertical axis $\hat{x}$, the drift direction, while the horizontal axis is a linear combination of the $\hat{y}$ and $\hat{z}$ directions. The two particles produced in this neutrino interaction are schematically represented with the blue and the orange line, matching their signatures across the three planes.

## 3D reconstruction

A TPC with a single wire plane would not be able to perform a 3D reconstruction; two wire planes could be enough but can still contain degeneracy, while three retain enough information for a proper 3D reconstruction.

MicroBooNE relies on three different reconstruction frameworks: Pandora[199], WireCell, and a Deep-Learning-based method. I will introduce the first one as it is employed in the rest of this thesis. Pandora algorithmically reconstructs a 3D image of the detector in terms of 3D objects called PFParticles, or simply reconstructed particles. The objects are reconstructed hierarchically through the process schematically illustrated in fig. 6.8. The deconvolved signals on the wires are fitted with Gaussian shapes to identify 2D hits[293], or simply hits: charge depositions at a given time on a given wire. Nearby 2D hits are clustered in 2D clusters, and clusters on different planes that share the same time span are matched to obtain a PFParticle. A PFParticle is defined by its 3D points, called SpacePoints, and its vertex. A PFParticle can be further reconstructed as a track or as a shower, which is the next chapter's topic. Eventually, PFParticles are arranged in a hierarchical structure, where each particle can have a Parent and multiple Daughters. For example, the electron produced by the muon decay at rest, often called Michel electron, is successfully reconstructed and is orga-nized as the *daughter* of the PFParticle associated with the muon.

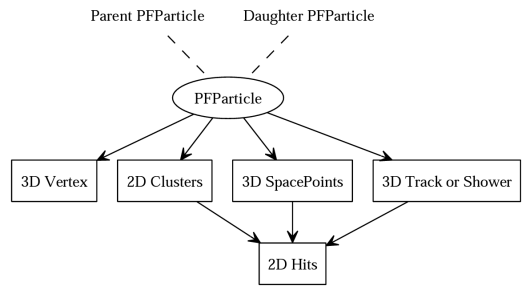

**Figure 6.8:** The Pandora pattern recognition framework reconstructs the event in a set of objects hierarchically arranged. 2D hits are organized into 2D clusters associated with each PFParticle, with an associated 3D vertex, a set of 3D space points, and a track or a shower object. Each PFParticle is part of a hierarchy of particles: it can be associated with a parent and multiple daughters. This figure is taken from[199].

The other two reconstruction techniques in MicroBooNE rely on deep learning and computer vision algorithms, treating the data like images[200,201], and on the tomographic WireCell approach[202], which reconstructs charge depositions in three dimensions.

# 7

# Identifying neutrinos: tracks and showers

We see neutrinos through the particles emerging from their interactions. In order to tell the type of neutrino involved in the interaction and the process that occurred, it is essential to identify the type of particle producing the signature. Metastable and stable particles with significant electromagnetic interaction, either charged particles or photons, leave a signature in the TPC which allows their reconstruction and identification. These particles are commonly classified as tracks or showers, depending on their signature. Tracks are characterized by straight lines, while showers by a more scattered profile. Tracks are the signature or particles that lose energy primarily through ionization, such as muons, charged pions, or protons. Showers are the images of electromagnetic showers produced by particles that lose energy primarily through bremsstrahlung and pair production, like electrons and photons. The classification between tracks and showers is performed by the Pandora reconstruction framework by means of a multivariate analysis that accounts for the shape of the charge depositions. In most MicroBooNE analyses, PFParticles with scores $<$ 0.5 are classified as showers, while PFParticles with scores $>$ 0.5 are classified as tracks. Noteworthy, at typical MicroBooNE energies, between hundreds of MeV to few GeV, hadrons do not produce hadronic showers. However, thanks to the high resolution, if a LArTPC like MicroBooNE were to be operated at larger energies, the different particles produced in hadronic showers could be reconstructed individually.

The central topic of this chapter is particle identification (PID), which is the determination of the particle type given its calorimetric measurement. It is performed by studying the profile of the ionization density $dE/dx$ along the particle trajectory, which is obtained from the measurement of the deposited charge and the reconstructed trajectory.

Tracks exhibit a peak in the ionization density (Bragg peak) when close to stopping, which differs in shape for different masses, allowing the distinction between particle types. Interactions of neutrinos in the GeV energy range in liquid argon lead to comparable rates of muons, charged pions, and protons, making the classification between these particle species important. In section 7.3, we will focus on the binary classification problem of distinguishing muons from protons. As pions and muons have very similar masses, the calorimetric separation of these two particle species is extremely challenging. Attempts to distinguish pions from muons rely on the identification of the decay products and are not yet very effective. Moreover, while calorimetric measurements are independent of the sign of the charge, this is not true for decay products, as $\pi^-$ and $\mu^-$ can form bound states and undergo weak interactions with the argon nuclei. Kaons are instead very rare (approximately 0.1% of the events in MicroBooNE are predicted to contain a kaon), and their classification is beyond the scope of this work.

For showers, we instead look at the beginning of the track, and LArTPCs provide two leverages. First, if we can identify the interaction vertex, we expect an electron shower to be attached to the vertex, while a photon shower would be detached by a distance of the order of the conversion distance, which is about 26 cm in LAr. Second, the beginning of the electromagnetic shower produced by an electron should be compatible with the ionization density of a single MIP particle, while the case of a photon shower should be compatible with the ionization density of two MIP particles, as the shower starts with the conversion of the photon into an $e^+e^-$ pair.

The left panel of fig. 7.1 shows the collection plane projection of a muon neutrino charged current ($\nu_\mu CC$) interaction, in which two track-like particles are produced. These tracks are classified as one proton and one muon, the two most common track-like particles in MicroBooNE, differentiated by the different amounts of energy deposited per unit length at any given point in their trajectories. The right panel shows the collection plane of an event where three different electromagnetic showers are produced. By measuring the ionization density at the beginning of the shower, and the distance between the vertex and the shower start point, these particles are identified as one electron and two photons, compatible with the signature of a $\nu_e CC$ $\pi^0$ event.

In this chapter, we discuss calorimetric reconstruction and its subtle angular effects. We perform high statistics data/Monte Carlo comparisons of calorimetric variables, resulting in a complete overview of the accuracy of the modeling of the calorimetry in MicroBooNE. In turn, we develop a tailored correction of the simulation as a function of geometrical variables and a new method for track classification that overcomes the limitations of the previous state-of-the-art. This new method is generally applicable to any present and future liquid argon time projection chamber with a wire readout and some of these ideas can be extended to pixel readout too. The code to compute the new

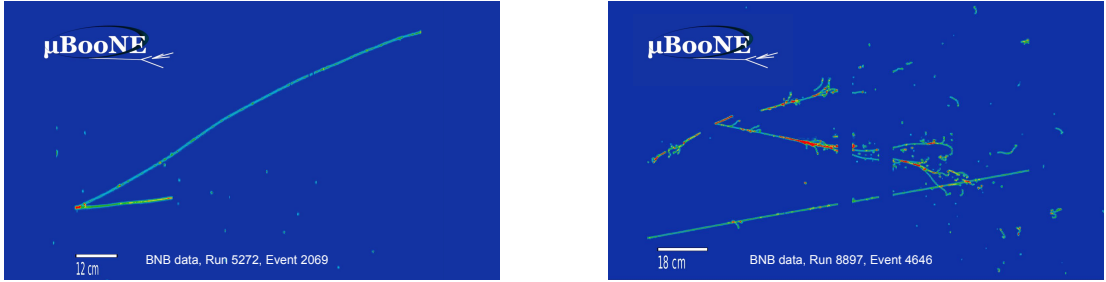

**Figure 7.1:** Examples of raw images of a neutrino interaction in MicroBooNE as recorded on the collection plane. The deposited charge (color scale) is shown as a function of wire number (x-axis) and time (y-axis). Left: Example of a $\nu_\mu CC$ interaction with two tracks in the final state, identified as one proton and one muon. Right: Neutrino interaction containing three electromagnetic showers, which are identified as one electron and two photons, compatible with the signature of a $\nu_e CC$ $\pi^0$ event.

experimental quantity used for track classification and to evaluate its performance is open source, and can be found on GitHub.[*]

We also discuss shower classification, based both on vertex-shower distance and on the measurement of the ionization density at the beginning of the shower, and attempts to improve it in the future.

## 7.1 Calorimetric reconstruction and angular effects

This section provides a deep understanding of the calorimetric reconstruction and its dependence on the angle the particle was traveling when depositing the charge.

### Coordinate system, pitch, d$Q$/d$x$ and d$E$/d$x$

Building up from the discussion in section 6.4, we now introduce the relevant coordinate system for calorimetric reconstruction. The coordinate system is plane-dependent, considering the way a given wire-plane projects the 3D charge into 2D pictures. The relevant directions are the drift direction $x'$, common to all planes, the direction parallel to the wires $y'$, and the one perpendicular to the wires $z'$. We can summarize the coordinate transformation as

$$x' = x, y' = \begin{cases} 1/2y + \sqrt{3}/2z & \text{plane = U} \\ 1/2y - \sqrt{3}/2z & \text{plane = V} \\ y & \text{plane = Y} \end{cases}, z' = \begin{cases} -\sqrt{3}/2y + 1/2z & \text{plane = U} \\ \sqrt{3}/2y + 1/2z & \text{plane = V} \\ z & \text{plane = Y} \end{cases} \cdot \quad (7.1)$$

---

[*] github.com/nfoppiani/calorimetry_likelihood.

Using these coordinates, we extract a measurement of $\Delta x$, $i.e.$ the distance over which the charge measured in a hit was deposited, which is essential to estimate the ionization density $dE/dx$. Assuming the particle trajectory is locally straight, we obtain $\Delta x$ as

$$\Delta x = 0.3 \,\text{cm} / |\cos(\gamma)|, \quad (7.2)$$

where 0.3 cm is the wire spacing and $\gamma$ is the three-dimensional angle between the local direction of the track and the vector that connects adjacent wires (also called wire-pitch direction), as illustrated in the right panel of fig. 7.2. The angle $\gamma$ ranges between 0 and 180 degrees, while $\Delta x$ takes values between 0.3 cm and infinity. $\gamma$ can be understood as the zenith angle of the coordinates in eq. (7.1), and together with the azimuthal angle $\phi'$, can be measured as:

$$\gamma = \arccos\left(u'_z\right) \quad and \quad \phi\prime = \arctan\left(u'_y/u'_x\right) \quad (7.3)$$

where $\hat{u}' = (u'_x, u'_y, u'_z)$ is the unit vector describing the local direction in the rotated frame of the wire-plane under consideration.

Because the calorimetric reconstruction in the TPC is symmetrical for particles going away or towards the anode, up or down, downstream or upstream the beam direction, $i.e.$ $u'_x \rightarrow -u'_x$, $u'_y \rightarrow -u'_y$, and $u'_z \rightarrow -u'_z$ that can be performed independently, we consider only one octant of the full angular space, taking the absolute value of $\hat{u}' = (|u'_x|, |u'_y|, |u'_z|)$ for eq. (7.3). Furthermore, while not directly affecting the measurement of $dE/dx$, the probability distribution of $dE/dx$ depends on $\phi'$, as discussed in the next subsection.

From this measurement, we can extract $dQ/dx$ as $\Delta Q/\Delta x$. This quantity can be converted to a measurement of $dE/dx$ using a non-linear function, shown in the right panel of fig. 7.2, which accounts for the recombination of electron and ions [294,295].

The local charge density is measured as $dQ/dx = \Delta Q/\Delta x$, and is converted to $dE/dx$ accounting for the recombination of electrons and ions, using the non-linear function shown in the right panel of fig. 7.2, as taken from [295]. This function is based on the Inverse Modified Box Model, and can be parametrized as follows:

$$dE/dx = \frac{\exp(\beta * W_{ion} * dQ/dx) - \alpha}{\beta} \tag{7.4}$$

where $\alpha = 0.93, \beta = 0.212/(\rho E), \rho = 1.396$ g/cm³, and $E = 0.273$ kV/cm, and $W_{ion} = 23.6 \times 10^{-6}$ MeV/e⁻.

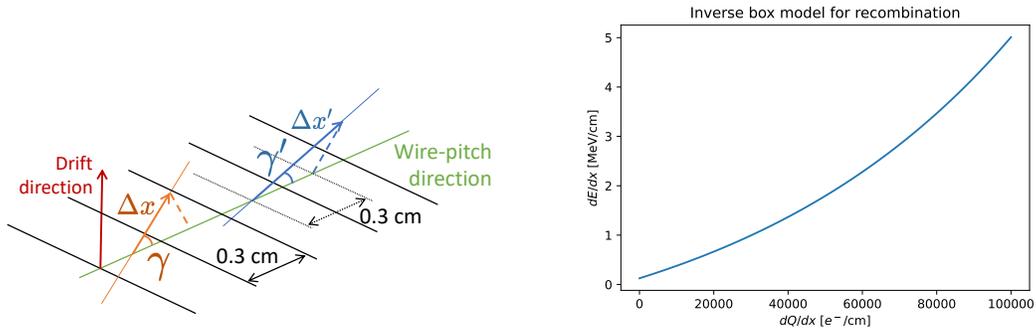

**Figure 7.2:** Left: A sketch of the relevant directions and angle in the calorimetric reconstruction. The orange and blue arrows represent two possible particle trajectories with different gamma angles and different local pitch values, represented by their lengths. Black solid lines represent wires, spaced $0.3$ cm apart. The dashed lines are perpendicular to the wire-pitch direction, and make evident the connection between the angle $\gamma$ and $\Delta x$, through eq. (7.2). Right: the recombination model used by MicroBooNE allows to convert $dQ/dx$ in $e^-$/cm into $dE/dx$ in MeV/cm. This graph shows a plot of eq. (7.4).

## ACPT selection

Anode or Cathode Piercing Tracks (ACPT) constitute a golden sample to perform detailed calorimetric studies. This is a high-purity sample of cosmic muons which cross either the cathode or the anode. While for most cosmic rays it is not possible to know the exact location along the x-axis, as it is degenerate with the time at which they cross the detector, by matching these tracks with the induced flash of light, it is possible to measure their spatial location and assure they cross detector, avoiding contamination with stopping muons. In Data Beam OFF, they are selected in the beam window, while in the overlay, they are simulated in the beam windows using Corsika, overlaid, as usual, with events recorded with the unbiased trigger. ACPT candidates are selected by matching a flash in the beam window with a crossing muon candidate. The selection requires downward going tracks, with a length of at least 20 cm, which has an efficiency of about 97% in both data and simulation. These samples have been used effectively for multiple calorimetric studies, as they provide a high statistics sample of through-going relativistic muons, which are, with very good approximation, minimum ionizing particles, with a constant average $\mathrm{d}E/\mathrm{d}x$. Figure 7.3 compares the data and Monte Carlo for the geometrical properties of these tracks. Discrepancies are at the ten percent level, which is why we need to carefully account for them in the following studies.

These tracks are employed in the rest of the section to study angular effects and to derive tailored corrections for the calorimetric reconstruction in the Monte Carlo simulation.

## Angular effects in calorimetric reconstruction

While a calibration applied early on in the analysis pipeline accounts for non-uniformity in space and between data and Monte Carlo[294], which are typically related to small discrepancies in the gains of the digital-to-analog converter units, the calorimetric reconstruction is non-isotropic, *i.e.* it exhibits a dependency on the angle. When measuring calorimetric information in a LArTPC, the fact that the charge is drifted along a particular direction (drift direction) and projected on wire planes with different orientations makes the calorimetric reconstruction angle-dependent. Both $dE/dx$ and the precision with which it is measured depend on the direction of the ionization trace left by the particle, even in a "perfect detector", absent of detector effects and angle-dependent detector response non-uniformity. The dependence appears primarily through the angle $\gamma$, and because it relates directly to the local pitch through a bijective function (eq. (7.2)), $\gamma$ and local pitch can be used interchangeably.

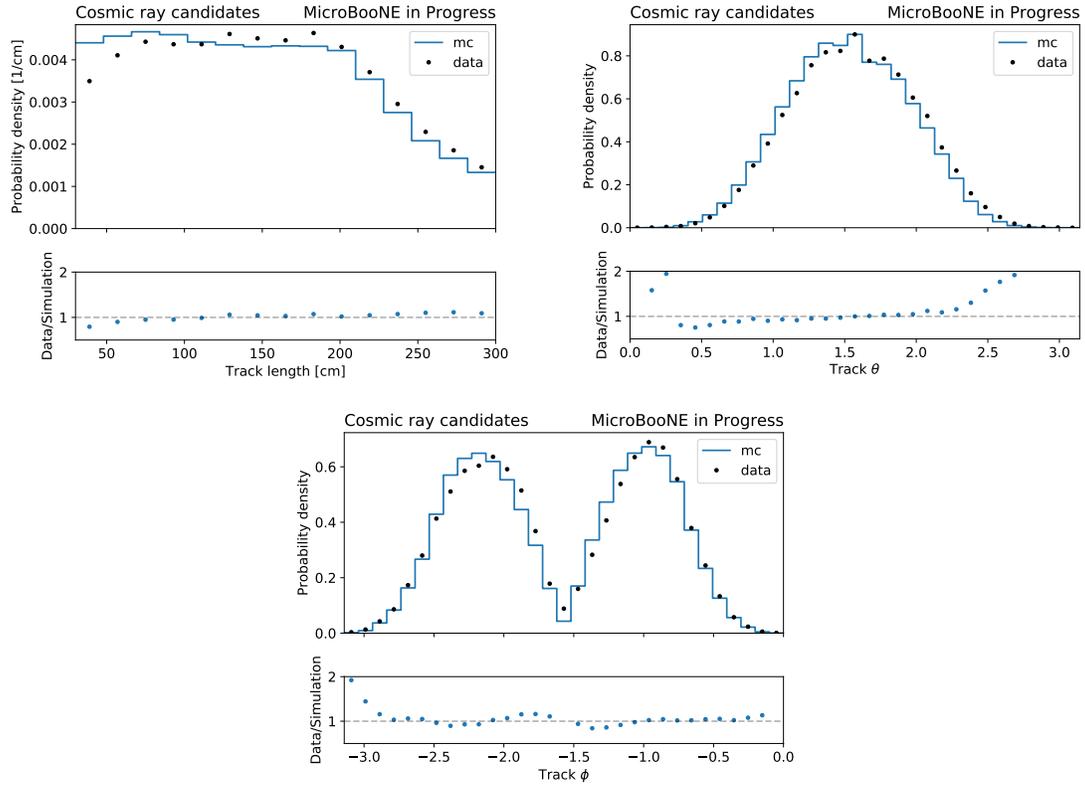

**Figure 7.3:** Distributions of geometrical properties (length, zenith angle $\theta$, azimuthal angle $\phi$) for selected anode cathode piercing tracks (ACPT), comparing the data with the simulation.

Measured $dE/dx$, even with a perfect detector, is angle-dependent because of intrinsic statistical fluctuations in particle energy loss, which impact the probability density function of measurements when averaged over different travel distances (different local pitch). A d$E$/d$x$ distribution, typically described by a Landau function for small local pitch values [297], becomes narrower at larger local pitch, and its most probable value moves to higher d$E$/d$x$ values while its average remains constant. This general geometrical effect applies to all LArTPCs, and it is shown in the left panel of fig. 7.4. We expect variations of the order of 10% while moving from 0.3 cm to 1 cm, as illustrated in orange on the right panel of fig. 7.4. This effect is present in MicroBooNE ACPT data, as seen in the blue line, although with a significant difference. In fact, the reconstruction chain, from the signal induced on the wires to the measured ionization density, affects the shape of the distribution, causing a significant shift in the peak value. As a result, the shape of the d$E$/d$x$ distribution, and not just the peak, also depends on the local pitch. Figure 7.5 illustrates these effects as measured with MicroBooNE ACPT data. The left plot shows the shape of the d$E$/d$x$ distribution for different small local pitch values, where the peak of the distributions shifts towards larger d$E$/d$x$ values for larger local pitch, varying by 4% in the low local pitch range (between 0.3 cm and 0.7 cm). The width of the distributions increases at a larger local pitch because the finite precision introduced by the detector smears out the distribution more than the predicted shrinking induced by geometrical effects.

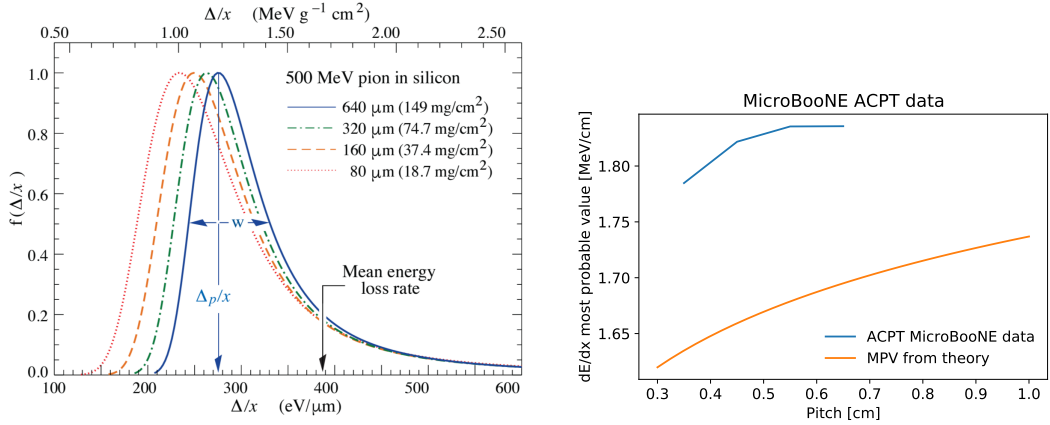

**Figure 7.4:** Left: the distribution of d$E$/d$x$ depends on the local pitch, as statistical fluctuations average out over longer distances, plot taken from [296]. Right: the most probable value of the d$E$/d$x$ distribution increases with the pitch as predicted by the theory (orange), and measured in MicroBooNE using cosmic ray data (blue).

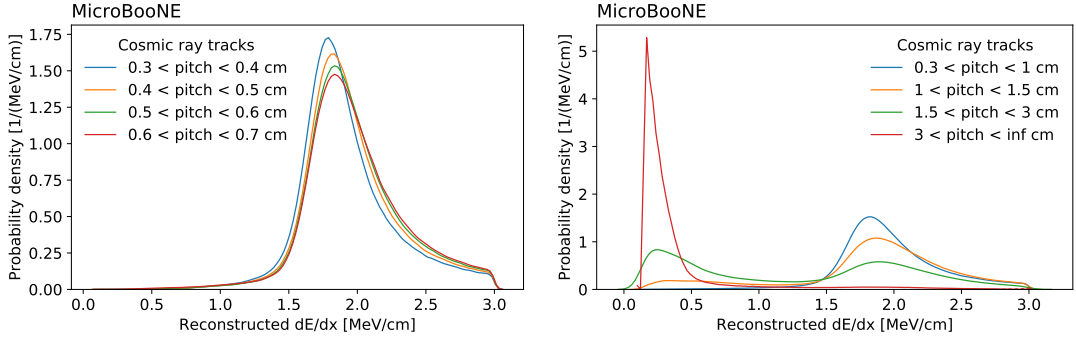

**Figure 7.5:** Normalized distributions of $dE/dx$ for different local pitch in the low-local pitch regime (left) and over the entire local pitch spectrum (right), as measured on the collection plane in a sample of cosmic muons tracks in the Micro-BooNE data.

The right plot shows analogous distributions integrated over a wider range of local pitch values, illustrating a major change in the shape of the distributions at very large local pitch. The distributions for larger local pitch values show a second peak at smaller $dE/dx$ values, populated by particles traveling parallel to the drift direction or the wire orientation.

It is well documented that the shape of the signals induced on the wires by the drifting charge, in ADC counts vs time, depends on the particle direction [298,299], appears very different at lower local pitch ($< 0.7$ cm) compared to larger local pitch ($> 0.7$ cm), impacting hit reconstruction and making measurements more precise at lower local pitch. Figure 7.6 shows how the shapes of the signals on the first induction plane varies for different value of $\theta_{xz}$ and $\theta_{y}$, which describes the angular direction in a different coordinate system. While it is still not fully understood how this effect translates to the distributions of the reconstructed $dE/dx$, it is clear how signals induced by particles traveling at large pitches differ from the most typical signals. The cartoon in fig. 7.7 illustrates how particles traveling parallel to a single wire ($\phi' \sim 90$ deg) induce high amplitude signals because all the charge is deposited at the same time on the same wire, while particles traveling orthogonal to the wires ($\phi' \sim 0$ deg) induce signals which are long in time. For the latter case, the Gaussian hit-finding process is not sufficient. Therefore, such signals are fit by a sum of Gaussian shapes, for which the overall number of hits, their positions, and their amplitudes are free parameters. The width is fixed to reduce the degeneracy of the problem. The total deposited charge is therefore segmented into multiple hits, leading to an underestimation of the hit charge and resulting in smaller $dE/dx$ values. These small and somehow non-physical values of $dE/dx$ encountered at large local pitch are only mildly correlated with the true d$E$/d$x$, bringing little additional information and degrading the particle identification performance.

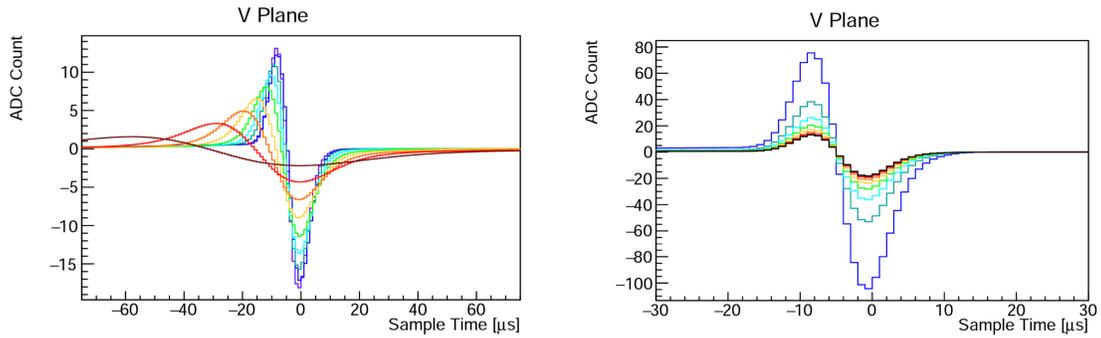

**Figure 7.6:** The shape of the signals induced on the wires depend on the direction of the particle at the moment of the charge deposition, in two qualitatively different ways for the two angles that describe the particle's direction. Plots taken from [298].

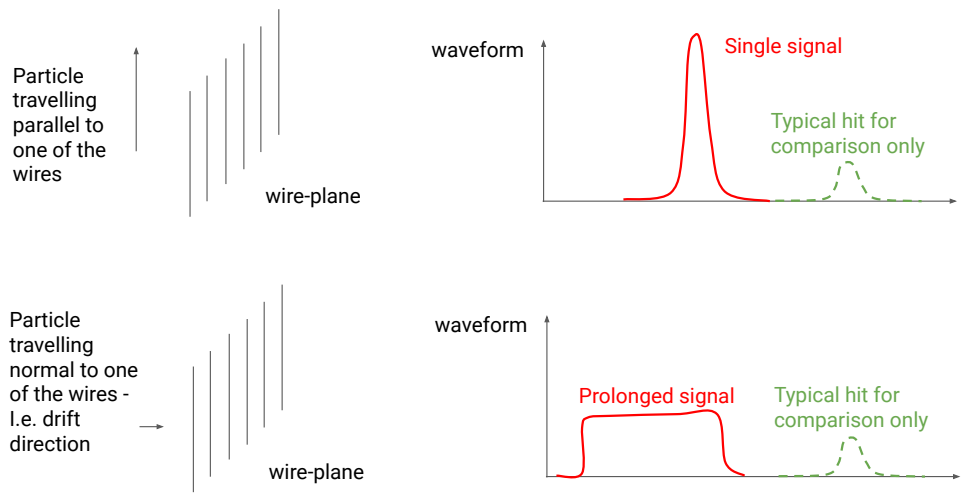

**Figure 7.7:** Signals generated at high local pitch induced unusual signals on the wires, with different shape depending if the high local pitch is obtained by traveling parallel to the wires or in the normal direction.

Despite being quantitatively evaluated with MicroBooNE data, this effect applies more generally to LArTPCs with wire-plane detectors and has two main consequences. First, because individual charge depositions are the basic ingredients for track classification, if wrongly simulated, this angular effect introduces systematic biases in the selection of tracks as a function of the angle. Second, comparing the measured d$Q$/d$x$ value, drawn from such distributions, to the mean value obtained from the Bethe-Bloch function results in a sub-optimal classification, especially in angular regions where these effects are the most prominent. The next two sections tackle these two questions.

## 7.2 Re-calibration

Good data/simulation agreement in the d$E$/d$x$ distributions is essential in order to ensure that PID-related cuts have the same impact on data and simulation.

The re-calibration is a correction applied to the simulation in order to match the data in the d$Q$/d$x$ distribution. It is called re-calibration because it is applied on top of the normal calibration, tailored to the Pandora eLEE analysis. Plans to extend it to the standard analysis framework are undergoing.

In this re-calibration, we compare the distribution of d$Q$/d$x$ between data and simulation using a sample of ACPT in bins of angular variables and derive multiplicative factors to be applied to the simulation to better match the data. We compare the distribution of d$Q$/d$x$ rather than d$E$/d$x$ as the re-calibration tries to address data/simulation discrepancies associated with measurements of drifting charge that are independent of possible effects of recombination mismodeling. As the measured d$E$/d$x$ is a bijective function of the measured d$Q$/d$x$, the only difference between a correction factor on d$Q$/d$x$ or d$E$/d$x$ arises from the non-linearity of this function.

The distributions seen in fig. 7.5 are the basis for this correction. However, only considering the dependence on the angle $\gamma$, is an approximation, as it encapsulates most, but not all the angular dependence. For example, fig. 7.7 sketches two different trajectories at a very high local pitch, which however induce very different signals. Including the azimuthal angle $\phi'$, that, together with $\gamma$, uniquely describes the 3D trajectory, breaks this degeneracy, and separates the two peaks in the d$E$/d$x$ distribution seen at large local pitch in fig. 7.5. Thanks to the large sample size of the ACPT dataset, we take into account both the local pitch and the azimuthal angle $\phi'$.

### Comparing the data with the simulation

The first step is to compare the data with the Monte Carlo in the distribution of d$Q$/d$x$ at the hit level.

We divide the local pitch-$\phi'$ plane into rectangular bins, separately per each wire plane. The binning (five bins in pitch and six bins in $\phi$) is defined as follows:

- pitch: $[0.3, 0.4, 0.7, 1., 1.5, 30]$ cm

- $\phi$: $[0, 0.26, 0.52, 0.79, 1.05, 1.31, 1.57]$

Figure 7.8 shows the comparison in some of the bins. The data is shown in black dots, the Monte Carlo in blue, while the orange line refers to the Monte Carlo after applying the correction, and will be discussed later. Statistical uncertainties on these quantities are not shown for simplicity and can be inferred by the level of fluctuations seen between points. Systematic uncertainties are not available for the ACPT sample, but they are not relevant for the discussion here. Detector uncertainties are implemented as new samples with variations in the detector simulation or reconstruction, while the purpose of this work is to improve the central value of the simulation. The first three rows show hits from the collection plane. The peak moves to lower values and broadens while moving to the larger values of the local pitch. When fixing the local pitch and changing $\phi'$ from low values (tracks on the horizontal plane) to large ones (tracks on the vertical plane) the agreement between the data and the Monte Carlo improves (second row). This pattern is less present when observing very large values of the local pitch ($>$ 1.5 cm), as illustrated in the third row. However, the relative height of the peaks at low and high d$Q$/d$x$ values changes drastically. Most of the discrepancy seems to be due to larger smearing induced by the detector resolution in the data rather than in the simulation. Eventually, the last row on this graph compares the same bin in local pitch and $\phi'$ among the three different wire planes. While the two induction planes show a consistent behavior, both the magnitude and the direction of the discrepancy, differ significantly between induction and collection planes.

## Fit for scaling factors

We perform a binned maximum-likelihood fit in each of the local pitch-$\phi'$ bins in order to extract the best multiplicative correction factor $\alpha$:

$$(\mathrm{d}Q/\mathrm{d}x)^{corrected} = \alpha(pitch, \phi)\mathrm{d}Q/\mathrm{d}x. \tag{7.5}$$

Figure 7.9 illustrates the likelihood scan for some of these fits. The full sample has not been divided into train and test sets, meaning that all the entries are used to derive these corrections. However, it has been verified that there are no temporal differences within the samples and that by splitting the sample, only some of the coefficients vary mildly because of statistical fluctuations.

The effect of the correction is visually shown in fig. 7.8 with the orange lines: the re-calibration improves the agreement significantly. When residual disagreement is still present, it typically signals a different width of the peak(s) between the data and the simulation and would require fitting an additional smearing factor, or a more complex unbinned correction.

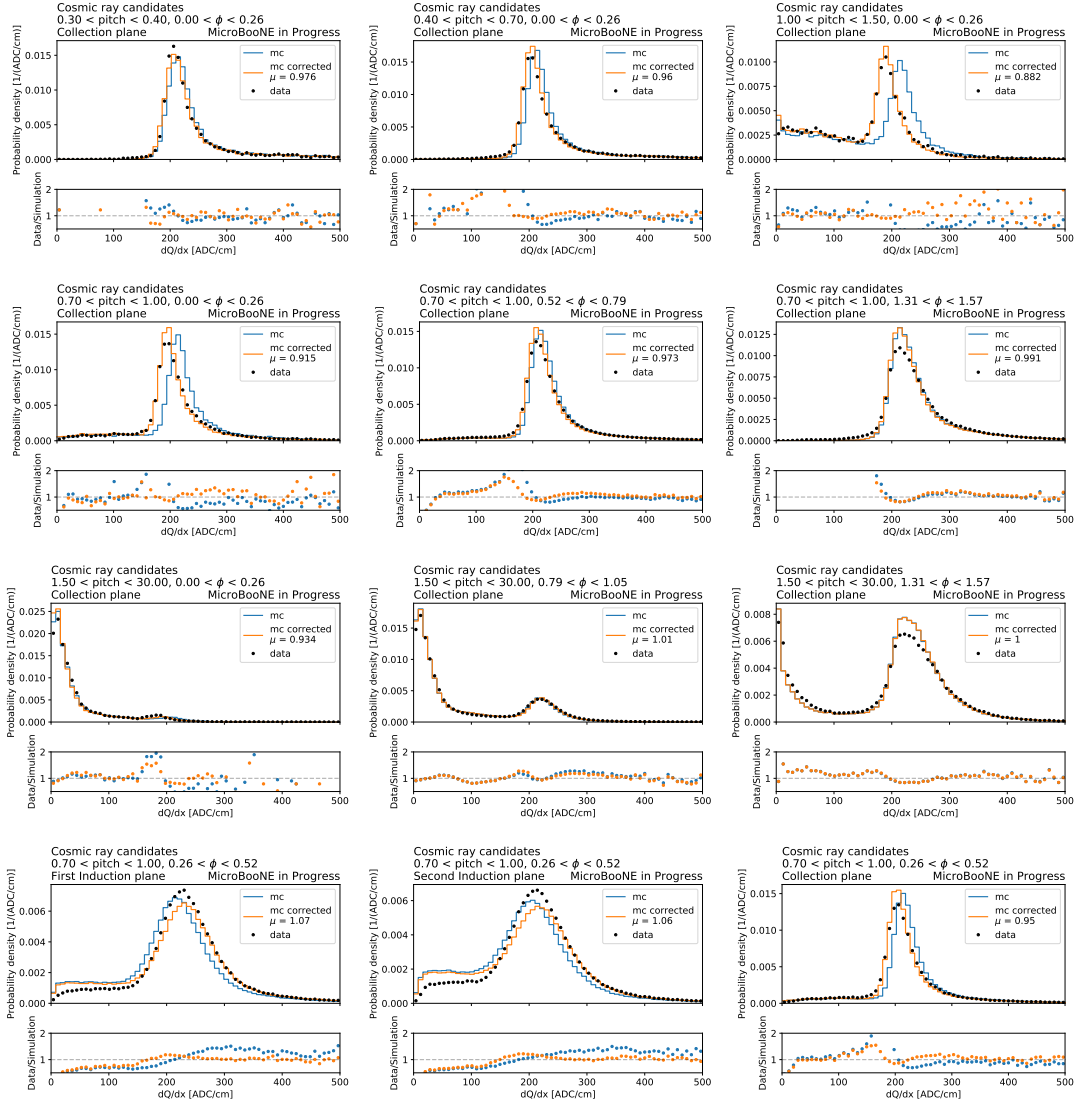

**Figure 7.8:** Examples of $\mathrm{d}E/\mathrm{d}x$ distributions, comparing the data (black dots) with the simulation before (blue) and after (orange) applying the re-calibration, for selected ACPTs, in bins of local pitch and $\phi'$. First row: collection plane hits, fixed local pitch at small value, varying $\phi'$. Second row: collection plane hits, fixed $\phi'$, varying local pitch. Third row: collection plane hits, fixed local pitch at large value, varying $\phi'$. Fourth row: the three different wire planes, fixed local pitch and $\phi'$.

## Re-calibration summary tables

The result of these fits is summarized in Figure 7.10 in the form of three annotated 2D plots, showing the re-calibration factor in each local pitch-$\phi'$ bin, for the three wire planes. The fourth plot illustrates the re-calibration factor if derived without binning in $\phi'$, by fitting a single factor per bin in local pitch. Roughly speaking, these factors should be comparable to the average of the factors from the table, weighted with the relative abundance of particles in each $\phi'$ bin. Remarkably, the factors are positive for the induction planes and negative for the collection plane. While averaging around $2 - 3\%$ correction for the collection plane, the discrepancies on the induction planes reach even 10% values at the larger local pitch. This study also justifies why, until now, most analyses doing calorimetric reconstruction in LAr, even outside of the MicroBooNE, were reluctant to use the induction planes.

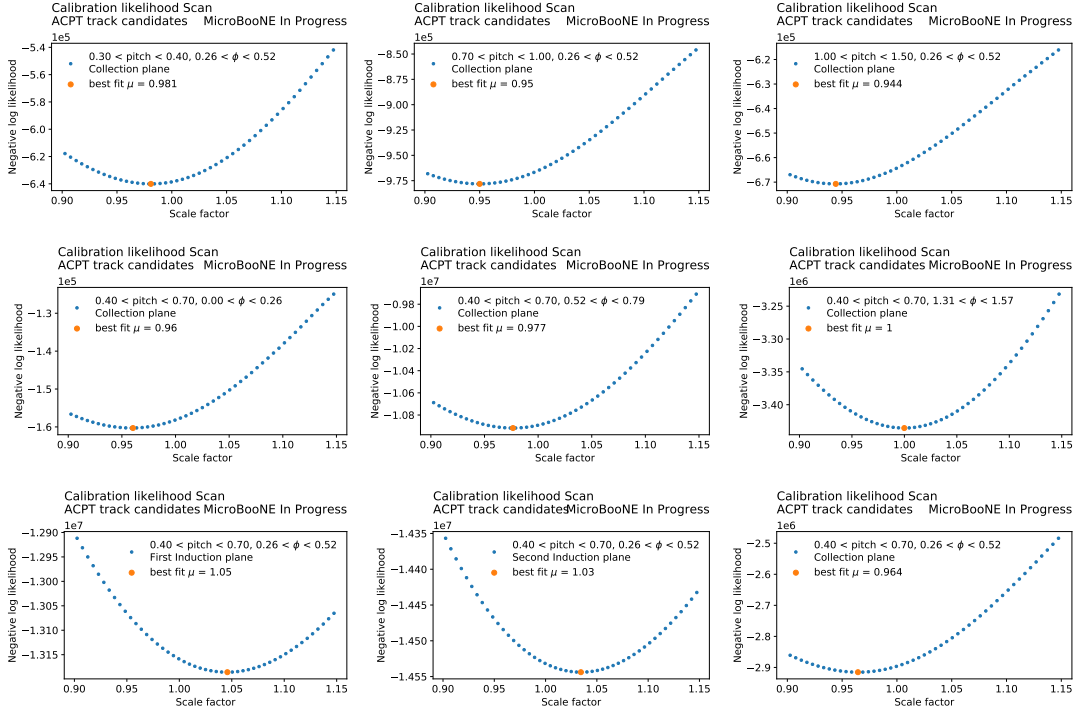

**Figure 7.9:** Likelihood scans for the best multiplicative factors, scaling $\mathrm{d}E/\mathrm{d}x$ measured in the simulation in order to match the data. The best fit point is shown in orange.

Figure 7.11 illustrates the same improvement by showing the distribution of $\mathrm{d}Q/\mathrm{d}x$ integrated over all values of local pitch and $\phi'$, with remarkable improvement across all three planes.

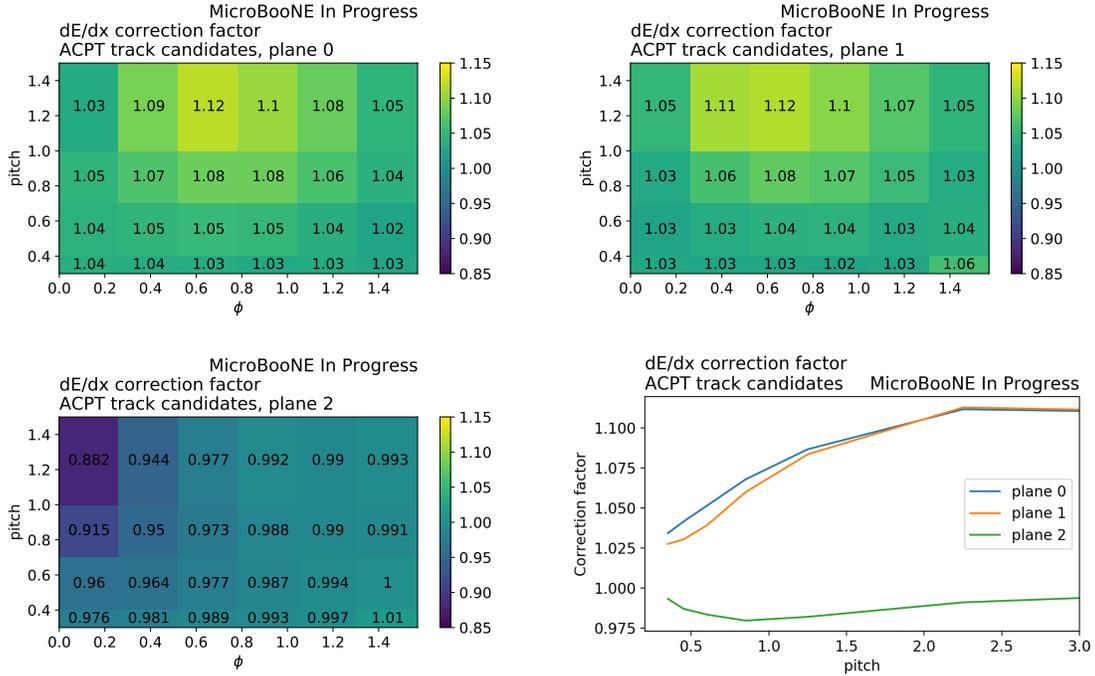

**Figure 7.10:** Summary of the re-calibration coefficients for each bin of local pitch and $\phi'$, for the first induction plane (top left), second induction plane (top right), and collection plane (bottom left). Re-calibration coefficients derived in bins of local pitch only (bottom right) for all three wire planes.

We acknowledge that this method has some limitations, and future work might improve above and beyond. The first limitation comes from being a binned correction, developed with especially coarse bins. An unbinned correction requires a larger effort, although it might be implemented using machine learning techniques, and overcome the coarseness of the bins, which is dictated by the limited statistics of the samples. Despite improving the agreement in every bin, the correction is not perfect as it lacks an additional smearing effect. A more complex correction, perhaps including smearing or a non-parametric correction, might be performed in a future iteration of this study. Eventually, the discrepancies can in principle depend on additional variables, such as the position in the detector. Including additional variables in a binned correction is not possible while preserving the necessary statistics needed for effective re-calibrations.

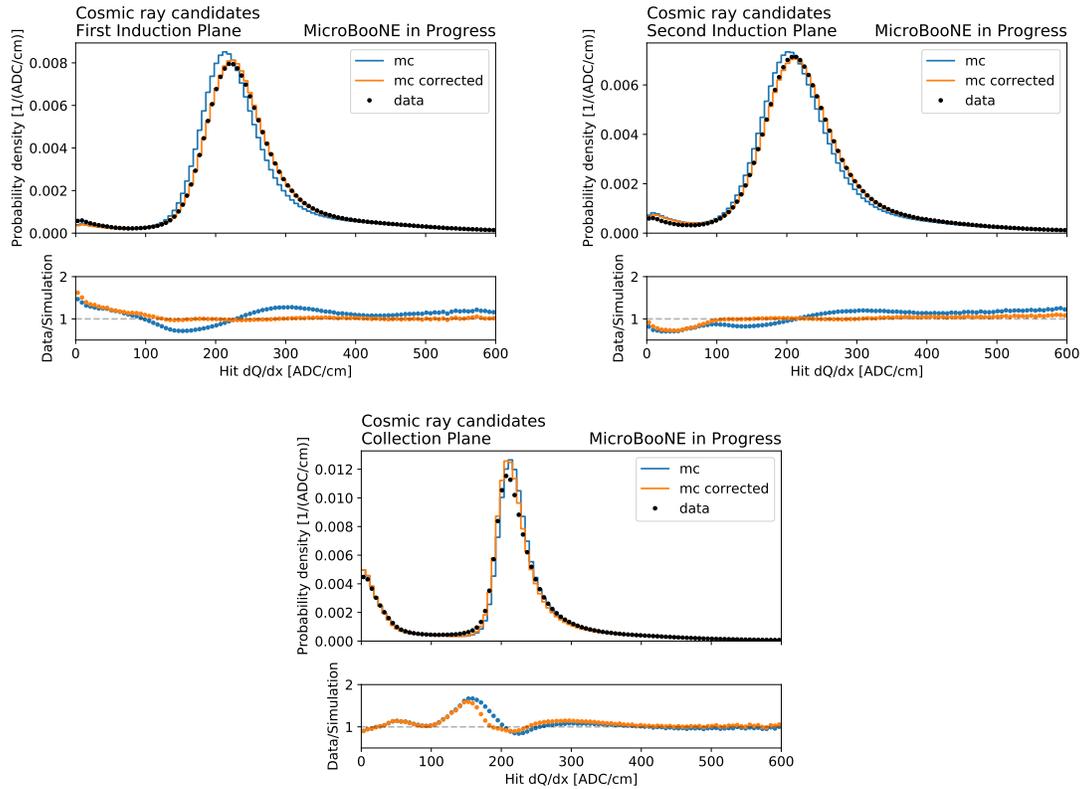

**Figure 7.11:** d$E$/d$x$ distribution integrated over all directions, comparing the data (black dots) with the simulation before (blue) and after (orange) applying the re-calibration for the three wire planes.

## 7.3 IDENTIFYING TRACKS - A NEW METHOD

The second task is to account for the angular effect in the identification of tracks. Identification of tracks relies on the d$E$/d$x$ profile as a function of the residual range, the distance of given energy deposition within the track from the endpoint of the track itself. This work describes the limitations of current methods, and how a deeper understanding of the calorimetric reconstruction in Micro-BooNE led to a new and improved method for track classification.

Typical particle identification methods condense calorimetric information into a test statistic used to distinguish different particle species. The $dE/dx$ profile is compared with the expectation for different particle hypotheses to choose the hypothesis that best matches the data. The hypothesized $dE/dx$ profile is computed by integrating the Bethe-Bloch function for a given particle mass and charge.

A common test statistic is the $X^2$-like variable under the proton hypothesis, defined as

$$X^2_{proton} = \sum_i \frac{(dE/dx_i - dE/dx^{theory}(rr_i | m = m_{proton}))^2}{\sigma^2} \tag{7.6}$$

where $i$ runs on the hits along a specific cluster, $rr_i$ is the residual range for hit $i$, $dE/dx^{theory}$ could either be the mean value of the most probable value for a specific pitch, and only its residual range dependency is considered, and $\sigma$ is a typical resolution value for $dE/dx$, sometimes taken as 10%. This approach is unsatisfactory for a series of reasons. For example, there are many arbitrary parameters, like what expected value from theory should be picked, or what mass should be considered. If we want to distinguish protons from muons, we could consider both the proton or the muon hypothesis, or even the kaon hypothesis, but it is not clear which one should give the highest separation power. Moreover, a constant value for $\sigma^2$ does not capture the complexity of the $dE/dx$ distribution, which depends on many different parameters. Eventually, we can obtain this test statistic individually for each plane, but the collection plane is typically used because of the better resolution. However, the induction planes can provide important information, especially when the signals on the collection plane do not carry important information, *e.g.* at high pitch. To summarize, despite utilizing the relevant physics, this approach is suboptimal because it is not derived in a principled way. It fails to properly account for the intrinsic statistical nature of $dE/dx$, as well as the angular dependencies, which also affect the shape of the d$E$/d$x$ distribution. Moreover, as local pitch values are different on different readout planes, combining the three wire-plane measurements ensures that the calorimetric information provided by the LArTPC in the entire $4\pi$ solid angle is fully leveraged for the purpose of performing particle identification.

## The d$E$/d$x$ probability density function

The d$E$/d$x$ probability density function (PDF) for each particle type is the basic ingredient of the likelihood calculation. In principle, the average d$E$/d$x$ as a function of the residual range could be estimated by integrating the Bethe-Bloch function before applying detector reconstruction. However, a complete characterization of the d$E$/d$x$ distribution requires an analytic description of the intrinsic fluctuations of the ionization energy loss and the effect of the detector reconstruction. This is challenging given the very long computational time required and because there is no straight-forward way to derive an analytic description of the detector reconstruction. The d$E$/d$x$ PDF is instead estimated from the MicroBooNE simulation which incorporates all the described effects, as demonstrated in the data/simulation comparisons in section 7.4.

The PDF is estimated through a three-dimensional histogram of d$E$/d$x$, residual range, and local pitch. The binning is defined as follows:

- d$E$/d$x$: $[0, 0.5, 1, 1.5, 2, 2.5, 3, 3.5, 4, 4.5, 5, 5.5, 6, 6.5, 7, 7.5, 8, 9, 10, 12, 15, 20, 25, 30, 35, 40, 45, 50, \infty]$ MeV/cm

- residual range: $[0., 2, 4, 7, 10, 15, 20, 30, 50, 100, 300, \infty]$ cm

- local pitch: $[0.3, 0.6, 1, 1.5, 3, 30, \infty]$ cm,

to have a number of entries large enough so that statistical fluctuations in the estimate of the PDF are not important while being fine enough to distinguish all the features of the PDF. The his-togram is normalized so that for each combination of values of residual range and local pitch, the integral of the $dE/dx$ distribution sums to one, providing an estimate of the conditional PDF $p(dE/dx|\text{residual range, local pitch})$ that is not informed by the underlying kinematics of tracks. This procedure is repeated for each plane and for two particle species, namely muons and protons. The histograms are filled with hits associated with well-reconstructed tracks produced in simulated neutrino interactions. These tracks are required to be complete, meaning that more than 90% of the true deposited charge is reconstructed. They are also required to be pure, meaning that more than 90% of their reconstructed charge was deposited by a single particle. The tracks are also re-quired to be contained within a fiducial volume, where both the start and end points are at least 20 cm away from the boundaries of the TPC. Moreover, each particle is required to have no associated daughter, to reject particles that undergo hard-scattering. Lastly, for each plane, we exclude the first and last hit from the calculation. Although they could provide additional important information, especially when it comes to the last hit, the charge measured could be significantly underestimated with respect to the expectation. This happens when the particle stops without covering the entire space over which the wire measuring the last hit integrates the charge. In such a case, the effective *Deltax* where the charge was deposited would be smaller than inferred, but there is no way to determine that.

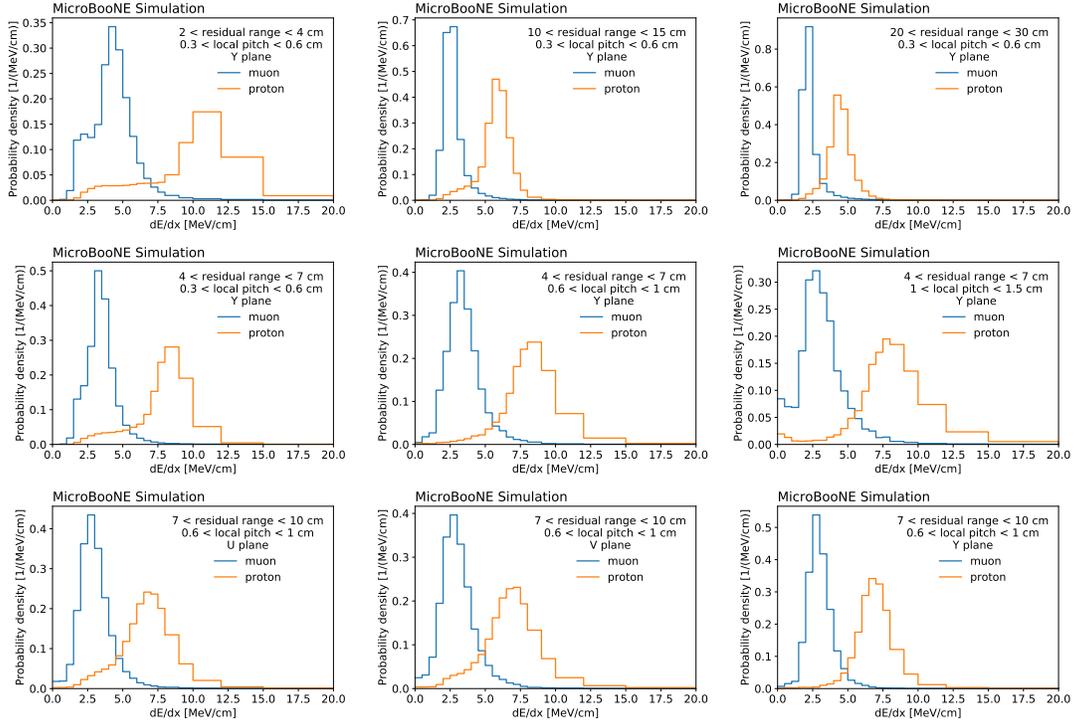

**Figure 7.12:** Expected $dE/dx$ distributions for muon (blue) and proton (orange) hits. The top row shows distributions on the collection plane, with a fixed value of local pitch, and three different values of the residual range. The middle row shows distributions on the collection plane, with a fixed value of the residual range, and three different values of local pitch. The bottom row shows distributions on the three wire planes, with a fixed value of residual range and local pitch. As expected, the peak $dE/dx$ value reduces at a higher residual range, passing from $11\text{ MeV/cm}$ to $5.75\text{ MeV/cm}$, and to $4.25\text{ MeV/cm}$ across the three bins under consideration for protons.

The $dE/dx$ PDF is visualized through three series of examples in fig. 7.12: in each row only one of the three parameters (residual range, local pitch, and plane, respectively) is varied while keeping the other two fixed. The PDF changes considerably, showing, for example, a reduction of the $dE/dx$ of the peak value at a higher residual range and an increase of the width at a higher local pitch, justifying the need for such a construction as a function of these three variables.

In order not to bias the evaluation of the method, only 90% of the available data is used to construct the PDF, while the remaining 10% is kept as test set to study the performance of the method.

### The likelihood ratio test statistic as PID score

Using the PDF previously constructed, the likelihood of any particle hypothesis can be computed for each reconstructed track. Interactions of neutrinos in the GeV energy range in liquid argon lead to comparable rates of muons, charged pions, and protons, making the classification between these particle species important. However, this work focuses only on the binary classification problem of distinguishing muons from protons. As pions and muons have very similar masses, the calorimetric separation of these two particle species is not addressed in this paper, and pion tracks will appear as muon-like by means of this algorithm. Kaons are instead very rare (approximately 0.1% of the events in MicroBooNE are predicted to contain a kaon) and are omitted in this work.

The likelihood for a track is computed starting from the single-hit-likelihood:

$$\mathcal{L}_{\text{hit}}(\text{type}|\text{plane}, dE/dx, \text{rr}, \text{local pitch}) = p(dE/dx|\text{type}, \text{plane}, \text{rr}, \text{local pitch}), \qquad (7.7)$$

where $p$ stands for the PDF, type refers to muon or proton and rr stands for residual range. The local pitch is measured locally, and it is generally different for each hit associated with the same track, because of changes in track trajectory due to multiple Coulomb scattering. The plane is included because the PDFs are significantly different for the different planes. The single-plane-likelihood is computed by taking the product of the single-hit-likelihood for each hit on a given plane:

$$\mathcal{L}_{\text{plane}}(\text{type}|\text{plane}, \{dE/dx\}_{i=1,\dots,N}, \{\text{rr}\}_{i=1,\dots,N}, \{\text{local pitch}\}_{i=1,\dots,N}) = \\ \prod_{i=1}^{N} \mathcal{L}_{\text{hit}}(\text{type}|\text{plane}, dE/dx_i, \text{rr}_i, \text{local pitch}_i), \quad (7.8)$$

where $i = 1, \dots, N$ indexes the hits on the plane under consideration. The three-plane-likelihood, which is the likelihood for the entire track, is then computed as the product of the single-plane-likelihoods for the three wire planes.

The likelihood defined this way is an approximation as it neglects correlations between the charge measured on different wires and planes. However, fluctuations of the hit charge are in general correlated among wires on different planes, as they record the same charge through different projections. Moreover, the induced charge on neighboring wires and correlated noise introduce additional correlations between the charge recorded on different wires on the same plane[298,299]. Modeling these correlations is in general complex, as they depend on the geometry on a track-by-track basis. Neglecting such correlations and using an approximation of the likelihood makes the method less optimal, and may result in a loss of separation power. A possible discrepancy between the data and the simulation for the values of $dE/dx$, which are the inputs of the PID method, could introduce a systematic bias. Moreover, someone might argue that this method relies more strongly on the simulation than previous methods, making it more sensitive to systematic biases due to an imperfect simulation. While it is true that systematic biases emerge from an imperfect simulation, any test statistic is making use of the same data, with the same discrepancies. For example, it is not correct to argue that this method relies on the simulation while the $X^2_{proton}$ on the theory, implying the latter is less sensitive to systematic biases. The reason is that once defined, any test statistic is simply a deterministic function of the observables, calorimetric measurements in this case. As long as the same function is applied to both the data and the Monte Carlo, the test statistic will only propagate the discrepancies from the observables. However, some test statistics, like this method, make use of more information from the observables than others, like $X^2_{proton}$, and might therefore result in stronger systematic biases than other test statistics[*]. However, as shown in the plots in section 7.4 where the data and the simulation are compared, there is no evidence of significant systematic biases. The Monte Carlo is accurate thanks to the precise simulation of the signals induced on the wires[298,299], and to the tailored recalibration of the $dQ/dx$ distribution in angular bins, as discussed in section 7.2. And lastly, the impact of possible systematic biases is evaluated through a study of detector systematic uncertainties, as discussed later in the following subsection.

The likelihood is then used to compute the likelihood ratio test statistic, which is employed to perform the classification task:

$$T = \mathcal{L}(\text{muon})/\mathcal{L}(\text{proton}), \tag{7.9}$$

where either a single-plane-likelihood (eq. (7.8)) or the three-plane-likelihood can be considered.

The binary classification problem of distinguishing protons from muons has the likelihood ratio as the most powerful statistical test, as proven by the Neyman-Pearson lemma. It provides the largest classification efficiency for any given value of the misidentification rate.

---

[*]A trivial example is a test statistic which assigns a fixed value independently of the observable, therefore exhibiting no difference between data and Monte Carlo.

For computational purposes, in the rest of article, instead of $T$, the PID score $\mathcal{P}$ will be considered, defined as:

$$\mathcal{P} = \frac{2}{\pi} \arctan\left(\log(T)/100\right). \tag{7.10}$$

Computing the logarithm of $T$ is convenient as it reduces to a sum of log-likelihoods rather than a product of likelihoods. This bijective non-linear transformation of $T$ does not change the separation power of the method, but it constrains the value of the PID score $\mathcal{P}$, otherwise unbounded, between $-1$ and $1$, making it easier to display.

## Implementation and lookup tables

An essential part of the implementation of this method lies in storing the PDF values and defining rules to access them at the analysis stage, possibly in a fast way.

First, we rearrange the calculation of eq. (7.9) using the log-likelihood ratio of each hit:

$$\log(T) = \log \mathcal{L}(\text{muon}) - \log \mathcal{L}(\text{proton}) = \tag{7.11}$$

$$= \sum_{i=0}^{N_{hits}} \log p(\mathrm{d}E/\mathrm{d}x | \text{muon}, \text{plane}_i, \text{rr}_i, \text{pitch}_i) \tag{7.12}$$

$$- \sum_{i=0}^{N_{hits}} \log p(\mathrm{d}E/\mathrm{d}x | \text{proton}, \text{plane}_i, \text{rr}_i, \text{pitch}_i) = \tag{7.13}$$

$$= \sum_{i=0}^{N_{hits}} \left( \log p(\mathrm{d}E/\mathrm{d}x | \text{muon}, \text{plane}_i, \text{rr}_i, \text{pitch}_i) - \log p(\mathrm{d}E/\mathrm{d}x | \text{proton}, \text{plane}_i, \text{rr}_i, \text{pitch}_i) \right) = \tag{7.14}$$

$$= \sum_{i=0}^{N_{hits}} t(\mathrm{d}E/\mathrm{d}x | \text{plane}_i, \text{rr}_i, \text{pitch}_i)) \tag{7.15}$$

where $t$ is the log-likelihood ratio for a single hit, $i$ runs over all hits associated with the particle, either on a single plane or on multiple planes. As a result, we only need to store the difference between the logarithm of the probability density function of protons and muons.

Some problems can arise in this process. For example, let's consider a hit $\{dE/dx_i, rr_i, pitch_i\}$ that ends up in a bin in which the probability density function has no muon entries. It can happen because of the limited Monte Carlo statistic. As a result $p(hit_i|muon) = 0$ while $p(hit_i|proton) \neq 0$, which produces $t(hit_i) = \log p(hit_i|proton) - \log p(hit_i|muon) = \infty$. This would be problematic, as one hit would unequivocally determine the classification of the particle. Therefore, we decided that every bin for which either the proton or the muon PDF contains no entry, gets a default value $t = 0$, effectively neglecting the contributions of the hits in those bins to the classification.

The $t$ values are stored through a lookup table, which enhances the speed of the calculation at the inference stage. It differs from a hierarchical structure, like a dictionary, in the sense that accessing each element is guaranteed to be $\mathcal{O}(1)$. Figure 7.13 illustrates this concept with a cartoon. The values are stored in a long array, with a rule that determines where the values for a given bin in residual range or local pitch begin and end. They are ordered so that the first value is the left-most bin in $dE/dx$, in the first bin of local pitch and residual range. The subsequent x numbers are the other bins in $dE/dx$. Then we switch to the following bin in local pitch and repeat for all local pitch bins keeping fixed the first bin of the residual range. We eventually switch to the second bin in the residual range and start and iterate the entire process. If we wanted to extend to a larger number of parameters, it would require only a trivial addition of one variable. Finding the correct bin where a value belongs is $\mathcal{O}(n_{bins})$, not as powerful as $\mathcal{O}(1)$. However, if we used a dictionary of dictionaries of dictionaries, the full processes would require $\mathcal{O}(n_{bins}^3)$.

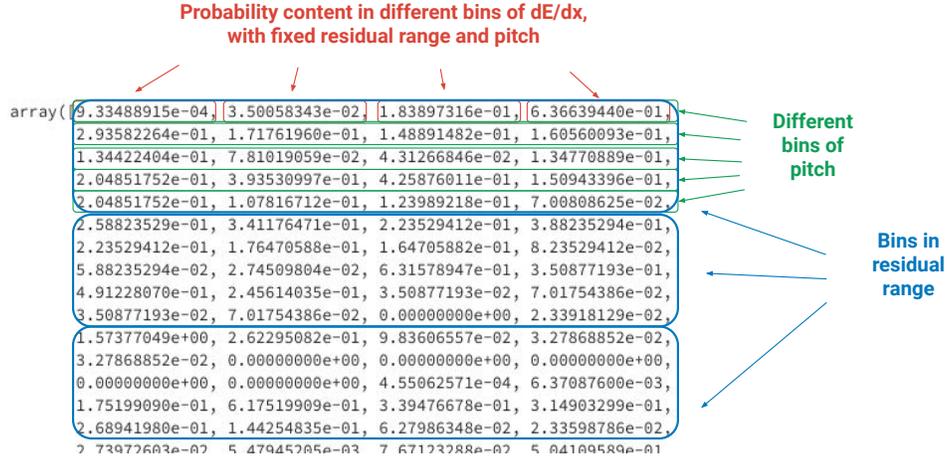

**Figure 7.13:** Cartoon showing the structure of a lookup table for this problem, in an exemplified case with four bins in d$E$/d$x$ (red), five bins in local pitch (green), and three bins in residual range (blue).

## Performance of the particle identification

The performance of $\mathcal{P}$ is evaluated on a test sample of more than 20000 simulated protons and muons, selected in the same manner as in section 7.3, with a small difference. To better mimic a real case, we drop all the requirements which rely on the ground truth, namely the cut on the completeness and purity. The sample contains inclusive neutrino interactions from the Booster Neutrino Beam (BNB). Figure 7.14 shows the distribution of the new test statistic for muons, protons, and for particles associated with cosmic ray muons. Muons populate the right side of the plot, protons on the left side, and the population around zero is composed of particles for which there is too little information to classify them. The population of protons at positive $\mathcal{P}$ is mostly related to protons that undergo a hard scattering before stopping, and they are a subject of the next subsection. Particles associated with cosmic ray muons span the entire spectrum. Indeed, these particles are not simulated, but rather measured by the detector with Beam OFF, and superimposed to simulated

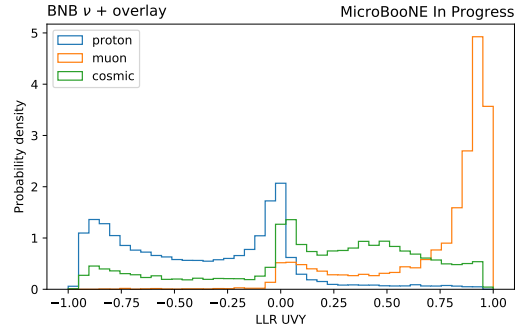

**Figure 7.14:** Distribution of $\mathcal{P}$ for protons (blue), muons (orange), and particles associated to cosmic rays (green), selected from the test set, using the same criteria as in ...

neutrino interactions. While it is impossible to know the type of these particles at truth level, $\mathcal{P}$ could help us classify them into muons from protons. A receiver operating characteristic (ROC)

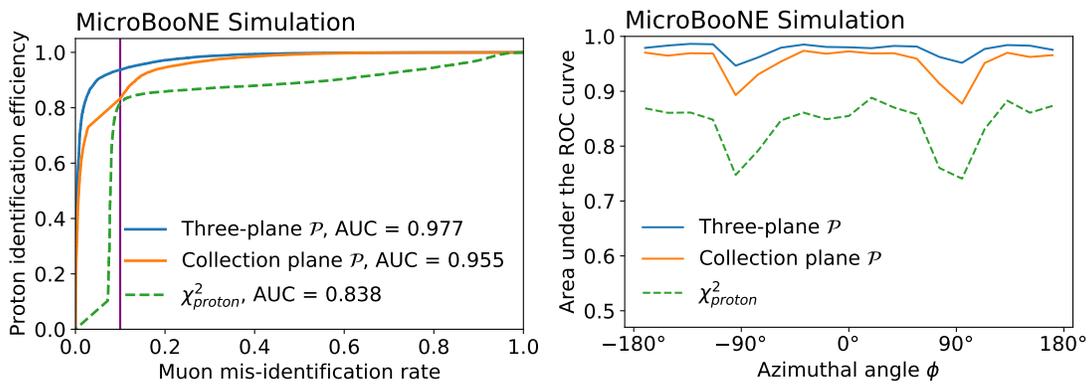

**Figure 7.15:** Comparison of the proton/muon separation power of different PID scores. The left plot shows the ROC curves on the entire sample, and the right plot shows the area under the curve (AUC) in bins of the track angle $\phi$. The blue curves refer to the proposed PID score $\mathcal{P}$ using three planes, the orange refers to the collection plane only $\mathcal{P}$, whereas the green curves show the $\chi^2$ test with respect to the proton hypothesis. The purple vertical line on the left plot slices the curves at a muon misidentification rate of $0.1$, comparing the proton identification efficiencies of the three methods at the same working point.

curve is calculated from the test statistics distributions for the two particle types and shown in the left plot of fig. 7.15 for the three-plane $\mathcal{P}$, for the collection plane only $\mathcal{P}$, and for the $\chi^2$ test with respect to the proton hypothesis, which represents the previous state of the art [2.94]. The latter quan-tity, computed by comparing the data with the expectation from the Bethe-Bloch theory, has been used in several previous MicroBooNE analyses and it is shown here as a reference for comparison. The ROC curves show the proton efficiency as a function of the muon misidentification rate, which is bounded between 0 and 1. For a given method, every possible cut value between $-1$ and 1 corresponds to a point on the ROC curve. The performance is quantified at the working point with 10% of the muons misidentified as protons: the three-plane $\mathcal{P}$ provides 93.7% efficiency at selecting protons compared to 83.4% for the collection plane only $\mathcal{P}$ and 81.6% for the $\chi^2$ test with respect to the proton hypothesis. An overall measure of the separation power is defined using the area under the ROC curve (AUC). When this metric is equal to 1, the variable allows perfect separation at any working point, whereas a value of 0.5 represents a random guess. The three-plane $\mathcal{P}$ scores an AUC of 0.977 compared to 0.955 for the collection-plane only $\mathcal{P}$ and 0.838 for the $\chi^2$ test with respect to the proton hypothesis. The robustness of the quoted performance is tested against detector systematic uncertainties. The performance is evaluated on a series of simulations with a modified detector response to assess the detector systematic uncertainty. Figure 7.16 shows the ratio of the ROC curves for the different simulated detector responses (more details in section 8.4) relative to the central value (blue line in fig. 7.15). This leads to an uncertainty on the proton efficiency of 1.2% at 10% muon misidentification rate. The uncertainty on the AUC is 0.002 units for the nominal value of 0.976. This uncertainty is dominated by the modeling of electron-ion recombination. The statistical uncertainty on the efficiency and AUC determination is negligible. The right plot of fig. 7.15 shows the AUC in bins of the azimuthal angle $\phi$ of the track, which describes the direction of the track on the plane orthogonal to the beam direction: $\phi = 0°, \pm 180°$ identifies the drift direction while $\phi = \pm 90°$ identifies to the vertical direction. Both plots illustrate an overall improvement of the separation power with respect to the $\chi^2$ test with respect to the proton hypothesis, and mitigation of the dependence of the performance on the track angle. Combining the three planes improves the separation power in every angular region, especially for vertical tracks

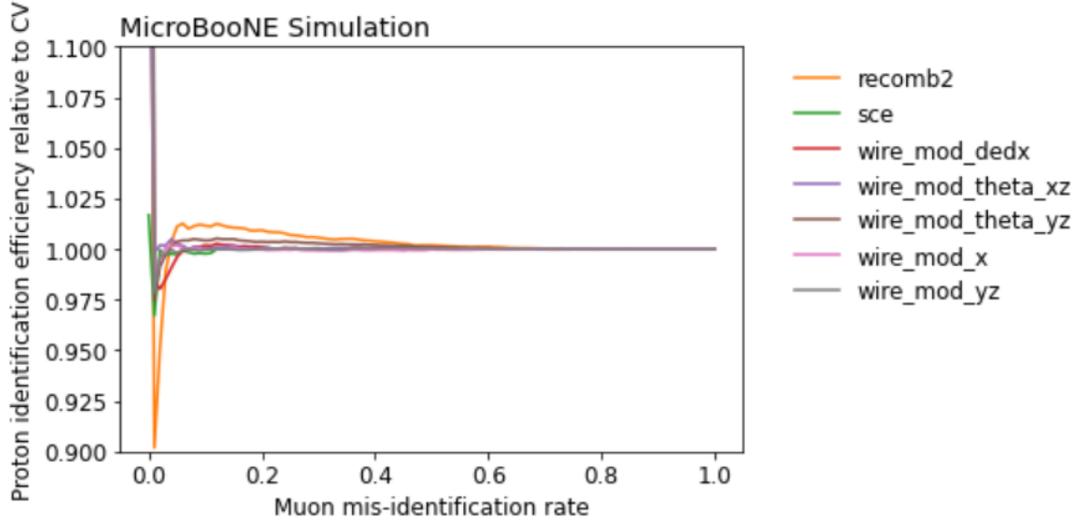

**Figure 7.16:** Systematic uncertainties associated with a mismodeling of the detector have a minimal impact on the selection efficiency of tracks based on the three-plane $\mathcal{P}$. The ratio of the ROC curves obtained with samples with different detector responses (one per line) to the central value (blue line in fig. 7.15) is of the order of a few percent across the whole spectrum, aside from the first couple of points at very low muon misidentification rate, where the selection efficiency is very small and relative fluctuation can be significant.

($\phi \sim \pm 90°$), where the collection plane is the least effective.

## 7.4 Identifying Tracks - physics applications

The following analyses were developed using data collected by MicroBooNE with the BNB during winter and spring 2016. This data amounts to $4.8 \times 10^{19}$ protons on target (POT). This data, in which neutrino interactions are present, is labeled as DATA Beam ON. The prediction comes from a combination of the simulation of neutrino interactions and data collected out of the beam windows, labeled as DATA Beam OFF. Even in events where a neutrino interaction is present, $\mathcal{O}(10)$ cosmic rays cross the detector on average. Instead of being simulated, cosmic ray waveforms are acquired out of the beam window and overlaid with simulated neutrino interactions.

The first test performed is to verify if the result obtained in the simulation in section 7.3 holds also with neutrino data. Tracks are selected by requiring track-score > 0.5, a measurement of the like-liness of a reconstructed particle to be a track, with values ranging from 0 for shower-like particles to 1 for track-like particles. Track-score is provided by the Pandora reconstruction. Tracks are also required to be reconstructed within 5 cm from the vertex, and to be contained within a fiducial volume, defined as the set of points that are at least 20 cm apart from every side of the TPC. Figure 7.17 shows the distribution of the PID score for these tracks, comparing the data (black cross) with the simulation (stacked colored histogram). Protons, reconstructed with a low $\mathcal{P}$, populate the left side of the distribution. These are well separated from lighter particles, such as muons and pions, which populate the region at larger values of $\mathcal{P}$. Tracks associated with cosmic rays are distributed along the whole spectrum, as they can be induced by cosmic muons or by protons kicked out of the argon nuclei. A peak at $\mathcal{P} \sim 0$ is also present. These are short tracks, for which there is too little information to discriminate between the two hypotheses. In fact, $\log(T)$ is additive for each hit: the longer the track, the more hits, the larger $\log(T)$ and thus $\mathcal{P}$ can be. The simulation reproduces the shape of the data, confirming the performance studied in the simulation. Two ad-ditional plots (fig. 7.18) illustrate that $\mathcal{P}$ correctly identifies the Bragg peaks, in good agreement with the theoretical prediction. Fully contained tracks, with track-score $> 0.8$ and collection plane *local pitch* $< 1$ cm are selected in beam data events. The 2D distributions $\mathrm{d}E/\mathrm{d}x$ vs residual range on the collection plane, for muon-like tracks with *pid* $> 0.5$, and proton-like tracks with $\mathcal{P} < -0.5$, are plotted on the left and right of fig. 7.18, respectively. The two Bragg peaks are clearly visible and distinct. This is possible because of the track local pitch requirement: selecting hits with a small local pitch ensures $\mathrm{d}E/\mathrm{d}x$ is measured properly, resulting in physical and meaningful values. The solid and dashed red lines show the theoretical prediction of the most probable value of the d$E$/d$x$ distribution for the extremes of the range of local pitch under consideration. The core of the data distribution lies between the two bands, demonstrating good calorimetric reconstruction for small local pitch.

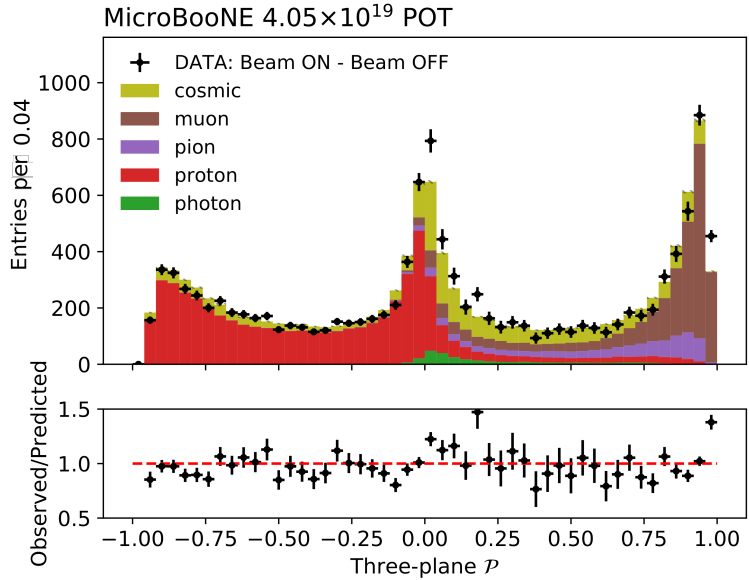

**Figure 7.17:** Distribution of the three-plane $\mathcal{P}$ for neutrino-induced tracks selected in data and simulation. The data (black cross) shows the difference between the DATA Beam ON and the DATA Beam OFF, in order to remove the contribution from non-beam events. The simulation (stacked histogram) is normalized to the same number of events observed in the data, and it is broken down for different particle species. In the case the particle selected in the simulation is an overlaid cosmic, it is assigned to the category "cosmic". The uncertainties shown on the data points are the expected statistical uncertainties from Poisson counting.

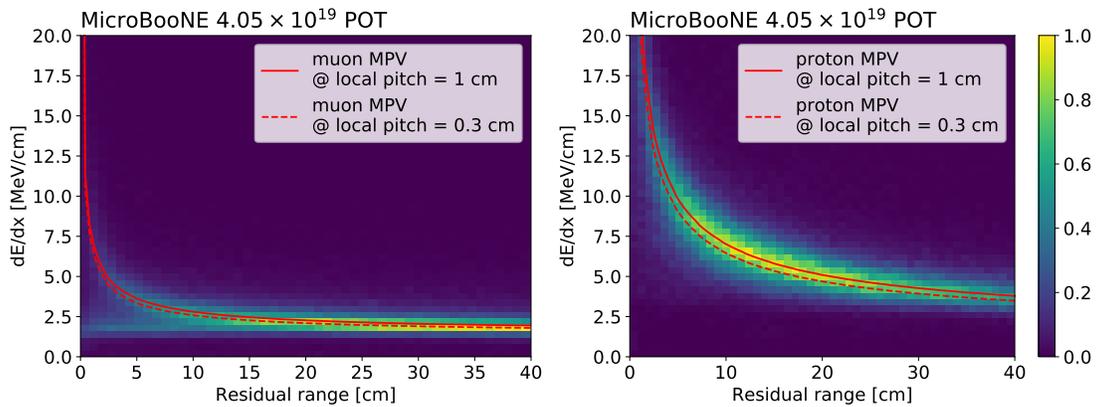

**Figure 7.18:** Collection-plane $\mathrm{d}E/\mathrm{d}x$ vs residual range profile for well reconstructed, contained, and low local pitch tracks in data, identified as muon (proton) candidates on the left (right) plot. The profiles are compared to the most probable values (MPV) as expected by the theory for the extremes of the range of local pitch under consideration (red lines). The two plots are normalized to the maximum value in order to share a common color scale.

## Large collection-plane-local pitch tracks identified with the two induction planes

The following example illustrates the efficacy of combining the calorimetric measurements performed with the three wire planes. Figure 7.19 shows the 2D distribution of d$E$/d$x$ and residual range measured on the U, V, and Y planes, for proton candidate tracks with large collection plane local pitch. Proton candidates are required to be fully contained and to have a track-score > 0.8. Proton-likeness is required through $\mathcal{P} < 0.5$. The collection plane local pitch is required to be > 1 cm: such tracks lie on the plane orthogonal to the beam, traveling in directions where the calorimetric reconstruction is subject to large distortion. For this set of tracks, only the induction planes exhibit the expected Bragg peak: combining the three planes recovers the separation power by correctly classifying protons whose calorimetric reconstruction is not accurate on one or more views.

## Exclusive $\nu_\mu$ selection

To further illustrate the separation power of $\mathcal{P}$, a general $\nu_\mu CC$ selection targeting contained events is performed, and the selected events are classified into different exclusive channels. Events are selected similarly to the procedure in [300], adding a containment requirement for all tracks reconstructed in the event by requiring the start and end points of each track to lie inside the fiducial volume, as described in section 7.4. First, a muon candidate is chosen as the highest $\mathcal{P}$ track among the tracks attached to the vertex longer than 10 cm. The top plot of fig. 7.20 shows the distribution

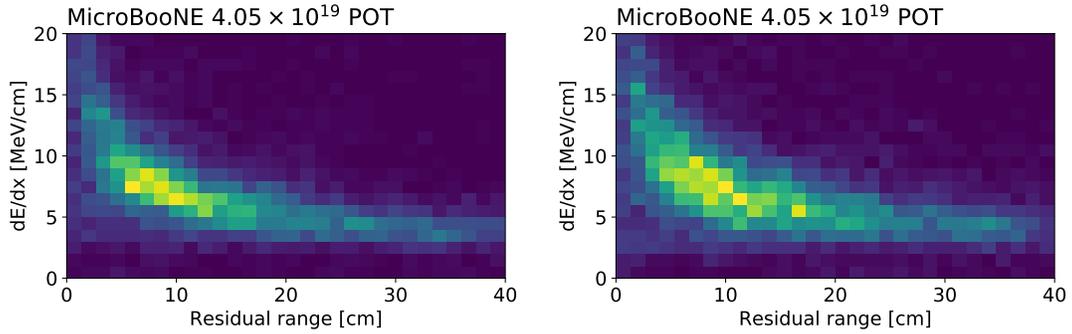

**(a)** First induction plane (U).　　　　　　　**(b)** Second induction plane (V).

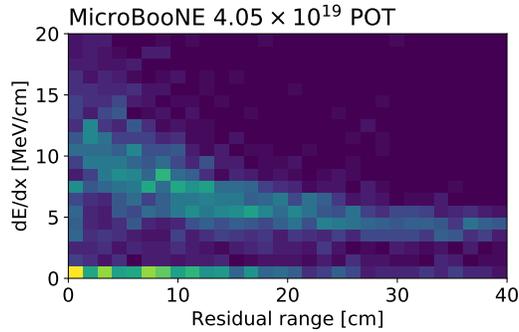

**(c)** Collection plane (Y).

**Figure 7.19:** 2D distribution of $\mathrm{d}E/\mathrm{d}x$ and residual range measured on the three wire planes for tracks identified as proton candidates in the data, with large collection plane local pitch.

of $\mathcal{P}$ for muon candidates, showing a good separation between muon and proton tracks. Selecting only events with a candidate with $\mathcal{P} > 0.2$ rejects most of the proton background, ensuring a pure selection of $\nu_\mu CC$ interactions.

Among the $\nu_\mu CC$ candidates, events with one additional reconstructed track (two-track events) are selected. If correctly reconstructed, they result predominantly from $\nu_\mu CC$ interactions with either one proton and no pions ($\nu_\mu \mathrm{CC}0\pi1\mathrm{p}$) or one pion and no protons ($\nu_\mu \mathrm{CC}1\pi0\mathrm{p}$) in the final state. In general, the former case predominantly (but not solely) results from quasi-elastic interactions while the latter is largely produced by the decay of a $\Delta$ resonance. The PID score of the second track (bottom left plot in fig. 7.20) separates the two cases, with $\nu_\mu CC0\pi 1p$ populating the left side while the $\nu_\mu CC1\pi 0p$ are located at positive values, because pions are indistinguishable from muons with this variable. By considering the events with $\mathcal{P} \leq 0$, we obtain a sample of contained $\nu_\mu CC0\pi 1p$ interactions with 61% purity and 40% efficiency. By applying the reverse cut, $\mathcal{P} > 0$, we have a background rejection of 98%, which provides the basis for a selection of contained $\nu_\mu CC1\pi 0p$ interactions. For this signature, the large cosmic ray background requires additional tailored background rejection. With a similar methodology, events with two additional reconstructed tracks (three-track events), are selected. Events with two protons and no pions in the final state ($\nu_\mu CC0\pi 2p$), predicted to be mainly induced by meson-exchange current interactions and final state effects, can be distinguished from events with one proton and one pion in the final state ($\nu_\mu CC1\pi 1p$), produced by a resonance decay. Because the presence of a proton, identified by a large negative $\mathcal{P}$, is common to the two cases, the track with the largest PID score among the two additional tracks (bottom right plot in fig. 7.20) is used to discriminate between $\nu_\mu CC0\pi 2p$, on the left, and $\nu_\mu CC1\pi 1p$, on the right. The cut $\mathcal{P} \leq 0$ provides a sample of contained $\nu_\mu CC0\pi 2p$ interactions with 54% purity and 24% efficiency, while the reverse cut, $\mathcal{P} > 0$, selects a sample of contained $\nu_\mu CC1\pi 1p$ interactions with 25% purity and 34% efficiency. In both cases, the background rejection is over 99.5%, emphasizing the difficulty of these selections, which could further benefit from additional cut variables.

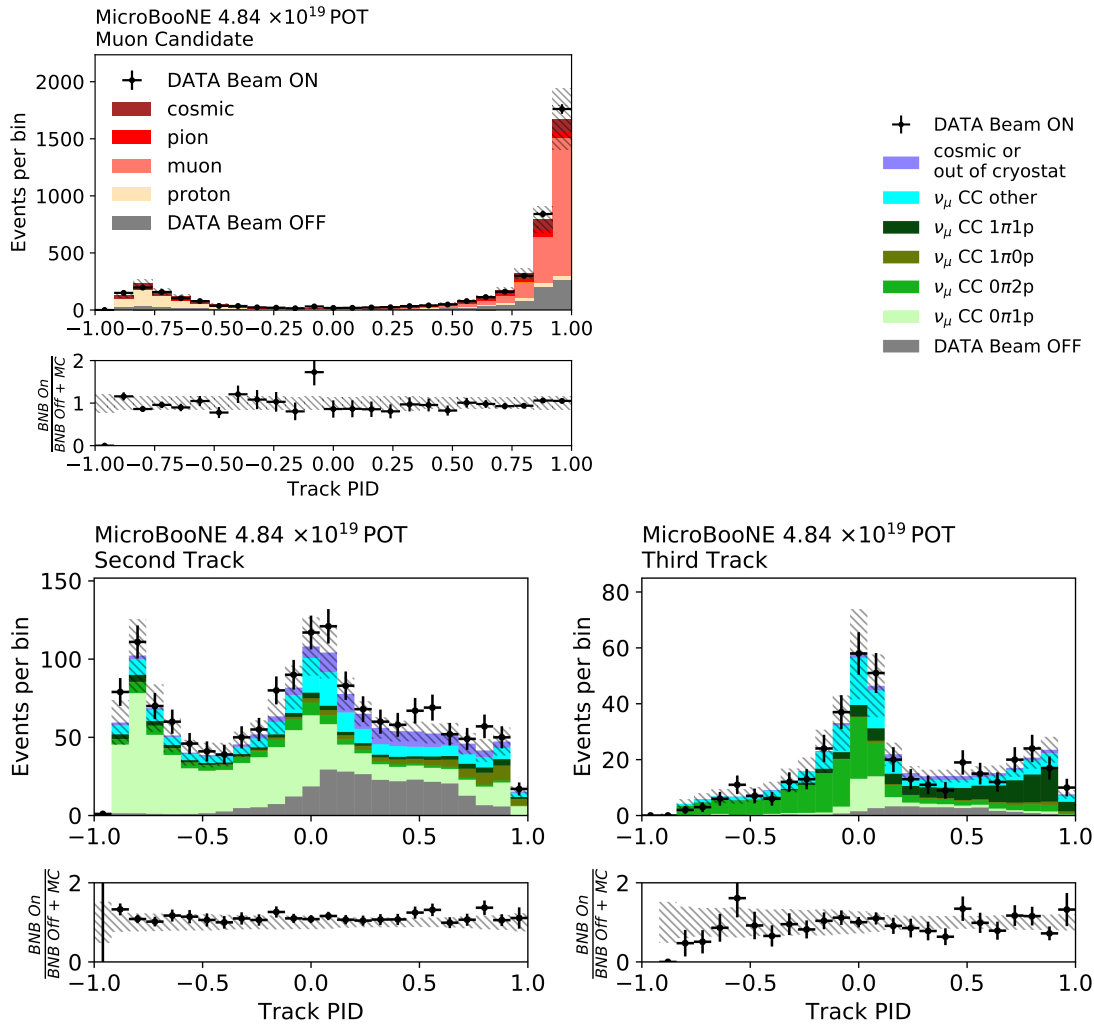

**Figure 7.20:** PID score distributions for the muon candidate track (top), the second track in events with two reconstructed tracks (bottom left), and the third track in events with three reconstructed tracks (bottom right). The DATA Beam ON (black cross) is compared with the prediction based on the sum of the simulation of neutrino interactions (stacked histogram) and DATA Beam OFF (gray bars). The selections are based on reconstructed quantities, while the prediction is broken down into different categories based on true information. In the first plot, the different colors correspond to different particle types, while in the other two they correspond to different final states. The uncertainties shown on the data points are the expected statistical uncertainties from Poisson counting, while the hashed patches on the stacked histogram illustrate systematic uncertainties in the prediction related to the simulation of the neutrino flux and interaction model.

Demonstrating the classification of exclusive $\nu_\mu CC$ final states is a novel result for liquid argon and stems from the potential of the new PID score. Future analyses will build on these examples, eventually leading to precise and detailed neutrino cross-section measurements.

## RE-INTERACTING PROTONS

Calorimetric-based classification assumes, and thus works best, with protons that stop through ionization, forming a clear Bragg peak. However, especially for momentum larger than 1 GeV, protons have a significant chance to undergo hadronic hard-scattering, resulting in broken tracks, which are harder to reconstruct. In most cases, only the first segment is reconstructed, and its d$E$/d$x$ profile does not present a Bragg peak, as the end of the track is the hard scattering.

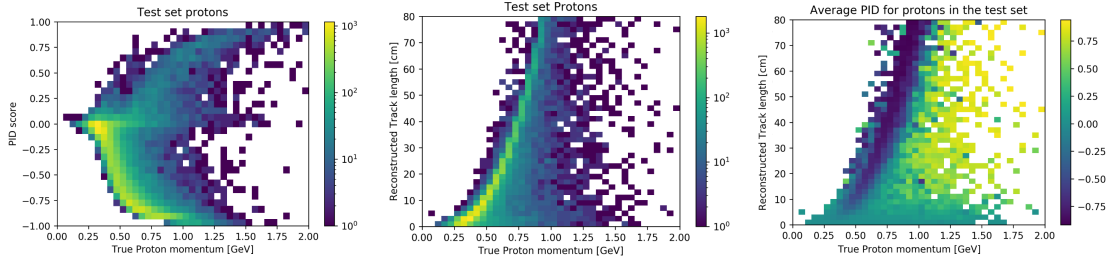

**Figure 7.21:** Left: $\mathcal{P}$ has a non-trivial dependency on the true proton momentum. Center: 2D distribution of the reconstructed track length vs the true proton momentum. Right: average $\mathcal{P}$ in bins of track length and true proton momentum.

The right plot in fig. 7.21 shows the $\mathcal{P}$ as a function of the true proton momentum, for protons in the test set. For very small momentum, the track does not produce enough hits to identify it. For larger momentum, most entries present a negative score, close to $-1$. However, there is a population in the opposite direction, which prefers the muon hypothesis. This is understandable if we consider that the profile of a broken proton when missing the true Bragg peak, will likely be more compatible with the Bragg peak of a muon rather than the one of a proton. The second plot shows the reconstructed track length versus the true proton momentum. The proton momentum is measured from the length of the track, so the vertical axis is equivalent to the reconstructed proton momentum. Most particles lie along the main line, with a spread induced by the detector resolution. However, there is a population of entries for which the reconstructed track length is significantly smaller than the expected one for their proton momentum, which can be explained by hard scatterings breaking the full expected track. If we plot the average $\mathcal{P}$ for the protons in each bin instead of the counts, as shown in the third plot, we see that, along the main line and for track length $\leq \sim 10$ cm, the average $\mathcal{P}$ is negative, while for the population of protons undergoing hard scattering, it is positive. For-tunately, at MicroBooNE energies, they rarely constitute a significant population. However, they might impact specific analyses or future experiments, and this work provides some foundation to develop tailored tools to identify them.

## 7.5 Distinguishing Showers

Calorimetric measurements can also be applied to showers to distinguish showers initiated by photons from the ones initiated by electrons.

### Shower objects

When a particle is classified as a shower, a shower object is constructed with important quantities regarding the dispersion of the charge along the main axis of the shower, such as the opening angle and the charge radius. However, the behavior that distinguishes electrons and photons is what happens in the first few centimeters of the shower, where it is indistinguishable from a track.

For this reason, we run the track reconstruction at the beginning of showers, producing a track object associated with the *trunk* of the shower. When the shower starts branching out, the track fitter follows one branch until the shower is too complicated for the main branch to be resolved. We can then obtain measurements of the local $dE/dx$ and direction and apply calorimetric tools similar to the ones developed for tracks.

There are two main handles to classify electron-induced from photon-induced showers, schematically illustrated in fig. 7.22. The first method relies in measuring the ionization density at the beginning of the shower, which results from a single electron in the first case, and from a collimated $e^+e^-$ pair in the second case, producing a value more or less double. The second handle requires measuring a gap between the vertex of the neutrino interaction, which is identified by other particles produced in the interaction, and the beginning of the shower. While there is no gap for an electron-induced shower, a significant gap is the footprint of a photon that traveled some distance before converting into an $e^+e^-$ pair.

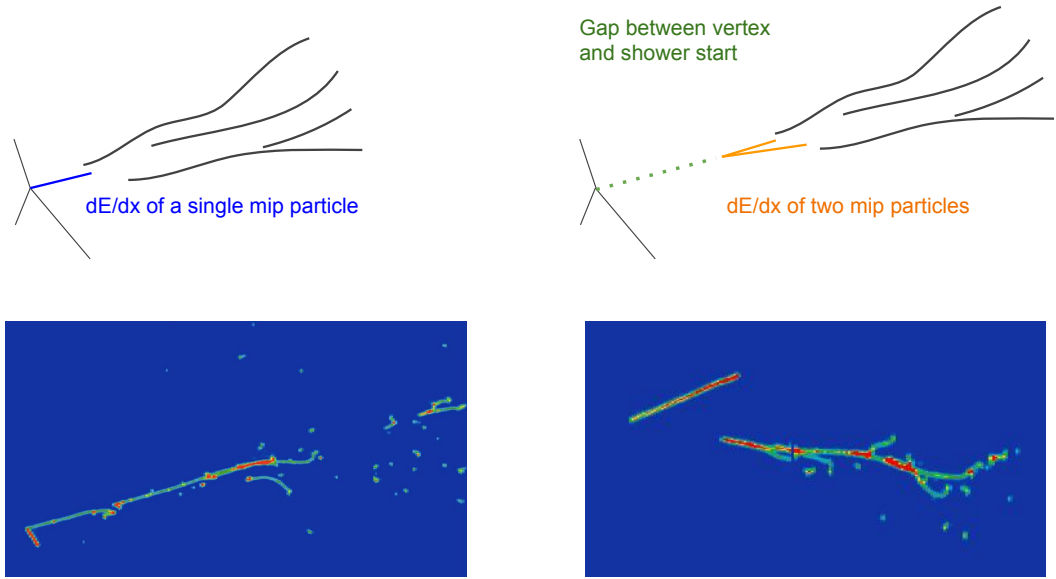

**Figure 7.22:** Electron- and photon-induced showers differ because of the gap between the vertex of the neutrino interaction, and because of the $dE/dx$ at the beginning of the shower, which correspond to a single electron in the first case, and to a collimated $e^+e^-$ pair in the second case. These features are exemplified by the two representative event displays.

## d$E$/d$x$ at the beginning of the shower

Electromagnetic showers in liquid argon are somewhat different depending on the energy. The left plot in fig. 7.23 shows the true energy distribution for an unbiased sample of reconstructed electromagnetic showers in the MicroBooNE simulation, originated by electrons or photons. In the following studies, in order to isolate physics effects from misreconstruction, we are considering only well-reconstructed showers, for which the purity at the hit level is larger than 90%, the shower start point is reconstructed in the fiducial volume and within 2 cm from the true start point. Photons are typically produced in $\pi^0$ decay and have typical energy of tens of MeV, while electrons result from $\nu_e$CC interactions and have typical energy of hundreds of MeV. While the shower length depends only logarithmically on the initial energy, the point of conversion varies significantly. As an example, the right plot in fig. 7.23 shows the distribution of the $\mathrm{d}E/\mathrm{d}x$ in the first 4 cm of the shower for photon-induced showers with different energies. Both characteristic peaks, at 2 MeV/cm and 4 MeV/cm, are present, although the relative heights change with energy. Low energy photons tend to be more electron-like from an experimental point of view. First, the Compton scattering cross section is still larger than the pair production cross section up to tens of MeV. Figure 7.24 shows the probability that a photon undergoes pair production as a function of its energy, for different materials. To a good approximation, all the other interactions in this energy range result from Compton scattering. In this case, the electromagnetic shower is induced by a single electron, thus indistinguishable from a genuine electron from a neutrino interaction. Second, the chance of producing low-energy electrons or positrons, that would stop after a very short distance and contribute only to a few hits, is not negligible for low-energy showers. Unfortunately these effects play against searches for low energy electron showers. The same effect is better visualized in fig. 7.25, showing the 2D distribution of $\mathrm{d}E/\mathrm{d}x$, as measured on the collection plane, versus the distance from the beginning of the shower, for electrons, on the left, and photons, on the right. While with electron showers we measure a constant $dE/dx \sim 2\,\mathrm{MeV/cm}$, photon showers exhibit a short region at $dE/dx \sim 4\,\mathrm{MeV/cm}$ before most of the hits enter the $2\,\mathrm{MeV/cm}$ region. Surprisingly, this re-gion is only $4-5$ cm long, while one might expect it to be of the order of the photon conversion length, about 29 cm in Argon [302]. In reality, it is relevant to know the distance after which at least one among the electron and the positron survives without bremsstrahlung, which is half of the interaction length for a single electron or positron. While this explains only partially why the region is $4-5$ cm long, it leaves room open for further exploration of the micro-physics happening in MicroBooNE. We further decompose the right plot in fig. 7.25 in energy bins, as shown in fig. 7.26.

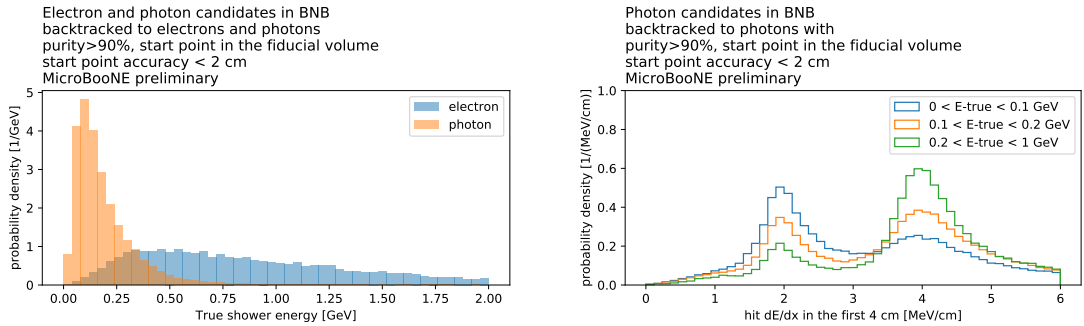

**Figure 7.23:** Left: distribution of the ground truth energy for showers initiated by photons (orange), and electron (blue). Right: normalized $\mathrm{d}E/\mathrm{d}x$ distribution of the hits within 4 cm from the shower start point, for photon-induced showers, in different energy bins.

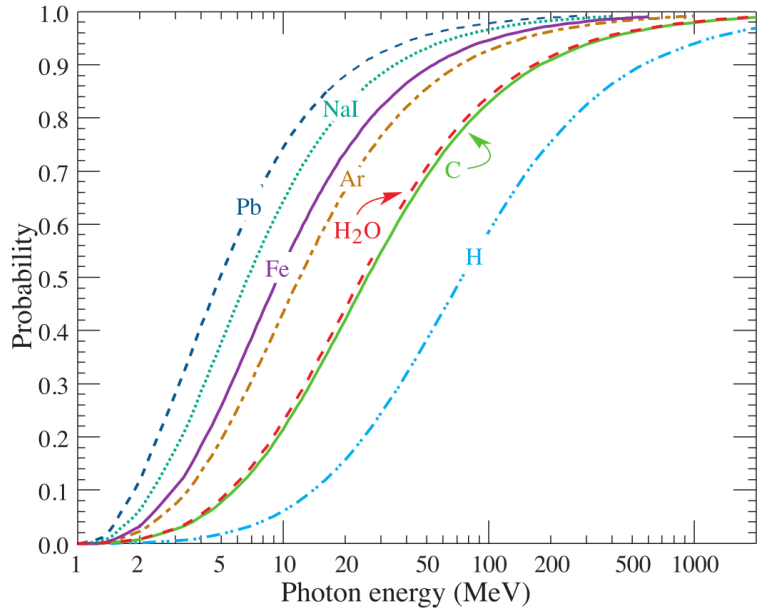

**Figure 7.24:** The probability that a photon undergoes pair production grows with energy, with a shape depending on the material the photon propagates in: the lighter the material, the less likely pair production with respect to other competing processes, which, in this energy region, are dominated by Compton scattering. This figure is taken from[301] (Fig. 33.17).

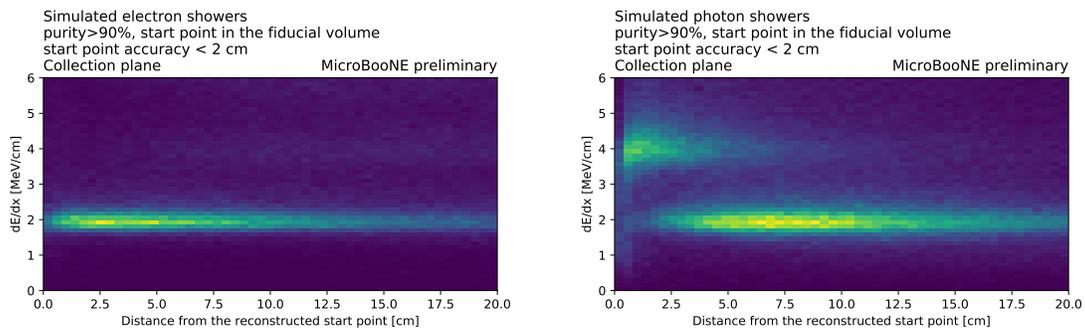

**Figure 7.25:** Hit $dE/dx$, as measured on the collection plane, versus distance from the reconstructed start point, for well reconstructed showers induced by electrons (left) and photons (right). While electrons showers contain hits mostly around $2\,\mathrm{MeV/cm}$, as produced by a single MIP particle, at least at the beginning of the shower, photons contain hits at $4\,\mathrm{MeV/cm}$, as produced by two super-imposed MIP particles.

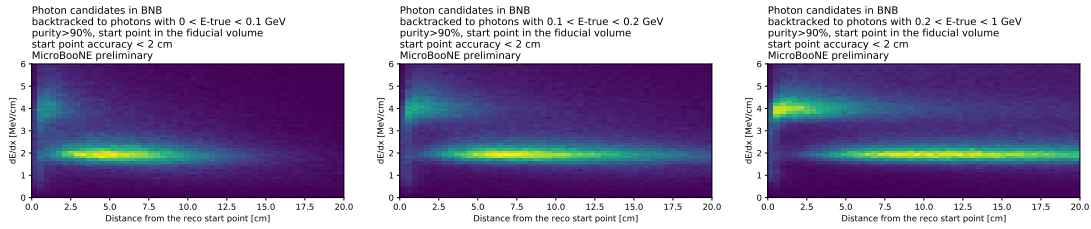

**Figure 7.26:** The distance after which a well reconstructed photon shower stop containing hits at $4\,\mathrm{MeV/cm}$ depends on the energy of the photon initiating the shower. The left plot in fig. 7.25 is split into three different energy bins: $0-100\,\mathrm{MeV}$ (left), $100-200\,\mathrm{MeV}$ (center), and $200-1000\,\mathrm{MeV}$ (right). At low energy, only the first 2 cm contain hits compatible with two MIP particles. This region extends to more than 5 cm at larger energies.

Showers at very low energy present only a few photon-like hits, only in the first $2-3$ cm. The region extends to longer distances when considering more energetic showers.

From this study we define the test statistic employed to distinguish electrons from photons as the median of the $\mathrm{d}E/\mathrm{d}x$ measurements in a range $\bar{d}$ from the conversion point

$$< \mathrm{d}E/\mathrm{d}x_{\bar{d}} > = med_{d_i < \bar{d}}\{\mathrm{d}E/\mathrm{d}x_i\}, \tag{7.16}$$

and we will consider $\bar{d} = 4$ cm, as obtained from the right plot of fig. 7.25 The three plots in fig. 7.27 illustrates the 2D distribution of the $< \mathrm{d}E/\mathrm{d}x_{\bar{d}} >$ for $\bar{d} = 4$ cm as a function of the shower energy, across the three different planes. While this technique provides a low rate of false-positive electrons misidentified as photons, the rate of false negatives, photons that are electron-like, is considerable, especially at low energy. This feature does not change much across wire planes, aside from producing sharper distributions around $2\,\mathrm{MeV/cm}$ and $4\,\mathrm{MeV/cm}$ with the collection plane, thanks to its higher resolution. Analogous plots for electrons would not be exciting, as it is unlikely for an actual single MIP particle to produce the equivalent energy deposition of two MIP particles.

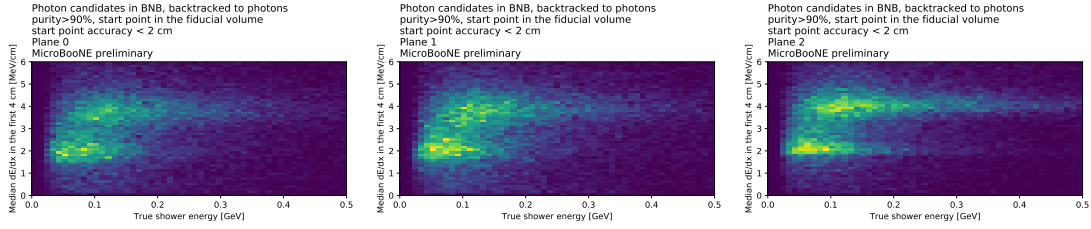

**Figure 7.27:** 2D distribution of the median $dE/dx$ in the first 4 cm and shower energy at truth level, for well-reconstructed photon showers across the three wire planes. While the two induction planes show similar distributions, the collection plane better resolves the two horizontal lines at $2\,\mathrm{MeV/cm}$ and $4\,\mathrm{MeV/cm}$, respectively. At low energy, most photon showers show only a peak at $2\,\mathrm{MeV/cm}$, in contrast with the behavior at larger energies, where the peak at $4\,\mathrm{MeV/cm}$ is predominant.

Lastly, the plot on the left in fig. 7.28 illustrates the distribution of the median $dE/dx$ on the first 4 cm, as measured by the collection plane, for the electron neutrino selection discussed in section 8.3. Events induced by an electron neutrino are shown in green: as they result in an electron in the final state, these events mostly populate the peak at $2\,\mathrm{MeV/cm}$. The main background of this selection, resulting from events with $\pi^0$ in the final state (light blue), makes up most of the peak at $4\,\mathrm{MeV/cm}$, as the selected showers are induced by final state photons in this case. However, while most electrons are in the first peak, and only the tail contaminates the second, a non-negligible fraction of photon showers ends up in the peak at $2\,\mathrm{MeV/cm}$, in agreement with the previous discussion. Moreover, the agreement between the data and the simulation is within the uncertainties, justifying its use in the selections targeting electron neutrinos. Together with the track identification score $\mathcal{P}$, this is one of the essential experimental quantities employed in these selections.

## Vertex - start point distance

The second vital handle to distinguish electrons and photons relies on reconstructing the whole event rather than the property of a single particle. While electrons release energy through ionization as soon as they are produced, photons are invisible until they Compton scatter or convert into an $e^+e^-$ pair. For this reason, the distance between the beginning of the shower and the vertex of the event, also known as conversion distance, is a good handle to distinguish electron-induced from photon-induced showers. The plot on the right in fig. 7.28 shows the distribution of the conversion distance in a sideband of the electron neutrino selection (section 8.3), enriched with events containing $\pi^0$ (light blue). This plot allows a characterization of the typical conversion distance, of the order of 10 cm, and shows the power of this variable to isolate events with electrons in green, which are only present in the first few bins. Finally, the simulation reproduces the data remarkably, making this experimental quantity reliable for the selection.

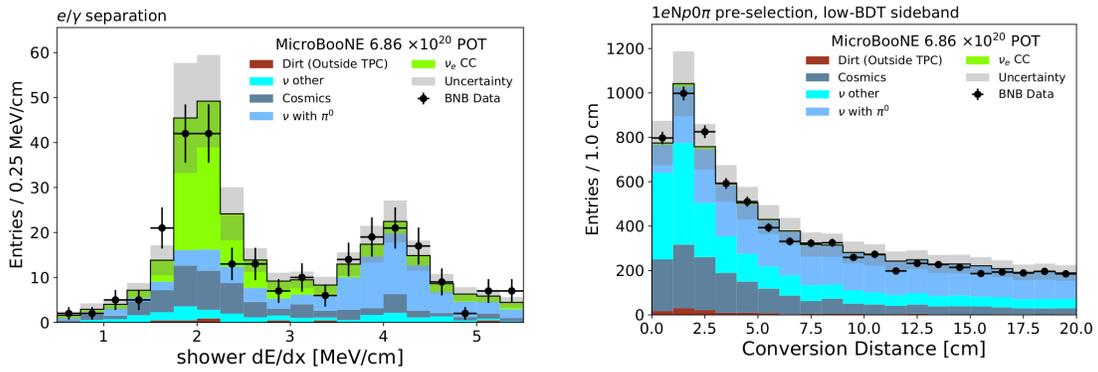

**Figure 7.28:** LArTPCs provide two main handles to distinguish electron-induced from photon-induced showers: the energy deposition at the beginning of the shower and the estimated photon conversion distance. The first is quantified through the median of the $dE/dx$ values along the initial 4 cm of the shower trunk (left plot), while the second one is by taking the distance between the vertex of the neutrino interaction and the start point of the shower (right plot). Both variables distinguish electrons (green) from photon showers, primarily present in events with $\pi^0$ (light blue), and show an excellent agreement between the data and the simulation.

We perform a preliminary study on improving the calorimetric classification of showers by adapting the technology developed for tracks in section 7.3 to showers. In this case, the local pitch still plays a role, as the angular effects in calorimetric reconstruction are analogous to track-like particles, as they do not depend on the nature of the particle that released energy. However, the residual range is not meaningful: instead the range, *i.e.*the distance from the start point, is helpful, as demonstrated by fig. 7.25. Figure 7.29 shows PDFs of $dE/dx$ for electrons and photons, analogous to fig. 7.12. In the first row, we fix the distance from the start point between 1 cm and 2 cm and vary the local pitch, while in the second row, we report examples with local pitch between 0.3 cm and 0.6 cm for different values of the distance from the start point. As observed for tracks, the peaks are sharper at smaller local pitch and smear out while moving at larger local pitch. The peak at 4 MeV/cm for photons is predominant at a small distance from the start point, becoming fainter further away, as seen in fig. 7.26.

Figure 7.30 illustrates the distribution of two classification variables: the median $dE/dx$ in the first 4 cm on the collection plane, as defined in eq. (7.16), and the new $\mathcal{P}$ variable obtained by applying the same methodology discussed in section 7.3 to the shower case, for showers originated by electrons and photons. A significant fraction of showers does not have any associated collection plane cluster, making the classification impossible if relying only on the collection plane. On the contrary, $\mathcal{P}$ always shows meaningful values, peaking close to $-1$ for photons and $+1$ for electrons. The performance is further evaluated by computing the different methods' ROC curves and the AUC. The left plot in fig. 7.31 shows the ROC curves for the new methods employing the collection plane only, the two induction planes only, and all three wire planes, in comparison with the median $dE/dx$ on the collection plane. The new method outperforms the well-established median calculation even when using only the collection plane, especially at working points that guaran-tee a small number of misidentified showers. The same information is further illustrated in the right plot, which shows the AUC in bins of the actual value of the shower energy. The improvement is remarkable over the entire energy spectrum, but especially in the medium energy range. While showing encouraging results, the method requires further validation using data; therefore, it is not employed in the rest of the analysis. However, this work sets the ground for an improved electron/photon classification which future iterations of the searches for electron neutrinos, both in MicroBooNE and other LArTPCs, will benefit from.

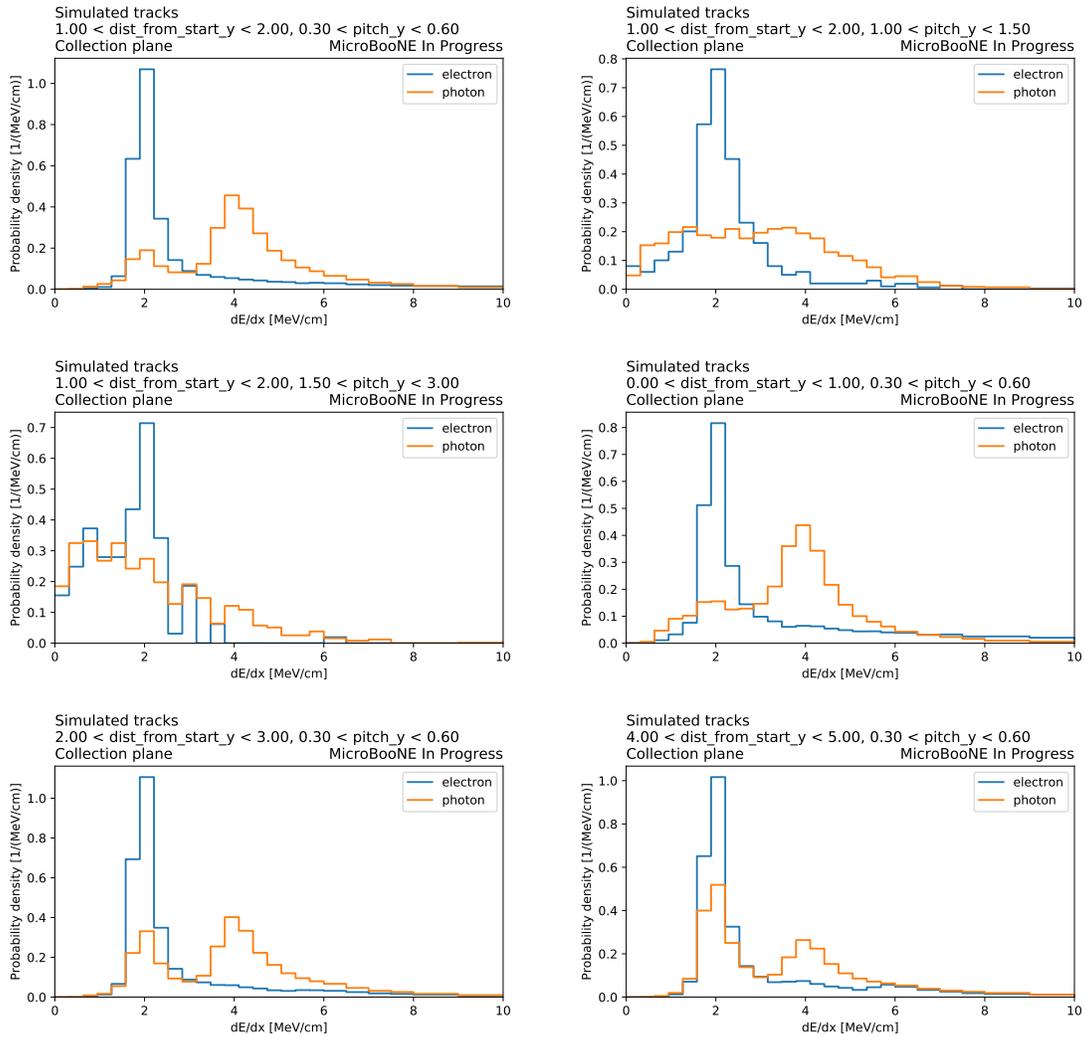

**Figure 7.29:** Analogously to fig. 7.12, we built the expected d$E$/d$x$ distributions for electron (blue) and photon (orange) hits. All plots are for the collection plane only. The first row shows the same bin in distance while varying the local pitch, while the second row has a fixed local pitch, and the distance from the start varies.

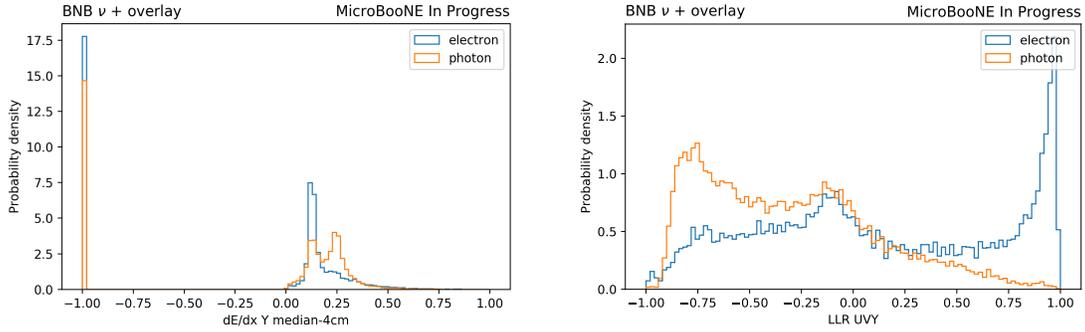

**Figure 7.30:** Both the median $\mathrm{d}E/\mathrm{d}x$ over the first 4 cm on the collection plane (left) and the likelihood based $\mathcal{P}$ (right) provide important information for classifying showers into electrons (blue) and photon (orange). Both distributions are normalized to the same area. While the median $\mathrm{d}E/\mathrm{d}x$ over the first 4 cm shows an unphysical peak at negative values, which contains all particles that do not have an associated cluster on the collection plane, the new $\mathcal{P}$ does not suffer this problem, separating electrons at positive values and photons at negative values.

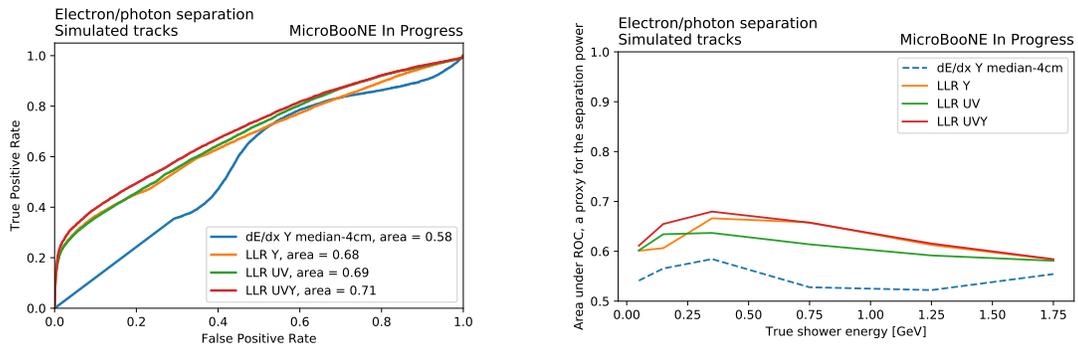

**Figure 7.31:** In analogy to fig. 7.15, we compare the performance of the new score $\mathcal{P}$ with the median $\mathrm{d}E/\mathrm{d}x$ in the first 4 cm using ROC curves and the AUC metric. The plot on the left shows the ROC curves integrated over the entire energy spectrum, while the plot on the right shows the AUC binned in the true shower energy. The new method outperforms the median $\mathrm{d}E/\mathrm{d}x$, especially in specific regions of misidentification rate and shower energy, even employing only the collection plane.

# 8

## The quest for electron neutrinos

THIS CHAPTER DESCRIBES A SEARCH FOR AN EXCESS of electron neutrino charged current interactions ($\nu_e$CC) over the predicted interaction rate of the $\nu_e$ component of the beam. It is performed by targeting final states without pions and using the Pandora reconstruction framework, as described in section 6.4. The signal model is constructed in order to benchmark sensitivity to potential new physics that results in an excess of $\nu_e$CC interactions that could explain the MiniBooNE excess. The experimental signature MiniBooNE observed, assuming that the excess of electromagnetic events is induced by final states electrons, should contain no pion, as pions would produce an additional visible signature in the MiniBooNE detector. Given MicroBooNE's capability to distinguish single protons, the analysis is split into the $1eNp0\pi$ and $1e0p0\pi$ channels, requiring the presence or the absence of protons, respectively. The selection of events employs a preselection, common to both channels, followed by multivariate analyses based on boosted decision trees (BDT) and tailored for each targeted final state. Since electron neutrino interactions are only 0.5% of the total neutrino interactions, in turn, overwhelmed by the cosmic ray background, these analyses aim at a very pure selection, compromising in terms of signal efficiency. Different sources of systematic uncertainties are considered, associated with the neutrino flux, cross-section modeling, or detector simulation. This analysis is performed with data collected between February 2016 and July 2018, summing up to a total exposure of $6.86 \times 10^{20}$ POT, which corresponds to the first three operation run periods. In order to minimize the risk of biasing the analysis result, the analysis strategy follows a blind scheme. Only a small fraction of the data, about $4.54 \times 10^{19}$ POT POT from Run 1 and $9.43 \times 10^{18}$ from Run 3[*], was available for the development of the analysis.

A combined analysis employs the $\nu_\mu CC$ analysis as a way to constrain and reduce the impact of systematic uncertainties in the $\nu_e$ channels, a procedure known as $\nu_\mu$ constraint. The physical interpretation and consequences of this analysis are discussed in chapter 9.

## 8.1 INTERACTION PROCESSES AND FINAL STATES

LArTPCs are powerful detectors because they allow the reconstruction of individual final state particles and the full characterization of the event, bringing neutrino detector technology closer to collider detectors. Depending on the momentum transferred between the neutrino and the nucleus, the interaction can undergo different processes and produce very different final states. Figure 8.1 shows a plot of the muon neutrino-nucleus charged current cross section as a function of the incoming neutrino energy $E_\nu$, split for different interaction processes and integrated over the momentum transferred to the nucleus, as taken from [303]. The cross section is normalized to the neutrino energy, as at first-order approximation, it grows linearly with the neutrino energy. Although the muon neutrino-nucleus charged current cross section is a specific case, the features of this plot are the same as for neutral current interactions or scattering of electron neutrinos.

---

[*]The main difference between Run1 and Run3 consists of better treatment of the noise of the electronics and the installation of a cosmic ray tagger (CRT) which improves the suppression of cosmic ray background.

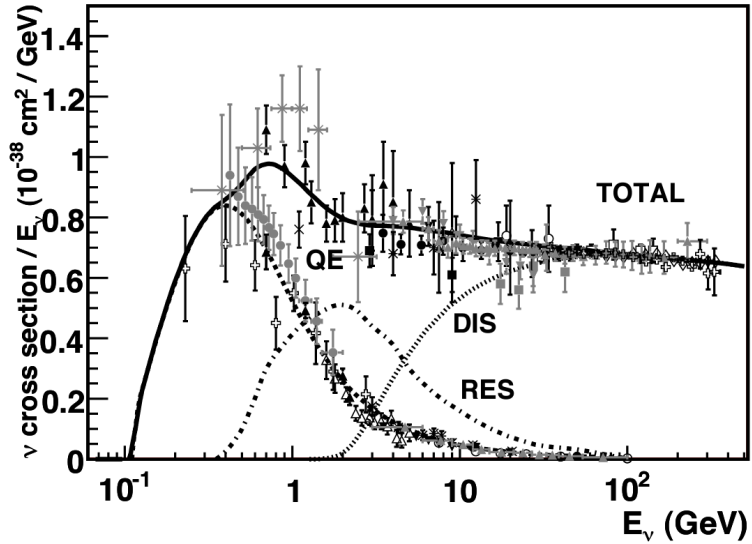

**Figure 8.1:** The charged current, neutrino-nucleus cross section, for muon neutrinos, normalized by and as a function of the incoming neutrino energy. Figure taken from [303].

At energies $E_\nu \lesssim 1$ GeV, most interactions are quasi elastic (QE). The momentum transferred is enough to resolve and scatter individual nucleons within the nucleus. The nucleus can be considered free, as the typical binding energy is of the order of 10 MeV, much smaller than the typical momentum transferred. The interaction can be approximated as elastic between the neutrino and the nucleon, which can be considered free. The quasi comes from the fact the final state particles are different than the initial state ones, as the neutrino converts into a lepton, while a neutron converts into a proton or vice-versa. However, the mass difference between initial and final state particles is much smaller than the typical momentum transferred, making the elastic approximation good enough when it comes to the theoretical calculation.

At intermediate energies $E_\nu \sim 1-2\,\text{GeV}$, the interaction between the neutrino and a nucleon can result in resonance excitations (RES). The most common one is $\Delta$, with a mass of about 1232 MeV, which, depending on the charge, can decay to neutral or charged pions. Moreover, the neutrino can scatter against a correlated nucleon-nucleon pair, a process called meson exchange current (MEC). As a result the scattering $\nu_l + np \rightarrow l + 2p$ is possible, which results in a two-protons final state.

At larger energies, the neutrino can exchange enough momentum to resolve individual quarks within each nucleon, a process called Deep inelastic scattering (DIS). These interactions result in the fragmentation of the nucleus and the formation of multi-hadron final states. Because of the large energy required, these interactions are rare in MicroBooNE, both with the BNB and the NuMI beam. Most interactions are QE, with a significant contribution from RES.

However, this description of different scattering processes and final state particles is an approximation valid only for light nuclei. Particles produced in the nucleus travel through the nuclear medium before exiting. During this process, they can re-interact with the nucleons and gluons, undergoing final state interactions (FSI). FSIs are more important with larger nuclei and can significantly affect the visible final state particles, for example, by changing the number of protons and pions that effectively exit the nucleus and are visible in the detector. FSI modeling is far from perfect, motivating analyses that are robust against mismodeling, for example, by measuring inclusive final states or a series of complementary exclusive final states.

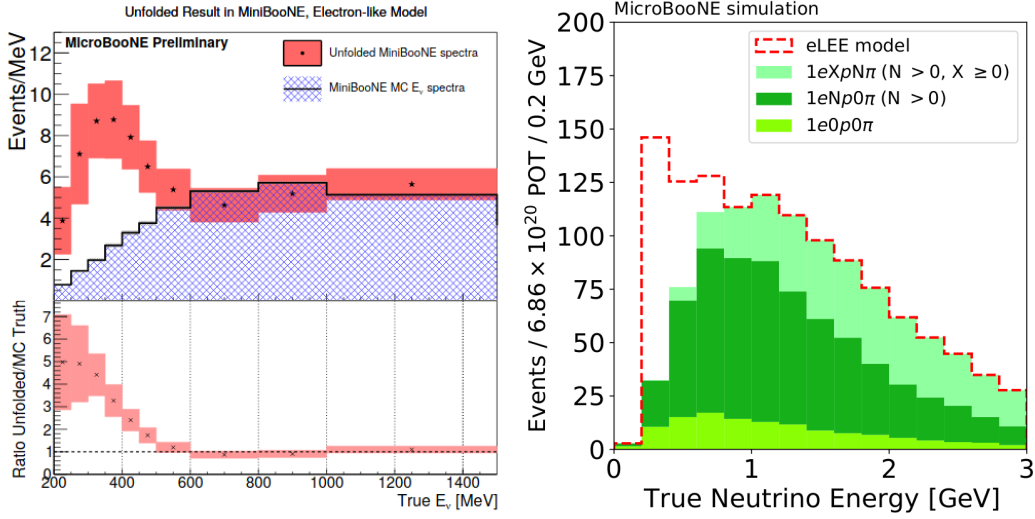

**Figure 8.2:** Left: the signal model employed in the eLEE search, as a function of the true neutrino energy, is obtained by rescaling the intrinsic flux by a factor varying between ∼ 1 at $E \sim 600$ MeV to ∼ 5 at $E \sim 200$ MeV. Right: relative contributions of the different final states to the total $\nu_e$CC rate, as a function of the true neutrino energy. The white area below the dashed red line represents the contribution from the eLEE signal model.

## 8.2 Signal model

The signal model is constructed empirically in order to benchmark the sensitivity to new physics models that could explain the MiniBooNE excess with the presence of additional particles in the beam. Future iterations of this analysis will look more carefully at a light sterile neutrino model, including not just $\nu_e$ appearance but also $\nu_e$ and $\nu_\mu$ disappearance. This model assumes that the MiniBooNE excess is induced by an additional component of $\nu_e$ in the beam,. It is indistinguishable from the intrinsic $\nu_e$ component of the beam, aside from the true energy spectrum. The signal is constructed starting from the data published by MiniBooNE in the 2018 result, using the following process, and it is shown in the left plot of fig. 8.2. The data containing the excess is subtracted from all backgrounds except for the intrinsic $\nu_e$ component of the BNB. Using MiniBooNE's electron neutrino energy smearing matrix, it is unfolded to obtain the best estimate of the true $\nu_e$ energy spectrum. The ratio between this spectrum (red histogram) and the intrinsic $\nu_e$ component of the BNB flux at MiniBooNE (blue histogram) provides energy-dependent weights that can be used to scale up the intrinsic $\nu_e$ component of the BNB flux at MicroBooNE to the signal expectation, as shown in the right plot of fig. 8.2. The filled histogram shows the intrinsic $\nu_e$ component, split into different final states, identified by the three different shades of green. The red line illustrates the expectation under the MiniBooNE LEE model, obtained by rescaling the filled histogram with the weights shown in the ratio plot of the plot on the left. As expected, most events fall in the low energy region, where most interactions are QE, thus producing either one or no visible proton in the final state. Systematic uncertainties on this signal related to the unfolding are not estimated, as they would require a combined analysis between MiniBooNE and MicroBooNE to properly account for all the correlations in the flux and cross-section modeling and, therefore, beyond the scope of this analysis. Finally, although this signal model is unrealistic, it captures the correct order of magnitude of the excess observed by MiniBooNE and propagates it to MicroBooNE. If the excess consists of events with one electron in the final state, which are caused by the presence of some additional particles in the beam, the sensitivity to this signal should be comparable with the sensitivity to a specific new physics model that explains the excess with this same final state. To account for this reasoning and in order to not fully bias the selection towards the low-energy region, the analysis targets the intrinsic component of beam as signal, and uses the MiniBooNE unfolded LEE signal only for estimated the significance of a possible excess. This strategy explains why this analysis is often referred to as a measurement of the intrinsic $\nu_e$ component of the beam.

## 8.3   Event Selection

The selection of neutrino candidates relies on the Pandora reconstruction framework, as discussed in section 6.4 and on the tools developed for particle identification, as illustrated in chapter 7.

Because MicroBooNE is located on the surface of the Earth, it is continuously crossed by a large number of cosmic rays, mostly muons. Their rate overwhelms the rate of neutrino interactions by several orders of magnitudes. On average, about six thousand cosmic rays cross the detector every second, while about one in a thousand beam spills lead to a neutrino interaction in the active volume of the TPC.

This fact produces two types of backgrounds: cosmic rays crossing the detector *in time* with the 1.6 μs beam spill, creating prompt scintillation light that triggers the recording of the event, and cosmic rays *out* of the beam spill, but within the 4.6 ms drift window, in events triggered by a genuine neutrino interaction. The first category is a few times larger than the rate of neutrino interactions and consists of events where no neutrino interaction is present. The second background is present in every event triggered by a neutrino interaction, as the TPC data always contains $\mathcal{O}(10)$ cosmic rays that can mimic and shade the actual neutrino interaction.

All cosmic backgrounds are estimated from the data. Cosmic rays *in time* with the beam spill are measured using the same trigger as beam data, but without the beam, and thus called Beam OFF. Cosmic rays *out* of the beam spill are measured with an unbiased trigger over a 4.6 ms window, and overlapped to simulated neutrino interactions.

In order to tackle these two backgrounds, the preselection stage combines optical and TPC information in order to cluster charge into different *slices* and select the one which is most likely to come from a neutrino interaction. Events with no slice compatible with a neutrino interaction are rejected as cosmic *in time* background.

For the selected neutrino slice, the hierarchical Pandora reconstruction discussed in section 6.4 is performed, resulting in a list of reconstructed particles. The selection for the $1e\mathrm{N}p0\pi$ and $1e0p0\pi$ channels is performed by computing high-level variables and combining them into multi-variate classifiers to reject the backgrounds in the most effective way.

In the first step of the preselection, Pandora groups the reconstructed particles into different *slices*, which are meant to cluster all the particles that belong to the same interaction. For example, all the final state particles from a neutrino interaction should be grouped in the same slice, as well as a cosmic muon that stops in the detector with the relative Michel electron resulting from its decay. The *Neutrino ID* tool (sometimes called Slice ID) allows the selection of the neutrino candidate slice among all the ones available, as illustrated in the sketch in fig. 8.3. First, all the *obvious cosmic rays* are removed. These are muons that cross the entire detector without producing any significant vertex-like features that are removed and are identified by looking at the geometry of their trajectories. Among the remaining slices, a combination of optical and geometrical information is employed. Compatibility between the flash that triggered the event and each slice is required by comparing the barycenter of the light measured by each PMT and the barycenter of the reconstructed charge. Moreover, a *topological score* assesses how alike the slice is to a neutrino interaction by exploiting features like kinks and vertices in the geometry of the charge. In most cases, no slice passes both criteria, resulting in a rejection of cosmic rays in time with the beam. In the other cases, the slice with the best match with the optical flash is selected as the neutrino candidate. The efficiency to select the correct neutrino slice, up to some small contamination from cosmic rays, is about 83% for both $\nu_\mu$ and $\nu_e$ interactions. The preselection efficiency also depends on the energy of the neutrino interaction, being about 60% at energies below 250 MeV, and grows with energy reaching a stable plateau around 1 GeV. Eventually, the presence of the one contained electromagnetic shower with reconstructed energy larger than 70 MeV is required to complete the preselection stage. Starting from Run 3 onward a cosmic ray tagger (CRT) was installed around the detector. If hits are present on these scintillation panels in time with the beam and can be matched to the reconstructed neutrino slice, the event is rejected as it is likely to come from a cosmic ray. Exceptions have been developed

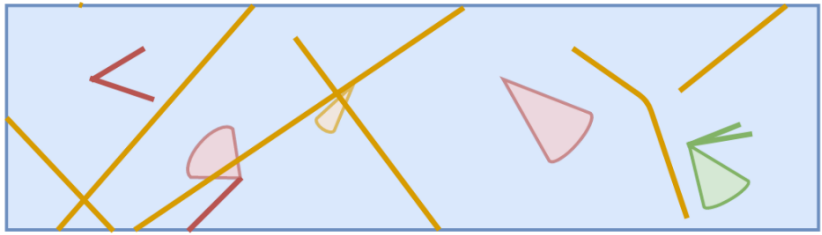

(a) Typical event with multiple interactions isolated by Pandora in `slices`. Candidate neutrino interactions are shown in green and red. Obvious cosmics are shown in orange.

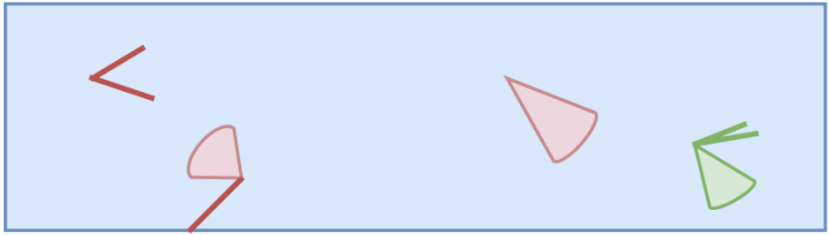

(b) Event after the removal of `obvious cosmics` tagged geometrically by Pandora.

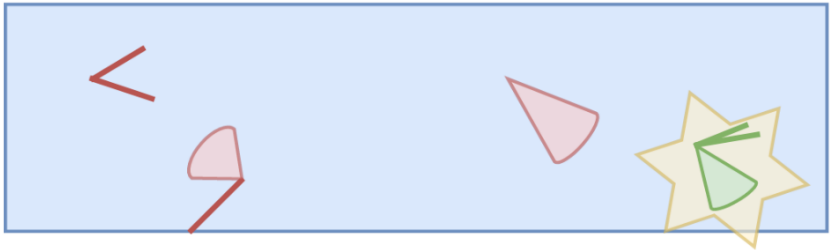

(c) Implementation of the `SliceID` tool to isolate possible candidate $\nu$ interactions. The selected candidate is highlighted, and shown in green.

**Figure 8.3:** A typical event contains several cosmic-ray interactions and might contain one neutrino interaction, clustered in *slices* by the Pandora reconstruction framework (first sketch). Slices that are very likely to originate from cosmic rays are tagged by Pandora using geometric information (second sketch). The Neutrino ID tool combines optical and geometrical information to isolate the slice which is most likely to contain a neutrino interaction (third sketch). The cartoon is taken from [304].

for uncontained events, typically $\nu_\mu CC$ interactions at high energy, which would produce hit only on one side of the CRT. This tool provides a slightly larger efficiency for Run 3 with respect to Run 1.

Energy reconstruction is an essential ingredient of the analysis, as it is the main handle to distinguish the signal, which is expected to hide at low energy. While the basic ideas of energy reconstruction are common among most analyses, the specific definitions are tailored for this analysis. Shower energy is estimated by summing up all the deposited charge on the collection plane and converting it to reconstructed energy with a constant factor assuming a flat recombination model. By performing a study using electron showers selected at the truth level, it turns out that this quantity needs to be corrected by 20% to account for the bias of all deposited energy below the threshold to form a hit. This bias is flat in reconstructed energy, as it is induced by some missing charge at a later stage in the shower development. Track energy is obtained by converting the range to the initial momentum with a specific assumption of the particle type: proton in the $1e$N$p0\pi$ analysis and muon in the $\nu_\mu$CC selection. The reconstructed neutrino energy, or total deposited energy, is the sum of the energy of all reconstructed particles and can be compared to the total visible energy, *i.e.* the energy deposited into visible final state particles at truth level. Figure 8.4 illustrates the 2D distribution of reconstructed versus true energy for showers, tracks, and electron neutrinos. The typical energy resolution for electron showers is of the order of 15%, for proton tracks varies between 4% at 100 MeV and 1% at 200 MeV, while for muons is of the order of 3%.

From this point onwards, the two channels $1e$N$p0\pi$ and $1e0p0\pi$ split into fully orthogonal selections. Figure 8.5 shows two examples of the interaction targeted by the two channels. The showery signature is evident, and it is accompanied by a proton track in the first case.

## $1e$N$p0\pi$ selection

This channel targets neutrino interactions resulting in one electron, at least one proton, and no other particle in the final state. It is the most sensitive for the signal model, as most electron neutri-nos at low energy result in one electron and one proton in the final state.

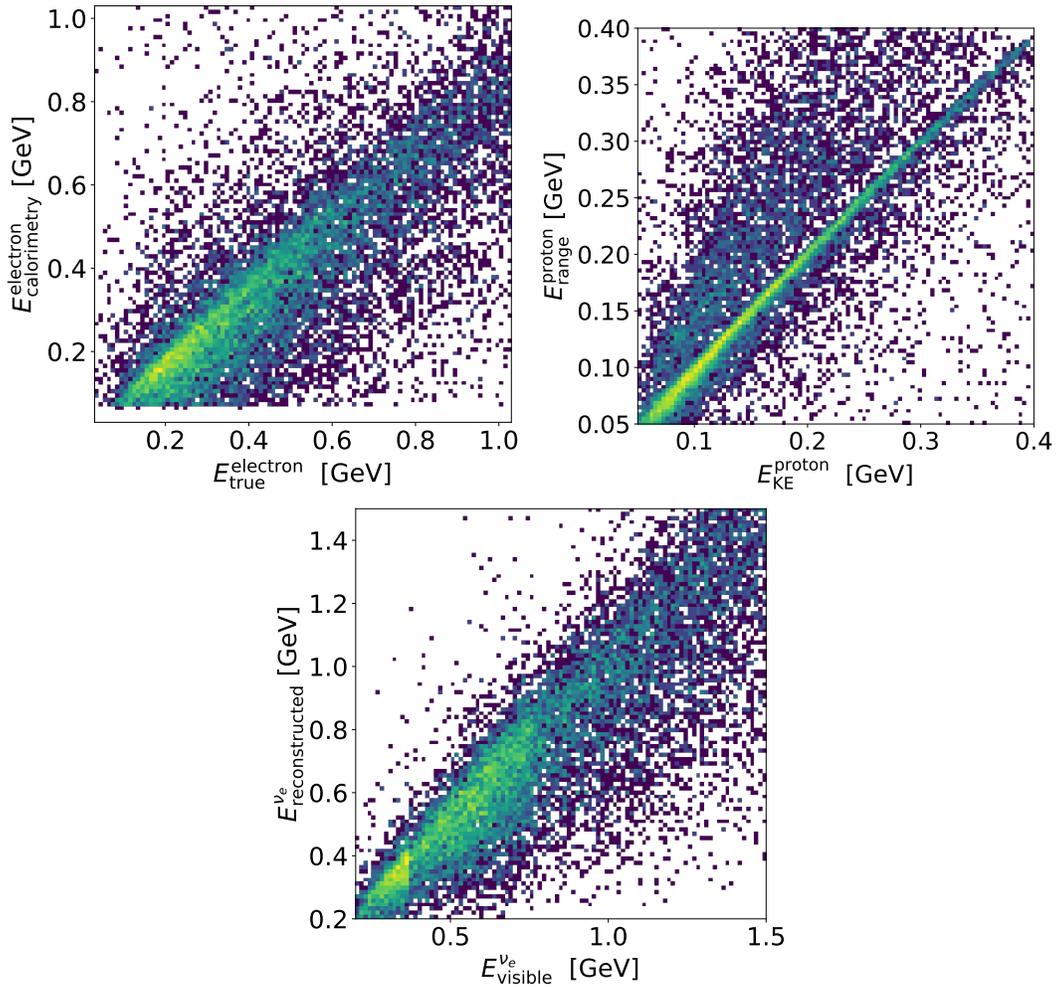

**Figure 8.4:** 2D distributions of the reconstructed (vertical axis) and true (horizontal axis) energy for electrons (left), proton tracks (middle), and electron neutrino events (right). For electrons, the energy is obtained calorimetrically as the sum of the deposited charge. For protons, it is extracted by converting the range into initial momentum. For the neutrino, it is reconstructed by summing the calorimetric energy for showers and the energy obtained from the range with the proton assumption for all tracks. It is compared with the visible energy rather than the true neutrino energy, which is the energy of the neutrinos spread over visible final state particles (excluding neutrons and particles below the detection threshold).

In order to complete the preselection requirements, we require the presence of at least one contained track to target the protons in the final state and that the reconstructed energy of the shower is larger than 70 MeV, which suppresses the background from Michel electrons, resulting from decays of muons. Lastly, 90% of the total charge is required to be contained within the fiducial volume, defined as points that are at least 20 cm away from every wall of the active volume of the TPC. This cut ensures that the energy reconstruction is quite accurate, removing events with significant missing charge, which are reconstructed at lower energy, creating a background for the eLEE signal model. Figure 8.6 shows the reconstructed energy spectrum after pre-the full $1eNp0\pi$ preselection. The agreement between the data and the simulation is within the systematic uncertainties, which will be better discussed in section 8.4. At this stage, about 25% of the background is dominated by cosmic rays, mostly in time with the beam, while $\nu_e$CC$0\pi Np$ constitutes about 1% of the total sample, while the signal constitutes only about 0.3% of the total selection.

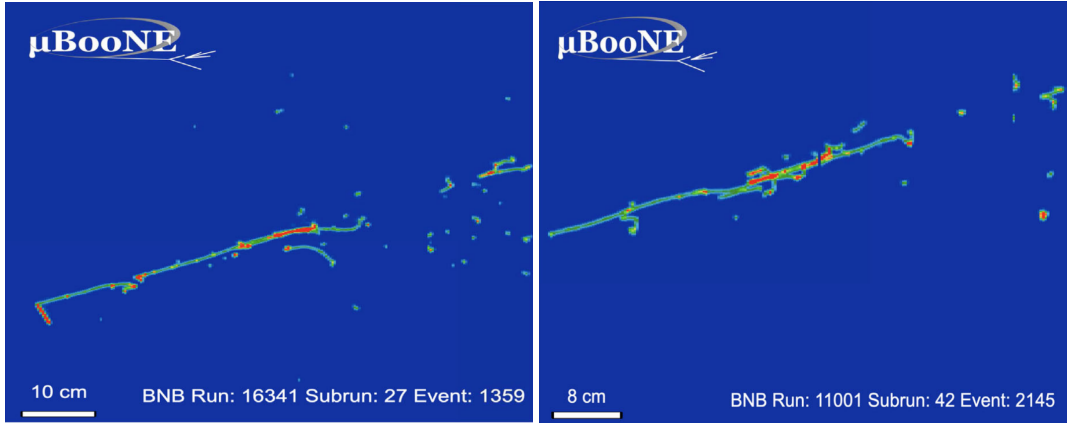

**Figure 8.5:** Examples of a $1eNp0\pi$ (left) and $1e0p0\pi$ (right) interactions, as measured by the collection plane. The two events look almost identical aside from the presence of the proton track in the first case.

The next steps in the analysis target the interpretation of the events a $1eNp0\pi$ final state. We require the presence of exactly one contained shower, which is also going to be called electron candidate in the following. Removing events with two or more showers significantly reduces the background induced by events with $\pi^0$. Some events in which a second shower is present and particularly well aligned with the most energetic shower are very likely to be induced by $\nu_e$CC interactions at high energy, where the reconstruction fails to cluster all the charge in a single object. Those events are re-interpreted as single-shower events, and information about the second shower is included in the analysis. Moreover, among all tracks, the longest, and thus most energetic, track is taken as the main proton candidate, which should reconstruct the final state proton in clean CCQE interactions.

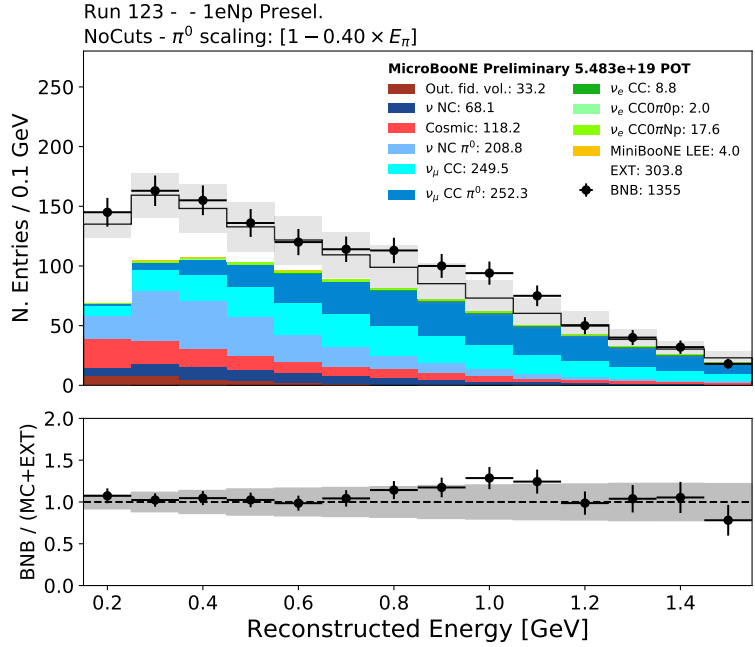

**Figure 8.6:** Reconstructed neutrino energy after preselection for the 1eN$p$0$\pi$ channel.

The analysis strategy proceeds as follows: two Boosted Decision Trees (BDT) are trained in order to distinguish the 1eN$p$0$\pi$ signal from the backgrounds. The first one targets explicitly background induced by events with $\pi^0$, while the second one targets the other backgrounds, mostly containing muons, either cosmic or from $\nu_\mu CC$ interactions. The two BDTs are designed to combine the information contained in all the experimental variables, which alone would not be enough for a high-purity selection, into scores that can be used for a more powerful rejection of the two main backgrounds. The BDTs are trained using the following variables:

- CosmicDirAll3D: Dot product between shower start and space points associated to tracks flagged as cosmic.

- CosmicIPAll3D: Closest distance between shower start and space points associated to tracks flagged as cosmic.

- trkshrhitdist2: Minimum distance between leading shower and longest track clusters in 2D.

- trkfit: Fraction of the 3D space points successfully fitted with the shower track-fitter algorithm.

- trkpid: Proton-muon $\mathcal{P}$ particle identification.

- tksh distance: Distance between the leading shower and longest track start points in 3D.

- subcluster: Number of isolated 2D segments of charge associated with a reconstructed shower on all three planes.

- tksh angle: Angle between the leading shower and longest track directions.

- shrmoliereavg: Average angle between the shower's direction and its 3D space points.

- shr score: Pandora SVM track/shower score for the leading shower.

- shr tkfit dedx max: Median dE/dx on the plane with most number of hits in $[0, 4]$cm trunk segment.

- secondshower Y nhit: Number of hits on each plane of the largest cluster associated with the recovered 2nd shower.

- secondshower Y vtxdist: 2D distance from vertex for the largest 2D cluster associated with the recovered 2nd shower in each plane.

- hits ratio: Ratio between hits from showers and the total number of hits in the slice.

- secondshower Y dot: Dot product between the vector connecting the vertex to the closest hit in the cluster and the charge-weighted cluster direction w.r.t. closest hit in a cluster.

- anglediff Y: 2D angle difference in each plane between the 2nd shower and the 1st shower cluster (cluster direction defined as the charge-weighted direction of cluster w.r.t. vertex).

Second-shower variables have been constructed to reject a special class of misreconstructed $\pi^0$ events, where one shower is reconstructed but not included with the neutrino slice. Locating single clusters near the neutrino interaction and likely emerge from the vertex of the interaction allows additional rejection of the $\pi^0$ background.

The signal for both BDTs is defined as $1eNp0\pi$ events with reconstructed energy smaller than 800 MeV. Dedicated samples are used for training, not to reduce the limited sample size used for the analysis. The signal comes from intrinsic $\nu_e$ samples simulated separately in the range $0 - 400$ MeV and $400 - 800$ MeV, in order to have a larger statistic at low energy. Beam OFF (EXT) events are taken from the NuMI stream. Tailored background samples enhancing the different backgrounds, for example, $\nu_\mu 0\pi^0$ for the non-$\pi^0$ BDT, are generated using truth-level filters. BDTs, training, and inference are implemented using the XGBoost package[305]. Figure 8.7 illustrates the relative importance of the different training variables in terms of the *gain*, which is the improvement in accuracy brought by a given feature. As expected, variables related to the electron candidate and the track PID provide the most gain. Remarkably, the relative contribution of the different variables to the two BDTs is quite similar. Figures 8.8 and 8.9 shows the distribution of the BDT score for the non-$\pi^0$ and the $\pi^0$ BDTs, respectively. The low score and low purity regions are shown on the left, while the high score regions, with larger purity, are shown on the right. Similarly to fig. 8.6, the plots are performed using the Run 1 and Run 3 open data. The full selection requires a combination of loose cuts and cuts on the BDT scores. These loose cuts are applied in order to remove events that have at least one feature which is significantly incompatible with the $1eNp0\pi$ signature. To further remove photon showers, the distance between the electron candidate and the proton candidate start point is required to be smaller than 6 cm, and the median $dE/dx$ at the beginning of the shower needs to be larger than 0.5 MeV/cm and smaller than 5.5 MeV/cm. Moreover, the impact parameter, defined as the minimum distance between the neutrino candidate vertex and the closest cosmic rays, is required to be larger than 10 cm, which removes misreconstructed delta-ray showers. At the same time, the longest track is required to have track-pid $\mathcal{P} < 0.02$ in order to remove tracks compatible with muons as well as misreconstructed proton tracks. In order to remove $\nu_\mu$ background, we require that the electron candidate makes up at least 50% of the total number of hits, that its estimated Moliere angle is larger than $9°$, and that it contains at least four different subclusters. Moreover, when considering the electron candidate fitted as a track, the fraction of the 3D space points successfully fitted with the track-fitter algorithm is required to be smaller than 0.65, meaning that fitting it with a track model results in a poor fit, and its length when fit as a track need to be smaller than 300 cm. The cosine of the angle between the electron and the proton candidate directions should be larger than $-0.9$, which removes misreconstructed events in which a true shower is split in a reconstructed track and a reconstructed shower going in opposite directions. Lastly, we apply cuts of 0.67 and 0.7 on the $\pi^0$ and non-$\pi^0$ scores, respectively. Figure 8.10 shows the predicted spectrum for the neutrino reconstructed energy, using Run 1, 2, and 3 simulated data. The sample is very pure, aside from some neutrino background, peaking around 500 MeV in reconstructed energy. With this selection, we predict about nine excess events in the analysis from the signal model.

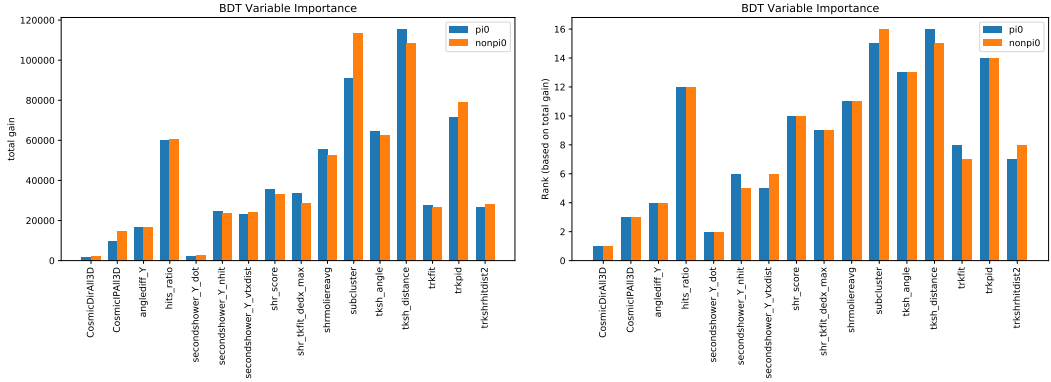

**Figure 8.7:** BDT variables important in terms of the *gain*, the improvement in accuracy brought by any given variable. Left: the total gain value is the sum of the gain across all branches of the tree. Right: Rank, from lowest ($= 1$) to highest ($= 15$) based on the total gain value.

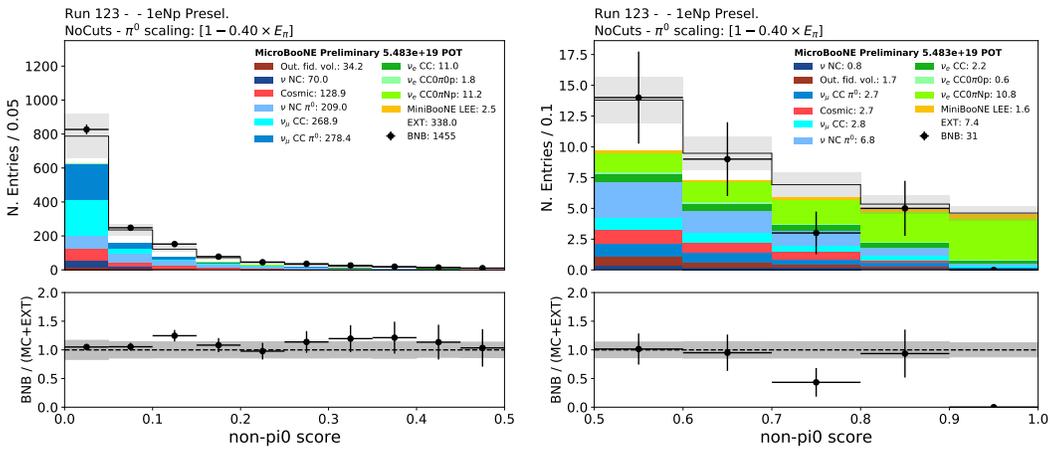

**Figure 8.8:** Classification score for the non-$\pi^0$ BDT, after the $1eNp0\pi$ preselection. The left plot shows the first half, mostly dominated by background, while the plot on the right shows the signal-enriched region.

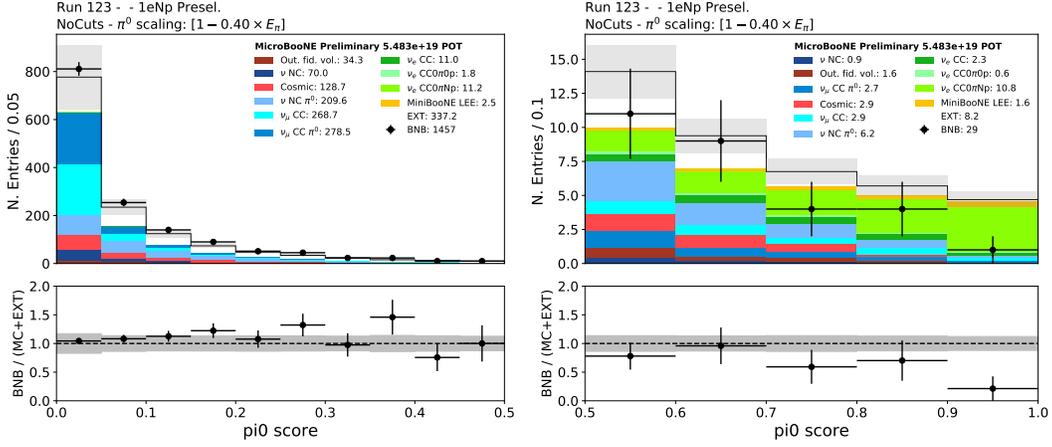

**Figure 8.9:** Classification score for the $\pi^0$ BDT, after the $1eNp0\pi$ preselection. The left plot shows the first half, mostly dominated by background, while the plot on the right shows the signal-enriched region.

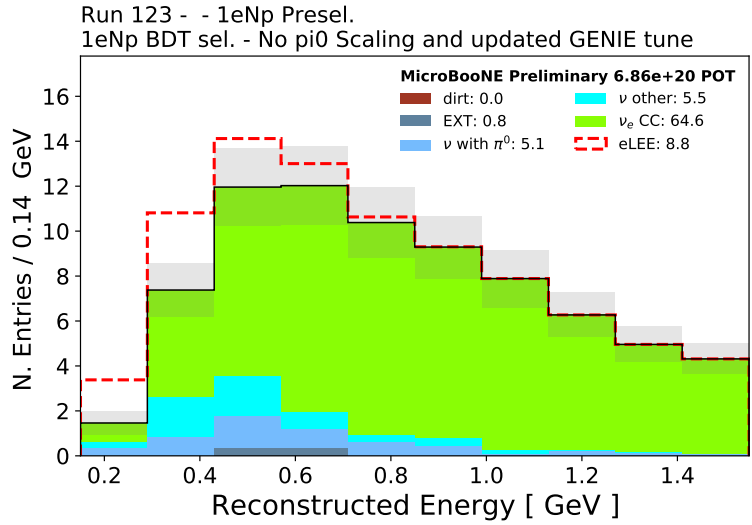

**Figure 8.10:** Simulated spectrum in reconstructed neutrino energy after the full BDT selection, using Run 1, 2, and 3 Monte Carlo. The selection is very pure, with only a small neutrino-induced background. Under the signal model hypothesis, we predict the presence of about nine excess events.

Efficiency and purity as a function of the neutrino energy are shown in fig. 8.11, at different selec-tion stages. The final signal efficiency turns out to be around 5% at low energy, peaking at 15% at energies close to 1 GeV. The contribution to the small efficiency comes mostly at the preselection stage, where the efficiency in the lowest energy bin is only 40%. Further selection reduces the efficiency further but moves the purity from 1% at preselection to the $60 - 80\%$ range. The main limitation causing the efficiency degradation is caused at the reconstruction level. In fact, if the reconstruction was perfect, the efficiency and purity would be both very large, as there is no other process that produces an electron shower and a proton track in the final state. However, the accuracy of the vertex reconstruction is the main limiting factor, as particles are reconstructed as coming out of the vertex: the whole reconstruction and interpretation of the neutrino slice are significantly degraded when the vertex is misplaced by a large amount, which is often the case. Studies performed using samples in which the correct location of the vertex is artificially injected in the reconstruction show improvement of the overall efficiency at preselection by more than 50%. While future iterations of this analysis will benefit from a more accurate reconstruction, improvements in the vertex placement are beyond the scope of this work.

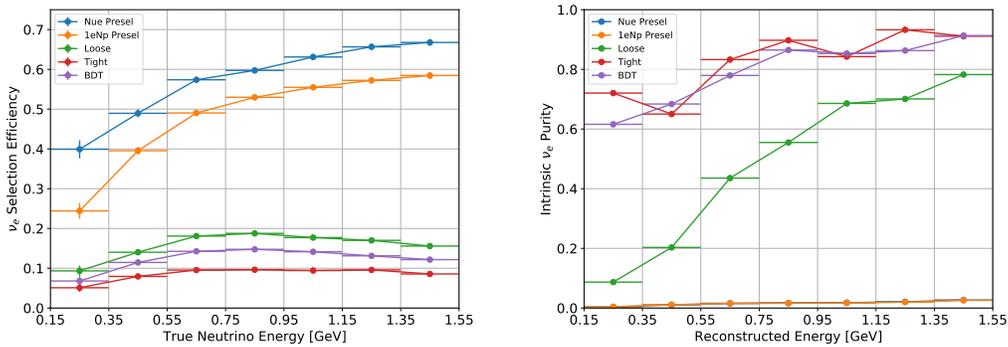

**Figure 8.11:** Left: efficiency of the $1e\mathrm{N}p0\pi$ analysis as a function of the true neutrino energy at different selection stages. Right: purity of the $1e\mathrm{N}p0\pi$ analysis as a function of the reconstructed neutrino energy at different selection stages. The BDT selection stands for the cuts on the two BDT scores on top of the loose cuts. "Tight" refers to a cut-based selection, an alternative to the BDT one, which provides a similar level of purity at the cost of a larger efficiency. The LEE signal model is not considered in these plots.

A final remark is about track multiplicity, as tracks play a very important role in this channel. Figure 8.12 shows the estimated joint and conditional distributions of the number of reconstructed contained tracks versus the number of protons above the threshold at the true level on the left and right plots, respectively. The distribution is obtained at reconstructed energy larger than 1.05 GeV, where multi-track events are more likely to occur, for true $\nu_e\text{CC}0\pi Np$ interactions, at loose selection cuts. While most events lie along the diagonal, there is a 10% chance of reconstructing two tracks in events with one proton only or a 20% chance of reconstructing one track in events with two protons at the true level. Given this study, we expect $\mathcal{O}(10\%)$ migration between signal regions with a different number of tracks, which justifies the choice of the analysis channel targeting $1e0p0\pi$ analysis.

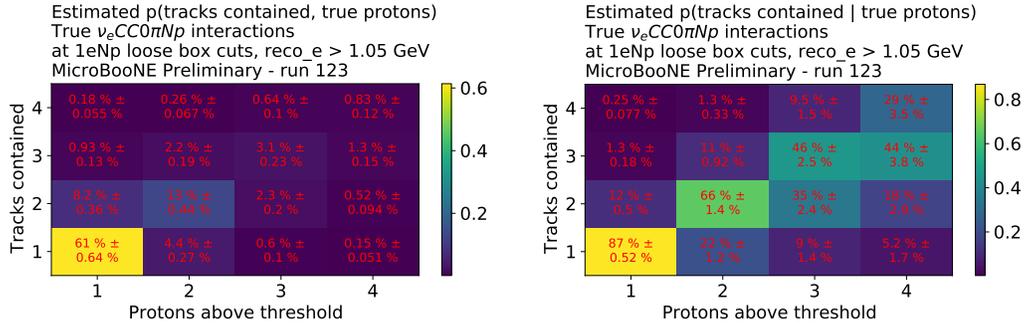

**Figure 8.12:** Number of reconstructed tracks versus the number of protons above the threshold at truth level, in the 1$e$N$p$0$\pi$ selection at loose selection and the high energy region. It shows the estimated joint distribution on the left and conditional on the right, where the content of each vertical slice adds up to one.

## 1$e$0$p$0$\pi$ SELECTION

The 1$e$0$p$0$\pi$ selection proceeds similarly to the 1$e$N$p$0$\pi$ selection, using, however, a fully orthogonal sample. No reconstructed track is allowed in the event, and thus no track identification variable can be employed in the BDT. All variables rely on shower properties, reconstructed cosmic rays, and the presence of a second shower not clustered in the event. The analysis in this channel is more difficult than the 1$e$N$p$0$\pi$, as additional tracks typically allow a more precise vertex reconstruction, which in turn results in a more correct event reconstruction. The plot on the left in fig. 8.13 shows the distribution of the BDT score used for the selection, when loose cuts are applied. The signal is enhanced at values close to 1: only events with a BDT score larger than 0.72 are selected, a value that allows a significant reduction of the background while retaining a decent efficiency. Most backgrounds are induced by events with $\pi^0$, where one of the showers is missing, and the other is identified as a single electron, which cannot be rejected by looking at the shower-vertex distance.

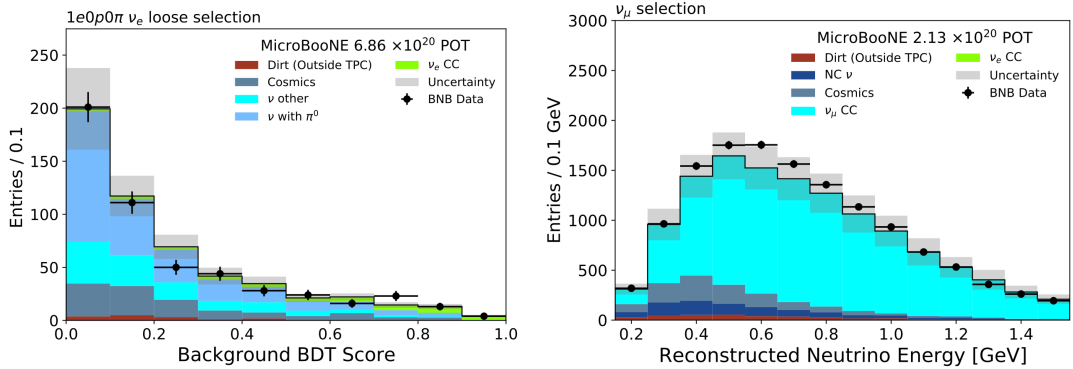

**Figure 8.13:** Left: Distribution of the BDT score for the $1e0p0\pi$ selection with loose cuts applied. The signal (green), while tiny, is enhanced at high values of the score, while the background, mostly containing events with $\pi^0$ (light blue), peaks at small values. Right: While showing agreement between the data and simulation within systematic uncertainties (section 8.4), the reconstructed energy spectrum in the $\nu_\mu CC$ selection emphasizes how the simulation underestimates the data. Using this data, the prediction for the $1eNp0\pi$ and $1e0p0\pi$ channels is adjusted through the "$\nu_\mu$ constraint" procedure (section 8.5).

## $\nu_\mu CC$ SELECTION

The $\nu_\mu CC$ selection is developed to study and correct systematic biases in the simulation. The central target is a pure selection with high energy resolution. For this reason, fully contained events containing a well reconstructed and identified muon are selected. This selection follows closely the one developed to study exclusive channels in section 7.4. The plot on the right of fig. 8.13 shows the distribution of the reconstructed neutrino energy, comparing the data with the simulation, dressed with systematic uncertainties (gray patches). The simulation describes the data within the uncertainties. However, there is a clear bias, as the simulation underestimates the data in most bins, which is likely due to a mismodeling of the neutrino flux and the cross-section model.

This selection effectively allows a reduction of the impact of systematic uncertainties correlated between $\nu_e$ and $\nu_\mu$ events (section 8.4), like flux and cross-section uncertainties, when all selections are employed in a combined analysis (section 8.5).

## 8.4 Systematics uncertainties

The predicted energy spectra in the different channels are subject to several types of systematic uncertainties. Each piece of the simulation carries its uncertainty, independent of the other. The ones considered in the analysis are the neutrino flux, the cross-section model used to simulate the rate and the final states of neutrino interactions, the propagation of the particles in the detector with attention to hadronic re-interactions, the simulation of the detector, in terms of light yield and shape of the signals, and, lastly, statistical uncertainties induced by the finite simulated sample size.

### The covariance matrix formalism

Uncertainties are dealt with using the covariance matrix formalism. This method relies on the assumption that uncertainties are Gaussian distributed and on the concept of propagation of uncertainties. Systematic uncertainties can be described by a series of nuisance parameters $\alpha$ and their associated prior distributions $p(\alpha)$. For example, in the case of the cross-section systematic uncertainties, about forty nuisance parameters define the neutrino generator, which can be varied within some uncertainty, estimated through a fit to previous data [306]. While they impact inference on the physical parameters, these nuisance parameters are not of physical interest. The expected number of entries $\nu_i$ in bin $i$ of the reconstructed energy spectrum of one of three analysis channels is a function of the nuisance parameters $\nu_i = \nu_i(\alpha)$. We can *marginalize* over $\alpha$ and assume that the uncertainty over $nu_i$ will be approximately Gaussian and can therefore be described by the mean $\nu_i^{\mathrm{CV}}$ and the covariance matrix $\Sigma_{ij}$, where CV stands for "Central Value." The mean $\nu_i^{\mathrm{CV}}$, or central value, is estimated using the simulation and fixing $\alpha$ to its most likely or mean value. The covariance matrix $\Sigma_{ij}$ is estimated in different ways depending on the type of uncertainty.

For the case or flux, cross-section, and hadronic re-interaction uncertainties, they are estimated through a *multi-universe* or *multi-sim* approach. Each universe $k$ contains a set of parameters $\alpha_k$ drawn from $p(\alpha)$, which is often assumed Gaussian. The flux and cross-section reweighing tools provide weights $w_k$, which can be applied to the simulated events to compute the expectation $\nu_i^k$ in the universe $k$. The covariance matrix can be estimated as

$$\Sigma_{ij} = \frac{1}{N_{univ}} \sum_{k=1}^{N_{univ}} (\nu_i^{\text{CV}} - \nu_i^k)(\nu_j^{\text{CV}} - \nu_j^k), \tag{8.1}$$

where $N_{univ}$ is the number of simulated universes. In this analysis, we simulate 100 universes.

For the case of detector uncertainties, we do not have the capability of reweighting events; we can only generate an entirely new sample by varying some parameters defining a different detector response. These variations are one-sided, meaning that $\alpha$ is varied to $\alpha'$, and the difference is considered as uncertainty. This approach is likely to provide a conservative estimate of the uncertainty. In addition, only the diagonal elements of the covariance matrix are considered, neglecting possible correlations and leading to a more conservative estimate. In this case, the covariance matrix is estimated as

$$\Sigma_{ii} = \sum_{k=1}^{N_{vars}} (\nu_i^{\text{CV}} - \nu_i^k)(\nu_i^{\text{CV}} - \nu_i^k), \tag{8.2}$$

where $N_{vars}$ is the number of variations considered.

Lastly, regarding finite statistics of the simulation, we consider a diagonal covariance matrix:

$$\Sigma_{ii}^{limstat,\eta} = \begin{cases} \nu_i^{\eta} & \text{if } \nu_i^{\eta} > 0, \\ 1.4 * w^{\eta} & \text{if } \nu_i^{\eta} = 0 \end{cases}, \tag{8.3}$$

for sample, $\eta$ (general BNB Monte Carlo, $\nu_e$ Monte Carlo, $\pi^0$ Monte Carlo, data Beam OFF), and $w^{\eta}$ is its specific scaling to much the number of POT or software triggers in the data.

The total uncertainty is obtained by summing up the covariance matrices from the different uncertainty sources. Figure 8.14 illustrate the fractional systematic uncertainty in all bins in recon-structed energy for the three different channels, considering only diagonal elements. The overall uncertainty budget is around 20-30% and tends to grow at low energy, negatively impacting the sensitivity to the signal model. The correlations between different selections, however, help improve the sensitivity. This procedure, often called $\nu_\mu$ constraint, is described in section 8.5.

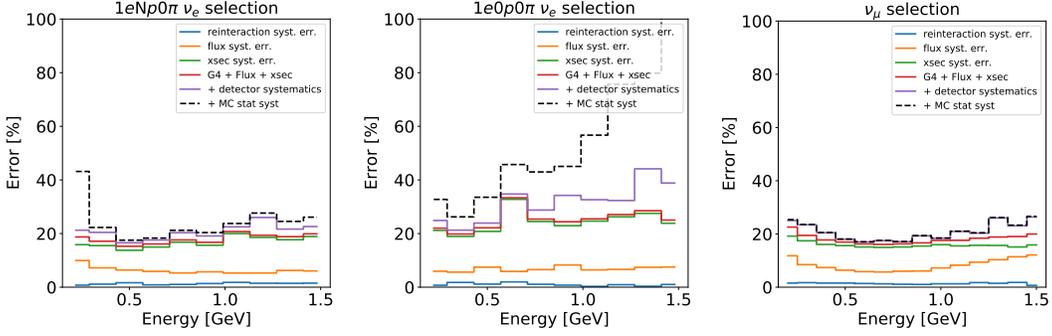

**Figure 8.14:** Fractional systematic uncertainties in each bin in reconstructed energy in the three analysis channels. Only the diagonal elements are shown, neglecting correlations, and providing a more conservative picture. The red line corresponds to the sum of hadronic re-interaction (blue), flux (orange), and cross-section (green) uncertainties. The purple line is obtained by adding detector systematic uncertainties to the red line. Eventually, the dashed black line adds the systematic uncertainties induced by finite Monte Carlo statistic on top of the purple line. Every line is obtained by summing up the relevant covariance matrices and taking the square root of the diagonal elements divided by the expected bin content in the central value of the simulation.

## Flux, cross-section, hadronic re-interaction systematic uncertainties

The procedure to estimate uncertainties related to the model of the flux, cross section, and particle propagation in the detector with attention to hadronic re-interaction closely follows the official MicroBooNE procedure.

Flux uncertainties depend on uncertainties in the current flowing in focusing horns, in the cross section for proton-Beryllium scattering, and the model of the propagation and re-interaction of hadrons within the Beryllium target. Thirteen parameters characterize them, randomly varied in a multi-universe fashion.

Cross-section uncertainties depend on the parameters in the GENIE Monte Carlo generator. GENIE v3 has been tuned to match the T2K near detector data, adjusting normalization of CCQE and MEC processes, resulting in an overall good agreement when looking at $\nu_\mu CC$ interactions, as seen in section 8.3. The uncertainties are characterized by 44 parameters sampled following the multi-universe procedure, plus 11 parameters for which only one-sided or two-sided variations are performed. These parameters are mostly related to the initial state momentum distribution of nucleons, scattering form factors, normalization and angular distributions for different processes, and propagation of particles in the nuclear medium that result in FSI.

The final cluster of uncertainties is related to the Geant4 [307] cross-section values used to simulate hadronic re-interactions of hadrons, mainly protons, and pions, with argon nuclei. These hard scatterings result in misclassified events. For example, the Bragg peak of a proton might not be visible, resulting in its classification as a muon instead, as discussed in section 7.4. Three parameters describe these uncertainties: the cross sections for protons, $\pi^+$, $\pi^-$. The three parameters are varied simultaneously in 1000 universes. This contribution is sub-dominant with respect to the flux and cross-section model, as shown in fig. 8.14. Figure 8.15 shows the correlation matrix, as obtained from the covariance matrix, for flux (left) and cross-section (right) systematic uncertainties. Most bins present high correlations within the same channel and different channels. These correlations are essential in decreasing the impact of the systematic uncertainties on the final sensitivity. Interestingly, most correlations are positive and significant, implying that most of the systematic effects result in a change in the overall normalization rather than a shape difference at fixed normalization.

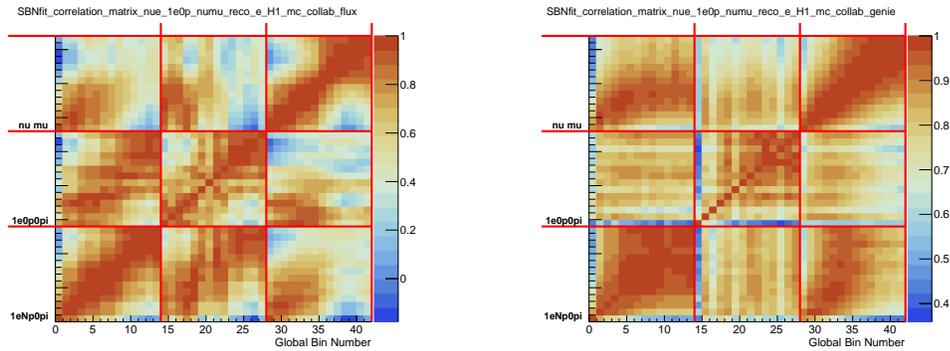

**Figure 8.15:** Correlation matrices for flux (left) and cross-section (right) systematic uncertainties. The matrix is read as a 2d plot, with increasing bin numbers going up and right. The first ten bins refer to the $1eNp0\pi$ channel, the next ten bins to the $1e0p0\pi$ channel, and the last 14 bins to the $\nu_\mu CC$ selection.

### DETECTOR SYSTEMATIC UNCERTAINTIES

Detector systematic uncertainties are obtained through single-sided parameter variations used in the data simulation. There are four different categories:

- Waveform variations, covering the uncertainty in the simulation of the charge deposition, drift, and measurement. These variations are obtained by profiling hit variables, like charge, amplitude, and width, as a function of the location in the detector, the angle with respect to the wire direction $\theta_{xz}$, with respect to the drift direction $\theta_{yz}$, and the ionization density $dE/dx$. They are obtained by comparing the simulation with the data using cosmic rays as rescaling factors for the simulation to match the data, which, by construction, covers the difference between the data and the simulation.

- Light response variations, where different models for the light propagation are used, can change the yield of triggered events and flash properties.

- Electric field map variations, where a different electric field map, obtained using cosmic-ray muons rather than laser measurement, is employed while simulating charge drift.

- Recombination variations, obtained by varying the parameters of the Modified Box Model (eq. (7.4)), to cover discrepancies observed with protons at high $dE/dx$.

Here we will discuss the effect of waveform variations, the most impactful ones, on calorimetric variables, which are the ones affected the most by these uncertainties. Figure 8.16 shows the selection efficiency of the cuts employed in the analysis on the proton/muon id $\mathcal{P}$ (left), and the electron/photon separation d$E$/d$x$ (right), as a function of the true proton or electron energy, for 1$e$N$p$0$\pi$ events selected at truth level. The different variations do not change the efficiency signifi-cantly, aside from the "Recomb" variation, which produces a sizable effect at large proton energies. The final impact on the analysis is, however, small, mainly because of the small proton population at large energy.

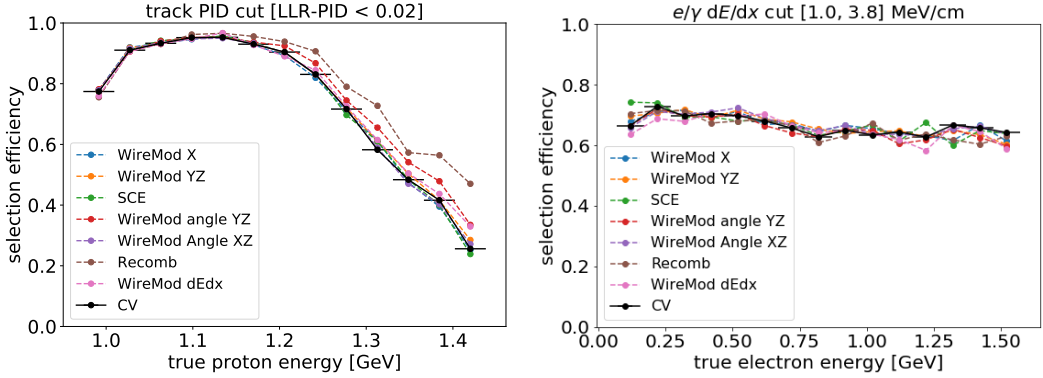

**Figure 8.16:** Selection efficiency of the particle identification cuts on the track $\mathcal{P}$ (left) and shower d$E$/d$x$ (right), as a function of the true proton and electron energy, respectively. The different lines show the same quantity using the different detector systematic samples.

The coverage of systematic uncertainties, with emphasis on recombination uncertainties, is validated through detailed calorimetric studies, performed using a high-purity sample of reconstructed protons, selected as in section 7.4, comparing the data with the central value and with the sample with the waveform variation profiled in d$E$/d$x$. The six plots in fig. 8.17 show an example of the comparison of the d$Q$/d$x$ distribution, performed in bins of residual range and pitch, differently for every plane. We show distributions for the residual range between 5 cm and 10 cm and pitch between 0.4 cm and 1 cm for different planes on different rows. This bin contains hits at large d$Q$/d$x$ because of the proton-like selection and the small residual range. In this region, we expect larger mismodeling, as the simulation relies more heavily on the correct simulation of the recombination process. Remarkably, the discrepancy between the data and the central value of the simulation is significantly corrected after applying this Waveform variation.

The three plots in figure fig. 8.18 summarize the information displayed in the $dQ/dx$ data/simulation comparison, by showing a 2-dimensional histogram of $dQ/dx$ and residual range, for the data, the central value of the simulation, and the sample with waveform variations profiled in $dE/dx$. To reduce possible biases due to the underlying distributions, these histograms are produced in bins of local pitch, normalized as estimates of $p(dQ/dx\,|\text{residual range})$, rather than $p(dQ/dx, \text{residual range})$. This procedure ensures that the probability of every vertical slice sums up to unity, making it less sensitive to the underlying track length distribution. The curve showing the theoretical most probable value is superimposed over the histogram to facilitate comparing the data and the simulation. The qualitative conclusion derived earlier, *i.e.* the sample with waveform variations profiled in $dE/dx$ improves the accuracy of the simulation, remains unchanged even after looking at these 2-dimensional plots. These studies also indicate that the analysis is robust against $dE/dx$ mismodeling.

## 8.5  Constraint of the systematic uncertainties

Because systematic uncertainties are correlated among different analysis channels, as shown in section 8.4, a combined analysis reduces their impact on the sensitivity to the eLEE model. The combination can be visualized through a procedure often called "$\nu_\mu$ constraint." The $1e\text{N}p0\pi$ and $1e0p0\pi$ selections are sensitive to the eLEE model, while the prediction in the $\nu_\mu CC$ selection does not depend on the eLEE hypothesis. We can then condition the prediction in the $1e\text{N}p0\pi$ and $1e0p0\pi$ selections on the observed spectrum in the $\nu_\mu CC$ selection. Let's dive into the details. The central value of the prediction and the full covariance matrix defines a multivariate Gaussian density over the bin contents in the reconstructed energy spectra of the three analysis channels. Let's represent

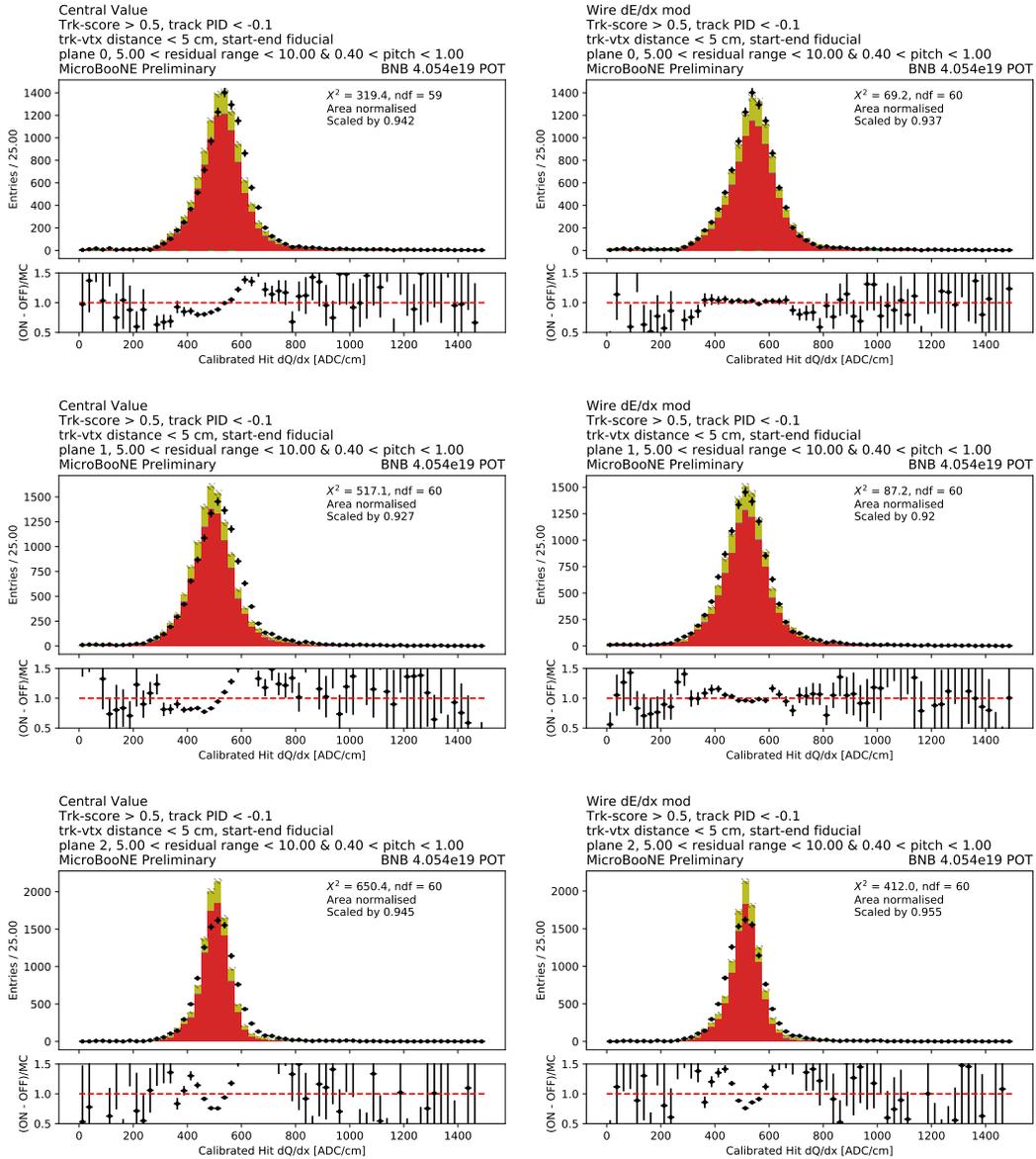

**Figure 8.17:** The sample with waveform variations profiled in $\mathrm{d}E/\mathrm{d}x$ (right) improves the accuracy of the simulation with respect to the central value (left) in the $\mathrm{d}Q/\mathrm{d}x$ distribution, in a bin with small residual range, between 5 cm and 10 cm, and small pitch, between 0.4 cm and 1 cm, and for the three wire planes, one per row. Hits come from proton-like tracks selected as in section section 7.4 The red component of the histogram corresponds to proton hits, while the brown to neutrino-induced muons.

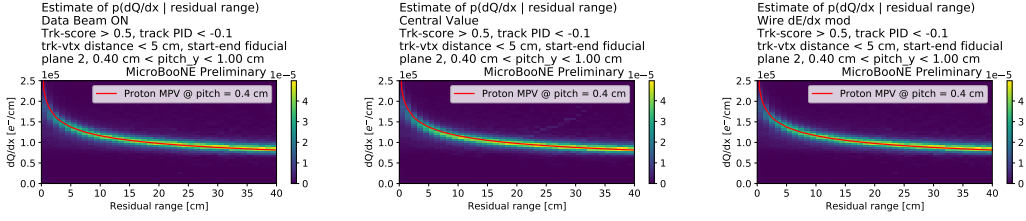

**Figure 8.18:** 2-dimensional histogram of $\mathrm{d}Q/\mathrm{d}x$ and residual range for the same contained proton selection in data, simulation central value, and the sample with waveform variations profiled in $\mathrm{d}E/\mathrm{d}x$, respectively. Only hits with local pitch between 0.4 cm and 1 cm are used for this comparison, and every vertical slice is normalized to one to reduce the dependence of these comparisons on the track kinematics. The red theoretical curves help the eye compare the data and the two simulations.

the central value and the covariance matrix in the following block form, and the

$$\nu^{CV} = \begin{pmatrix} \nu_e & \nu_\mu \end{pmatrix} \tag{8.4}$$

$$\Sigma^{syst} = \begin{pmatrix} \Sigma_{ee} & \Sigma_{e\mu} \\ \Sigma_{e\mu}^T & \Sigma_{\mu\mu} \end{pmatrix}, \tag{8.5}$$

where the subscript $e$ refers to all $1e\mathrm{N}p0\pi$ and $1e0p0\pi$ bins in reconstructed energy, while the subscript $\mu$ refers to all bins of the $\nu_\mu CC$ selection in reconstructed neutrino energy. Suppose we observe $n_\mu^{\mathrm{obs}}$ in the $\nu_\mu CC$ selection. Then, the $1e\mathrm{N}p0\pi$ and $1e0p0\pi$ bin contents are described by a multivariate Gaussian:

$$p(n_e | n_\mu = n_\mu^{\mathrm{obs}}) \sim \mathcal{N}(\nu_e^{\mathrm{constrained}}, \Sigma_{ee}^{\mathrm{constrained}}), \tag{8.6}$$

with parameters

$$\nu_e^{\mathrm{constrained}} = \nu_e + \Sigma_{e\mu}\Sigma_{\mu\mu}^{-1}(n_\mu^{\mathrm{obs}} - \nu_\mu) \tag{8.7}$$

$$\Sigma_{ee}^{\mathrm{constrained}} = \Sigma_{ee} - \Sigma_{e\mu}\Sigma_{\mu\mu}^{-1}\Sigma_{e\mu}^T. \tag{8.8}$$

We note that the correction to the central value is proportional to the difference between the observed bin content and the central value. At the same time, the reduction of systematic uncertainties only depends on the structure of the covariance matrix. We can think about this procedure as reducing a 2D Gaussian distribution to one slice identified by selecting the observed value on the axis corresponding to the $\nu_\mu CC$ selection. These formulas are a well-known fact within Bayesian inference with multivariate normal; see for example [308]. This procedure, strongly relying on the Gaussian approximation of systematic uncertainties, ensures that every test statistic computed using eq. (8.4) or using eq. (8.7) produces the same result. Figure 8.19 shows the fractional uncertainties in each bin content for the three analysis channels, before and after the constraint for the $1eNp0\pi$ and $1e0p0\pi$ channels. The uncertainty reduction varies from 10% at the lowest energy to about 50% in the medium energy range. Given the agreement observed with the $\nu_\mu CC$ selection illustrated in the plot on the right in fig. 8.13, the shift in the central value of the $1eNp0\pi$ and $1e0p0\pi$ is relatively small, of the order of 10%. This procedure leads to a stronger sensitivity to the eLEE model, ensuring that statistical rather than systematic uncertainties dominate the analysis.

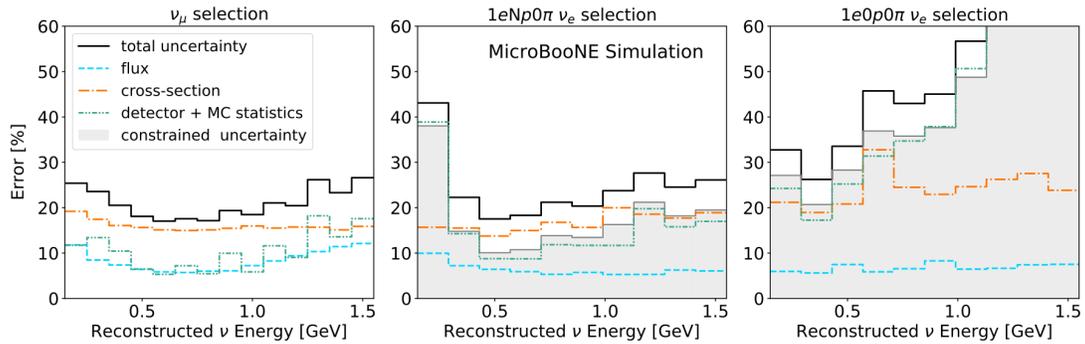

**Figure 8.19:** Fractional systematic uncertainties in bins of the reconstructed energy in the three analysis channels. After observing the $\nu_\mu CC$ data, the uncertainties in the $1eNp0\pi$ and $1e0p0\pi$ channels are reduced, from the black line to the gray shaded line. The constrained uncertainties cannot be further split into the different components, as this method relies on the Gaussian approximation of the bin content observables and does not distinguish different sources of uncertainties.

# 9

# Interpreting the results

After characterizing the full eLEE analysis over the simulation and validating it with a small fraction of the data, the full dataset, consisting of Run 1, 2, and 3, is unblinded. It corresponds to data collected between February 2016 and July 2018, corresponding to a total exposure of $6.86 \times 10^{20}$ POT. The unblinding procedure relies on first identifying a signal region, events at low energy with high $\nu_e$ purity. A set of sidebands with progressively lower energy and larger purity is unblinded in steps to make sure possible problems can be solved before accessing the region with the highest chance of hosting a signal.

After unblinding the data, its physical implications are drawn through a combined analysis of the three channels. Statistical tests to assess the significance of excluding the eLEE signal models are performed, both as a single hypothesis test and as a composite test where confidence intervals for a scaling factor of the signal are extracted.

Eventually, this data could be further interpreted in terms of the light sterile neutrino model, and possibly more complex models.

## 9.1 Unblinding the data

In order to unblind the data in the signal region, a series of sidebands with progressively lower energy and a larger signal-to-background ratio has been defined. This section describes the procedure for the $1e\mathrm{N}p0\pi$ channel. An analogous procedure has been carried out for the $1e0p0\pi$ channel, while the results of both analyses are discussed in chapter 9. We identified three regions of progressively lower energy and three regions of progressively higher event ID, the identification score, or equivalently the signal-to-background ratio. This is shown schematically in the left plot of fig. 9.1. While the vertical axis is the reconstructed neutrino energy, the horizontal one requires a combination of BDT scores, as well as preselection and loose selection cuts. The energy cuts are defined as:

- High energy: 1.05 GeV $< E_{reco} <$ 2.05 GeV

- Medium energy: 0.75 GeV $< E_{reco} <$ 1.05 GeV

- Low energy: 0.05 GeV $< E_{reco} <$ 0.75 GeV,

while the regions at different event ID are defined as:

- Low event ID: preselection cuts, exactly one contained shower, and $0 < \pi_0$ score $< 0.1$ or $0 < \text{non } \pi_0$ score $< 0.1$

- Medium event ID: preselection cuts, exactly one contained shower, and $(0.1 < \pi_0$ score $< 0.67$ or $0.1 < \text{non } \pi_0$ score $< 0.7)$ and $(\pi_0$ score $> 0.1)$ and $(\text{non } \pi_0$ score $> 0.1)$

- High event ID: loose-selection cuts, and $\pi_0$ score $> 0.67$ and non $\pi_0$ score $> 0.7$.

These regions define the far sideband as events at high energy or events at low event ID, and the near sideband as medium energy or medium event ID but not in the far sideband. The signal region consists of events at low energy and high event ID. The significant correlation between the two BDT scores, as shown in fig. 9.1, justifies the choice of combined cuts. For the low and medium event ID, the loose selection cuts have been relaxed to the preselection cuts plus the requirement of precisely one contained shower. This cut relaxation ensures larger sideband statistics, allowing for more informative data/simulation comparisons. Requiring a single contained shower leaves the sample with two or more contained showers as an additional sideband, enriched of events with two showers, mainly produced by $\pi^0$.

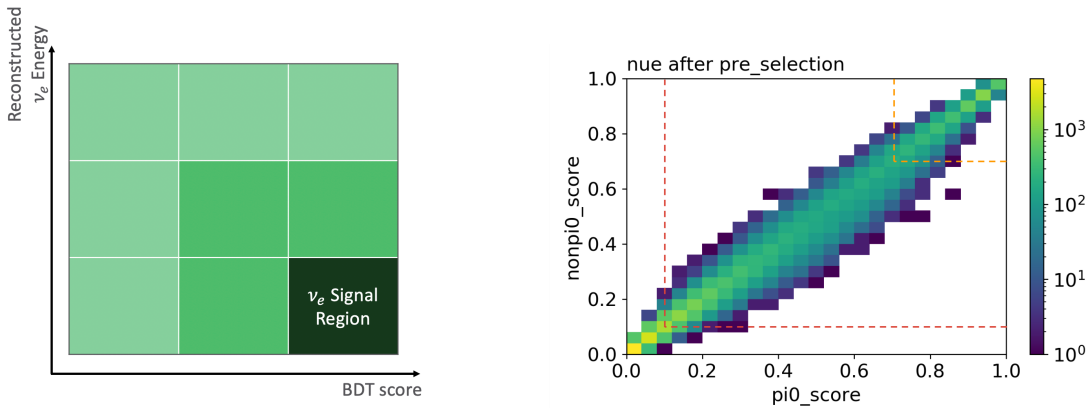

**Figure 9.1:** Left: schematic of the sideband structure. Right: correlation among the two BDT scores employed in the $1e\mathrm{N}p0\pi$ analysis, in the $\nu_e$CC sample.

The following studies have been carried out to test the effectiveness of these sidebands. The underlying true neutrino energy and three kinematic variables, namely the energy of the leading shower chosen as the electron candidate, the length of the track chosen as the proton candidate, and the angle between these two objects, have been studied. The four plots in figure fig. 9.2 show the distributions of these four variables in the high PID region for the different energy regions, corresponding to the three boxes in the last column of the schematics in the left plot of figure fig. 9.1. The histograms show the expected distributions for the 1eNp0π events, selected with truth level information in the $\nu_e$ sample, scaled up to $6.86 \times 10^{20}$ POT. Analogously, the four plots in figure fig. 9.3 show similar distributions for events in the low energy, comparing the three different PID regions. The three histograms correspond, this time, to the three boxes in the bottom row of the schematics shown in the left plot of figure fig. 9.1.

The take-home message of these studies is that the 1eNp0π events in the three energy regions at high PID cover a similar phase space in the kinematic variables, despite being at different energies. This means that the resolution of our reconstructed energy is smaller than the size of the sidebands. Additionally, the 1eNp0π events in the different PID regions at low energy are very similar and cover similar phase space in the kinematic variables.

## 2+ SHOWERS SIDEBAND

An additional sideband enriched of events with two or more showers has been identified. This sideband is particularly useful to study $\pi^0$ backgrounds in a region that is more similar to what is expected in the signal region. This sideband is defined by preselection cuts, with no track requirement, and requires at least two showers contained. As no track requirement is applied, this sideband is useful for both the 1eNp0π and 1e0p0π selections. The three plots in fig. 9.4 show the reconstructed

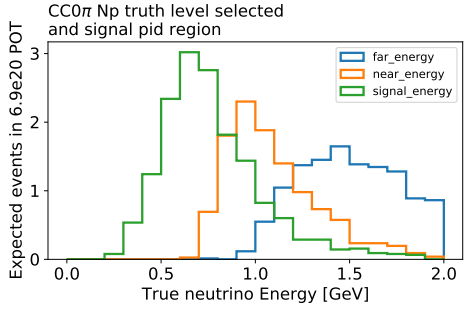

**(a)** True neutrino energy spectrum.

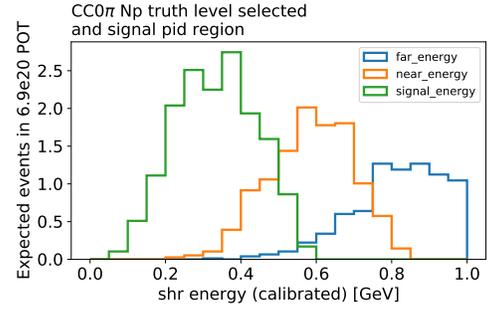

**(b)** Distribution of the reconstructed energy of the shower chosen as electron candidate.

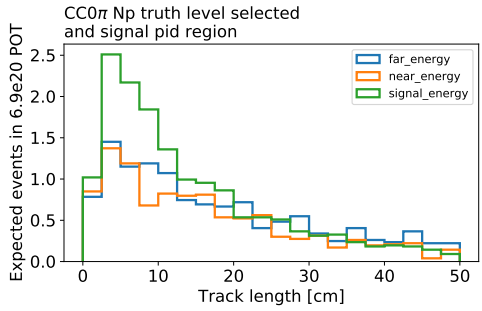

**(c)** Distribution of the reconstructed length of the track chosen as proton candidate. This is a bijective function of the reconstructed track energy as the energy is extracted from the range.

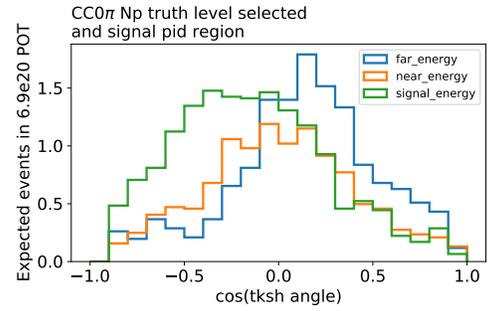

**(d)** Distribution of the cosine of the angle between the main shower and the main track.

**Figure 9.2:** Expected distributions of $1eNp0\pi$ events scaled up to $6.86 \times 10^{20}$ POT. All the plots are for the high PID region, and the three different colored histograms identify the different regions in reconstructed energy.

neutrino energy, the shower $dE/dx$, and the track identification score in this sideband after applying very loose selection cuts. The data agrees with the simulation within uncertainties, demonstrating the accuracy of modeling background events with showers that fall close to the signal box as defined by the PID cuts.

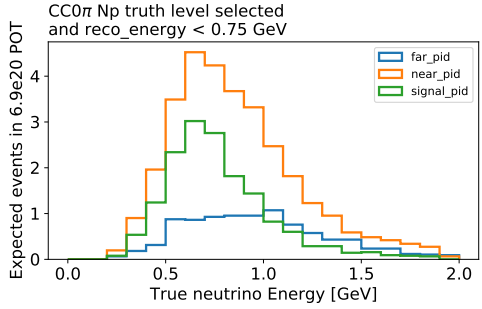

**(a)** True neutrino energy spectrum.

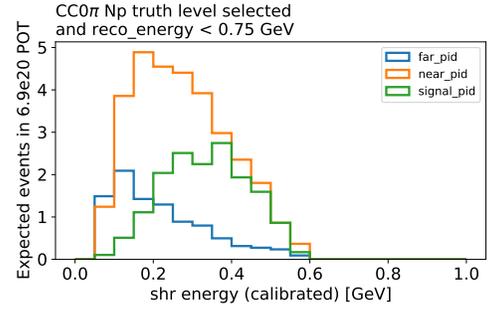

**(b)** Distribution of the reconstructed energy of the shower chosen as electron candidate.

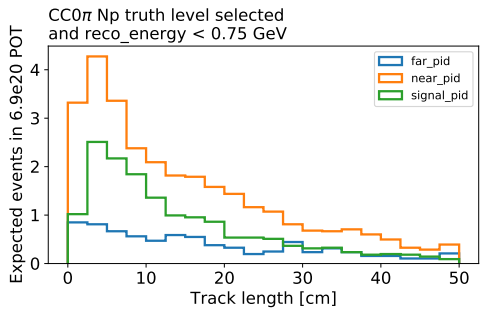

**(c)** Distribution of the reconstructed length of the track chosen as proton candidate. This is a bijective function of the reconstructed track energy as the energy is extracted from the range.

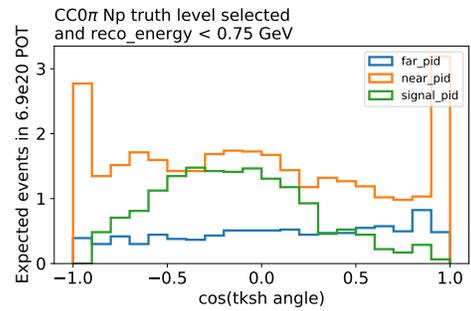

**(d)** Distribution of the cosine of the angle between the main shower and the main track.

**Figure 9.3:** Expected distributions of $1eNp0\pi$ events scaled up to $6.86 \times 10^{20}$ POT. All the plots are for the low energy region, and the three different colored histograms identify the different regions in PID.

### Far sidebands at high energy

While providing no sensitivity to the eLEE model, the far sideband demonstrates the accuracy of the simulation in essential regions of the phase space. For example, the high purity region at high energy provides insight into modeling electron neutrino interactions. Figure 9.5 shows the distribution of the two BDT scores employed in the $1eNp0\pi$ selection, at loose cuts. The agreement is within uncertainties, demonstrating an overall agreement between the data and the simulation in the entire selection process. We can observe distributions of the properties of the electron candidate in fig. 9.6.

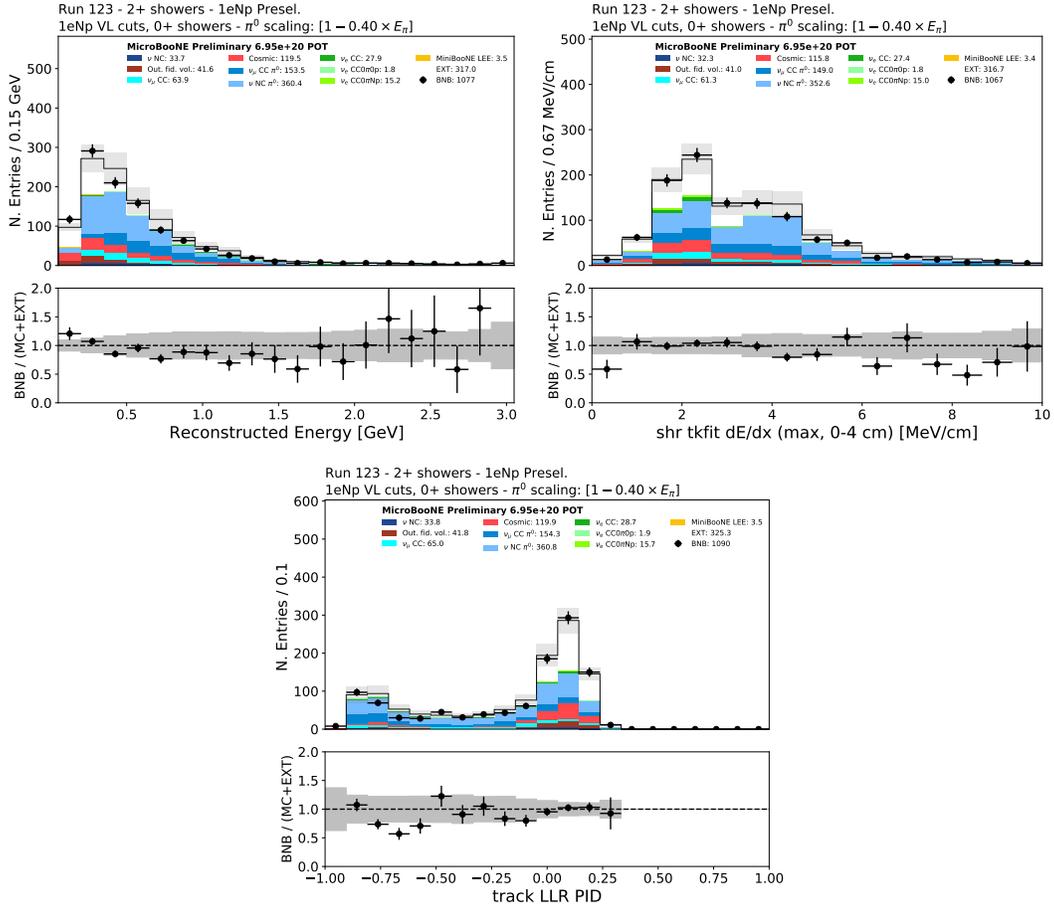

**Figure 9.4:** Selecting events with two or more showers allows the definition of high-statistics sideband, with background analogous to the signal region, mostly induced by $\pi^0$ events, but no recoverable signal. The simulation is accurate, as demonstrated by the agreement between the data and the simulation shown in the reconstructed energy spectrum (left), the shower $dE/dx$ (middle), the track particle identification score (right).

A systematic bias is present: the prediction mildly overestimates the data, which is believed to be connected to the overall deficit of events observed in the entire energy spectrum, as shown in a later section.

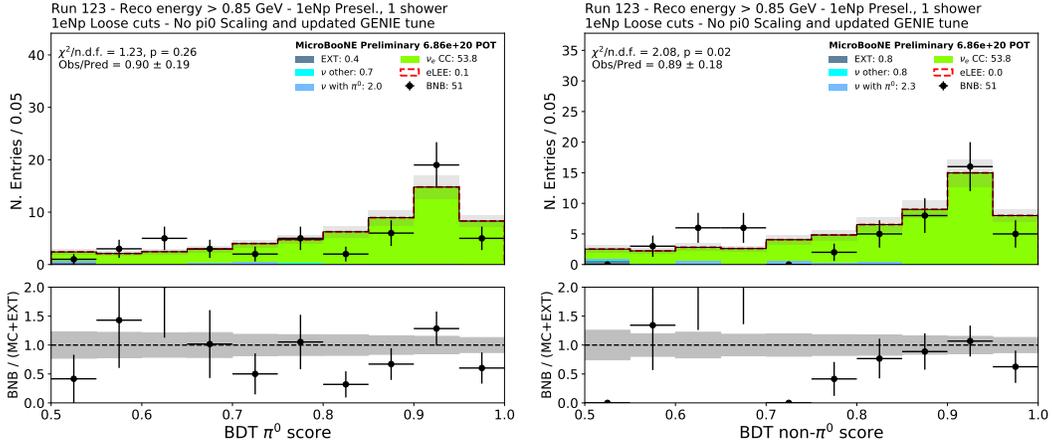

**Figure 9.5:** Both BDT scores ($\pi^0$ score on the left, and non-$\pi^0$ score on the right) employed in the $1eNp0\pi$ selection shows good agreement between the data and the simulation in the far sideband at high energy.

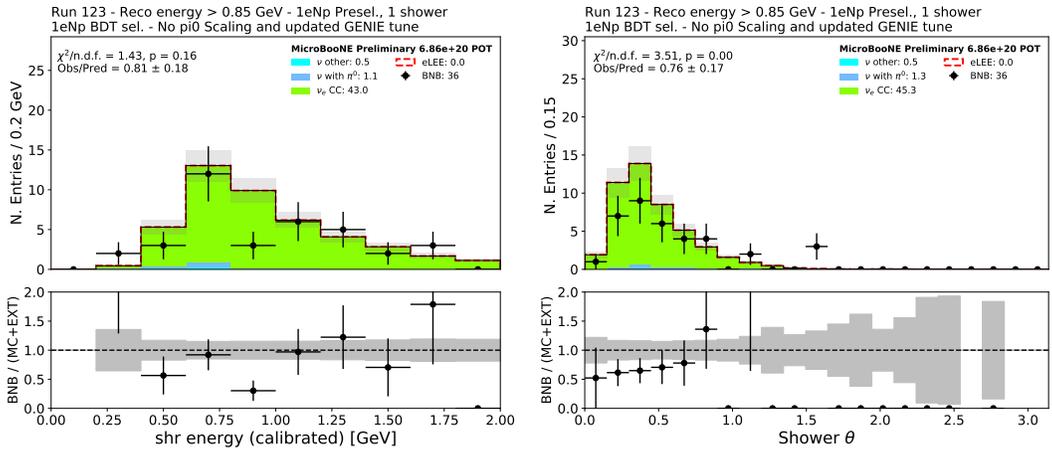

**Figure 9.6:** The distributions of shower properties, total calorimetric energy on the left and angle with respect to the beam on the right, show an overall deficit of events with respect to the prediction after applying the full BDT selection in the high energy region of the far sideband.

## Far sideband at low event ID

Another interesting region in the far sideband is at low event ID. This is another background-enriched region, which provides additional insights on the background modeling, especially when it comes to background containing photon showers induced by events with $\pi^0$. Figure 9.7 shows the spectrum of reconstructed energy, track-shower distance, and track pid score, with only preselection cuts applied. It provides a genuinely high-statistic data/simulation comparison, demonstrating accurate modeling of the background, and further validating the conclusions from fig. 9.4.

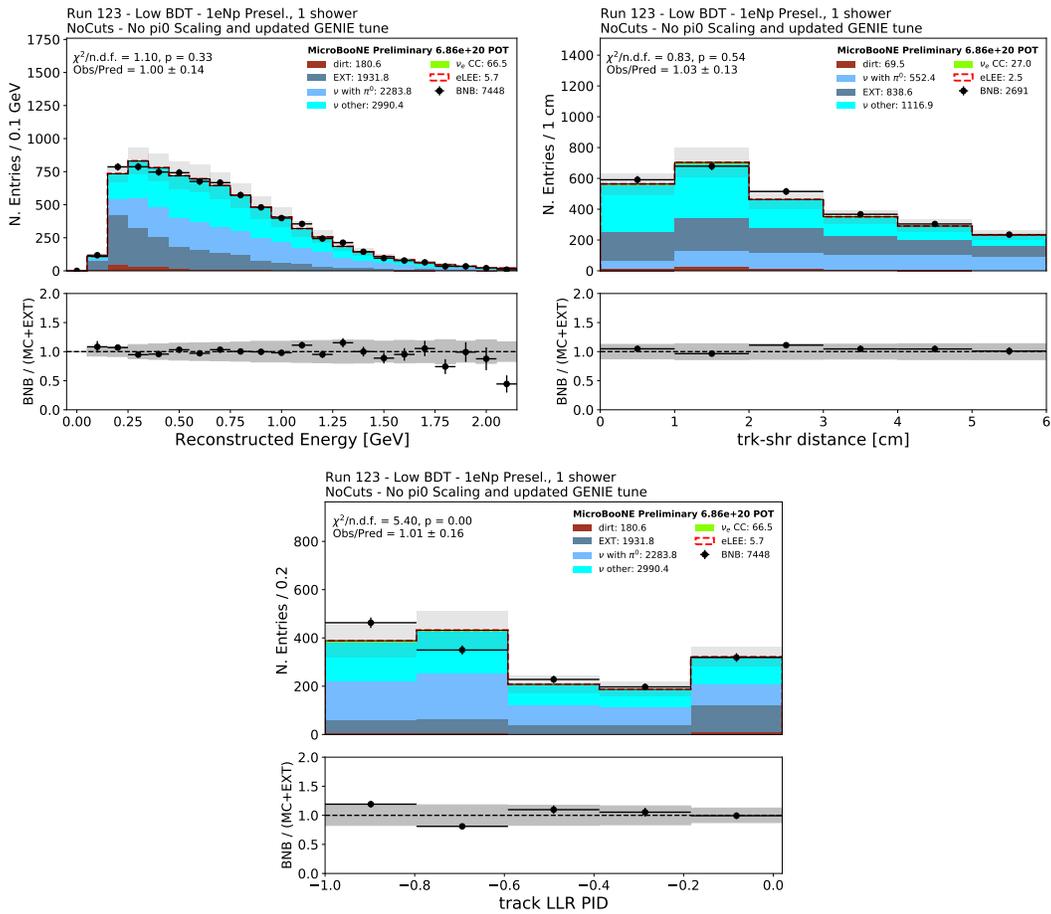

**Figure 9.7:** The far sideband at low event ID, or low purity, provides high statistics data/Monte Carlo comparisons in background-enriched sidebands. The accuracy of the simulation is demonstrated by the reconstructed energy spectrum (left), the track-shower distance (middle), and the track identification score (right).

The near sideband was unblinded after the conclusions drawn from the previous studies with the far sideband and the 2+ showers sideband. This sideband has a more substantial fraction of signal events, although still too small to provide any sensitivity to the signal model. However, it allows the validation of signal and background modeling in adjacent regions to the signal box. Figure 9.8 illustrates a summary of these studies. The first row shows plots of the shower $dE/dx$ and the track-shower distance at preselection cuts, at medium event ID and medium energy, respectively. The second row illustrates the distribution of the BDT score and ratio of shower hits to the total hits in the slice with loose cuts, again at medium event ID and medium energy, respectively. While the agreement between the data and the simulation at preselection is remarkable, the selection at loose cuts provides further insights into the trend already seen: the prediction underestimates the data in regions at higher purity, in particular at large values of the BDT scores. However, this behavior is not attributable to any specific problem and might have a specific physics explanation.

## SIGNAL REGION

The signal region was unblinded on June 24th, 2021, at 8 am CT, and the results were announced, together with the other eLEE analyses, on Wednesday October 27th, 2021, at 10 am CT. Figure 9.9 shows the full reconstructed energy spectrum, including the signal region, for the $1eNp0\pi$ and $1e0p0\pi$ channels. The first remarkable observation is that the data visually agrees with neither the intrinsic neutrino-only prediction nor the eLEE model. The prediction overestimates the data in the $1eNp0\pi$ channel and underestimates it in the $1e0p0\pi$ channel. The implications of this energy spectrum are further explored in the following few sections.

Within the $1eNp0\pi$ signal region, we further explore the features of the deficit of events by looking at the distribution of the angle with respect to the beam and proton kinetic energy, as show in

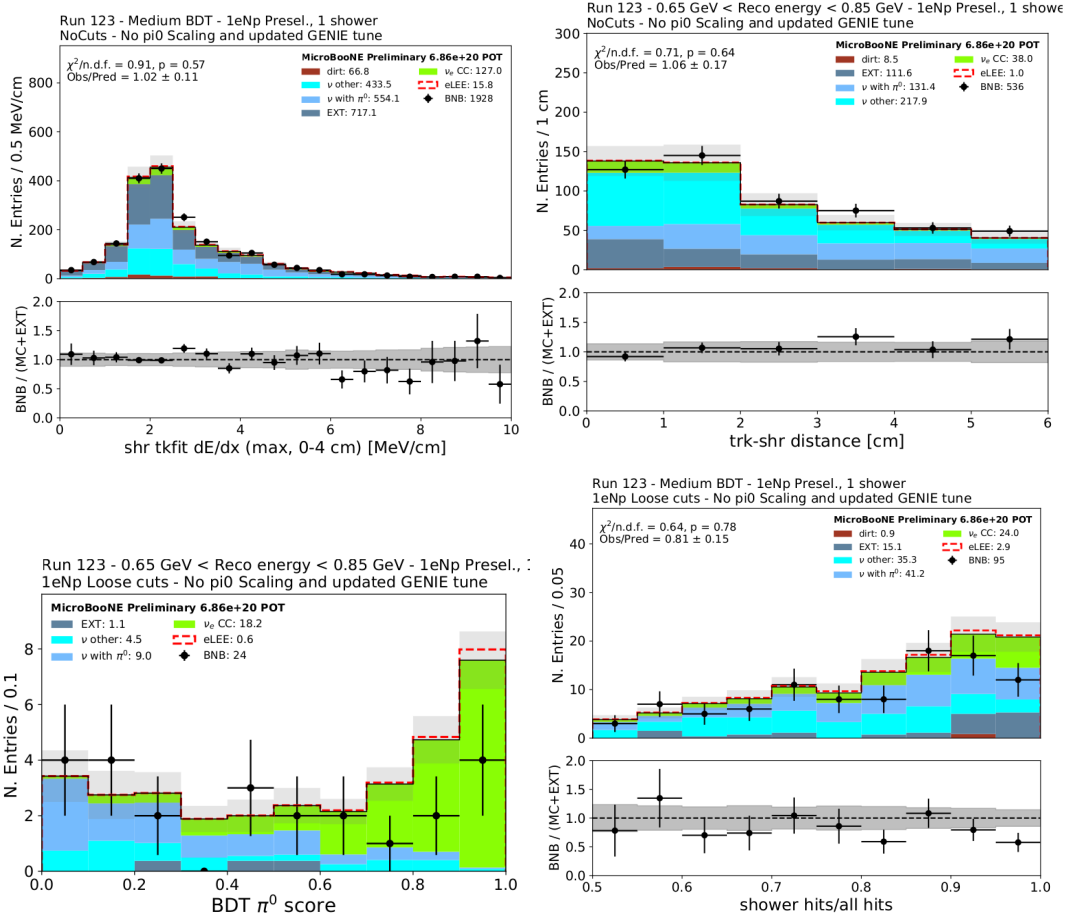

**Figure 9.8:** The near sideband allows data/Monte Carlo comparisons in regions where the signal is visible, but with no statistical sensitivity. The left column shows distributions in the medium event ID region, while the right one is in the medium energy range. The first row illustrates distributions at preselection, while the second one at loose selection. The four variables shown are, in clockwise order, the shower $dE/dx$, the track-shower distance, the ratio between the hits associated with the shower and the total hits in the slice, and the $\pi^0$ BDT score. The simulation accurately describes the data within the uncertainty at preselection but underestimates the data in the highest purity regions at loose selection.

fig. 9.10. While in the proton kinetic energy distribution the deficit seems to be constant over the entire spectrum, the disagreement between the data and the simulation seems to be concentrated in the forward direction.

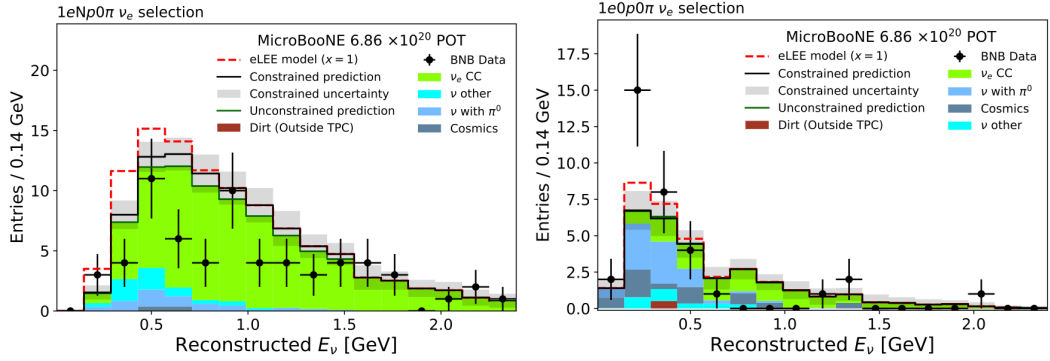

**Figure 9.9:** The full reconstructed energy spectra at high event ID in the $1e\text{N}p0\pi$ (left), and $1e0p0\pi$ (right) selections. The prediction overestimates the data in the first channel, while underestimating it in the second one.

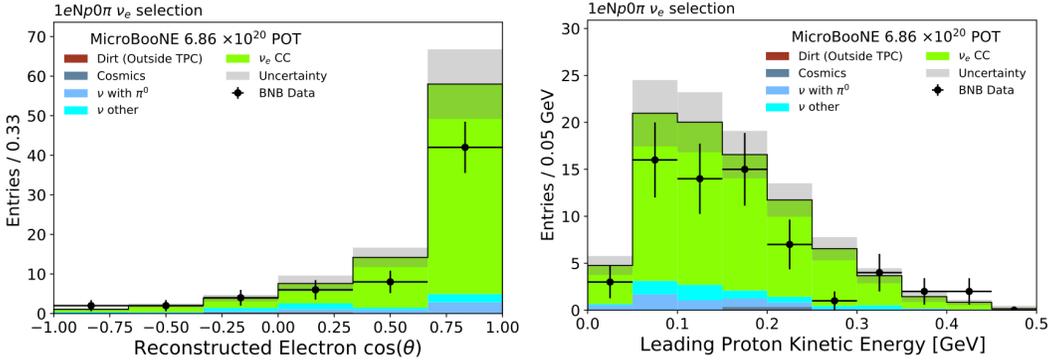

**Figure 9.10:** Two variables provide further insights on the features of the deficit of events observed in the signal region of the $1e\text{N}p0\pi$ channel: the angle with respect to the beam (left) and the proton kinetic energy (right).

## 9.2 Statistical test of new physics

We perform three different statistical tests for quantifying the presence of new physics. First, we quantify how well the prediction matches the data by performing a simple test of goodness of fit. Second, we perform a simple hypothesis to distinguish between the two hypotheses, with or without the eLEE model. Third, we expand this simple hypothesis test to a composite test in which the signal has a fixed shape, but its total normalization is allowed to float by a scaling factor $\mu$. We extract the best fit value and a confidence interval for $\mu$.

The common element to these three tests is the Combined Neyman-Pearson (CNP) $\chi^2$ test statistic [309], defined as:

$$\chi^2 = \sum_{i,j}(n_i - \nu_i)\Sigma_{ij}^{-1}(n_j - \nu_j) \tag{9.1}$$

$$\Sigma_{ij} = \Sigma_{ij}^{\text{stat CNP}} + \Sigma_{ij}^{\text{syst}}, \tag{9.2}$$

with $i$ and $j$ running over the bins, and $n$ refers to the observed, while $\nu$ to the predicted bin content. The statistical covariance matrix is defined as

$$\Sigma_{ij}^{\text{stat CNP}} = 3/(1/n_i + 2/\nu_i)\delta_{ij}, \tag{9.3}$$

in order to cancel at first order the bias introduced by using the Gaussian approximation of statistical uncertainties in place of the Poisson likelihood, while remaining compatible with the covariance matrix formalism.

## Modeling of electron neutrinos

The agreement between the observed data and the prediction is obtained using the $\chi^2$ test statistic with the central value prediction of the simulation. No assumption on the distribution of the $\chi^2$ test statistic is performed. Instead, toy experiments are drawn in the following way. First, a central value is sampled according to the distribution described by the systematic covariance matrix. Secondly, the bin contents are drawn from a Poisson distribution with a mean equal to that central value. The data appeared to be consistent with the standard model prediction with p-values of 0.182, 0.126, and 0.098 for the $1e\text{N}p0\pi$ channel, $1e0p0\pi$ channel, and the combination of the two.

The simple hypothesis test is performed to quote the p-value to exclude MiniBooNE's explanation based on the presence of additional electron neutrinos in the beam, according to the eLEE model. The most powerful test-statistic for a simple hypothesis test is the likelihood ratio, according to the Neyman-Pearson lemma, which is approximated with the difference in $\chi^2$ defined as in eq. (9.1):

$$\Delta\chi^2 = \chi^2(H_0) - \chi^2(H_1), \tag{9.4}$$

where $H_0$ refers to the standard model only, and $H_1$ to the standard model plus eLEE model. The choice of hypothesis between $H_0$ and $H_1$ impacts the predicted central value and the covariance matrix used for the $\chi^2$ calculation. We compute p-values with respect to the expected distribution of $\Delta\chi^2$ assuming $H_1$ is true. The sensitivity is obtained by taking the median value of the $\Delta\chi^2$ under $H_0$.

While the $1e\mathrm{N}p0\pi$ channel observes an underfluctuation of the data, resulting in a preference of $H_0$ over $H_1$ ($\Delta\chi^2 = -3.89$, p-value $= 0.285$, sensitivity $= 0.06$), the overfluctuation of the data in the $1e0p0\pi$ channel produces a preference for $H_1$ ($\Delta\chi^2 = 3.11$, p-value $= 0.984$, sensitivity $= 0.249$). While the combination of the two analysis channels is dominated by the smaller uncertainty of the $1e\mathrm{N}p0\pi$ channel, it produces an intermediate result ($\Delta\chi^2 = -0.58$, p-value $= 0.748$, sensitivity $= 0.049$) which shows no clear preference between the two hypotheses. Figure 9.11 illustrates the distribution of the $\Delta\chi^2$ test statistics for the $1e\mathrm{N}p0\pi$ channel, the $1e0p0\pi$ channel, and the combination of the two, under the hypotheses $H_0$ and $H_1$. Vertical lines identified the data and the median values, which are the quoted sensitivities. In the $1e0p0\pi$ channel, the data falls in the tail of both distributions. While there is a preference for the eLEE hypothesis, the analysis is not powerful enough to distinguish between the two hypotheses, with a sensitivity of 0.25.

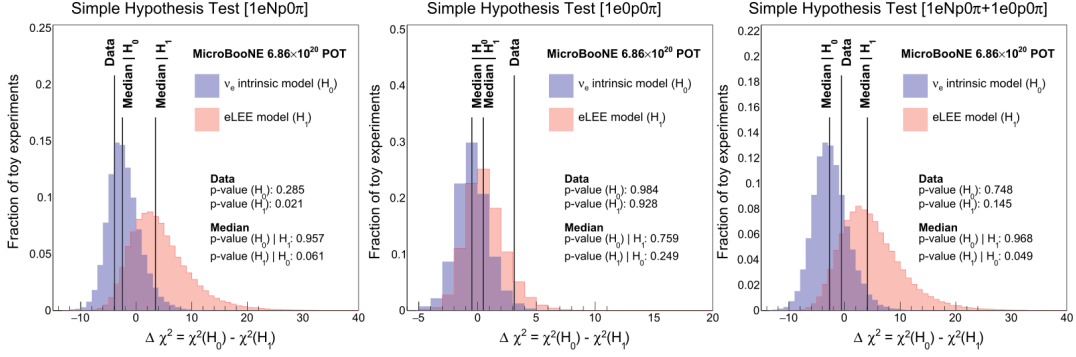

**Figure 9.11:** The simple hypothesis test to distinguish the $\nu_e$ intrinsic model only ($H_0$, blue) and the $\nu_e$ intrinsic plus eLEE signal model ($H_1$, red) relies on computing the distribution of $\Delta\chi^2$ under the two models. The exercise is repeated for the $1e\text{N}p0\pi$ channel only (left), $1e0p0\pi$ channel only (middle), and the combination of the two (right). The data observation is identified by a vertical line, together with the median, or expected sensitivity, to reject one of the two models assuming the other one is true. Expected and observed p-values are also quoted, both to reject $H_0$ and $H_1$. However, in the final result we only quote the significance to exclude $H_1$.

## Signal strength measurement

The last test of the eLEE model is performed by fitting the signal strength $\mu$, defined as the scaling factor of the total normalization of the eLEE signal. $\mu = 0$ refers to the SM only expectation, or $H_0$, while $\mu = 1$ to the eLEE model as defined in section 8.2, or $H_1$. Any other value indicates the presence of a signal compatible in shape with the eLEE model but with a different normalization. The procedure is based on the Feldman-Cousin approach [310]. First, the best value $\mu_{\text{fit}}$ is obtained by minimizing the following $\chi^2$:

$$\chi^2(n|\mu) = \sum_{i,j}(n_i - (m_i^{SM} + \mu m_i^{eLEE}))\Sigma_i^{-1}j(\mu)(n_j - (m_j^{SM} + \mu m_j^{eLEE})) \qquad (9.5)$$

$$\Sigma_{ij} = \Sigma_{ij}^{\text{stat CNP}}(\mu) + \Sigma_{ij}^{\text{syst}}. \qquad (9.6)$$

Second, the confidence interval is defined using the ordering rule based on the likelihood ratio, which is approximated by the difference in $\chi^2$,

$$R(x|\mu) \sim \Delta\chi^2(n|\mu) = \chi^2(n|\mu) - \chi^2(n|\mu_{\text{best-fit}}), \tag{9.7}$$

where $n$ is a generic observation of the bin contents, and $\mu_{\text{bestfit}}$ is the value that minimizes eq. (9.5). The procedure requires drawing a set of toy experiments for each value of $\mu$ at the truth level and extracting the distribution of $\Delta\chi^2(n|\mu)$ for each value of $\mu$. The confidence interval at a given confidence level $c$ is the set of all $\mu$ for which the observed $\Delta\chi^2(n^{\text{data}}|\mu)$ is smaller than the $c$-quantile of the distribution of $\Delta\chi^2(n|\mu)$ obtained with the toy experiments. We report results with for a confidence level (CL) of 90%, meaning $c = 0.9$. Figure 9.12 illustrates the $\chi^2$ scan as a function of $\mu$ for the $1e\text{N}p0\pi$ and $1e0p0\pi$ analysis channels, and their combination, while table 9.1 summarizes the results. The sensitivity is computed by taking the data equal to the Asimov dataset, which is the dataset with bin-content values equal to the predicted values for $\mu = 0$. The $1e\text{N}p0\pi$ analysis observed a best-fit point of $\mu = 0$, with an upper limit of 0.82 at 90% CL, lower than the sensitivity, equal to 1.16. In contrast, the $1e0p0\pi$ analysis shows a two-sided interval between 1.13 and 15.01, with the best fit value of 4 and an expected upper limit of 3.41. Despite the value $\mu = 0$ being excluded at 90% CL, the confidence interval is wide, justifying the low sensitivity of this channel alone. As expected, combining the two channels results in a wider confidence interval than the 1eNp0π channel alone. We exclude all values of $\mu$ smaller than 1.57, mildly ruling out the eLEE signal model under study.

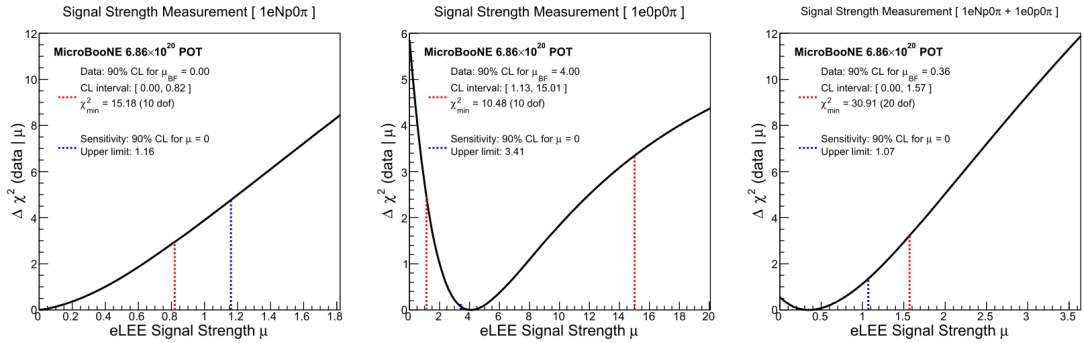

**Figure 9.12:** The signal strength $\mu$ is extracted through a scan of $\Delta\chi^2(\text{data}|\mu)$. The expected upper limit at 90% CL, obtained with the Asimov dataset, assuming $\mu = 0$ is drawn with a blue line. The observed confidence interval at 90% CL is obtained through the Feldman-Cousin procedure and indicated with red lines. The exercise is repeated for the $1e\text{N}p0\pi$ channel only (left), $1e0p0\pi$ channel only (middle), and the combination of the two (right).

| Channel | $\mu_{\text{best−fit}}$ | 90% CI on $\mu$ | 90% expected upper limit on $\mu$ |
|---|---|---|---|
| 1eNp0π | 0.00 | [0.00, 0.82] | 1.16 |
| 1e0p0π | 4.00 | [1.13, 15.01] | 3.41 |
| 1eNp0π + 1e0p0π | 0.36 | [0.00, 1.57] | 1.07 |

**Table 9.1:** Summary of the signal strength measurement in terms of best-fit value, 90% confidence interval, and 90% expected upper limit (sensitivity).

## COMPARISON WITH THE OTHER ANALYSES

The other analyses targeting electron neutrinos in MicroBooNE, namely the CCQE one, targeting the 1e1p0π final state and using deep-learning-based reconstruction, and the inclusive $\nu_e$CC analysis based on the WireCell reconstruction framework, observe results compatible with this work. Figure 9.13 summarizes and compares the data from all analyses. The left plot shows the ratio between the observed and the predicted number of events with no eLEE model in their respective signal regions. The standard model expectation of one is shown with the corresponding systematic uncertainty, while the red line identifies the expected ratio if the eLEE model was true. The observed ratio appears consistently below one for all analyses aside from the 1e0p0π channel, although consistent with one within systematic uncertainties. The right plot illustrates the result in terms of the signal strength $\mu$ for the different analyses. The blue band around one represents the expectation $\mu = 1$ with systematic uncertainties in case the eLEE signal model was true. The red lines illustrate the expected upper limits, while the black lines show the observed confidence intervals at the 1 and $2\sigma$ level. All analyses aside from the $1e0p0\pi$ channel report an upper limit, varying between $\sim 0.3$ and $\sim 1$, consistently and consistent with the ratio of the observed and expected number of events described before.

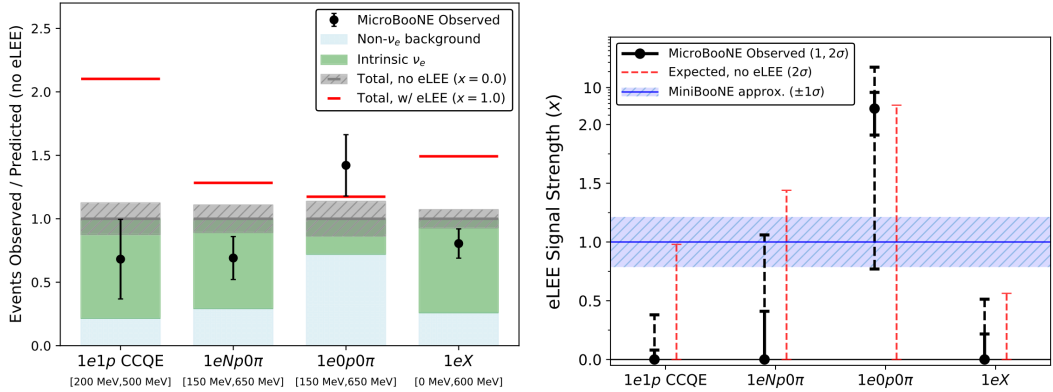

**Figure 9.13:** The different eLEE analyses report results consistent with each other within uncertainties. Most analyses report a null result, aside from the $1e0p0\pi$ channel, which results in a preference for the eLEE signal, although with weak significance. The ratio between the number of observed and predicted events in the signal regions is below one in most analyses (left), and it is typically incompatible with the expectation under the eLEE model. When extracting the signal strength, all analyses but the $1e0p0\pi$ selection report an upper limit (right), often stronger than the expected one because of the underfluctuation of the data with respect to the prediction.

### 9.3 Interpretation and implications

The most important implication of this result is the preference of the standard model with respect to simple explanations of the MiniBooNE anomaly in terms of an additional component of $\nu_e$ in the beam. Although fig. 9.13 seems to show that this explanation is ruled out at more than $2\sigma$, the current eLEE model does not include systematic uncertainties in the unfolding procedure. It has been pointed out that by introducing uncertainties in the unfolding process, the significance reduces, and the upper limit increases [208].

On top of this simplistic model invoked as an explanation of the MiniBooNE anomaly, Micro-BooNE's data allows tests of more realistic models, like the presence of one light sterile neutrino on top of the three active ones. The observation of an underfluctuation of the data with respect to the prediction in selections enriched with electron neutrinos can be interpreted as no electron neutrino appearance and a hint of electron neutrino disappearance. A complete study requires exploring the entire parameter space with two mixing angles and one oscillation frequency. Preliminary results show mild evidence of electron neutrino disappearance [311], or simply an exclusion in parameter space [208] as systematic uncertainties wash out any proof of oscillations. The second study points out that, despite the increase in significance due to the underfluctuation of the data, the inclusive analysis cannot exclude the interesting part of the best-fit region to MiniBooNE data (plot on the right in fig. 9.14). However, preliminary results from the MicroBooNE collaboration show that the combination of BNB and NuMI data can place essential constraints in the parameter space, carving out significant portions of the best-fit regions to LSND data (plot on the left in fig. 9.14). Further analyses are expected to be released soon, with a more refined analysis and a larger dataset.

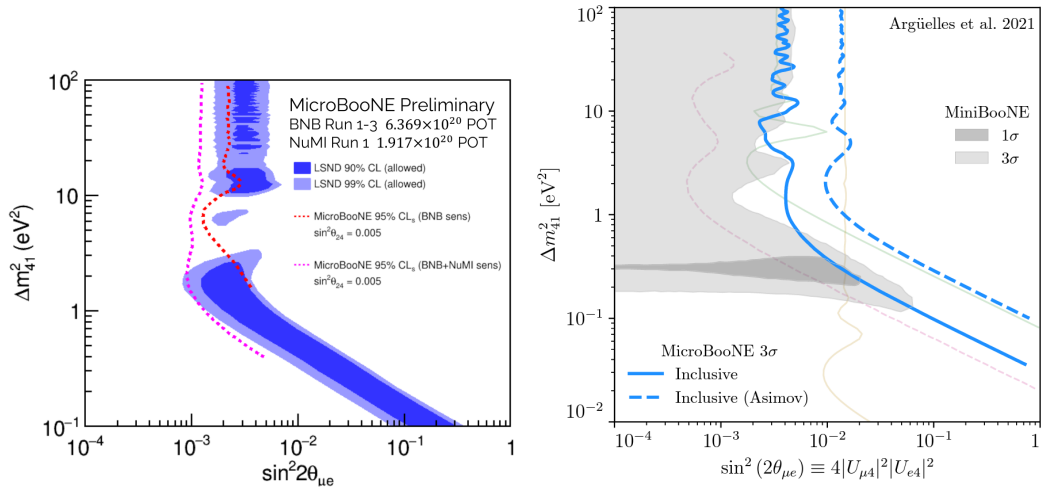

**Figure 9.14:** the inclusive analysis based on the WireCell reconstruction framework can be further interpreted in terms of the light sterile neutrino model, with three active neutrinos and one light sterile neutrino. The parameter space of interest is defined by $\Delta m_{14}^2$ and $\sin^2 2\theta_{e\mu}$, profiling (or marginalizing in the case of Bayesian analysis) over the other free parameter. The left plot, taken from [312], shows the result obtained using the $CL_s$ method, using BNB data only (red) and the combination of BNB and NuMI data (purple). The combination can exclude most of the best-fit regions to LSND data. The right plot shows the exclusion at $3\sigma$ obtained by recasting the inclusive analysis with BNB data only. This plot emphasizes that the result is significantly stronger than the sensitivity (dashed line) because of the smaller number of observed events with respect to the expectation. Despite this stronger significance, the line does not carve out a significant portion of the MiniBooNE best-fit region.

The data excess in the $1e0p0\pi$ channel is a little mysterious, although not particularly significant at this stage. One straightforward interpretation is in terms of migration of events from $1e\text{N}p0\pi$ channel: a mismodeling of the track simulation and reconstruction at low energy would result in fewer events with low energy protons in the $1e0p0\pi$ channel than observed in the data. However, the excess is observed in a region with low $\nu_e$CC purity, and, in the absence of further data, it is interpreted as an overfluctuation of the data.

*Imagine that right here, right now, dark matter scientists*
*are holding a conference about what they call dark matter.*
*A scientist illustrates their theory of dark matter, described*
*by a gauge group $SU(3)_C \times SU(2)_L \times U(1)_Y$ with sponta-*
*neous symmetry breaking and many other complex features,*
*what would the other scientists say? They would probably*
*start laughing, saying it is too complicated because nature is*
*undoubtedly simple.*

Christopher Stubbs

# 10

# Epilogue

I WISH I COULD OPEN THIS THESIS AGAIN IN A FEW DECADES with the awareness the short baseline anomaly puzzle has been solved. Indeed, the work described in this thesis can be framed in the general evolution of the short baseline neutrino anomaly puzzle over the past five years. Despite new experimental measurements and theoretical insights, something remained constant: the puzzle is still far from being solved.

From the experimental side, some results grew more robust, and some became weaker. Thanks to more accurate input data to the flux calculations and a new generation of experiments capable of sampling the event rate at different distances, the reactor antineutrino anomaly seems to be significantly reduced and likely disappeared. At the same time, the BEST experiment confirmed the long-lasting gallium anomalies (GALLEX and SAGE) with an even larger significance. This anomaly was observed at distances and energies similar to reactor antineutrinos but using sources producing electron neutrinos. These two results are in strong tension if interpreted under the light sterile neutrino hypothesis. The new light sterile neutrino search in IceCube confirmed the lack of evidence for muon neutrino disappearance. This analysis is especially interesting because it tests the hypothesis by looking for a resonant shift to the sterile state induced by the matter effects of atmospheric neutrinos passing through the Earth. However, some people interpreted the closed contour at 90% CL and the agreement with the SM with a p-value of 8% ($1.4\,\sigma$ to reject the SM) as mild hints for light sterile neutrinos. And while the latest iteration of the MiniBooNE analysis confirmed the excess and its significance, and proved to be related to the beam, MicroBooNE made public the first, long-awaited iteration of the short baseline searches, of which part III discusses a part of a broader collaboration effort. Both the single photon search [207] and the electron neutrino searches [203,204,205,206] reported null results. However, these results are not sensitive enough to fully exclude the LSND and MiniBooNE best-fit regions. The lack of an excess in the electron neutrino searches has been used as evidence against the MiniBooNE observation. However, this statement is invalid because of the difficulties in making model-independent comparisons. Indeed, the backgrounds in the two analyses are very different, with MicroBooNE measuring a much purer spectrum of electron neutrinos. If interpreted under the light sterile neutrino model, MicroBooNE instead seems to observe a mild hint of electron neutrino disappearance, washed out by systematic uncertainties. Moreover, MiniBooNE and MicroBooNE use different models for generating neutrino interactions. Although tuned on their respective data, re-interpretation of this data with the same neutrino-interaction model would be highly insightful.

Across the theoretical landscape, a mindset shift thoroughly established the difference between short baseline anomalies, meaning the mere experimental results, and their interpretation under the light sterile neutrino hypothesis, concepts too often mixed and tangled. This mindset shift instilled the idea that short baseline anomalies might be the result of many different physical effects rather than a single new physics effect. Many more model proposals flourished: they are based on heavy sterile neutrinos, light mediators, and dark portals that could explain at least some of the anomalies. These models, some of which we studied in part II, are exciting because the light sterile interpretation requires a workaround to accommodate the many null results, as discussed in section 2.4. It is moreover not particularly well-motivated theoretically, while heavy sterile neutrino models could explain the origin of neutrino masses, and dark sectors could involve the particles that make up dark matter.

Even though the short baseline anomaly puzzle is complex and easily changeable, I still believe, as I did when I started this journey, that it is one of the most interesting open problems in particle physics these days. New experiments, theoretical frameworks, and tools are coming up, making prospects look very exciting.

From the experimental point of view, I believe we need bold projects capable of testing the LSND and MiniBooNE results with large sensitivities and are ideally able to distinguish between different interpretations. Indeed, many low-sensitivity results, either anomalous or in agreement with the standard model, made the picture only blurrier and more confusing. I am looking forward to the results of JSNS[2] [313,314], which will replicate the LSND experiment with a much larger statistic, lower background, and a better understanding of systematic uncertainties. Using two liquid scintillator detectors at different baselines, with sections loaded with gadolinium to improve the neutron capture rate, it will uniquely test all the LSND allowed parameter space. The IsoDAR experiment at YemiLab [315,316], South Korea, will also provide high-sensitivity results. A new exper-imental approach allows the production of a large flux of $^8$Li, which beta decays producing $\bar{\nu}_e$ with energies between 3 and 13 MeV. Paired with a 2.3 kton liquid scintillator detector, the experiment will collect 1.6 million inverse beta decays, at distances $L \sim 1 - 10$ m. Such an experiment would be sensitive to several oscillation wiggles for the preferred values from the BEST experiment, making conclusive statements about electron antineutrino disappearance. It will also be able to differentiate the vanilla model from models with more than one sterile neutrino or where sterile neutrinos decay on distances of the order of the oscillation wavelength. Lastly, testing the oscillation hypothesis with coherent neutrino-nucleus scattering will also achieve strong sensitivities. These measurements would sample the total flux of neutrinos, being sensitive to a possible disappearance rate. Additionally, as for the case of the Short Baseline Neutrino program, multiple detectors could sample the neutrino rate at different distances along the same beamline to provide measurements of the oscillation wiggles, should the light sterile neutrino hypothesis be proven correct.

On the other hand, new and larger experiments will not be enough without significant technological and tool development work. For example, the MicroBooNE experience taught us that liquid argon technology still requires more technical work to fully express its potential, achieving the high reconstruction and selection efficiencies that the technology could be capable of. The work done on tracks and shower classification, discussed in chapter 7, is an example of this direction: although ionization and charge deposition was already understood decades ago, there was still room to improve the modeling of the detector measurement and reconstruction of the charge, and drastically increase selection efficiencies. However, because it is more specific and of narrower interest to the community, this type of research is often not rewarded as much as measurements of physical observables.

Similarly, improving the computational and statistical tools would benefit current and future experiments. Because the datasets are becoming larger and larger, computational time plays a crucial role. For example, the SBND experiment will have a significant pile-up of neutrino interactions, meaning that a substantial fraction of the events will contain two or more neutrino interactions.

Such event rates are orders of magnitudes larger than previous generation experiments. Reducing times from days to hours or even minutes is often possible thanks to new libraries and programming languages, multiplying the number of possible analyses and limits that can be cast.

Statistical methods also improved significantly, showing limitations of the approximate calculations used in the past. A clear example comes from the many reactor experiments, where it was shown that Wilks' theorem could inflate the results significantly [317,76]. Another example is Neutrino4, whose statistical treatment was criticized because it can artificially create wiggles and bolster the significance of the result [82].

If properly implemented, these improvements would also contribute to better treatments of systematic uncertainties, resulting in larger sensitivities. Many recent experimental results are limited by systematic uncertainties or have comparable systematic and statistical uncertainties. Better tools like multi-universe approaches, such as the SnowStorm method used in IceCube [276] to handle detector systematics, will dramatically improve the impact of detector systematic uncertainties on the final results.

From the phenomenological side, I believe that critically re-reading, re-understanding, and re-interpreting the critical experimental results, like LSND and MiniBooNE, would prove beneficial. Many of these results became stories repeated many times at conferences and workshops to the point that I would not be surprised if someone noticed details that the community has overlooked. For example, when scrutinizing results from the PS191 experiment from the 1980s, we discovered that the constraints on the heavy neutrino parameter space were overestimated by a factor of $\sim 6$, as discussed in chapter 4. Similarly, looking more broadly at all experimental measurements can result in constraints on new-physics models that the experiment was not even designed to test. An example is the novel test of possible new physics explanations of the KOTO anomaly performed by MicroBooNE [197], or the recent constraints on dark sector models obtained using data from the CHARM experiment [318,236].

The phenomenology panorama will also tremendously benefit from improvements in the simulation and analysis tools. I am looking forward to the latest developments brought by the Achilles generator [319], which will extend MadGraph to neutrinos and dark sector models, and by LeptonInjector and LeptonWeighter [320], that allow simulations and weighting of neutrino events in arbitrary geometries, for arbitrary fluxes and cross sections. The statistical tools to explore and set constraints on large parameter spaces also require more development. The work discussed in chapter 5 allowed us to interpolate predictions across the parameter space in a fast way, and to easily cast constraints on any slice of the parameter space. If interfaced with posterior sampling and a more accurate simulation, it has the potential to cast limits on many different models from the same dataset, speeding up the rejection of the ones that turned out to be excluded. In this sense, I also believe that setting up more precise standards on how to set and compare limits *e.g.* Bayesian or frequentist, profiled or marginalized, etc. will clear up some of the confusion that often emerges when comparing different papers. Lastly, better inputs from experimental collaborations will be highly beneficial. It would be constructive if the community would agree on a standardized way to share and release the data for every analysis and measurement. It would be extremely important if this release contains the Monte Carlo simulation with the list of events, together with reconstruction and selection efficiencies as a function of the causal variables: *i.e.* the $e^+e^-$ selection efficiency depends on the invariant mass of the pair through the momenta and the opening angle, and extrapolating it to different models or different invariant mass values is difficult without knowing how the efficiency varies as a function of momenta and angles.

While some of the anomalies might indeed be induced by new physics, some others might be caused by wrong estimates of systematic uncertainties. For this reason, developments in cross-section calculations and predictions are essential. New input data from beta decay measurements, together with a strong effort to update the calculations of the reactor antineutrino flux, cleared up the confusing panorama of reactor antineutrino measurements that started ten years ago. Analo-gous efforts for the gallium electron capture cross section and the flux of the chromium and argon sources might shed new light on the radiochemical anomalies. Similarly, improvements in the accelerator neutrino cross sections for charged current and photon production and a deeper understanding of final state interactions, the interactions final state particles undergo while exiting the nucleus, could bring new insights into the MiniBooNE and LSND anomaly. However, we need to acknowledge that most of these estimates are connected to well-understood neutrino physics: for example, the gallium electron capture cross section is the same that provided measurements of the solar neutrino flux, while cross sections for producing resonances decaying into single photons in the nucleus are tightly related to similar and well measured nuclear processes.

Circling back to the research illustrated in these pages, these null results for interpretations of the anomalies based on heavy or light sterile neutrinos are two additional pieces to the short baseline anomaly puzzle. These results and the full development of the field over the past few years taught us that if new physics lies behind the short baseline anomalies, it is certainly not described by a simple model, and we do not have any good reason why this should be the case. Although the picture looks still uncertain, pursuing the study of short baseline anomalies is extremely relevant, especially while many other experiments in the lab find agreement with the Standard Model with unprecedented precision. Perhaps, some new physics will be discovered, which might unveil hints towards detecting dark matter, understanding baryogenesis, and solving the hierarchy problem, currently the most important open questions in fundamental physics. But even if new physics does not lie behind the short baseline anomalies, there is certainly something to understand.